\documentclass[11pt,a4paper]{book}

\usepackage{fullpage}
\usepackage{natbib}
\usepackage{graphicx}
\usepackage{appendix}
\usepackage{url}
\usepackage{bm}
\usepackage{bbm}
\usepackage[slovene,english]{babel}

\def\Z{{\mathbbm{Z}}}
\def\>{\rangle}
\def\<{\langle}
\def\ket#1{|#1\>}
\def\bra#1{\<#1|}
\def\braket#1#2{\< #1 | #2 \>}
\def\bracket#1#2#3{\< #1 | #2 | #3 \>}
\def\ave#1{\left\< #1\right\>}
\def\ma#1{{\rm \mathbf{#1}}}
\def\tr{{\,{\rm tr}}}
\def\ii{{\rm i}}
\def\Md{M_\delta(t)}
\def\Fn{F(t)}
\def\Fr{F_{\rm R}(t)}
\def\Fp{F_{\rm P}(t)}
\def\oFn{\bar{F}}
\def\oC{\bar{C}}
\def\oCr{\bar{C}_{\rm R}}
\def\oCp{\bar{C}_{\rm P}}
\def\oV{\bar{V}}
\def\oG{\bar{\Gamma}}
\def\Vres{{V_{\rm res}}}
\def\fn{f(t)}
\def\aave#1{\left\<\! \left\< #1\right\> \! \right\>}
\def\vec#1{{\bm{#1}}}
\def\sumn#1{\sum_{\vec{m}\neq\vec{0}}{#1}}
\def\tre#1{\tr_{\rm e}{\lbrack #1 \rbrack}}
\def\trc#1{\tr_{\rm c}{\lbrack #1 \rbrack}}

\usepackage{fancyhdr}
\makeatletter 
\def\cleardoublepage{\clearpage\if@twoside \ifodd\c@page\else%
\hbox{}%
\thispagestyle{empty}
\newpage%
\if@twocolumn\hbox{}\newpage\fi\fi\fi} 
\makeatother

\makeatletter  
\renewcommand*\@pnumwidth{2.4em}
\makeatother

\begin{document}

\selectlanguage{english}

\pagenumbering{roman} \setcounter{page}{1}

\begin{titlepage} 
\begin{center} 
\thispagestyle{empty} 
\large{University of Ljubljana\\
Faculty of Mathematics and Physics\\
Physics Department}\\
\vspace{4cm}
\textsf{\Large Marko \v Znidari\v c}\\
\vspace{1.5cm}
\textsf{\huge \bf STABILITY OF QUANTUM DYNAMICS}\\
\vspace{0.9cm}
\large{Dissertation}\\
\vspace{6cm}
\begin{flushright}
\parbox{95mm}{
\large{advisor: Prof.~Toma\v z Prosen\\\\
external advisor: Prof.~Thomas H. Seligman}\\
}
\end{flushright}
\vspace{3.5cm}
\normalsize{Ljubljana, 2004}
\end{center}

\thispagestyle{empty}
\end{titlepage} 

\thispagestyle{empty}
\hbox{}\newpage

\chapter*{
}
\pagenumbering{roman} \setcounter{page}{1}
\vspace{-2cm}
\centerline{\sf \Large Abstract}
\vspace{0.5cm} 
\begin{center}
\begin{minipage}{125mm}
The stability of quantum systems to perturbations of the Hamiltonian is
studied. This stability is quantified by the fidelity, an overlap
of an ideal state obtained by the unperturbed evolution, and a perturbed
state obtained by the perturbed evolution, both starting from the same
initial state. Dependence of fidelity
on the initial state as well as on the dynamical properties of the
system is considered. In particular, systems having a chaotic or regular classical limit are analysed. The fidelity decay rate is
given by an integral of the correlation function of the perturbation
and is thus smaller the faster correlation function decays. Quantum
systems with a chaotic classical limit can therefore be more stable than
regular ones. If the perturbation can be written as a time derivative
of another operator, meaning that the time averaged perturbation vanishes, fidelity freezes at a
constant value and starts to decay only after a long time inversely
proportional to the perturbation strength. In composite systems stability of entanglement to
perturbations of the Hamiltonian is analysed in terms of purity. For regular systems purity decay is shown to be independent of
Planck's constant for coherent initial states in the semiclassical limit. The accelerated
decoherence of macroscopic superpositions is also explained. The theory of fidelity decay is applied to the stability of
quantum computation and an improved quantum Fourier transform
algorithm is designed and shown to be more stable against random perturbations.\\\\
{\bf Keywords:} quantum stability, fidelity, purity, quantum chaos,
decoherence, entanglement, quantum computation, quantum Fourier transformation,
the kicked top, Jaynes-Cummings model.\\\\
\newcommand{\pacslabel}[1]{\mbox{\sf{#1}}\hfil}
\newenvironment{pacs}{
\begin{list}{}{
\renewcommand{\makelabel}{\pacslabel}
\setlength{\itemsep}{0pt}\setlength{\parskip}{0pt}
\setlength{\parsep}{0pt}
\setlength{\leftmargin}{49pt}
\setlength{\labelwidth}{43pt}\setlength{\topsep}{0pt}}}
{\end{list}}
{\bf PACS numbers:}
\begin{pacs}
\item[03.65.Sq] Semiclassical theories and applications
\item[03.65.Yz] Decoherence; open systems; quantum statistical
  methods
\item[03.67.Lx] Quantum computation
\item[03.67.Mn] Entanglement production, characterization, and manipulation
\item[03.67.Pp] Quantum error correction and other methods for protection
against decoherence
\item[05.45.-a] Nonlinear dynamics and nonlinear dynamical systems
\item[05.45.Mt] Quantum chaos; semiclassical methods
\item[05.45.Pq] Numerical simulations of chaotic systems
\item[42.50.Ct] Quantum description of interaction of light and
  matter; related experiments
\end{pacs} 
\end{minipage}
\end{center}

\chapter*{}
{\Large \sf Acknowledgement}
\vspace{2cm} 
\begin{center}
\begin{minipage}{105mm}
\sf
I would like to thank my advisor Prof.~Toma\v z Prosen for giving me
the opportunity to work in his group, guiding me and showing me the
way. It has been a great pleasure to work with him, sharing the
excitement of new discoveries. Initially small project gradually grew,
eventually surpassing all our expectations and resulting in the present thesis.\\\\ 
Thanks also go to Prof.~Thomas H.~Seligman for cooperation on parts of
this thesis and for warm hospitality during
my stay in Cuernavaca. In particular, he has been a very careful
reader of the final manuscript, pointing out english
mistakes and confusing statements.
\end{minipage}
\end{center}

\tableofcontents


\chapter{Introduction}
\label{ch:intro}
\pagenumbering{arabic} \setcounter{page}{1}
\begin{flushright}
\baselineskip=13pt
\parbox{90mm}{\baselineskip=13pt
\sf In science one tries to tell people, in such a way as to be understood by everyone, something that no one ever knew before. But in poetry, it's the exact opposite.}\medskip\\
---{\sf \itshape Paul Dirac}\\\vspace{20pt}
\end{flushright}

Irreversibility of macroscopic behaviour has attracted attention ever since Boltzmann introduced his $H$-theorem in a seminal paper in 1872~\citep{Boltzmann:72}, together with the equation bearing his name. The $H$-theorem is a precursor of what is today known as the second law of thermodynamics, stating that entropy can not decrease in the course of time. The problem was how to reconcile this ``arrow of time'' with the underlying {\em reversible} microscopic laws. How come that macroscopic systems always seem to develop in one direction whether the underlying dynamics is symmetric in time? This so called reversibility paradox is usually attributed to Josef Loschmidt. He mentioned it briefly at the end of a paper published in 1876~\citep{Loschmidt:76} discussing the thermal equilibrium of a gas subjected to a gravitational field, in an attempt to refute Maxwell's distribution of velocities for a gas at constant temperature. He also questioned Boltzmann's monotonic approach towards the equilibrium. Discussing that, if one would reverse all the velocities, one would go from equilibrium towards the initial non equilibrium state, he concludes with : {\em ``...Das ber\" uhmte Problem, Geschehenes ungeschehen zu machen, hat damit zwar keine L\" osung...''}. Bolzmann was quick to answer Loschmidt's objections in a paper from 1877~\citep{Boltzmann:77}, pointing out the crucial importance of the initial conditions and of the probabilistic interpretation of the second law. The fact that for macroscopic systems we always observe a definite ``arrow of time'' is a consequence of a vast majority of the initial conditions representing an equilibrium state. If we choose the initial state at random it will almost certainly evolve according to the second law of thermodynamics. Of course this probabilistic interpretation still does not solve the problem. Resolution lies in the initial condition, chosen to be a macrostate representing a very small part of the phase space, i.e. having a low entropy. So, provided we start with a low entropy state, the above probabilistic arguments can be used~\citep{Lebowitz:99} to explain the second law. The question of the initial low entropy state of say the whole universe still remains. For a popular account of this ``cosmological'' subject see~\citet{Penrose:89}.   
\par
The first written account of the reversibility paradox is actually not due to Loschmidt but due to William Thompson (later Lord Kelvin), although it is possible that Loschmidt mentioned the paradox privately to Boltzmann before. Boltzmann and Loschmidt became good friends while working at the Institute of Physics in Vienna around 1867 (directed at the time by the Slovenian physicist Jo\v zef Stefan). In 1869 Boltzmann moved to Graz, but returned to Vienna in the period 1873-1876. For a detailed biography of Boltzmann see~\citet{Cercignani:98}. In a vivid paper~\citep{Thompson:74} from 1874 Thompson gave a very modern account of irreversibility. When discussing the heat conduction and the equalization of temperature he says: {\em ``...If we allowed this equalization to proceed for a certain time, and then reversed the motions of all the molecules, we would observe a disequalization. However, if the number of molecules is very large, as it is in a gas, any slight deviation from absolute precision in the reversal will greatly shorten the time during which disequalization occurs... Furthermore, if we take account of the fact that no physical system can be completely isolated from its surroundings but is in principle interacting with all other molecules in the universe, and if we believe that the number of these latter molecules is infinite, then we may conclude that it is impossible for temperature-differences to arise spontaneously...''}. The interesting question is then, how short is this {\em short disequalization time}? The quantity in question is nothing but the fidelity or the Loschmidt echo as it is sometimes called. We evolve the system forward in time with the {\em unperturbed} evolution, then backward in time with the {\em perturbed} evolution, and look at the overlap with the initial state. It can be considered in classical mechanics as well as in quantum mechanics. The decay time of the fidelity will then be the time of disequalization in question. Despite its importance for thermodynamics the fidelity was not considered until some years ago, with motivation coming from quantum rather than classical theory.
\par
Quantum theory is arguably the greatest achievement in 20th century
physics. There are estimates~\citep{Tegmark:01} that up to $30\%$ of
the gross national product of the US relies on quantum devices. Alone
the semiconductor industry is of enormous importance. Still, all these
quantum devices do not manipulate individual quanta but rather exploit
macroscopic phenomena involving many particles. Experiments involving
individual quantum systems became possible only in the '80 with the
progress made in e.g. manipulation of cold atoms in traps, single
electron devices, entangled photons etc.. This was so to say the
experimental birth of what is now called quantum information
theory~\citep{Nielsen:01}. Quantum information theory married quantum
mechanics with information theory and with computer science. It deals
with means of processing and transmitting information, and by using
quantum systems can achieve things not feasible in any classical
way. For instance, one can teleport a quantum state, or perform secure
communication over a public channel or do a quantum computation. By
using quantum resources to do a computation one is able for instance
to factorize a number in a polynomial time, which is presently not
known to be possible by classical computer. Also, quantum computers
are very efficient in simulating other quantum systems, answering the
problem posed by~\citet{Feynman:82}. Namely, he asked whether it is
possible to build a computing machine whose size will grow only
linearly with the size of a quantum system simulated on it. With
classical computers this is not possible since the size of the Hilbert
space needed merely to describe the system grows exponentially with
the number of particles. We do not know yet if it is possible to build
a quantum computer that will achieve that goal, but on paper quantum
computer will be able to do the trick. We say on paper because
presently one is able to perform laboratory computations only on a few
qubit (less than 10) quantum computer. The main obstacle are errors in
the evolution, either due to unwanted coupling with the environment or
due to internal errors. Therefore, the main goal is to build a stable
quantum computer, resistant to such perturbations. The usual benchmark
for stability is fidelity and therefore one ought to understand the
behaviour of fidelity in different situations to know hot to maximise it. Yet again, the original push to study fidelity came neither from quantum information theory nor from thermodynamics, but from the field of quantum chaos.
\par
The exponential instability of classical systems is a well
known and much studied subject. As the underlying laws of nature are
quantum mechanical the obvious question arises how this ``chaoticity''
manifests itself in quantum systems whose classical limit is chaotic. The field
of quantum chaos mainly dealt with stationary properties of
classically chaotic systems, like spectral and eigenvector
statistics. Despite classical chaos being defined in a dynamical
way it was easier to pinpoint the ``signatures'' of classical chaos in
stationary properties~\citep{Haake:91,Stockmann:99}. There were not
as many studies of the dynamical aspects of quantum evolution in chaotic
systems, some examples being studies of the reversibility of quantum
evolution~\citep{Shepelyansky:83,Casati:86}, the dynamical
localization~\citep{Fishman:82,Grempel:84}, energy
spreading~\citep{Cohen:00}, wave-packet evolution~\citep{Heller:91}. Classical instability is usually defined as an exponential separation of two nearby trajectories in time. In quantum mechanics the state of a system is completely described by a wave function and so one could be tempted to look at the sensitivity of quantum mechanics to variations of the initial wave function. But quantum evolution is unitary and therefore preserves the dot product (i.e. the distance) between two states and so there is no exponential sensitivity with respect to the variation of the initial state. This, at first sight perplexing, conclusion has been reached because we compared two different things. Classical mechanics can also be stated in terms of a Liouiville propagation of phase space densities and this is also unitary. If we want to compare quantum and classical mechanics we have to compare them on the same footing. Quantum mechanics is a probabilistic theory, the wave function just gives the probabilities of measurement outcomes. Therefore we should also formulate classical mechanics in a probabilistic way as a propagation of probability densities in phase space. The idea, first proposed by~\citet{Peres:84}, was to study not the sensitivity to the variation of the initial condition but with respect to the variation of the {\em evolution}. He compared two slightly different evolutions starting from the same initial state -- the quantum fidelity. For classical systems fidelity gives the same exponential sensitivity to perturbations of the evolution as to perturbations of initial conditions; the two things are equivalent. The quantum fidelity though can behave in a very different way, as we will see in the present work.
\par
The fidelity lies at the crossroad of three very basic areas of physics: thermodynamics, quantum information theory and quantum chaos. The features of the fidelity turned out to be very interesting, as one would expect for such a crossroad.    


\section{Historical Overview}
\label{sec:hist}
 \subsection{Quantum Fidelity}
The quantum fidelity $F(t)$, being the square of the overlap of the state $\ket{\psi(t)}=U_0(t)\ket{\psi(0)}$ obtained by the unperturbed evolution $U_0(t)$ and the state $\ket{\psi_\delta(t)}=U_\delta(t)\ket{\psi(0)}$ obtained by the perturbed evolution $U_\delta(t)$,
\begin{equation}
F(t):=|\braket{\psi(t)}{\psi_\delta(t)}|^2=|\bracket{\psi(0)}{U^\dagger_0(t) U_\delta(t)}{\psi(0)}|^2,
\end{equation}
has been first used as a measure of quantum stability
by~\citet{Peres:84}, see also his book~\citep{Peres:95}. The fidelity
can also be interpreted as an overlap of the initial state and the
{\em echo} state obtained after forward unperturbed evolution followed
by a backward perturbed evolution, i.e. after evolution with
$U_\delta^\dagger(t) U_0(t)$. Quite generally we can imagine
unperturbed evolution being governed by an unperturbed Hamiltonian
$H_0$ and the perturbed evolution by slightly perturbed Hamiltonian $H_\delta=H_0+\delta V$,
with $\delta$ a dimensionless perturbation strength and $V$ the perturbation
operator. Peres reached non-general conclusion that the decay
of fidelity is faster and has a lower asymptotic value for chaotic than for regular classical dynamic. As we will see the general
situation can be exactly the opposite. Non decay of the fidelity for
regular dynamics in Peres's work was due to a very special choice of the initial
condition, namely that a coherent wave packet was placed in the centre of a
stable island. Such a choice is special in two ways, first the centre
of an island is a stationary point and second the number of
constituent eigenstates of the initial state is very small. After
Peres's work the subject lay untouched for about a decade. In
1996~\citet{Ballentine:96} numerically studied a quantity similar to
fidelity. Instead of perturbing the backward evolution (i.e. the
Hamiltonian), they took the same backward evolution but instead
perturbed the state after forward evolution by some instantaneous
perturbation, like shifting the whole state by some $\delta x$. They
also looked at the corresponding classical quantity. The conclusion
they reached was that for chaotic dynamics quantum stability was
much higher than the classical one, while for regular dynamics the two
agreed. All these results were left mainly
unexplained. ~\citet{Gardiner:97,Gardiner:98} proposed an experimental
sheme for measuring the fidelity in an ion trap. Somehow related to the studies of stability was also a work by~\citet{Schack:92,Schack:93,Schack:96}, where they studied how much information about the environment is needed to prevent the entropy of the system to increase. Fidelity studies received new impetus by a series of NMR experiments carried out by the group of Pastawski.
\par
In NMR echo experiments are a standard tool. The so called spin echo
experiment of~\citet{Hahn:50} refocuses free induction decay in
liquids due to dephasing of the individual spins caused by slightly
different Larmour frequencies experienced due to magnetic field
inhomogeneities. By an appropriate electromagnetic pulse the Zeeman
term is reversed and thus the dynamics of {\em non-interacting} spins
is reversed. The first real {\em interacting} many-body echo
experiment was done in solids by~\citet{Rhim:70}. Time reversal,
i.e. changing the sign of the interaction, is achieved for a dipolar
interaction whose angular dependence can change sign for a certain
``magic'' angle, that causes the method to be called magic
echo. Still, the magic echo showed strong irreversibility. Much
later,~\citet{Zhang:92} devised as sequence of pulses enabling a {\em
local} detection of polarisation (i.e. magnetic moment). They used a
molecular crystal, ferrocene ${\rm Fe}({\rm C}_5{\rm H}_5)_2$, in
which the naturally abundant isotope $^{13}{\rm C}$ is used as an
``injection'' point and a probe, while a ring of protons $^1{\rm H}$
constitutes a many-body spin system interacting by dipole forces. The
experiment proceeds in several steps: first the $^{13}{\rm C}$ is
magnetised, then this magnetisation is transfered to the neighbouring
$^1{\rm H}$. We thus have a single polarised spin, while others are in
``equilibrium''. The system of spins then evolves freely, i.e. spin
diffusion takes place, until at time $t$ the dipolar interaction is
reversed and by this also spin diffusion. After time $2t$ the echo is
formed and we transfer the magnetisation back to our probe $^{13}{\rm
C}$ enabling the detection of the {\em polarisation echo}. Note that
in the polarisation echo experiments the total polarisation is
conserved as the dipole interaction has only ``flip-flop'' terms like
$S^j_+S^{j+1}_-$, which conserve the total spin. To detect the spin
diffusion one therefore needs a local probe. With the increase of the
reversal time $t$ the polarisation echo -- the fidelity -- decreases
and Zhang {\em et al.} obtained approximately exponential decay. The
nature of this decay has been furthermore elaborated
in~\citet{Pastawski:95}. The group of Pastawski then performed a
series of NMR experiments where they studied in more detail the dependence of the polarisation echo on various parameters~\citep{Levstein:98,Usaj:98,Pastawski:00}. They were able to control the size of the residual part of the Hamiltonian, which was not reversed in the experiment and is assumed to be responsible for the polarisation echo decay. For small residual interactions they obtained a Gaussian decay while for a larger ones the decay rate saturated and was independent of the perturbation strength, i.e. of the size of the residual interaction. While there is still no complete consensus on the interpretation of these experimental results they triggered a number of theoretical and even more numerical investigations. We will briefly list them in chronological order.
\par
Using the semiclassical expansion of the quantum
propagator~\citet{Jalabert:01} derived a {\em perturbation
independent} fidelity decay for localised initial states and chaotic
dynamics, $F(t) \sim {\rm e}^{-\lambda t}$, also called a Lyapunov
decay due to its dependence on the Lyapunov exponent
$\lambda$. Perturbation independent fidelity decay obtained for strong
perturbations has also been studied numerically in a Lorentz
gas~\citep{Cucchietti:02}, the perturbation being in the mass of
the particle. In the same paper the authors also studied the
asymptotic fidelity saturation value which, for the strong perturbations
considered, is independent of the perturbation strength, but becomes
perturbation dependent for smaller perturbations (see
Section~\ref{sec:time_averaged}). Studying fidelity turned out to be
particularly fruitful in terms of the {\em correlation
function}~\citep{Prosen:02}, see also~\citep{Prosen:01}. For coherent
initial states, the fidelity of regular systems decays as a Gaussian
while for chaotic systems we have different regimes, the most
prominent being the perturbation {\em dependent} exponential
decay. The decay of quantum fidelity in  chaotic systems has also been
studied by~\citet{Jacquod:01} and by~\citet{Cerruti:02}. For
sufficiently small perturbations one gets a Gaussian decay (also
called a perturbative regime), for intermediate ones the so-called Fermi
golden rule regime of exponential, perturbation dependent decay, while
for still stronger perturbations we get the Lyapunov decay. All these
regimes, including the fidelity decay for regular dynamics, and for
different initial states, were carefully discussed using the
correlation function approach in~\citet{Prosen:02JPA}. Several
interesting results were obtained, perhaps the most surprising one
being that in a certain range of parameters we can, by {\em increasing}
chaoticity of the corresponding classical system, {\em increase}
quantum fidelity, i.e. improve the {\em stability} of quantum
dynamics. Different time and
perturbation scales were discussed as well as their dependence on the
number of degrees of freedom. It is well known that the quantization
of classical system is not unique, i.e. there are different quantizations leading to the same
semiclassics. ~\citet{Kaplan:02} compared the quantization ambiguity in chaotic and regular
systems, reaching a conclusion that in chaotic systems the quantization ambiguity is {\em
supressed} as compared with regular ones. ~\citet{Karkuszewski:02} connected the fidelity decay with the decay of the off-diagonal matrix elements of the reduced density matrix, and therefore with decoherence. They claimed that quantum systems, whose classical limit is chaotic, are particularly sensitive to perturbations due to small scale structures in their Wigner functions. Actually, what their results show, was just the dependence of the fidelity decay on the size of the initial state. Fidelity decay is faster for random initial states than for coherent ones and there are no quantum effects in the regime they studied. For related comments see also~\citep{Srednicki:01,Jacquod:02}. The transition between Fermi golden rule and the Lyapunov decay has been further considered in a Bunimovich stadium~\citep{Wisniacki:02}, see also~\citep{Cucchietti:02ly} for a study of Lyapunov decay. We should stress that the Lyapunov decay of quantum fidelity is purely a consequence of the quantum-classical correspondence. There is nothing ``quantum'' in it and can be explained in terms of the classical fidelity displaying the same perturbation independent decay as the quantum fidelity~\citep{Veble:04}. The fidelity decay can also be connected with the local density of states, although not in a straightforward manner~\citep{Wisniacki:02b}.
\par
Fidelity decay in mixed systems, having a coexisting regular and chaotic components, is not as well studied as in purely chaotic or regular situation. In~\citep{Weinstein:02} they studied the fidelity decay for initial states placed at the border between regular and chaotic regions and observed a power-law fidelity decay although over less than an order of magnitude.~\citet{Jacquod:03} studied averaged fidelity decay in regular systems for perturbations with a zero time average for which they predict a universal power-law decay with a power $3/2$. This does not agree with our findings, see discussion at the end of Section~\ref{sec:CIS_V0}. If the packet after an echo in the chaotic system drifts exponentially away from its position at the beginning, the fidelity can decay also in a super-exponential way~\citep{Silvestrov:03}. In the same paper the authors also considered the influence of different averaging procedures on the decay of short-time fidelity. From the correlation function approach one easily sees that the decay of the fidelity does not only depend on the unperturbed dynamics, say being regular or chaotic, but also on the perturbation. If the correlation function decays to zero sufficiently fast, we can get exponential decay also for regular unperturbed dynamics. Such is the case for instance if the perturbation is modelled by a random matrix~\citep{Emerson:02}, but note that such perturbation has no direct classical limit. Instead of the fidelity one can also study the Fourier transformation of the fidelity, see~\citep{Wang:02}. In~\citep{Wisniacki:03} it was studied how different perturbations of the billiard border influence the short-time fidelity decay. For an attempt to give a uniform approximation of the fidelity decay in the crossover regime between the Gaussian perturbative and the exponential Fermi golden rule decays using random matrix theory see~\citep{Cerruti:03,Cerruti:03a} and also~\citep{Gorin:04}. The asymptotic saturation level of the fidelity has been studied in~\citep{Weinstein:02un}, confirming results of~\citep{Prosen:02JPA}. Perturbative calculation of the fidelity decay in disordered systems by field theoretical method, i.e. diagrammatic expansion of Green's function, has been done by~\citet{Adamov:03}.
\par
Recently~\citet{Vanicek:03} devised an efficient numerical scheme for
a semiclassical evaluation of the quantum fidelity. The method
consists of transforming intractable (due to exponentially many
contributing orbits) semiclassical expressions in coordinate space into
an initial momentum space representation. A surprising quantum phenomenon of a prolonged stability, called freeze of fidelity, has been described in~\citet{Prosen:03} for regular systems. Later, it has been generalised to arbitrary dynamics, in particular to chaotic systems~\citep{Prosen:04}. The decay of fidelity in regular one dimensional systems has been studied in~\citep{Sanka:03} and they also numerically observed a very short correspondence between the classical and the quantum fidelity for initial states placed in a rotational part of phase space, where the average perturbation is zero. This is nothing but the freeze of fidelity, not present in classical fidelity. The Lyapunov regime, being of purely classical origin, and its borders of validity have been furthermore elaborated in~\citep{Cucchietti:03}, see also~\citep{Prosen:02JPA} for a detailed discussion of borders within which different regimes are valid. They also stressed the importance of noncommutativity of the limits $\hbar \to 0$ and $\delta \to 0$ in recovering the classical behaviour, as already explained in~\citep{Prosen:02JPA}. A nontrivial question, addressed by Hiller, Kottos, Cohen and Geisel~\citep{Hiller:04} concerns the optimal time of the unperturbed evolution, i.e. we fix the duration of the unperturbed evolution and seek the duration of the perturbed evolution for which the fidelity will be maximal. The quantum fidelity decay in weakly chaotic systems when one might not get the Lyapunov or the Fermi golden rule decay has been explored in~\citep{Wang:04}. The fidelity for various random matrix models has been analysed by~\citet{Gorin:04}, see also~\citep{Cerruti:03,Cerruti:03a}. For further random matrix results and their relevance for quantum computation see also~\citep{Frahm:04}. In~\citep{Iomin:04} the fidelity of a nonlinear time-dependent chaotic oscillator with respect to the time-dependent perturbation of its frequency has been studied analytically. Most of the studies so far focused on a few degrees of freedom systems, the exception being~\citep{Prosen:02}. For additional results on the stability of many-body systems see~\citep{Izrailev:04}.
\par
There has also been a large number of papers dealing with the
stability of quantum computation, i.e. studies of the fidelity for
specific quantum algorithms and perturbations, mainly relying on
numerical simulations. Some of these
include~\citep{Miquel:96,Miquel:97,Banacloche:98,Banacloche:99,Banacloche:00}.
The group of Berman studied in detail the stability of the Ising quantum computer, see for instance~\citep{Berman:01,Berman:02b,Berman:02} and references therein and also~\citep{Celardo:03}. The group of Shepelyansky studied the stability of many different quantum algorithms~\citep{Georgeot:00,Song:01,Benenti:02qc,Terraneo:03}, see also~\citep{Benenti:03qc} and~\citep{Bettelli:04}. The scaling of errors with various parameters is easily explained by our correlation function approach. 
\par
With the advances in experiments manipulating individual quantum
systems it has become possible to actually measure quantum
fidelity. Apart from NMR echo experiments already mentioned, one is
able to measure the quantum fidelity on a few qubit NMR quantum
computer~\citep{Weinstein:02qc}. Particularly promising candidates are
ion traps~\citet{Gardiner:97,Gardiner:98}, ultra cold atoms in optical
traps~\citep{Andersen:03,Schlunk:03}. Experiments with microwave
resonators in billiards are also under way~\citep{Schafer:03}.

 \subsection{Classical Fidelity}
\label{sec:class_fid}
Classical fidelity can be defined in an analogous way as quantum fidelity and has been first used in~\citep{Prosen:02JPA}. It is an overlap integral of two classical densities in phase space, obtained by unperturbed and perturbed evolutions,
\begin{equation}
F_{\rm clas}(t):=\int{\!{\rm d}\vec{x} \rho_0(\vec{x},t)\rho_\delta(\vec{x},t)},
\label{eq:Fclas_def}
\end{equation}
where $\rho_\delta(\vec{x},t)$ is the density in phase space obtained
by a perturbed evolution $\vec{\phi}_\delta(\vec{x})$, where
$\vec{\phi}_\delta(\vec{x})$ is a volume preserving flow in phase
space. The density at time $t$ can be obtained by backward propagating phase space point $\vec{x}$, $\rho_\delta(\vec{x},t)=\rho(\vec{\phi}^{-1}_\delta(\vec{x}),0)$. Note that in order for fidelity to be normalized to $1$, the classical density has to be square normalized, $\int{\!{\rm d}\vec{x} \rho^2(\vec{x},0)}=1$.
\par
Classical fidelity behaves markedly different from quantum
fidelity. For chaotic systems and localized initial states the
classical fidelity agrees with quantum only up to the short Ehrenfest
time, logarithmic in Planck's constant, when the quantum-classical
correspondence breaks down. For regular systems and if the time
average perturbation is nonzero though, classical fidelity follows
quantum fidelity. On the other hand, if the time averaged perturbation
is zero, quantum fidelity exhibits the so called freezing
(Chapter~\ref{ch:freeze}), while classical fidelity does not,
except in the non-generic case of a harmonic oscillator
(Section~\ref{sec:har_freeze}). The classical fidelity will depend on the stability of
orbits in phase space. Linear response will therefore depend
on the stability of the flow and this involves the derivatives of the
flow, a derivative being an unbounded operator. For chaotic systems
for instance, the derivatives grow exponentially in time due to orbit
separation, a simple consequence of the famous Lyapunov
instability. In a phase space picture, the classical dynamics can produce structures on an arbitrary small scales, even if we start from a smooth density. Resolution of the quantum mechanics on the other hand is limited by a finite Planck constant.
\par
We will give a brief overview of the known results about classical fidelity. This will help us to understand the differences with quantum fidelity which will be the object of study in the present work. The literature on classical fidelity is not nearly as extensive as on quantum fidelity. Linear response calculation has been done in~\citep{Prosen:02JPA}. Numerical results on the classical fidelity and its correspondence with the quantum fidelity in chaotic systems and in systems exhibiting diffusion have been presented in~\citep{Benenti:02}. Classical fidelity in regular and chaotic systems has also been theoretically discussed in~\citep{Eckhardt:03}. A detailed explanation of the asymptotic decay in chaotic systems has been given in~\citep{Benenti:03b} and a theoretical explanation of the Lyapunov decay for short times in~\citep{Veble:04}. Classically regular systems on the other hand have been worked out in~\citep{Benenti:03}. 
 
\subsubsection{Chaotic Systems}
The classical fidelity in chaotic systems will go trough different decay regimes as time increases. We will consider a localized initial state of size $\nu$. Starting from $t=0$ fidelity stays close to $1$ until time $t_\nu$~\citep{Benenti:02,Benenti:03b},
\begin{equation}
t_\nu \sim \frac{1}{\lambda}\ln{\frac{\nu}{\delta}},
\end{equation}
where $\lambda$ is the Lyapunov exponent. The time $t_\nu$ can be
thought of as the time in which the initial perturbation is
``amplified'' to the size of the initial packet and $\rho_\delta$
starts to differ from $\rho_0$. For a rigorous derivation of $t_\nu$
as well as for a discussion of multi-degree of freedom systems, where there is a cascade of times $t_\nu$, see~\citep{Veble:04}. After $t_\nu$ the so called Lyapunov decay $F(t)=\exp{\{-\lambda(t-t_\nu)\}}$ sets in and lasts until time $t_\delta$ determined by the spreading of the packet over the whole phase space. The Lyapunov decay has been explained by~\citet{Veble:04} and for systems with more than one degree of freedom there is a whole cascade of Lyapunov decays determined by the Lyapunov spectrum\footnote{In case of drift of packets one can get a super-exponential instead of exponential decay of fidelity. For sufficiently small times quantum fidelity will show the same phenomena, see~\citep{Silvestrov:03}.}. If we have a classical system with diffusion, such that the phase space is much larger in one direction, say $q,p \in [0,2\pi]\times [0,L]$, with $L \gg 2 \pi$, time $t_\delta$ will be given by~\citep{Benenti:03b} 
\begin{equation}
t_\delta \sim \frac{1}{\lambda} \ln{\frac{2\pi}{\delta}}.
\end{equation}
After $t_\delta$ fidelity will decay diffusively $F(t)\sim 1/\sqrt{D t}$ with the diffusion constant $D$, until the diffusive process reaches the phase space boundary also in $p$-direction, i.e. at $t_{\rm D}\sim L^2/D$. Note that this diffusive regime is present only if $L \gg 2\pi$. After $t_{\rm D}$ the asymptotic decay of classical fidelity begins. This asymptotic decay is determined by the largest Ruelle-Pollicott resonance~\citep{Benenti:03b}, i.e. the eigenvalue of the Perron-Frobenius operator, and is thus the same as the asymptotic decay of classical correlations. If there is a gap in the spectrum of the Peron-Frobenius operator, this decay will be exponential, otherwise it can be power law. Note that this asymptotic decay rate does not depend on the perturbation strength $\delta$ but only on the phase space size $L$. For large times fidelity will decay towards the asymptotic value $\bar{F}=F(t \to \infty)$ determined by the ratio of the initial packet size and the phase space size. In order to see the asymptotic regime of fidelity decay one has to look at $F(t)-\bar{F}$. Furthermore, for the fidelity at the end of the Lyapunov decay $t_\delta$ to be larger than $\bar{F}$ we must have $\delta > \bar{F}$.
\par
The decay rate for classical fidelity in chaotic systems, apart from the non-decaying ``shoulder'' until $t_\nu$, does not depend on the perturbation strength $\delta$ (decay time borders do depend though) and is therefore independent of the perturbation itself. This must be contrasted with quantum fidelity, which as we will see does depend strongly on perturbation strength and type.

\subsubsection{Regular Systems}
The classical fidelity decay for regular systems has been explained~\citep{Benenti:03} by studying changes in the action-angle variables caused by the perturbation. There is a competition between two contributions to the fidelity decay. It can decay as a consequence of the ballistic separation of the perturbed and unperturbed packets caused by different frequencies of the perturbed and unperturbed tori on which the initial packet is placed, or it can decay due to different shapes of unperturbed and perturbed tori. In the latter case the decay is caused by transitions of the perturbed packet between unperturbed tori. 
Therefore, if the perturbation predominantly changes the {\em frequency} of tori, 
the classical fidelity will exhibit a {\em ballistic} decay. The shape of this ballistic decay is determined by the shape of the initial packet. 
If the packet is a coherent state having a Gaussian shape, the decay will be Gaussian. On the other hand, if the perturbation predominantly changes 
the {\em shape} of tori the decay will be {\em algebraic} $F(t) \sim
1/(\delta t)^d$, in a system with $d$ degrees of freedom. Which type of decay one gets, depends on the shape of the perturbation but not on its strength $\delta$, provided it is small enough. The type of the decay can also depend on the position of the initial packet. 
\par
Note that the above results were derived under the assumption that the
angles across the packet are spread over $2\pi$,
i.e. $t>2\pi/(\Delta_j\partial\vec{\omega}/\partial\vec{j})$, where
$\Delta_j$ is the width of the initial packet in the action
direction. Incidentally, this time is equal to the time $t_1$
(\ref{eq:t1}) after which quantum fidelity freezes at a constant
value if the time averaged perturbation is zero (see Section~\ref{sec:denominator}). Furthermore, the change of the actions across the packet caused by the perturbation must be small, i.e. $\Delta_j \gg \delta$. This last condition translates for coherent packets into $\delta \ll \hbar^{1/2}$, also being the upper limit of validity of approximations used in the calculation of the quantum fidelity plateau (\ref{eq:plateauCS}).
\par
Comparing to quantum fidelity, the case of ballistic decay of classical fidelity corresponds to perturbations having a nonzero time average, discussed in Chapter~\ref{ch:general}. For such perturbations quantum fidelity agrees with the classical one under certain conditions. The case of the algebraic decay corresponds to the perturbation with a vanishing time 
average. In this case though, the quantum and the classical fidelity do not 
agree. What is more, quantum fidelity displays an intriguing new
feature called freezing (see Section~\ref{sec:denominator}) and decays
only on a much longer time scale $\sim 1/\delta^2$. Whereas only the
functional dependence of the classical fidelity decay changes
depending on the perturbation type, always decaying on a time scale
$\sim 1/\delta$, the decay of quantum fidelity drastically changes if one has a perturbation with a zero time average.

\subsection{Entanglement}
The literature on decoherence is exhaustive and we will here list only those more or less directly related to our work.
\par
Time independent perturbative expansion has been used
by~\citet{Kubler:73} to study the eigenvalues of the reduced density
matrix. A similar expansion was used much later for the
purity~\citep{Kim:96}. Note that these perturbative approaches are not
equivalent to our linear response expressions presented in Chapter~\ref{ch:coupling}
as they correspond only to the short time regime in which the correlation
function is constant. Among the early studies of entropy growth
and its relation to the chaoticity of the underlying system is the one
by~\citet{Alicki:96}.~\citet{Miller:99} observed a linear entropy
growth with the slope given by the Lyapunov
exponent. ~\citet{Gorin:02,Gorin:03} studied the purity decay for
random matrix models. Random matrix asymptotic value of the purity (or
of linear entropy) was later re-derived in~\citep{Bandy:02} and
numerically observed in chaotic systems.~\cite{Prosen:02spin} first
defined the purity fidelity, generalising the purity to echo
dynamics. The purity fidelity and its relation with the fidelity in
chaotic and regular systems was studied also in~\citep{Prosen:03evol}
using a correlation function approach which was used also
by~\citet{Tanaka:02} for studying chaotic systems. The connection
between the purity and the fidelity was put on a firm ground by a
rigorous inequality between the
two~\citep{Znidaric:03,Prosen:03corr}. The classical analog of
decoherence has been studied
in~\citep{Gong:03,Gong:03pra,Gong:03pra2}. Using a perturbative
approach the influence of the type of the perturbation and of the dynamics on
the quantum-classical correspondence were explored, see
also~\citep{Angelo:04} for a study of quantum-classical correspondence
of entanglement. The entanglement in weakly coupled composite
systems was studied in~\citep{Znidaric:03} as well
as~\citep{Fujisaki:03}, reaching the same conclusion, namely that the
increase of chaos can inhibit the production of
entanglement. Subsequently the inequality between fidelity and purity
was used in~\citep{Cucchietti:03I}. Entanglement production in a coupled baker's map has been studied in~\citep{Scott:03} and in coupled kicked tops in~\citep{Bandy:02b}. Similar semiclassical methods as for the evaluation of the fidelity have been used also for the purity~\citep{Jacquod:04}, predicting an exponential decay in chaotic systems, confirming prediction in~\citep{Znidaric:03}, and algebraic decay in regular systems. The predicted power of the algebraic decay though does not agree with our theory and numerical results presented in Section~\ref{sec:I}. The entanglement under echo situation during a quantum computation has been studied in~\citep{Rossini:03}. Entanglement in weakly coupled kicked tops has been studied also in~\citep{Rafal:04}, among other things also for random initial states.

\section{Outline}
In Chapter~\ref{ch:fidelity} we introduce the quantum fidelity and briefly review numerical models used for the illustration of the theory. The material in sections dealing with the average fidelity has been mainly published in~\citep{Prosen:02JPA} and~\citep{Prosen:03ptps}. The core of Chapter~\ref{ch:general} has been published in~\citep{Prosen:02JPA}, which is the main paper presenting the correlation function approach to fidelity. New, unpublished material consists of numerical demonstration of a Gaussian distribution of diagonal matrix elements in chaotic systems, relevant for a Gaussian perturbative decay of fidelity. New is also the discussion about the average fidelity in regular systems as well as the last part of Section~\ref{sec:time_scales} with the illustration of the fidelity decay in terms of Wigner functions and some additional figures hopefully clarifying the relations between different decay regimes. The material of Chapter~\ref{ch:freeze} has been published in~\citep{Prosen:03} and~\citep{Prosen:04}. We unified the approach in mixing and regular situation making the exposition of a regular case more concise. New material is the explanation of the average fidelity decay in regular systems with zero time average perturbation. About half of the contents of Chapter~\ref{ch:coupling} has been published in~\citep{Prosen:03evol,Znidaric:03,Prosen:03corr}. The new material consists of the purity fidelity and the reduced fidelity calculation in regular systems beyond linear response, most of the discussion about the Jaynes-Cummings model, the section describing the freeze in a harmonic oscillator and the last section explaining the accelerated decoherence of cat states. The application of the fidelity theory to the improvement of the quantum Fourier transformation has been published in~\citep{Prosen:01} and an extension to the Ising model of quantum computer, not presented in the present work, in~\citep{Celardo:03}.


\chapter{Fidelity}
\label{ch:fidelity}
\begin{flushright}
\baselineskip=13pt
\parbox{70mm}{\baselineskip=13pt
\sf Basic research is what I am doing when I don't know what I am doing.
}\medskip\\
---{\sf \itshape Werner von Braun}\\\vspace{20pt}
\end{flushright}

Quantum fidelity $F$ between two general density matrices $\rho$ and $\sigma$, representing either a pure or a mixed state, has been used by Uhlmann~\citep{Uhlmann:76},
\begin{equation}
F(\rho,\sigma)=\left(\! \tr{\sqrt{\rho^{1/2} \sigma \rho^{1/2}}} \right)^2.
\label{eq:Uhl_fid} 
\end{equation}
He called it a transition probability and the name fidelity was introduced by Jozsa~\citep{Jozsa:94}. It is symmetric with respect to the exchange of $\rho$ and $\sigma$ and in the case of one density matrix being a pure one, the general expression for fidelity simplifies into $F=\bracket{\psi}{\sigma}{\psi}$ and if both are pure it is
\begin{equation}
F(\psi,\varphi)=|\braket{\psi}{\varphi}|^2,
\label{eq:overlap}
\end{equation}
where obviously $\rho:=\ket{\psi}\bra{\psi}$ and $\sigma:=\ket{\varphi}\bra{\varphi}$. We will always deal with pure states and the latter definition (\ref{eq:overlap}) will be sufficient for us. Note that sometimes the name fidelity is used for a quantity without a square, i.e. for a fidelity amplitude. 
\par
We will study stability of quantum dynamics with respect to the perturbation of evolution and the quantity studied will be the fidelity between states obtained by {\em unperturbed} and {\em perturbed} evolutions, starting from the same initial state. Let us denote the initial state by $\ket{\psi(0)}$, and the states at time $\tau$ as
\begin{equation}
\ket{\psi(\tau)}=U_0(\tau) \ket{\psi(0)},\qquad \ket{\psi_\delta(\tau)}=U_\delta(\tau) \ket{\psi(0)},
\label{eq:psit}
\end{equation}
where $U_0(\tau)$ is a unitary propagator from time $0$ to time $\tau$ and $U_\delta(\tau)$ is a perturbed propagator. The overlap between the perturbed and unperturbed states, denoted by $F(\tau)$, will serve us as a criterion of stability,
\begin{equation}
F(\tau)=|\braket{\psi_\delta(\tau)}{\psi(\tau)}|^2=|f(\tau)|^2,\qquad f(\tau)=\bracket{\psi(0)}{M_\delta(\tau)}{\psi(0)},
\label{eq:F_def}
\end{equation}
where we introduced a complex fidelity amplitude $f(\tau)$ and the unitary {\em echo operator}
\begin{equation}
M_\delta(\tau)=U^\dagger_0(\tau) U_\delta(\tau).
\end{equation}
The fidelity as defined in (\ref{eq:F_def}) is a real quantity between $0$ for orthogonal states and $1$ iff the two states are equal (up to a phase) and is a standard measure of stability.
\par
Up to now we have not specified the propagators $U_0(\tau)$ and
$U_\delta(\tau)$ and to simplify the matter we will limit ourselves to
the case where a propagator for time $\tau$ can be written as a power
of some basic single-step propagator $U_0$. That is, we will use
time index $t$ measuring the number of basic units of duration
$\tau_0$, $\tau=t\tau_0$. We therefore write a single step propagator
$U_0 \equiv U_0(\tau_0)$ and similarly for the perturbed evolution
$U_\delta \equiv U_\delta(\tau_0)$. The propagator for $t$ steps is
now simply the $t$-th power of a basic propagator $U_0(\tau_0 t)=U_0^t$. These discrete time formalism allows us to treat two interesting cases at once: the case of time-independent Hamiltonian and the case when the Hamiltonian is time periodic function $H(\tau)=H(\tau+\tau_0)$ with a period $\tau_0$. The latter case includes the so-called kicked systems which will be used for the numerical demonstration of our theory. Note that the theory can be easily generalised also to time-dependent Hamiltonians, only notation becomes more cumbersome. For an example see Section~\ref{sec:QFT} describing the application of fidelity theory to the quantum Fourier transformation algorithm. The general perturbed evolution can be written in terms of the perturbation generator $V$ as
\begin{equation}
U_\delta=U_0 \exp{(-\ii V \delta \tau_0 /\hbar)},
\label{eq:Ud_def}
\end{equation}
where $\delta$ is a dimensionless perturbation strength. The above
equation (\ref{eq:Ud_def}) can be considered as a definition of a
hermitian operator $V$, given the unperturbed and perturbed one time
step propagators $U_0$ and $U_\delta$, respectively. In the case of
Hamiltonian dynamics with $H_0$ generating unperturbed evolution as
$U_0^t=\exp{(-\ii H_0 t \tau_0 /\hbar)}$ and with the perturbed
Hamiltonian of the form $H_\delta=H_0+H' \delta$ we have $V=H'+{\cal
O}(\tau_0 \delta)$. The difference between $H'$ and $V$ therefore goes
to zero in the limit of either small perturbation strength $\delta$ or
in the limit of small time step $\tau_0$. The latter limit, namely
$\tau_0 \to 0$, corresponds to the case of time-independent
Hamiltonians, where the time step $\tau_0$ can be chosen arbitrarily
small. From now on we will drop the irrelevant parameter $\tau_0$
(i.e. take a unit of time to be $\tau_0$) in all equations, so that the discrete time index $t$ has units of time. We will use the same letter $t$ for a
discrete time index as well as for a continuous time (limit $\tau_0
\to 0$) on few
occasions. Whether $t$ is a discrete time index or a continuous will
be clear from the context. Our definition of fidelity (\ref{eq:F_def}) is in discrete time formulation 
\begin{equation}
\Fn=|\bracket{\psi(0)}{\Md}{\psi(0)}|^2,\qquad \Md:=U_0^{-t} U_\delta^t.
\label{eq:Fn_def}
\end{equation}
The fidelity is just the expectation value of the echo operator. It can be equivalently expressed in terms of the initial pure density matrix $\rho(0)$ and the {\em echo} density matrix $\rho^{\rm M}(t)$ (sometimes refered to as the Loschmidt echo) obtained from $\rho(0)$ by evolving it with the echo operator $\Md$ 
\begin{equation}
\rho(0)=\ket{\psi(0)}\bra{\psi(0)},\qquad \rho^{\rm M}(t):=\Md \rho(0) M^\dagger_\delta(t),
\label{eq:rho_def}
\end{equation}
as
\begin{equation}
\Fn=\tr{\left[ \rho(0)\rho^{\rm M}(t) \right]},\qquad \fn=\tr{\left[ \rho(0)\Md \right]}.
\label{eq:Fn_trace}
\end{equation}
On several occasions we will be interested in the fidelity averaged over some ensemble of initial states. In such a case the density matrix $\rho(0)$ in the above definitions of fidelity and fidelity amplitude $\fn$ (\ref{eq:Fn_trace}) must be replaced by appropriate mixed density matrix. We will frequently use a uniform average over whole Hilbert space of dimension ${\cal N}$, $\rho(0)=1/{\cal N}$, see Section~\ref{sec:state_avg} for details.
\par
Using our definition (\ref{eq:Ud_def}) of perturbed dynamics in terms of the perturbation generator $V$, the echo operator can be rewritten in a more useful way by writing $V$ in the interaction picture\footnote{It can also be considered as the Heisenberg picture of the unperturbed dynamics.}, i.e. for an arbitrary operator $A$ its interaction picture $A(t)$ is  
\begin{equation}
A(t)=U_0^{-t} A U_0^t,
\label{eq:Heis_def}
\end{equation}
and so the echo operator is
\begin{equation}
\Md=\exp{(-\ii V(t-1) \delta/\hbar) } \cdots \exp{(-\ii V(0) \delta / \hbar)}={\cal T} \exp{\left( -\ii \Sigma(t) \delta/\hbar \right)},
\label{eq:Md_prod}
\end{equation}
where ${\cal T}$ is a time-ordering operator and $\Sigma(t)$ is the sum of operators $V(j)$,  
\begin{equation}
\Sigma(t):=\sum_{j=0}^{t-1}{V(j)}.
\label{eq:Sigma_def}
\end{equation}
In the case of the time-independent Hamiltonian one has $\Sigma(t)=\int_0^t{V(t'){\rm d}t'}$. The form of the echo operator (\ref{eq:Md_prod}) is nothing but the interaction picture of the perturbed propagator, familiar in quantum field theory. Methods of quantum field theory have been actually used by Adamov, Gornyi and Mirlin~\citep{Adamov:03} to calculate the fidelity in a disordered system. Remember that the time dependence of $V(t)$ comes from the interaction picture (\ref{eq:Heis_def}) and so $V(t)$ is time dependent even if original perturbation $V$ (\ref{eq:Ud_def}) is not, which is the case throughout our derivations. The same definition of fidelity, just by replacing density matrices with the densities in phase space, can be used for classical fidelity, see~\citet{Veble:04}.  
\par
The echo operator $\Md$ can we written as an exponential function of a single operator by using the Baker-Campbell-Hausdorff (BCH) formula
\begin{equation}
{\rm e}^A{\rm e}^B=\exp{\left(A+B+\frac{1}{2}[A,B]+\frac{1}{12}[A,[A,B]]+\frac{1}{12}[B,[B,A]] +\cdots \right)}.
\label{eq:BCH}
\end{equation}
Applying the BCH formula on the product form of $\Md$ (\ref{eq:Md_prod}) gives to the order $\delta^2$ in the argument of the exponential function
\begin{equation}
\Md=\exp{\left\{ -\frac{\ii}{\hbar} \left(\Sigma(t) \delta+\frac{1}{2}\Gamma(t) \delta^2 + \cdots  \right) \right\}},
\label{eq:Md_exp}
\end{equation}
with the hermitian operator $\Gamma(t)$ being
\begin{equation}
\Gamma(t)=\frac{\ii}{\hbar}\sum_{j=0}^{t-1}{\sum_{k=j}^{t-1}{[V(j),V(k)]}}.
\label{eq:Gamma_def}
\end{equation}
Both operators $\Sigma(t)$ and $\Gamma(t)$ standing in the expression for the echo operator have a well defined classical limit, provided $V$ has a classical limit. The classical limit of $\Gamma(t)$ can be obtained by replacing commutator with the Poisson bracket $\{\bullet,\bullet\}$, $(-\ii/\hbar)[\bullet,\bullet]\to \{\bullet,\bullet\}$. The BCH form (\ref{eq:Md_exp}) of the echo operator will be particularly useful in the case of regular dynamics. For now let us list three different possible fidelity decays depending on the behaviour of operator $\Sigma(t)$. For mixing dynamics the fluctuations of $\Sigma(t)$ give the dominant contribution, i.e. terms like $\ave{\Sigma^2(t)}$ grow linearly with time as $\sim t$. For regular dynamics second moment $\ave{\Sigma^2(t)}\sim t^2$ grows quadratically with time, corresponding to the existence of a nontrivial time-averaged perturbation $V(t)$. In certain cases we can have even $\ave{\Sigma^2(t)} \sim t^0$ which results in the so-called freeze of fidelity. Only in this last case is the fidelity decay caused by the operator $\Gamma(t)$.
\par
If we want the fidelity expressed as a power series to all orders in $\delta$ the BCH formula becomes too cumbersome. An easier approach is to just expand the product form of the echo operator $\Md$ (\ref{eq:Md_prod}), giving the fidelity amplitude
\begin{equation}
\fn=1+\sum_{m=1}^{\infty}{\frac{(-\ii)^m \delta^m }{m! \hbar^m} \sum_{j_1,\ldots,j_m=0}^{t-1}{{\cal T} \ave{V(j_1)V(j_2)\cdots V(j_m)}}},
\label{eq:fn_series}
\end{equation}
where we introduced the notation $\ave{\bullet}=\bracket{\psi(0)}{\bullet}{\psi(0)}$. One can see, that the fidelity is expressed in terms of $m$-point quantum correlation functions of the perturbation generator $V$. In the semiclassical limit one is able to replace quantum correlation functions with the classical ones and therefore the quantum fidelity is expressed in terms of classical quantities. One should keep in mind that the classical fidelity behaves distinctively different from quantum fidelity (see Chapter~\ref{sec:class_fid}) and therefore it is not obvious that quantum fidelity can be expressed in terms of classical quantities. Power series expansion of the classical fidelity for example does not yield a fruitful result (for chaotic systems) due to unboundedness of the classical operators and the classical fidelity can not be expressed in a similar way as the quantum fidelity (\ref{eq:fn_series}). 
\par
Fidelity $\Fn$ is now obtained by taking absolute value square of $\fn$ (\ref{eq:fn_series}). Only even orders in $\delta$ survive as odd orders just rotate the phase of the fidelity amplitude $\fn$, the lowest order being quadratic in $\delta$
\begin{equation}
\Fn=1-\frac{\delta^2}{\hbar^2}\sum_{j,k=0}^{t-1}{C(j,k)}+{\cal O}(\delta^4)=1-\frac{\delta^2}{\hbar^2}\left\{ \<\Sigma^2(t)\>-\<\Sigma(t)\>^2 \right\}+{\cal O}(\delta^4),
\label{eq:Fn_2order}
\end{equation}
with $C(j,k)$ being the quantum correlation function
\begin{equation}
C(j,k)=\ave{V(j) V(k)}-\ave{V(j)}\ave{V(k)}.
\label{eq:Cjk}
\end{equation}
The second order expansion of the fidelity (\ref{eq:Fn_2order}) is one of the central theoretical results. Although it is very simple and is just the lowest order expansion it contains most of the essential physics of fidelity decay. Furthermore, there is one very pragmatic reason why it is sufficient and higher orders are not needed. In all practical and experimental applications where the fidelity is the relevant quantity, one is mainly interested in a regime of high fidelity, i.e. of high stability. If the interesting range is say $\Fn>0.9$, higher orders will give only corrections of order $0.01$. To see this, let us denote the second order term with $x=(\delta /\hbar)^2 \sum_{j,k}{C(j,k)}$. From the expansion (\ref{eq:fn_series}) one can see that the term with $\delta^{2m}$ will be at most of the order $(\delta t/\hbar)^{2m} \sim x^{2m}$. So if $1-\Fn\approx x^2 \ll 1$ one can safely neglect higher orders in $\delta$, i.e. terms of order $x^{2m}$ with $m>1$, irrespective of the values of individual parameters like $\delta$, $\hbar$ or time $t$. The range of validity of the lowest order expansion in (\ref{eq:Fn_2order}) is limited only by the value of $1-\Fn$. Furthermore, in certain cases such as mixing dynamics (Section~\ref{sec:mixing}), regular dynamics (Section~\ref{sec:regular}) or in the so-called freeze of fidelity (Chapter~\ref{ch:freeze}) the series (\ref{eq:fn_series}) can be resummed to all orders in $\delta$ and one gets an expression for the fidelity valid in the whole range from $1$ to its asymptotic value. 
\par
Let us now discuss in more detail the lowest order term in the expansion of fidelity which can be in the case of continuous time-independent Hamiltonian written as
\begin{equation}
1-F(t)=\frac{\delta^2}{\hbar^2} \int_{0}^t\!\!{\int_0^t{C(t',t'')\,{\rm d}t'}\, {\rm d}t''}.
\label{eq:Ft_2order}
\end{equation}
This {\em linear response} expression is reminiscent of Green-Kubo like formulas. It says that the ``dissipation of quantum information'' $1-F(t)$ equals the double integral of the correlation function of the perturbation $V$. An interesting and somehow counterintuitive conclusion can be drawn from it, namely, the {\em smaller} the integral of time correlation function the higher fidelity will be. In the semiclassical limit the quantum correlation function approaches the classical one, provided the initial state $\ket{\psi(0)}$ also has a well defined classical limit, and one will see a fast decay of the correlation function for chaotic systems while for regular systems the correlation function typically will not decay. The double integral of the correlation function will therefore grow as $\propto t$ for chaotic systems and like $\propto t^2$ for regular systems. The fidelity will in turn decay {\em more slowly} for chaotic systems than for regular ones. Or in other words, the more chaotic the systems is, the slower decay of quantum fidelity it will have, i.e. the more stable it is to perturbations. This must be contrasted with the behaviour of the classical fidelity (see Section~\ref{sec:class_fid}) which is just the opposite. The more chaotic the system is, the faster classical fidelity will decay.     

 \section{Models for numerics}
We will always use numerical simulations to compare with the theoretical derivations. Two classes of systems will be used. One group will be the so-called kicked top models, describing dynamics of a spin ``kicked'' by some external field. In the case of a kicked top the Hamiltonian is time periodic and the propagator $U_0$ or $U_\delta$ represent a Floquet map over the period of one kick. The second example will be a Jaynes-Cummings system familiar in quantum optics. The Jaynes-Cummings model is a time-independent model with two degrees of freedom representing a harmonic oscillator (boson) coupled with a spin (fermion). In the following two subsections we will present both models and briefly describe their properties, while the actual numerical results will be presented in each chapter as needed.
\subsection{The Kicked Top}
The kicked top has been introduced by Haake, Ku\' s and Scharf~\citep{Haake:87} and has served as a numerical model in numerous studies since then~\citep{Haake:91,Shack:94,Fox:94,Alicki:96,Miller:99,Breslin:99}. In addition the kicked top might also be experimentally realizable~\citep{Haake:00}.
\par
In Chapter~\ref{ch:general}, discussing the fidelity decay for general perturbations, we will use a unitary one step propagator
\begin{equation}
U_0=U(\gamma,\alpha)=\exp{(-\ii \gamma S_{\rm y})} \exp{\left(-\ii \alpha \frac{S_{\rm z}^2}{2S} \right) },
\label{eq:KT_def}
\end{equation}
where $S_{\rm x,y,z}$ are standard spin operators $[S_k,S_l]=\ii \varepsilon_{klm}\,S_m$ and $\alpha,\gamma$ are two parameters determining dynamical properties. The propagator for $t$ steps is $U_0^t$. Half-integer (integer) spin $S$ determines the size of the Hilbert space ${\cal N}=2S+1$ and the value of the effective Planck constant $\hbar=1/S$. The perturbed propagator $U_\delta$ is obtained by perturbing the parameter $\alpha$
\begin{equation}
U_\delta=U(\gamma,\alpha+\delta),
\label{eq:KT_delta}
\end{equation}
so that the perturbation generator $V$ is
\begin{equation}
V=\frac{1}{2}\left( \frac{S_{\rm z}}{S} \right)^2.
\label{eq:KT_V}
\end{equation}
In the classical limit $S \to \infty$ the area preserving map corresponding to $U_0$ can be obtained from the Heisenberg equations for spin operators, $\mathbf{S}(1)=U_0^\dagger \mathbf{S}\, U_0$. The classical map is most easily written in terms of a unit vector on a sphere $\mathbf{r}=({\rm x,y,z})=\mathbf{S}/S$ as
\begin{eqnarray}
x' &=& \cos{\gamma}\, (x \cos{(\alpha z)}-y \sin{(\alpha z)})+z\sin{\gamma} \nonumber \\
y' &=& y \cos{(\alpha z)}+x\sin{(\alpha z)} \\
z' &=& z\cos{\gamma}-\sin{\gamma}\, (x \cos{(\alpha z)}-y \sin{(\alpha z)}) \nonumber.
\label{eq:KT_class}
\end{eqnarray}
The classical perturbation generator (\ref{eq:KT_V}) is simply~\footnote{We will use lowercase letters for classical quantities.} $V \to
v=z^2/2$. The angle $\gamma$ in the propagator is usually set to $\pi/2$
whereas we will use two different values, $\gamma=\pi/2$ and
$\gamma=\pi/6$. For these two values of $\gamma$ the classical
correlation function displays two different decays towards zero, a
monotonic decay for $\gamma=\pi/6$ and an oscillatory decay for $\gamma=\pi/2$. 
\par
The phase space of the classical map (\ref{eq:KT_class}) is regular for small values of $\alpha$, at $\alpha \sim 3$ (see e.g.~\citep{Peres:95}) most of tori disappear and for still larger $\alpha$ the system is fully chaotic.
\par
For $\gamma=\pi/2$ the unperturbed propagator $U_0$ (as well as the perturbed one) commutes with the operator $R_{\rm y}$ of a $\pi$ rotation around ${\rm y}$-axis, $R_{\rm y}=\exp{(-\ii \pi S_{\rm y})}$. In addition there is an antiunitary symmetry so that the Hilbert space is reducible to three invariant subspaces. Following notation in Peres's book~\citep{Peres:95} we denote them with EE, OO and OE with the basis states (here we assume $S$ to be even)
\begin{equation}
\begin{array}{lll}
{\rm EE :} & \ket{0},\left\{ \ket{2m}+\ket{-2m}\right\}/\sqrt{2} & {\cal N}_{\rm EE}=S/2+1 \\
{\rm OO :}& \left\{ \ket{2m-1}-\ket{-(2m-1)}\right\}/\sqrt{2} & {\cal N}_{\rm OO}=S/2  \\
{\rm OE :}& \left\{ \ket{2m}-\ket{-2m}\right\}/\sqrt{2},\left\{\ket{2m-1}+\ket{-(2m-1)} \right\}/\sqrt{2}\; & {\cal N}_{\rm OE}=S,
\end{array}
\label{eq:KT_subspaces}
\end{equation}
where $m$ runs from $1$ to $S/2$ and $\ket{m}$ are standard eigenstates of $S_{\rm z}$. For $\gamma \neq \pi/2$ the subspaces EE and OO coalesce as $R_{\rm y}$ is the only symmetry left and we have just two invariant subspaces. Except if stated otherwise the initial state will always be chosen from subspace OE (i.e. initial coherent state will be projected onto OE subspace) so that the size of the relevant Hilbert space will be ${\cal N}=S$.
\par
In Chapter~\ref{ch:freeze}, dealing with the so-called freeze of
fidelity, we will take a slightly different form of the propagator which will be presented in the mentioned chapter. Apart from the one dimensional kicked top we will also use a system of two coupled kicked tops to demonstrate the dependence of the fidelity decay on the number of degrees of freedom. The one step propagator for two coupled kicked tops is chosen to be
\begin{equation}
U_\delta=U_1(\gamma,\alpha)U_2(\gamma,\alpha) \exp{(-\ii (\delta+\varepsilon)V/\hbar)}.
\label{eq:2KT_def}
\end{equation}
Propagators $U_1$ and $U_2$ are the usual single kicked top propagators (\ref{eq:KT_def}) acting on the first and the second top, respectively, and the last term with operator $V$ is responsible for the coupling of strength $\varepsilon$ for unperturbed evolution and $\varepsilon+\delta$ for perturbed one. The operator $V$ will be left unspecified for now as we will use different $V$'s.
\par
The initial state $\ket{\psi(0)}$ used for fidelity evaluation will be either a random state with the expansion coefficients $c_m=\braket{m}{\psi(0)}$ being independent Gaussian complex numbers or a coherent state. A random state might be the most relevant for quantum computations for instance as it contains the most information and the states used in quantum computation are expected to be ``random''. Coherent state centred at the position $\mathbf{r}^*=(\sin{\vartheta^*}\cos{\varphi^*},\sin{\vartheta^*}\sin{\varphi^*},\cos{\vartheta^*})$ is given by expansion
\begin{equation}
\ket{\vartheta^*,\varphi^*}=\sum_{m=-S}^S{{2S \choose S+m}^{1/2} \cos^{S+m}{(\vartheta^*/2)}\sin^{S-m}{(\vartheta^*/2)}{\rm e}^{-\ii m \varphi^*}\ket{m}}.
\label{eq:SU2_coh}
\end{equation}
Equivalently, it can be written in terms of a complex parameter $\tau$ as
\begin{equation}
\ket{\vartheta^*,\varphi^*}=\frac{{\rm e}^{-\ii \varphi^* S}}{(1+|\tau|^2)^S}\exp{(\tau S_-)}\ket{S},\qquad \tau={\rm e}^{\ii \varphi^*}\tan{(\vartheta^*/2)},
\end{equation}
with $S_\pm=S_{\rm x}\pm \ii S_{\rm y}$. In the semiclassical limit of large spin $S$ the expansion coefficients of coherent state $c_m$ go towards $c_m\asymp \exp{(-S(m/S-z^*)^2/2(1-z^{*2}))}{\rm e}^{-\ii m \varphi^*}$. Coherent states have a well defined classical limit and this enables to compare the quantum fidelity for coherent initial states with the corresponding classical fidelity. The initial classical phase space density corresponding to a coherent state is~\citep{Fox:94}
\begin{equation}
\rho_{\rm clas}(\vartheta,\varphi)=\sqrt{\frac{2S}{\pi}} \exp{\{-S \lbrack (\vartheta-\vartheta^*)^2+(\varphi-\varphi^*)^2 \sin^2{\vartheta} \rbrack \}}.
\label{eq:rho_clasSU2}
\end{equation}
The above density is normalised as $\int{\!\rho_{\rm clas}^2 {\rm d}\Omega}=1$.

\subsection{The Jaynes-Cummings model}
\label{sec:Jaynes}
The Jaynes-Cummings model~\citep{Jaynes:63,Tavis:68}, see also \cite[p.~336]{Qoptics}, is a system of a coupled harmonic oscillator and a spin. It can be realized experimentally in the cavity electrodynamics experiments (QED) by sending a beam of atoms trough a cavity. The electromagnetic field in the cavity is quantized with the Hamiltonian $\hbar \omega a^+a$, and the spin degree of freedom $\hbar \varepsilon S_{\rm z}$ of atoms interacts with the electromagnetic field. The dominant interaction is a dipolar $\vec{d}\cdot \vec{E}$, with the monochromatic field $\vec{E}=\vec{\varepsilon} E_0 (a {\rm e}^{\ii k r}-a^+ {\rm e}^{-\ii k r})$. For two level atoms only ${\rm y}$-component of the dipole moment $\vec{d}$ is nonzero and is proportional to $S_{\rm y}=-\frac{\ii}{2}(S_+ - S_-)$. All this results in the Hamiltonian
\begin{equation}
H=\hbar \omega a^+ a+\hbar \varepsilon S_{\rm z}+\frac{\hbar}{\sqrt{2S}}\left\{ G\,(a S_+ + a^+ S_-)+G'\,(a^+ S_+ + aS_-) \right\},
\label{eq:JC_H}
\end{equation}
with boson lowering/raising operators $a,a^+$, $[a,a^+]=1$. Planck's constant is chosen as $\hbar=1/S$ so that the semiclassical limit is reached for $S \to \infty$. If we are close to a resonance $\omega=\varepsilon$ the rotating-wave approximation can be made in which the fast oscillating term $G'$ can be neglected and only the $G$-term is retained (oscillating with small $\omega-\varepsilon$). In our discussion we will predominantly focus on this situation of $G'=0$. In addition, frequencies $\omega$ and $\varepsilon$ are usually in a GHz regime, while the coupling $G$ is of the order of kHz and is therefore small. If either $G=0$ or $G'=0$ the model is integrable with an additional invariant being the difference or the sum of the spin and boson (harmonic oscillator) quanta. 
\par
The initial state will be always chosen as a product state of spin and boson coherent states. The spin coherent state is the same as above (\ref{eq:SU2_coh}), while the coherent state for a boson is
\begin{equation}
\ket{\alpha}={\rm e}^{\alpha a^+-\alpha^* a}\ket{0}=\exp{(-|\alpha|^2/2)}\sum_{k=0}^\infty{\frac{\alpha^k}{\sqrt{k!}}\ket{k}},
\label{eq:boson_coh}
\end{equation}
with $\alpha$ being a complex parameter determining the position of the coherent state and $\ket{k}$ is an eigenstate of operator $a^+ a$ having $k$ boson quanta. The classical density corresponding to the boson coherent initial state with a complex parameter $\alpha=\alpha_{\rm r}+\ii\, \alpha_{\rm i}$ is
\begin{equation}
\rho_{\rm class}(q,p)= \sqrt{\frac{2S}{\pi}}\exp{\left\{-S \left[(q-\sqrt{2/S}\, \alpha_{\rm r})^2+(p-\sqrt{2/S}\,\alpha_{\rm i})^2 \right] \right\}}.
\label{eq:rho_classbos}
\end{equation}
The normalisation is as usual $\int{\!\rho_{\rm class}^2 {\rm d}q\,{\rm d}p}=1$.\par
The classical Hamiltonian is obtained by taking the limit $S \to \infty$ and defining new canonical quantum operators, $q_1$ and $p_1$ for a boson and $q_2$ and $p_2$ for a spin 
\begin{equation}
a=\sqrt{\frac{S}{2}}(q_1+\ii p_1),\qquad \sqrt{1-p_2^2}\, {\rm e}^{\ii q_2}=S_+/S.
\end{equation}
They satisfy commutation relations $[q_1,p_1]=\ii \hbar$ and $[q_2,p_2]=\ii \hbar$. In the limit $S \to \infty$ they commute and can be replaced by the classical variables resulting in the classical Hamiltonian
\begin{equation}
H_{\rm class}=\frac{\omega}{2}(p_1^2+q_1^2)+\varepsilon \, p_2+G_+ \sqrt{1-p_2^2}\, q_1 \cos{q_2}-G_-\sqrt{1-p_2^2}\, p_1\sin{q_2}-\frac{\omega}{2S},
\label{eq:JC_class}
\end{equation}
with $G_\pm=G\pm G'$.

 \section{Average Fidelity}
Sometimes the average fidelity is of interest, i.e. the fidelity averaged over some ensemble of initial states. Such an average fidelity is also more amenable to theoretical treatment. Easier to calculate is the average fidelity amplitude $\fn$ which is of second order in the initial state $\ket{\psi}$ while the fidelity $\Fn$ is of fourth order in $\ket{\psi}$. We will show that the difference between the average fidelity amplitude and the average fidelity is semiclassically small.
\par
In a finite Hilbert space the fidelity will not decay to zero but will instead fluctuate around some small plateau value. The value of this plateau equals to a time averaged fidelity. For ergodic systems this time averaged value equals to the phase space averaged one.

\subsection{Time Averaged Fidelity}
\label{sec:time_averaged}
We want to calculate the value of fidelity in the limit $t\to \infty$ that is its asymptotic value for large time. For a finite Hilbert space size ${\cal N}$ the fidelity will start to fluctuate for long times due to a discreteness of the spectrum of the evolution operator. The size of this fluctuations can be calculated by evaluating a time average fidelity $\oFn$
\begin{equation}
\oFn=\lim_{m\to \infty} \frac{1}{m} \sum_{t=0}^m{\Fn}.
\label{eq:Fn_avg_def}
\end{equation}
This is easily done if we expand the initial state in eigenbasis of
the unperturbed propagator $U_0$ and denote the matrix elements between unperturbed and perturbed eigenstates by $P_{kl}$
\begin{equation}
U_0\ket{\phi_l}=\exp{(-\ii \phi_l)}\ket{\phi_l},\quad U_\delta\ket{\phi_l^\delta}=\exp{(-\ii \phi_l^\delta)}\ket{\phi_l^\delta},\quad P_{kl}=\braket{\phi_k}{\phi_l^\delta}.
\end{equation}
We denoted the eigenphases of unperturbed and perturbed one-step
propagator with $\phi_l$ and $\phi_l^\delta$, respectively. The matrix
$P$ is unitary and in the case when both eigenvectors can be chosen to
be real it is orthogonal. This happens if $U_0$ and $U_\delta$ commute
with an antiunitary\footnote{Antiunitary operator must satisfy
$\braket{T\psi}{T\phi}=\braket{\psi}{\phi}^*$.} operator
$T$ whose square is identity~\footnote{Square of an arbitrary
antiunitary operator is $T^2=\pm \mathbbm{1}$. Time reversal operator for
a system with spin is
$T=\exp{(-\ii \pi S_{\rm y})}K$, with a complex conjugation
operator $K$. If the system has an integer spin (or half-integer spin
and an additional rotational invariance symmetry) we have $T^2=\mathbbm{1}$.}. The fidelity amplitude can now be written
\begin{equation}
\fn=\sum_{lm}{(P^\dagger \rho)_{lm} P_{ml} \exp{(-\ii (\phi_l^\delta-\phi_m)t)}},
\label{eq:fn_P}
\end{equation}
with $\rho_{lm}=\bracket{\phi_l}{\rho(0)}{\phi_m}$ being the matrix elements of the initial density matrix in the unperturbed eigenbasis. To calculate the average fidelity $\oFn$ we have to take the absolute value square of $\fn$. Averaging over time $t$ we will assume that the phases are nondegenerate
\begin{equation}
\overline{\exp{(\ii(\phi_l^\delta-\phi_{l'}^\delta+\phi_m-\phi_{m'})t)}}=\delta_{m\,m'}\delta_{l\,l'}.
\label{eq:time_avg}
\end{equation}
This results in the average fidelity
\begin{equation}
\oFn=\sum_{ml}{|(\rho P)_{ml}|^2 |P_{ml}|^2}.
\label{eq:Fn_avg}
\end{equation}
The time averaged fidelity therefore understandably depends on the initial state $\rho$ as well as on the ``overlap'' matrix $P$. 
\par
For {\em small perturbation} strengths, say $\delta$ smaller than some critical $\delta_{\rm rm}$, the unitary matrix $P$ will be close to identity. Using $P \to \mathbbm{1}$ for $\delta \ll \delta_{\rm rm}$ (\ref{eq:Fn_avg}) gives us
\begin{equation}
\oFn_{\rm weak}=\sum_l{\rho_{ll}^2}.
\label{eq:Fnavg_weak}
\end{equation}
One should keep in mind that for the above result $\oFn_{\rm weak}$ we needed eigenphases to be nondegenerate, $\phi_l^\delta \neq \phi_m$ (\ref{eq:time_avg}), and at the same time $P\to \mathbbm{1}$. This approximation is justified in the lowest order in $\delta$, when off diagonal matrix elements are $|P_{ml}|^2 \propto \delta^2$. 
\par
On the other hand, for sufficiently {\em large} $\delta \gg \delta_{\rm rm}$ and complex perturbations $V$ one might assume $P$ to be close to a random matrix with independent real or complex matrix elements $P_{ml}$. Then we can average expression (\ref{eq:Fn_avg}) over a Gaussian distribution $\propto \exp{(-\beta {\cal N} |P_{ml}|^2/2)}$ of matrix elements $P_{ml}$, where we have $\beta=1$ for orthogonal $P$ and $\beta=2$ for unitary $P$. This averaging gives $\< |P_{ml}|^4 \>=(4-\beta)/{\cal N}^2$ and $\langle |P_{ml}|^2 \rangle=1/{\cal N}$ for the variance of $P_{ml}$ (brackets $\< \bullet \>$ denote here averaging over the distribution of matrix elements and not over the initial state). The average fidelity for strong perturbation can therefore be expressed as
\begin{equation}
\oFn_{\rm strong}=\frac{4-\beta}{\cal N}\sum_l{\rho_{ll}^2}+\frac{1}{\cal N}\sum_{l\neq m}{|\rho_{lm}|^2}.
\label{eq:Fnavg_strong}
\end{equation}
The point of crossover $\delta_{\rm rm}$ from weak
(\ref{eq:Fnavg_weak}) to strong (\ref{eq:Fnavg_strong}) perturbation regime is system dependent and can not be discussed in general apart from expecting it to scale with $\hbar$ similarly as a mean level spacing $\delta_{\rm rm} \sim \hbar^{d}$. We will discuss the value of $\oFn$ for three different initial states:
\begin{enumerate}
\item[(i)] 
First, let us consider the simplest case when the initial state is an {\em eigenstate} of $U_0$ say, $\rho=\ket{\phi_1}\bra{\phi_1}$ with matrix elements $\rho_{lm} = \delta_{l,1}\delta_{m,1}$. For {\em weak perturbations} this gives (\ref{eq:Fnavg_weak}) $\oFn_{\rm weak}=1$, therefore the fidelity does not decay at all. This result can be generalised to the case when $\rho$ is a superposition of a number of eigenstates, say $K$ of them, all with approximately the same weight, so that one has diagonal density matrix elements of order $\rho_{ll} \sim 1/K$, resulting in $\oFn_{\rm weak}\sim 1/K$. On the other hand, for {\em strong perturbation} $\delta\gg\delta_{\rm rm}$ we get $\oFn_{\rm strong}=(4-\beta)/{\cal N}$ for an initial eigenstate. Summarising, for an initial eigenstate we have time averaged values of fidelity
\begin{equation}
\oFn_{\rm weak}=1,\qquad \oFn_{\rm strong}=(4-\beta)/{\cal N}.
\label{eq:Fnavg_eig}
\end{equation}
With this simple result we can easily explain the numerical result of Peres~\citep{Peres:95} where no-decay of fidelity was found for a coherent initial state
sitting in the centre of an elliptic island, thus being a superposition of a very small number of eigenstates (it is almost an eigenstate). The behaviour in generic case may be drastically different as described in the present work.

\item[(ii)]
Second, consider the case of a {\em random pure} initial state $\ket{\psi}=\sum_m c_m\ket{\phi_m}$, giving $\rho_{ml}=c_m c_l^*$. The coefficients $c_m$ are independent random complex Gaussian variables with variance $1/{\cal N}$, resulting in averages $\< |\rho_{lm}|^2 \>=1/{\cal N}^2$ for $m\neq l$ and $\< \rho_{ll}^2 \>=2/{\cal N}^2$ (average is over Gaussian distribution of $c_m$). Using this in expressions for average fidelity (\ref{eq:Fnavg_weak}) and (\ref{eq:Fnavg_strong}) we get
\begin{equation}
\oFn_{\rm weak}=2/{\cal N},\qquad \oFn_{\rm strong}=1/{\cal N}.
\label{eq:Fnavg_random}
\end{equation}
For random initial state there is therefore only a factor of $2$ difference between finite size fluctuating plateau for weak and for strong perturbation. The result for weak perturbation agrees with the case (i) where we had $\oFn_{\rm weak} \sim 1/K$ if there were $K$ participating eigenvectors.
\item[(iii)]
Third, for a {\em uniform average} over the whole Hilbert space, i.e. taking a non-pure initial density matrix $\rho=\mathbbm{1}/{\cal N}$, we have
\begin{equation}
\oFn_{\rm weak}=1/{\cal N},\qquad \oFn_{\rm strong}=(4-\beta)/{\cal N}^2.
\label{eq:Fnavg_avg}
\end{equation}
As expected, the fluctuating plateau is the smallest for an uniform average over the whole Hilbert space and strong perturbation.
\end{enumerate}
\begin{figure}[h]
\centerline{\includegraphics{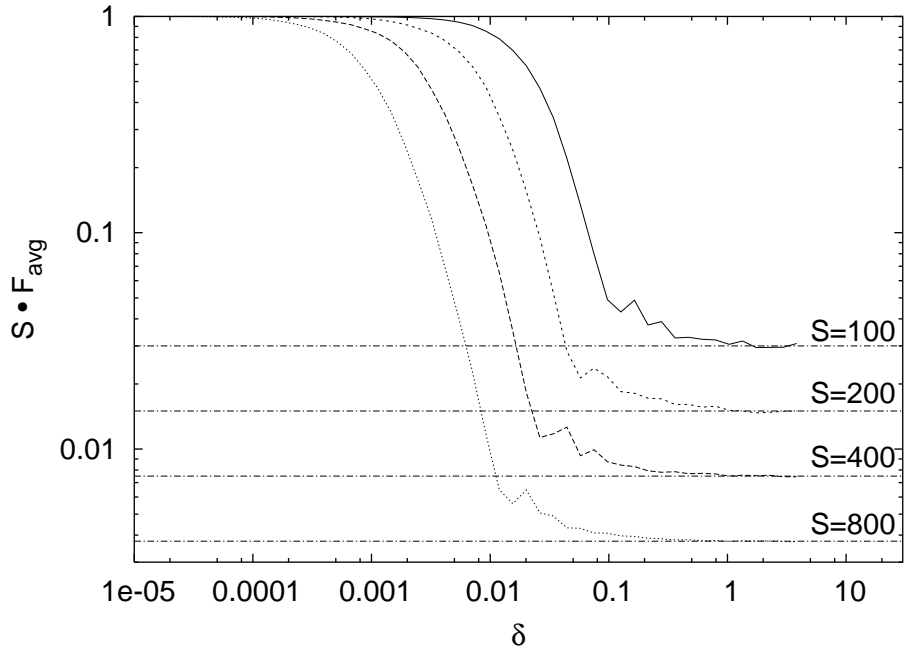}}
\caption{Dependence of time averaged fidelity (multiplied by the
Hilbert space size ${\cal N}=S$) $\oFn$ on $\delta$ is shown for a
chaotic kicked top system and Hilbert space average
$\rho=\mathbbm{1}/{\cal N}$, i.e. our case (iii). The transition from
weak to strong perturbation regime is seen
(\ref{eq:Fnavg_avg}). Horizontal full lines are the theoretical predictions $\oFn_{\rm strong}$ (\ref{eq:Fnavg_avg}), while the theoretical result for the weak regime corresponds to $1$.} 
\label{fig:avgF_chaotic}
\end{figure}
Observe that the average fidelity $\oFn$ (\ref{eq:Fn_avg}) is of
fourth order in matrix elements of $P$, the same as the inverse
participation ratio (IPR) of the perturbed eigenstates. Actually, in
the case of initial eigenstate, our case (i), the average fidelity
(\ref{eq:Fn_avg}) can be rewritten as $\oFn=\sum_m{|P_{1m}|^4}$,
exactly the IPR. The inverse of the IPR is a number between $1$ and
${\cal N}$ which can be thought of as giving the
approximate number of unperturbed eigenstates represented in the
expansion of a given perturbed eigenstate. For an average over the whole space, case (iii), we
have instead $\oFn=\sum_{l,m}{|P_{lm}|^4}/{\cal N}^2$, i.e. the
average IPR divided by ${\cal N}$. The time averaged fidelity is thus directly related to the
{\em localisation} properties of eigenstates of $U_\delta$ in terms of eigenstates of $U_0$. However, except for the pathological case of the initial state being a small combination of eigenstates of $U_0$ with weak perturbation, the
fidelity fluctuation is always between the limiting values $2/{\cal N}$, and $3/{\cal N}^2$. Therefore, the fidelity will decay only until it reaches the value of finite size fluctuations and will fluctuate around $\oFn$ thereafter. The time $t_\infty$ when this happens, $F(t_\infty)=\oFn$, depends on the decay of fidelity and will be discussed in subsequent chapters.
\par
\begin{figure}[ht]
\centerline{\includegraphics{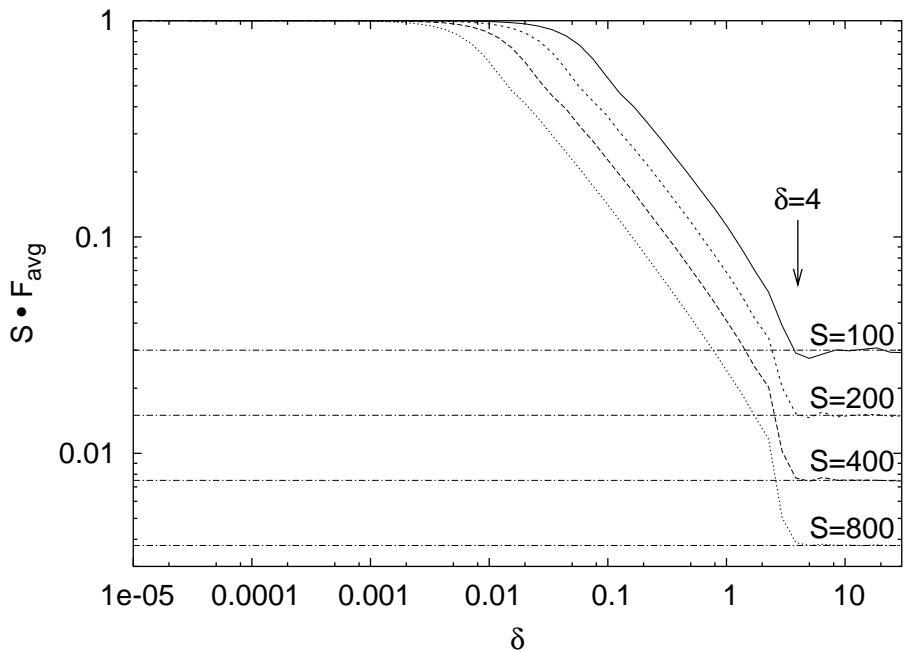}}
\caption{The same as Figure~\ref{fig:avgF_chaotic} but for a regular kicked top.}
\label{fig:avgF_regular}
\end{figure}

To illustrate the above theory we have calculated the average fidelity (\ref{eq:Fn_avg}) for a kicked top with a propagator (\ref{eq:KT_def}). As an initial state we used $\rho=\mathbbm{1}/{\cal N}$, i.e. the case (iii), where the Hilbert space size is determined by the spin value, ${\cal N}=S$ (OE subspace). We calculated the dependence of $\oFn$ on $\delta$ for two cases: a chaotic one for kicked top parameters $\alpha=30$, $\gamma=\pi/2$ shown in Figure~\ref{fig:avgF_chaotic} and a regular one for $\alpha=0.1$, $\gamma=\pi/2$ shown in Figure~\ref{fig:avgF_regular}. In both cases one can see a transition from the weak perturbation regime $\oFn_{\rm weak}=1/{\cal N}$ to the strong regime $\oFn_{\rm strong}=3/{\cal N}^2$ for large $\delta$. In the chaotic case the critical $\delta_{\rm rm}$ can be seen to scale as $\delta_{\rm rm} \sim \hbar=1/S$. In the regular situation, the strong perturbation regime is reached only for a  strong perturbation $\delta \sim 4$, where the propagator $U_\delta$ itself becomes chaotic. Namely, the transition from the regular to chaotic regime in the kicked top happens at around $\alpha=3$, see e.g.~\citep{Peres:95}. Still, if one defines $\delta_{\rm rm}$ as the points where the deviation from the weak regime starts (point of departure from $1$ in Figure~\ref{fig:avgF_regular}) one has scaling $\delta_{\rm rm}\sim 1/S$ also in the regular regime.  

\subsection{State Averaged Fidelity}  
\label{sec:state_avg}
In previous section we used the initial density matrix $\rho(0)=\mathbbm{1}/{\cal N}$ in expression (\ref{eq:Fn_trace}) for the fidelity amplitude to calculate the average fidelity over the whole Hilbert space. As the fidelity is of fourth order in the initial state $\ket{\psi(0)}$, whereas the fidelity amplitude is bilinear in $\ket{\psi(0)}$ the average fidelity is not equal to the quantity obtained by first averaging fidelity amplitude and subsequently squaring it. In general the difference between the two depends on the ensemble of states over which we average. 
\par
Let us look only at the simplest case of averaging over random initial
states, denoted by $\aave{\bullet}$. In the asymptotic limit of large
Hilbert space size ${\cal N}\to \infty$ the averaging is simplified by the fact that the expansion coefficients $c_m$ of a random initial state in an arbitrary basis become independent Gaussian variables with
variance $1/{\cal N}$. Quantities bilinear in the initial state, like the fidelity amplitude or the correlation function, result in the following expression
\begin{equation}
\aave{\psi |A|\psi}=:\aave{A}=\aave{c_m^* \, A_{ml}\, c_l}=\frac{1}{\cal N}\tr{A},
\label{eq:Avg_trace}
\end{equation}
where $A$ is an arbitrary operator. The averaging is done simply by means of a trace over the whole Hilbert space. For the fidelity $\Fn$ which is of fourth order in $\ket{\psi}$ we get
\begin{equation}
\aave{\Fn}=\aave{ c_m^* [\Md]_{ml} \, c_l\, c_p\, [\Md]_{pr}^* \,c_r^* }= \left| \aave{\fn} \right|^2+\frac{1}{\cal N}.
\label{eq:Avg_4order}
\end{equation}
The difference between the average fidelity and the average fidelity amplitude is therefore semiclassically small~\citep{Prosen:03ptps}.
\par
There are two reasons why averaging over random initial states is of
interest. First, in the field of quantum information processing this
are the most interesting states as they have the least structure,
i.e. can accomodate the largest amount of information. Second, for ergodic
dynamics and sufficiently long times one can replace expectation
values in a specific generic state $\ket{\psi}$ by an ergodic
average. If the system is ergodic on the whole space one can calculate
the correlation function $C(j,k)$ (\ref{eq:Cjk}) by means of a simple
trace (\ref{eq:Avg_trace}), so that it does not depend on the initial
state, or even calculate it classically in the leading semiclassical order, which greatly simplifies theoretical derivations~\footnote{In the case of autonomous systems the canonical or micro-canonical averaging should be used instead.}. For regular systems, where the decay of fidelity depends on the initial state, the ergodic averaging differs from the average in a specific initial state, although one can still be interested in the behaviour of the average fidelity. Such an averaging will be discussed in the section describing the decay of fidelity in regular systems. 
\par
We have seen that for sufficiently large Hilbert spaces there is no difference between averaging the fidelity amplitude or the fidelity or taking a single random initial state. For mixing dynamics the long time fidelity decay is independent of the initial state even if it is a non-random, wheras in the regular regime it is state dependent. For instance, the long time Gaussian decay (Section~\ref{sec:CIS_gen}) depends on the position of the initial coherent state. The fidelity averaged over this position of the initial coherent state might be of interest and will not be equal to the fidelity averaged over random initial states. We will discuss averaging over coherent states at the end of Section~\ref{sec:CIS_gen} describing long time fidelity decay for coherent initial states. 


\chapter{General Perturbation}
\label{ch:general}
\begin{flushright}
\baselineskip=13pt
\parbox{70mm}{\baselineskip=13pt
\sf An expert is a person who has made all the mistakes that can be made in a very narrow field.
}\medskip\\
---{\sf \itshape Niels Bohr}\\\vspace{20pt}
\end{flushright}

\section{Mixing Dynamics}
\label{sec:mixing}
Here we assume that the system is mixing such that the
correlation function of the perturbation $V$ decays sufficiently fast; this typically 
(but not necessarily) corresponds to globally chaotic classical motion.
Due to {\em ergodicity} we will assume the initial density matrix to be $\rho(0)=\mathbbm{1}/{\cal N}$, so that all averages over a specific initial state can be replaced by a full Hilbert space average, $\ave{.}=\tr(.)/{\cal N}$ (Section~\ref{sec:state_avg}). For any other initial state (e.g. in the worst case for the minimal wave packet -- coherent state) one obtains identical results on $\Fn$ for sufficiently long times\footnote{The exception might be systems with localized states.}, i.e. longer than the {\em Ehrenfest} time $t_{\rm E} \approx \ln(1/\hbar)/\lambda$ (for a
classically chaotic system with maximal Lyapunov exponent $\lambda$) needed for a minimal wave packet to spread effectively over the accessible phase space~\citep{Berman:78}. The state averaged quantum correlation function is homogeneous in time, i.e. $C(j,k)=C(k-j)$, so we simplify the second order linear response formula for the fidelity (\ref{eq:Fn_2order})
\begin{equation}
\Fn=1-\frac{\delta^2}{\hbar^2}\left\{t C(0) + 2\sum_{j=1}^{t-1}{(t-j)C(j)} \right\} + {\cal O}(\delta^4).
\label{eq:LR_mixing}
\end{equation}
\par
If the decay of the correlation function $C(j)$ is sufficiently fast, namely if its integral converges on a certain characteristic {\em mixing time} scale $t_{\rm mix}$, then the above formula can be further simplified. 
For times $t \gg t_{\rm mix}$ we can neglect the second term under the summation in (\ref{eq:LR_mixing}) and obtain a
linear fidelity decay in time $t$ (in the linear response)
\begin{equation}
\Fn = 1 - 2 (\delta/\hbar)^2 \sigma t,
\label{eq:linear_mixing}
\end{equation}
with the transport coefficient $\sigma$ being
\begin{equation}
\sigma = \frac{1}{2}C(0) + \sum_{j=1}^\infty{C(j)}=\lim_{t\to\infty}{\frac{\ave{\Sigma^2(t)}-\ave{\Sigma(t)}^2}{2t}}.
\label{eq:sigma_def}
\end{equation}
In the continuous time case $\sigma$ is just the integral of the correlation function. Note that $\sigma$ has a well defined classical limit obtained from the classical correlation function and in the semiclassical limit this classical $\sigma_{\rm cl}$ will agree with the quantum one.
\par
We can make a stronger statement in a non-linear-response regime if we make an additional assumption on the
factorisation of higher order time-correlations, $n-$point mixing. This implies that $2m$-point correlation
$\ave{V(j_1)\cdots V(j_{2m})}$ is appreciably different from zero for $j_{2m}-j_1 \to \infty$ only if all (ordered) time indices $\{j_k,k=1\ldots 2m\}$ are {\em paired} with the time differences 
within each pair, $j_{2k}-j_{2k-1}$, being of the order or less than 
$t_{\rm mix}$. Then we can make a further reduction, namely if $t\gg m t_{\rm mix}$ the terms in the expansion of the fidelity amplitude $\fn$ (\ref{eq:fn_series}) are
\begin{figure}[ht!]
\centerline{\includegraphics{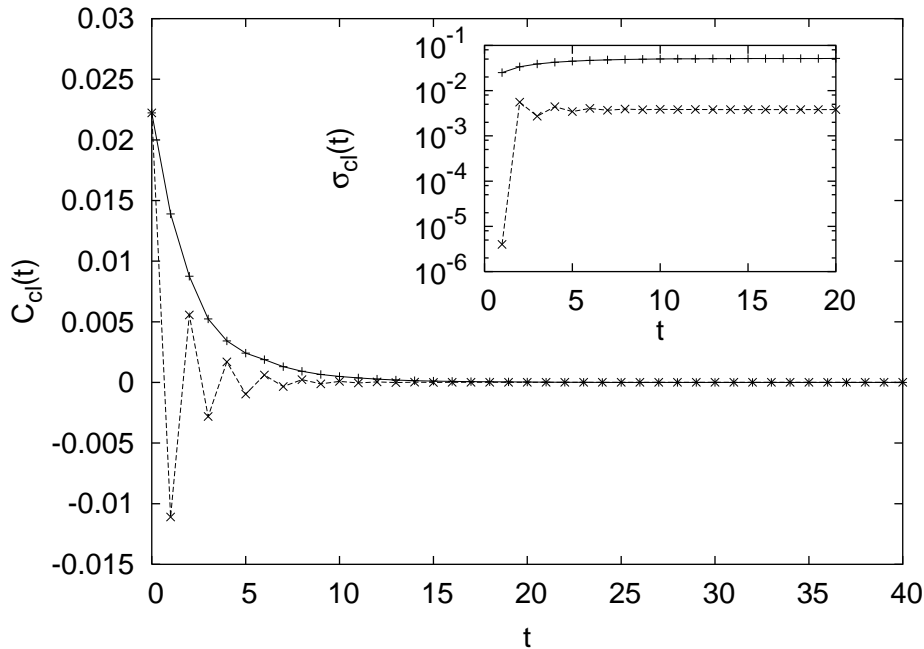}}
\caption{The classical correlation function (\ref{eq:Cjk}) of perturbation $V$ (\ref{eq:class_v}) for chaotic kicked top and $\gamma=\pi/6$ (top solid curve) and $\gamma=\pi/2$ (bottom broken curve). The finite time integrated correlation function is shown in the inset.}
\label{fig:class_cor}
\end{figure}
\begin{eqnarray}
&{\cal T}&\!\!\!\sum_{j_1,\ldots,j_{2m}=0}^{t-1}\!
\ave{V(j_1) V(j_2) \cdots V(j_{2m})}
\rightarrow \nonumber \\
\rightarrow &{\cal T}&\!\!\!\sum_{j_1,\ldots,j_{2m}=0}^{t-1}\!
\ave{V(j_1) V(j_2)}\cdots\ave{V(j_{2m-1}) V(j_{2m})}
\rightarrow
\frac{(2m)!}{m!2^m}(2\sigma t)^m.
\label{eq:factor}
\end{eqnarray}
The fidelity amplitude is therefore $\fn=\exp{(-\delta^2 \sigma t/\hbar^2)}$ and the fidelity is
\begin{equation}
\Fn=\exp{(-t/\tau_{\rm m})}, \qquad \qquad \tau_{\rm m}=\frac{\hbar^2}{2 \delta^2 \sigma_{\rm cl}},
\label{eq:Fn_mixing}
\end{equation}
with a mixing decay time-scale $\tau_{\rm m}={\cal O}(\delta^{-2})$ and a classical limit of the transport coefficient $\sigma_{\rm cl}$. 
We should stress again that the above result (\ref{eq:Fn_mixing}) has been derived under the assumption of
true quantum mixing which can be justified only in the limit ${\cal N}\rightarrow\infty$, e.g.
either in the semiclassical or the thermodynamic limit. Thus for the true quantum-mixing dynamics the fidelity will decay exponentially. The same result has been derived also by a quite different approach, using a Fermi golden rule~\citep{Jacquod:01,Cerruti:02}. That is why this regime of exponential fidelity decay is sometimes called a Fermi golden rule regime. 
\par
To numerically check the above exponential decay, we will use the kicked top (\ref{eq:KT_def}) with parameter $\alpha=30$, giving a totally chaotic classical dynamics. As argued before, one can calculate the transport coefficient $\sigma$ (\ref{eq:sigma_def}) by using the classical correlation function of the perturbation (\ref{eq:KT_V}),
\begin{equation}
V_{\rm cl}=v=\frac{1}{2}z^2.
\label{eq:class_v}
\end{equation}
We consider two different values of kicked top parameter $\gamma$,
namely $\gamma=\pi/2$ and $\gamma=\pi/6$. The classical correlation
functions can be seen in Figure~\ref{fig:class_cor}. The correlation
function (obtained by averaging over $10^5$ initial conditions on a
sphere) is shown in the main frame. The correlation functions have
qualitatively different decay towards zero for the two chosen $\gamma$'s. In the inset the convergence of classical $\sigma$ (\ref{eq:sigma_def}) is shown, where one can see that the mixing time is $t_{\rm mix} \sim 5$. The values of $\sigma_{\rm cl}$ are $\sigma_{\rm cl}=0.00385$ for $\gamma=\pi/2$ and $\sigma_{\rm cl}=0.0515$ for $\gamma=\pi/6$. 
\begin{figure}[h]
\centerline{\includegraphics{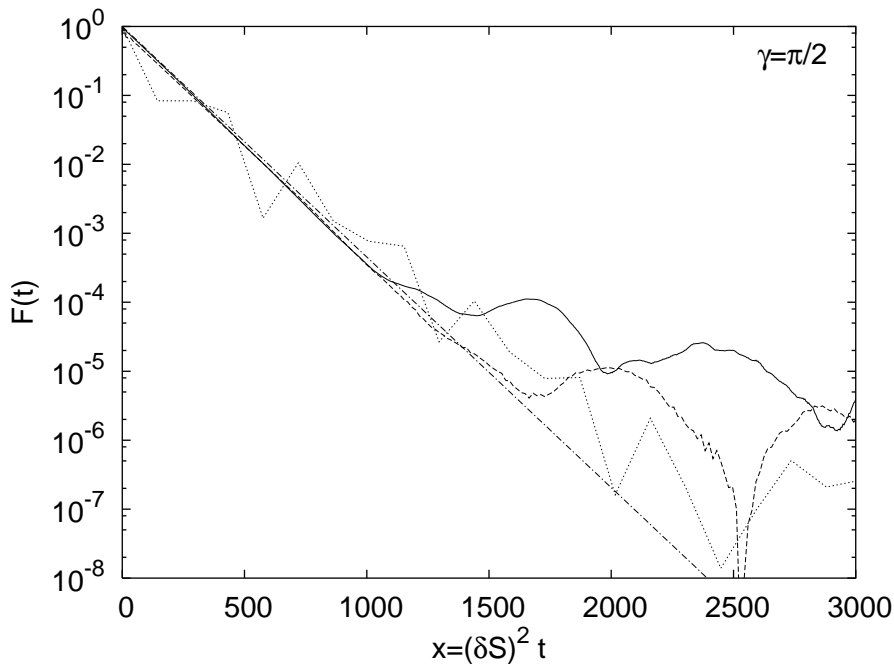}}
\caption{Quantum fidelity decay in the chaotic regime for
  $\gamma=\pi/2$ and three different perturbation strengths
  $\delta=5\times 10^{-4}$, $1\times 10^{-3}$ and $3\times 10^{-3}$
  (solid, dashed and dotted curves, respectively) is shown. The chain
  line gives theoretical decay (\ref{eq:Fn_mixing}) with the classically calculated $\sigma$ seen in Figure~\ref{fig:class_cor}.}
\label{fig:exp_pi2}
\end{figure}
These values are used to calculate the theoretical decay of fidelity $\Fn=\exp{(-\delta^2 S^2 2 \sigma_{\rm cl} t)}$ (\ref{eq:Fn_mixing}) which is compared with the numerical simulation in Figures~\ref{fig:exp_pi2} and \ref{fig:exp_pi6}. We used averaging over the whole Hilbert space, $\rho(0)=\mathbbm{1}/{\cal S}$ and checked that that due to ergodicity there was no difference for large $S$ if we choose a fixed initial state, say a coherent state. As fidelity will decay only until it reaches its finite size fluctuating value $\oFn$ (\ref{eq:Fnavg_avg}) we choose a large $S=4000$ in order to be able to check exponential decay over as many orders of magnitude as possible. In Figure~\ref{fig:exp_pi2} the decay of quantum fidelity is shown for $\gamma=\pi/2$. The agreement with the theory is excellent. Note that the largest $\delta$ shown corresponds to $\tau_{\rm m} \sim 1$ so the condition for $n-$point mixing $t\gg t_{\rm mix}$ is no longer satisfied. The agreement with theory is still good due to the oscillatory nature of the correlation decay (see Figure~\ref{fig:class_cor}) fullfiling the factorisation assumption (\ref{eq:factor}) on average.
\begin{figure}[h]
\centerline{\includegraphics{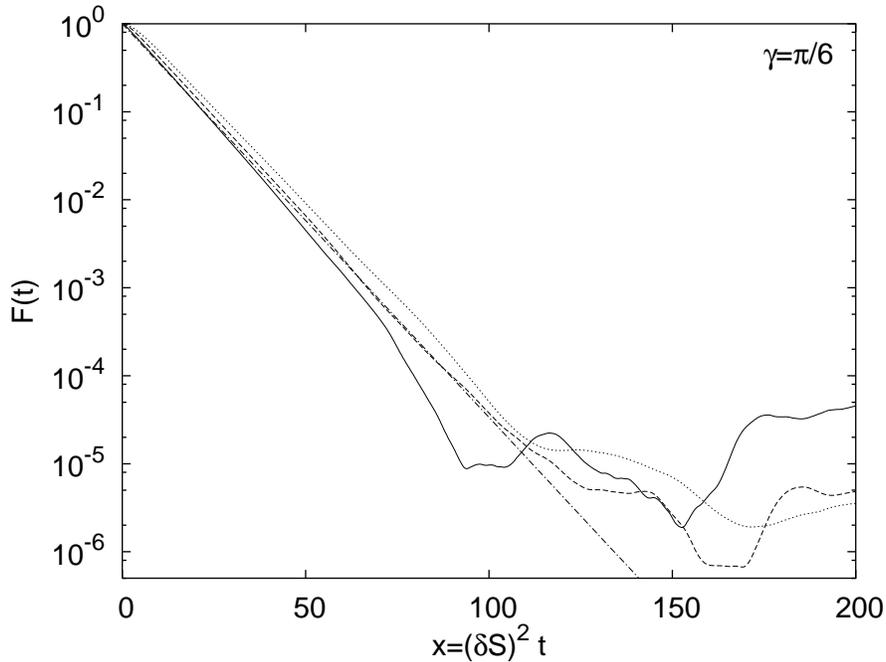}}
\caption{Similar figure as~\ref{fig:exp_pi2}, only for $\gamma=\pi/6$ and perturbation strengths $\delta=1\times 10^{-4}$, $2\times 10^{-4}$ and $3\times 10^{-4}$ (solid, dashed and dotted curves, respectively).}
\label{fig:exp_pi6}
\end{figure}
In Figure~\ref{fig:exp_pi6} for $\gamma=\pi/6$ a similar decay can be seen. In both cases fidelity starts to fluctuate around $\oFn$ calculated in the section~\ref{sec:time_averaged} for times larger than $t_\infty$.

\subsection{Long Time Behaviour}
\label{sec:long_time}
\begin{figure}[h]
\centerline{\includegraphics{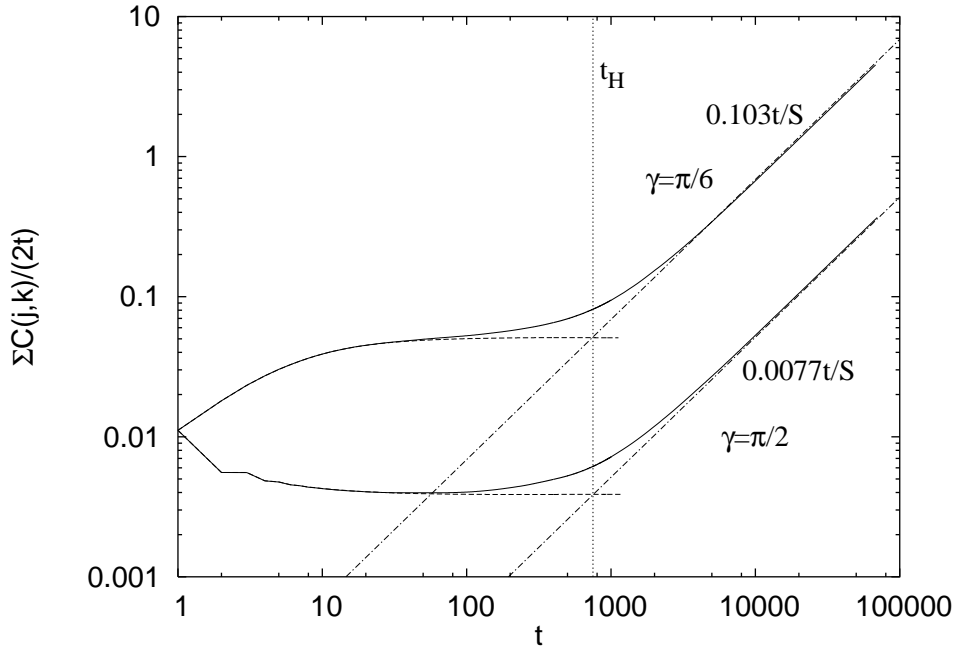}}
\caption{The finite time quantum correlation sum 
$\sigma(t)=\sum_{j,k=0}^{t-1} {C(j,k)/2t}$ (solid curves) 
together with the corresponding classical sum $\sigma_{\rm cl}(t) =
\sum_{j,k=0}^{t-1}{C_{\rm cl}(j,k)}/2t$ (dashed curves saturating at
$\sigma_{\rm cl}$ and ending at $t \sim 1000$) is shown for the chaotic kicked top. Quantum data are for a full trace $\rho=\mathbbm{1}/{\cal N}$ with $S=1500$. Upper curves are for $\gamma=\pi/6$ while lower curves are for $\gamma=\pi/2$. 
Chain lines are best fits for asymptotic linear functions corresponding to $\bar{C}t/2=0.0077t/S$ for $\gamma=\pi/2$ and $0.103 t/S$ for $\gamma=\pi/6$.}
\label{fig:cinf}
\end{figure}
We assumed that the quantum correlation function $C(j)$ decays to zero and its integral converges to $\sigma$. For a system having a finite Hilbert space of size ${\cal N}$, the correlation function asymptotically does not decay but has a non-vanishing plateau $\oC$ due to finite ${\cal N}$, similarly as we have a finite asymptotic value of the fidelity $\oFn$. This will cause the double correlation sum to grow quadratically with time. Because this plateau $\oC$ is small, the quadratic growth will overtake linear growth $2\sigma_{\rm cl} t$ only for large times. The time averaged correlation function $C(j,k)$ (\ref{eq:Cjk}) can be calculated assuming a nondegenerate unperturbed spectrum $\phi_k$ as
\begin{equation}
\oC=\lim_{t \to \infty}{\frac{1}{t^2}\sum_{j,k=0}^{t-1}{C(j,k)}}=\sum_k{\rho_{kk} (V_{kk})^2-\left( \sum_k {\rho_{kk} V_{kk} }\right)^2},
\label{eq:Cbar_def}
\end{equation}
where $\rho_{kk}$ are diagonal matrix elements of the initial density matrix $\rho(0)$ and $V_{kk}$ are diagonal matrix elements of the perturbation $V$ in the eigenbasis of the unperturbed propagator $U_0$. One can see that $\oC$ depends only on the diagonal matrix elements~\footnote{The case when diagonal matrix elements are zero is the subject of Chapter~\ref{ch:freeze}.}, in fact it is equal to the variance of the diagonal matrix elements. Since the classical system is ergodic and mixing, we will use a version of the {\em quantum chaos conjecture}~\citep{Feingold:86,Wilkinson:87,Feingold:89,Prosen:93,Prosen:94} saying that $V_{mn}$ are independent Gaussian random variables with 
a variance given by the Fourier transformation $S(\omega)$ (divided by ${\cal N}$) of the corresponding classical correlation 
function $C_{\rm cl}(j)$ at frequency $\omega=\phi_m-\phi_n$. On the diagonal we have $\omega=0$ and an additional factor of $2$
due to random matrix measure on the diagonal 
. Using $2 \sigma_{\rm cl} t=\sum_{j,k}^t{C(j,k)}=S(0) t$ we can write
\begin{equation}
\bar{C} = \frac{2 S(0)}{\cal N}=\frac{4\sigma_{\rm cl}}{\cal N}.
\label{eq:Cbar}
\end{equation}
Because of ergodicity, for large ${\cal N}$, $\oC$ does not depend on the statistical operator $\rho$ used in the definition of the correlation function, provided we do not take some non generic state like a single eigenstate $\ket{\phi_k}$ for instance. Note that Equation~(\ref{eq:Cbar}) is valid on a single quantum invariant subspace. If $U_0$ has symmetries, so that its Hilbert space is split into $s$ components of sizes ${\cal N}_j$, the average $\oC$ will be different on different subspaces, $\oC_j=4\sigma_{\rm cl}/{\cal N}_j$. Averaging over all invariant subspaces then gives 
\begin{equation}
\oC=\frac{4s \sigma_{\rm cl}}{{\cal N}},
\label{eq:Cbar_s}
\end{equation}
so that $\oC$ is increased by a factor $s$ compared to the situation with only a single subspace. The fidelity decay will start to be dominated by the average plateau (\ref{eq:Cbar})
at time $t_{\rm H}$ when the quadratic growth takes over, $\bar{C} t_{\rm H}^2 \approx 2\sigma_{\rm cl} t_{\rm H}$,
\begin{equation}
t_{\rm H}=\frac{1}{2}{\cal N} \propto \hbar^{-d},
\label{eq:np_def}
\end{equation}
which is nothing but the {\em Heisenberg time} associated with the inverse density of states. Again, if one has $s$ invariant subspaces, the Heisenberg time is $t_{\rm H}={\cal N}/2s$. This crossover time agrees with the result of~\citet{Cerruti:02} and for random matrix models with~\citet{Gorin:04}.
\par
In Figure~\ref{fig:cinf} we show numerical calculation of the correlation sum for the chaotic kicked top at $\alpha=30$. We compare the classical correlation sum (the same data as in Figure~\ref{fig:class_cor}) and quantum correlation sum. One can nicely see the crossover from linear growth of quantum correlation sum $2 \sigma_{\rm cl} t$ for small times $t<t_{\rm H}$, to the asymptotic quadratic growth due to correlation plateau $\oC$. In addition, numerically fitted asymptotic growth $0.103 t/S$ and $0.0077t/S$ nicely agree with formula for $\oC$, using ${\cal N}=S$ and classical values of transport coefficients $\sigma_{\rm cl}=0.0515$ and $0.00385$ for $\gamma=\pi/6$ and $\gamma=\pi/2$, respectively.
\par
\begin{figure}[h]
\centerline{\includegraphics[height=80mm,angle=-90]{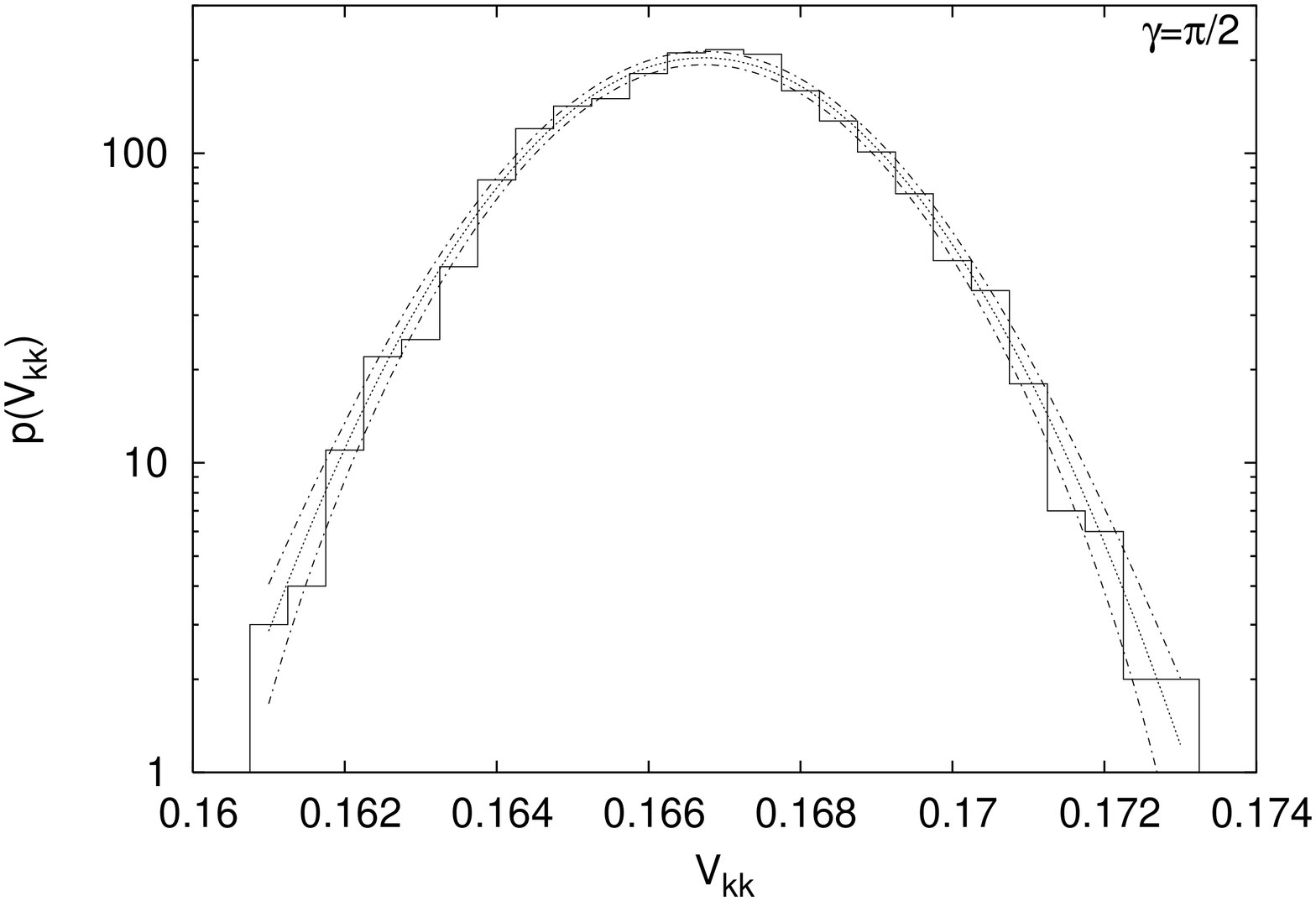}\includegraphics[height=80mm,angle=-90]{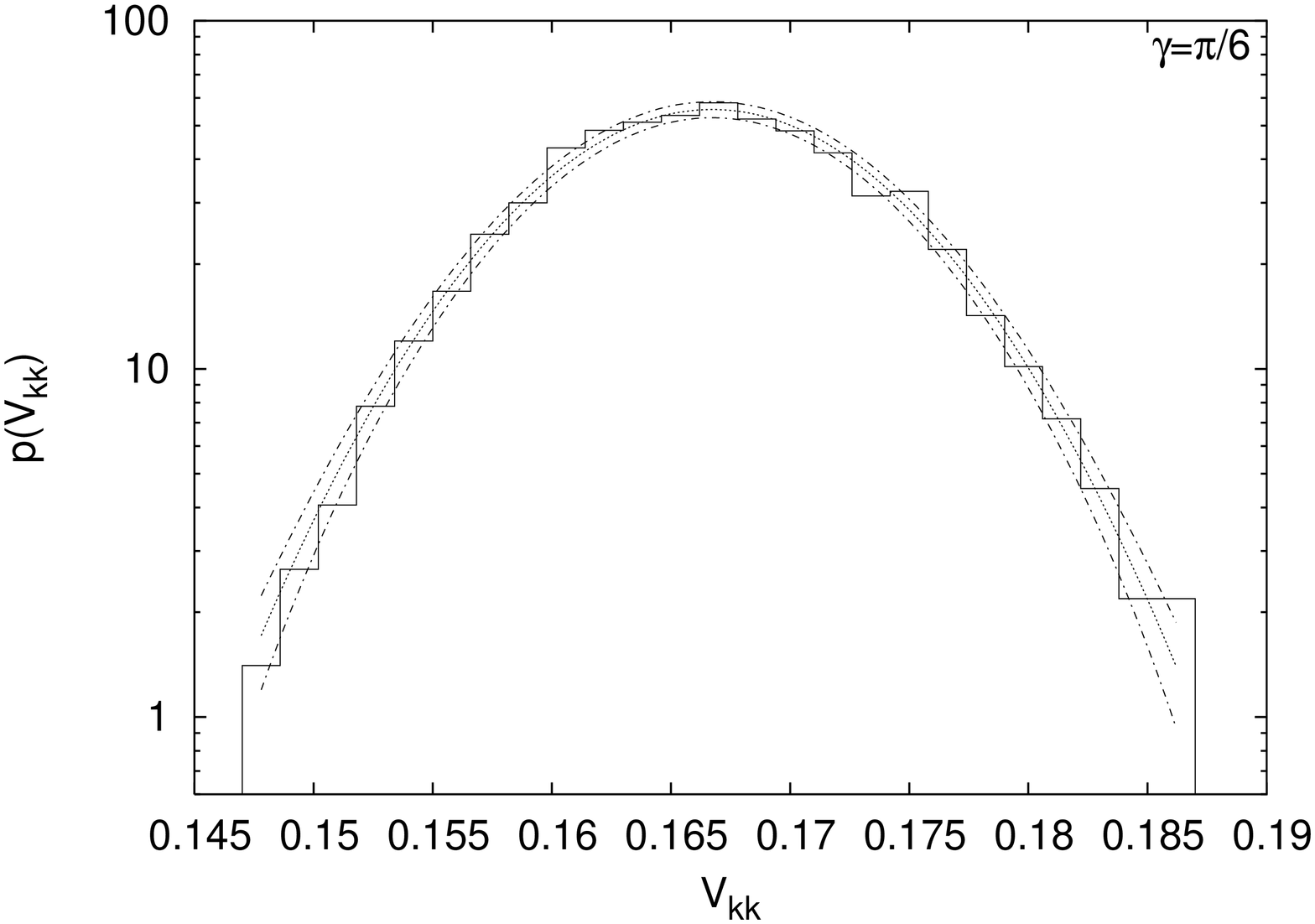}}
\caption{Histogram of the normalised distribution of the diagonal
  matrix elements $V_{kk}$ for the chaotic kicked top and $S=4000$ on
  OE subspace (\ref{eq:KT_subspaces}). The dotted line is the
  theoretical Gaussian distribution with the second moment $\oC$ and
  the two chain lines are expected $\sqrt{N_i}$ statistical deviations
  if there are $N_i$ elements in the $i-$th bin. Note the different
  $x$-ranges in two figures due to different $\sigma_{\rm cl}$ for the
  two chosen $\gamma$.}
\label{fig:Vkk}
\end{figure}
For times $t>t_{\rm H}$, and provided $\delta$ is sufficiently small, the correlation sum will grow quadratically and the linear response fidelity reads
\begin{equation}
\Fn=1-\frac{\delta^2}{\hbar^2} \frac{4\sigma_{\rm cl}}{\cal N} t^2.
\end{equation}
To derive the decay of fidelity beyond the linear response regime one
needs higher order moments of diagonal elements of perturbation
$V$. If we use the BCH form of the echo operator (\ref{eq:Md_exp}) and
discard the term involving~\footnote{in Chapter~\ref{ch:freeze} we
  will see that the size of $\Gamma(t)$ term grows at most linearly
  with time and so can be neglected because of $\delta^2$ prefactor.}
$\Gamma(t)$, we have the fidelity amplitude $\fn=\sum_k{\exp{(-\ii
    V_{kk} \delta t/\hbar)}}/{\cal N}$, where we choose an ergodic
average $\rho=\mathbbm{1}/{\cal N}$. In the limit ${\cal N}\to \infty$
we can replace the sum with an integral over the probability distribution of diagonal matrix elements $p(V_{kk})=p(V)$,
\begin{equation}
\fn=\int{\!{\rm d}V\,p(V)\exp{(-\ii V \delta t /\hbar)}}.
\label{eq:fn_fourier}
\end{equation}
For long times the fidelity amplitude is therefore a Fourier transformation of the distribution of diagonal matrix elements. For classically mixing systems the distribution is conjectured to be Gaussian with the second moment equal to $\oC$ (\ref{eq:Cbar}). This is confirmed by numerical data in Figure~\ref{fig:Vkk}. The mean value of diagonal matrix elements is perturbation specific and is for our choice of the perturbation (\ref{eq:KT_V}) $\sum_k{V_{kk}}/(2S+1)=(2S+1)(S+1)/12S^2$. From the figure we can see that the distribution is indeed Gaussian with the variance agreeing with the theoretically predicted $\oC=4\sigma_{\rm cl}/S$. The Fourier transformation of a Gaussian is readily calculated and we get a Gaussian fidelity decay
\begin{equation}
\Fn=\exp{\left(-(t/\tau_{\rm p})^2 \right)},\qquad \tau_{\rm p}=\sqrt{\frac{{\cal N}}{4 \sigma_{\rm cl}}} \frac{\hbar}{\delta}.
\label{eq:gaussian_mixing}
\end{equation}
In order to see a Gaussian fidelity decay for mixing systems the perturbation strength $\delta$ must be sufficiently small. If it is not, the fidelity will decay exponentially (\ref{eq:Fn_mixing}) to its fluctuating plateau $\oFn$ before time $t_{\rm H}$ when the Gaussian decay starts. Demanding that the mixing decay time $\tau_{\rm m}$ is smaller than $t_{\rm H}={\cal N}/2$ gives the critical perturbation strength $\delta_{\rm p}$,
\begin{equation}
\delta_{\rm p}=\frac{\hbar}{\sqrt{\sigma_{\rm cl}{\cal N}}}.
\label{eq:deltap}
\end{equation}
For $\delta<\delta_{\rm p}$ we will have a Gaussian decay (\ref{eq:gaussian_mixing}) otherwise the decay will be exponential (\ref{eq:Fn_mixing}), for details see Section~\ref{sec:time_scales}.
\par
\begin{figure}[ht!]
\centerline{\includegraphics{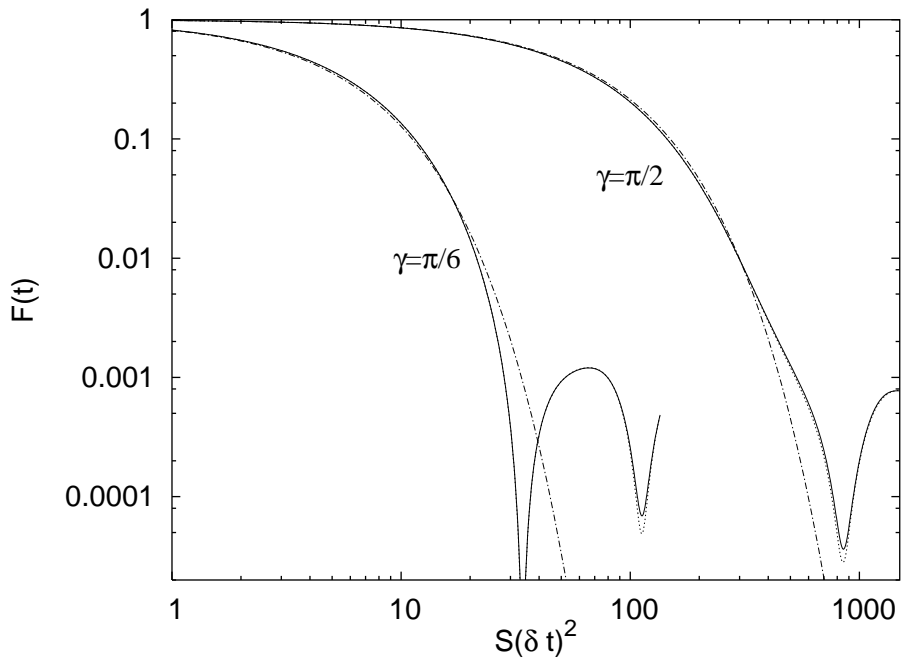}}
\caption{Quantum fidelity decay for $\delta<\delta_{\rm p}$ in the chaotic regime. For $\gamma=\pi/2$ data for $\delta=1\cdot 10^{-6}$ (solid curve) and $5\cdot 10^{-6}$ 
(dotted curve) are shown. For $\gamma=\pi/6$, $\delta=3\cdot 10^{-7}$ (solid) and $1\cdot 10^{-6}$ (dotted) are shown. 
Note that for both $\gamma$ the curves for both $\delta$ practically overlap. The chain curves are theoretical 
predictions (\ref{eq:gaussian_mixing}) with classically computed $\sigma_{\rm cl}$.}
\label{fig:j15kpert}
\end{figure}
Again we numerically checked the predicted Gaussian decay for the chaotic kicked top with $\alpha=30$ and $S=1500$ and a full trace average over Hilbert space. The results of numerical simulation, together with the theory are shown in Figure~\ref{fig:j15kpert}.
\par
The regime of Gaussian decay is sometimes referred to as the perturbative regime~\citep{Cerruti:02,Jacquod:01} because it can be derived using the lowest order perturbation theory. Writing first order corrections in phases $\phi_k^\delta=\phi_k+V_{kk} \delta /\hbar$ and for overlap matrix $P_{kl}=\delta_{kl}+{\cal O}(\delta)$ (\ref{eq:fn_P}) one gets the Fourier transform formula (\ref{eq:fn_fourier}). 

 \section{Regular Dynamics}
\label{sec:regular}
As opposed to mixing dynamics which was characterised by a linear growth of the double sum of the correlation function, the regular regime will typically exhibit  {\em quadratic} growth. By a regular regime we again refer to the behaviour of the correlation function. Typically, its double sum will exhibit quadratic growth for integrable or mixed (KAM) systems. In such case, we can define a time average correlation function, denoted by $\oC$
\begin{equation}
\oC=\lim_{t\to \infty}\frac{1}{t^2}\sum_{j,k=0}^{t-1}C(j,k).
\label{eq:Cbar_reg}
\end{equation}
For nonzero $\oC$ we will call this a ``regular regime''. Of course,
we have ${\cal N}$-dependent nonzero value of $\oC$ also in mixing systems (\ref{eq:Cbar}), as discussed in Section~\ref{sec:long_time}. Here we consider only systems where  $\oC$ exists in the limit of ${\cal N} \to \infty$, i.e. nonzero $\oC$ is a consequence of the dynamics and not of the finite Hilbert space size. Due to non-ergodicity all expectation values (like $\oC$) depend on the initial state, and we can not use ergodic average over the whole Hilbert space as for mixing dynamics. The time in which time averaging of the correlation function converges will be denoted by $t_{\rm ave}$. For times $t \gg t_{\rm ave}$ the linear response fidelity decays quadratically in time
\begin{equation}
\Fn=1-\frac{\delta^2}{\hbar^2}\oC t^2.
\label{eq:reg_lr}
\end{equation}
In contrast to the mixing regime one does not need any further assumptions in order to go beyond the linear response formula. Namely, for times $t \gg t_{\rm ave}$ we can define a time average perturbation $\oV$
\begin{equation}
 \oV=\lim_{t\to \infty}{\frac{1}{t}\sum_{k=0}^{t-1}V(k)}=\lim_{t\to\infty}{\frac{\Sigma(t)}{t}}.
\label{eq:Vbar_def}
\end{equation}
Observe that $\oV$ is by construction a constant of motion, $[\oV,U_0]=0$. In a mixing regime $\oV$ is trivial (in the limit ${\cal N}\to\infty$), i.e. proportional to the identity~\footnote{Actually it can be made zero by subtracting from $V$ the identity operator which only rotates a phase of the fidelity amplitude and does not affect the fidelity itself.}, whereas for regular dynamics nontrivial $\oV$ exists. The special case of $\oV=0$ will be considered in Chapter~\ref{ch:freeze}. In the case of nondegenerate spectrum of the unperturbed propagator $U_0$, the time average is simply the diagonal part of $V$ in the eigenbasis $\ket{\phi_k}$ of the unperturbed propagator, namely
\begin{equation}
\oV=\sum_k{V_{kk} \ket{\phi_k}\bra{\phi_k}},
\label{eq:Vbar_matrix}
\end{equation}
where as usual $V_{kk}=\bracket{\phi_k}{V}{\phi_k}$. Note that the average correlation function is
\begin{equation}
\oC=\< \oV^2 \>-\< \oV\>^2.
\label{eq:oC}
\end{equation}
For times $t \gg t_{\rm ave}$ the operator $\Sigma(t)$ is dominated by the linearly growing term $\oV t$ and one can neglect contributions not growing with time. The BCH form of the echo operator can therefore be simply written as
\begin{equation}
\Md=\exp{(-\ii \oV \delta t/\hbar )},\qquad t \gg t_{\rm ave},
\label{eq:Md_Vbar}
\end{equation}
where we neglected the term $\sim \Gamma(t)\delta^2$ as it becomes important only at large times $\sim 1/\delta^2$, whereas the first term will cause the fidelity to decay already in time $\sim 1/\delta$. The above form of the echo operator will be the main ingredient of theoretical calculation of the fidelity decay in the regular regime.
\par
As the operator $\oV$ commutes with the unperturbed propagator it is diagonal in the eigenbasis $\ket{\phi_k}$. In integrable systems the basis states $\ket{\phi_k}$ can be ordered in a very special way. Namely, there exist quantum numbers which are eigenvalues of canonical action operators having a very simple algebra. Using action-angle operators will make derivations easier and will furthermore enable us to use classical action-angle variables in the leading semiclassical order, thereby approximating quantum fidelity in classical terms. So before proceeding with the evaluation of $\fn$ for various initial states, let us have a look at action-angle operators. 

\subsection{Action-angle Operators}
\label{sec:actions}
Since we assume the classical system to be completely integrable (at least locally, by KAM theorem, in the phase space part of interest) we
can employ action-angle variables, $\{j_k,\theta_k,k=1\ldots d\}$, in $d$ degrees of freedom system. In the present section, dealing with the regular regime as well as in Section~\ref{sec:freeze_regular} describing quantum freeze of fidelity in regular systems, we shall use lowercase letters to denote {\em classical variables} and capital letters to denote the corresponding {\em quantum operators}. For instance, the quantum Hamiltonian will be given as $H(\vec{J},\vec{\Theta})$ whereas its classical limit will be written as $h(\vec{j},\vec{\theta})$.
\par
As our unperturbed Hamiltonian is integrable, it is a function of
actions only, i.e. $h_0=h_0(\vec{j})$. The solution of classical equations of motion is very simple,
\begin{eqnarray}
\vec{j}(t) &=& \vec{j}, \nonumber\\ 
\vec{\theta}(t) &=& \vec{\theta} + \vec{\omega}(\vec{j})t \pmod{2\pi}
\label{eq:class_eq}
\end{eqnarray}
with a dimensionless frequency vector
\begin{equation}
\vec{\omega}(\vec{j}) := \frac{\partial h_0(\vec{j})}{\partial\vec{j}}.
\end{equation}
The classical limit $v(\vec{j},\theta)$ of our perturbation generator $V$ can be written as a Fourier series
\begin{equation}
v(\vec{j},\vec{\theta}) = \sum_{\vec{m}\in\Z^d} v_{\vec{m}}(\vec{j}) 
e^{\ii\vec{m}\cdot\vec{\theta}},
\label{eq:fourier}
\end{equation}
where a multi-index $\vec{m}$ has $d$ components. The classical limit of the time-averaged perturbation $\bar{V}$ is $\bar{v} = v_{\vec{0}}(\vec{j})$, i.e. just the zeroth Fourier mode of the perturbation.
\par
In quantum mechanics, one quantises the action-angle variables using the famous
EBK procedure (see e.g.~\citep{Berry:77}) where one defines the action (momentum) operators $\vec{J}$
and angle operators $\exp(\ii\vec{m}\cdot\vec{\Theta})$ satisfying the canonical
commutation relations,
\begin{equation}
[J_k,\exp(\ii\vec{m}\cdot\vec{\Theta})] = \hbar m_k \exp(\ii\vec{m}\cdot\vec{\Theta}), \qquad k = 1,\ldots,d.
\label{eq:com}
\end{equation}
As the action operators are mutually commuting they have a common eigenbasis 
$\ket{\vec{n}}$ labelled by $d$-tuple of quantum numbers $\vec{n}=(n_1,\ldots,n_d)$,
\begin{equation}
\vec{J}\ket{\vec{n}} = \hbar (\vec{n} + \vec{\alpha})\ket{\vec{n}}
\label{eq:J_eigen}
\end{equation}
where $0 \le \alpha_k \le 1$ are the Maslov indices which are irrelevant for the leading order semiclassical approximation we will use. It follows from (\ref{eq:com}) that the angle operators act as shifts
\begin{equation}
\exp(\ii \vec{m}\cdot\vec{\Theta})\ket{\vec{n}} = \ket{\vec{n}+\vec{m}}.
\label{eq:shift}
\end{equation}
The Heisenberg equations of motion can be trivially solved in the leading semiclassical order by simply disregarding the operator ordering,
\begin{eqnarray}
\vec{J}(t) &=& {\rm e}^{\ii H_0  t/\hbar}\vec{J} {\rm e}^{-\ii H_0  t/\hbar} = \vec{J},\nonumber\\
{\rm e}^{\ii\vec{m}\cdot\vec{\Theta}(t)} 
&=& {\rm e}^{\ii H_0  t/\hbar}{\rm e}^{\ii\vec{m}\cdot\vec{\Theta}} {\rm e}^{-\ii H_0  t/\hbar} \cong 
{\rm e}^{\ii\vec{m}\cdot\vec{\omega}(\vec{J})t} {\rm e}^{\ii\vec{m}\cdot\vec{\Theta}},
\end{eqnarray}
in terms of the frequency operator $\vec{\omega}(\vec{J})$. Throughout this paper we use the symbol $\cong$ for 
'semiclassically equal', i.e. asymptotically equal in the leading order in $\hbar$.
Similarly, time evolution of the perturbation observable is obtained in the 
leading order by substitution of classical with quantal action-angle 
variables in the expression (\ref{eq:fourier})
\begin{equation}
V(t) = {\rm e}^{\ii H_0  t/\hbar}V {\rm e}^{-\ii H_0  t/\hbar} \cong
\sum_{\vec{m}} v_{\vec{m}}(\vec{J}) 
{\rm e}^{\ii\vec{m}\cdot\vec{\omega}(\vec{J})t}{\rm e}^{\ii\vec{m}\cdot\vec{\Theta}}.
\label{eq:Vt_quant}
\end{equation}
\par
Operator $\oV(\vec{J})$ is diagonal in the eigenbasis $\ket{\vec{n}}$ and therefore the expectation value of the echo operator (\ref{eq:Md_Vbar}) in the initial density matrix $\rho$ is
\begin{equation}
\fn=\sum_{\vec{n}}{\exp{\left(-\ii \delta t \oV(\hbar\{\vec{n}+\vec{\alpha}\})/\hbar \right)} D_\rho(\hbar \vec{n})},\qquad D_\rho(\hbar \vec{n})=\bracket{\vec{n}}{\rho}{\vec{n}}.
\label{eq:LDOS}
\end{equation}
For pure initial states $D_\rho$ is just $D_\rho=|\braket{\psi}{\vec{n}}|^2$.
This is still the exact quantum mechanical expression of the fidelity amplitude. Now we make a leading order semiclassical approximation by replacing quantum $\oV$ with its classical limit $\bar{v}$ and replacing the sum over quantum numbers $\vec{n}$ with the integral over classical actions $\vec{j}$. The replacement of the sum with the action space integral (ASI) is valid up to classically long times $t_{\rm a}$, such that the variation of the argument in the exponential across one Planck cell is small,
\begin{equation}
t_{\rm a}=\frac{1}{|\partial_{\vec{j}}\bar{v}|\delta}\sim \hbar^0/\delta.
\label{eq:regular_t*}
\end{equation}
Subsequently we will see that the fidelity decays on shorter times and so the approximation is justifiable. By denoting with $d_\rho(\vec{j})$ the classical limit of $D_\rho(\hbar \vec{n})$ we arrive at the fidelity amplitude
\begin{equation}
\fn \cong \hbar^{-d}\int{\!{\rm d}^d \vec{j} \exp{\left(-\ii \frac{\delta}{\hbar} t \bar{v}(\vec{j})\right)} \,d_\rho(\vec{j})}.
\label{eq:fn_ASI}
\end{equation} 
This expression will be used to evaluate the fidelity for different initial states.

   \subsection{Coherent Initial States}
\label{sec:CIS_gen}
We proceed to evaluate fidelity amplitude (\ref{eq:fn_ASI}) for coherent initial state centred at $(\vec{j}^*,\vec{\theta}^*)$ in action-angle phase space. The expansion coefficients of the initial density matrix $\rho=\ket{\vec{j}^*,\vec{\theta}^*}\bra{\vec{j}^*,\vec{\theta}^*}$ for a general coherent state in $d$ degrees of freedom system can be written as
\begin{equation}
\!\!\!\!\!\!\!\!\!\!\!\!\!\!\!\!\!\!
\braket{\vec{n}}{{\vec{j}^*,\vec{\theta}^*}} = 
\left(\frac{\hbar}{\pi}\right)^{d/4}
\!\!\!\left|\det\Lambda\right|^{1/4}
\exp\left\{-\frac{1}{2\hbar}(\vec{J}_{\vec{n}} - \vec{j}^*)\cdot\Lambda(\vec{J}_{\vec{n}}-\vec{j}^*) - 
\ii\vec{n}\cdot\vec{\theta}^*\right\},
\label{eq:coh}
\end{equation}
where $\Lambda$ is a positive symmetric $d\times d$ matrix of squeezing 
parameters and $\vec{J}_{\vec{n}}=\hbar(\vec{n}+\vec{\alpha})$ is an eigenvalue of operator $\vec{J}$ in eigenstate $\ket{\vec{n}}$. The classical limit of $D_\rho$ (\ref{eq:LDOS}) is therefore
\begin{equation}
d_\rho(\vec{j}) = (\hbar/\pi)^{d/2}\left|\det\Lambda\right|^{1/2}
\exp(-(\vec{j}-\vec{j}^*)\cdot\Lambda (\vec{j}-\vec{j}^*)/\hbar).
\label{eq:drho_coh}
\end{equation}
The ASI for the fidelity amplitude can now be evaluated by the stationary phase method. To lowest order in $\delta t$ the stationary point $\vec{j}_{\rm s}$, i.e. the zero of $[ -\ii \delta t \bar{v}-(\vec{j}-\vec{j}^*) \cdot \Lambda(\vec{j}-\vec{j}^*)]$, is at
\begin{equation}
\vec{j}_{\rm s} = \vec{j}^* + \frac{\ii t\delta}{2}\Lambda^{-1}\vec{\bar{v}}' + {\cal O}(\delta^{2}),
\end{equation}
where the derivative of the average perturbation is
\begin{equation}
\vec{\bar{v}}' = \frac{\partial\bar{v}(\vec{j}^*)}{\partial\vec{j}}. 
\label{eq:vbar_c}
\end{equation}
For $\delta t \ll 1$ the stationary point is simply at $\vec{j}^*$ and the fidelity is
\begin{equation}
\Fn=\exp{\left\{ -(t/\tau_{\rm r})^2\right\}},\qquad  \tau_{\rm r}=\frac{1}{\delta}\sqrt{\frac{2\hbar}{\vec{\bar{v}}'\cdot\Lambda^{-1}\vec{\bar{v}}'}}.
\label{eq:Fn_regcoh}
\end{equation}
The fidelity amplitude on the other hand has an additional phase,
\begin{equation}
\fn=\exp{\left\{ -(t/\tau_{\rm r})^2/2\right\}} \exp{\left(-\ii \bar{v}(\vec{j}^*) \delta t/\hbar\right)}.
\label{eq:fn_phases}
\end{equation}
Comparing the fidelity (\ref{eq:Fn_regcoh}) with the linear response formula (\ref{eq:reg_lr}) we see that the average correlation function for a coherent initial state is $\bar{C}=\frac{1}{2}\hbar(\vec{\bar{v}}'\cdot\Lambda^{-1}\vec{\bar{v}}')$.
\par
We have already briefly explained the behaviour of classical
fidelity in Section~\ref{sec:class_fid}, for details
see~\citep{Benenti:03}. For regular dynamics one gets
a ballistic decay of classical fidelity for perturbations predominantly changing
frequencies of tori or power-law decay for perturbations
predominantly changing the shape of tori. In the ballistic case,
provided one is allowed to replace the perturbation $v$ with its time average
$\bar{v}(\vec{j})$, one gets the same Gaussian decay of the classical
fidelity as for the quantum one (\ref{eq:Fn_regcoh}). Therefore, in the regime where $\bar{v}$ determines the decay
of classical fidelity, classical and quantum fidelity agree. From the
expression for the echo operator (\ref{eq:Md_Vbar}) one can in this
case interpret
the fidelity as the overlap between the initial coherent state and the
state obtained after the evolution with the Hamiltonian
$\delta\bar{v}$. Because $\bar{v}$ is a 
function of actions only, the classical equations of motion are very simple,
namely only the frequency of classical motion is changed by the amount
$\Delta \vec{\omega}=\vec{\bar{v}}'\,\delta$. This change in frequency
causes the ``echo'' packet $\Md \ket{\psi(0)}$ to move {\em
  ballistically} away from its initial position and as a consequence
fidelity decays. For coherent state the shape of the packet is
Gaussian and therefore fidelity decay is also Gaussian. For other
forms of localized initial packets the functional form of the fidelity
will be different but with the same dependence on the ballistic
separation ``speed'' $\vec{\bar{v}}'$. 
If $\bar{V}\equiv 0$ we obviously have
$\vec{\bar{v}}'=0$, this situation will be discussed in Chapter~\ref{ch:freeze} describing a freeze of fidelity.
\par
In one dimensional systems ($d=1$) another phenomena should be
observable. After long time the echo packet will make a whole
revolution around the torus causing the fidelity to be large
again. This will happen after the so-called beating time $t_{\rm b}$ determined by $\delta \bar{v}'(j^*) t_{\rm b}=2\pi$,
\begin{equation}
t_{\rm b}=\frac{2\pi}{\bar{v}' \delta}.
\label{eq:tb}
\end{equation}
This beating phenomena is particular to one dimensional systems as in general the incomensurability of frequencies will suppress the revivals of fidelity. 
\par
\begin{figure}[ht]
\centerline{\includegraphics{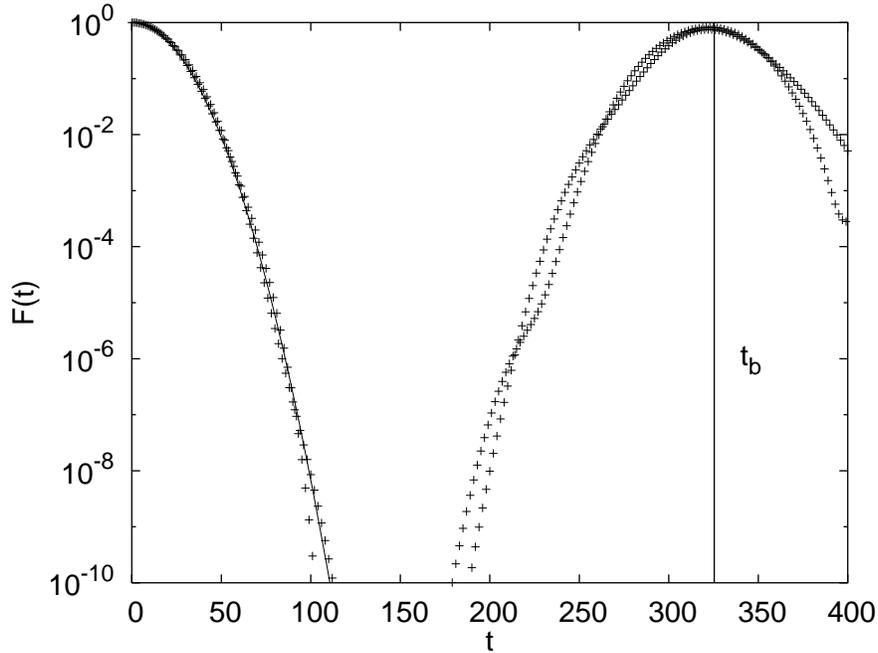}}
\caption{Quantum fidelity for a coherent state of the regular kicked top,
  $\alpha=0.1$, $S=100$, $\gamma=\pi/2$ and $\delta=0.0025$. The full
  curve is the theoretical decay (\ref{eq:reg_cohtheo}) and the
  vertical line is indicates the theoretical beating time (\ref{eq:tb}).}
\label{fig:regcoh}
\end{figure}
Let us now illustrate the above theory with numerical
experiment. We again take the kicked top (\ref{eq:KT_def}), but this
time with $\gamma=\pi/2$ and $\alpha=0.1$ resulting in regular classical dynamics. The perturbation is the same as for mixing situation, i.e. a perturbation in $\alpha$ resulting in a classical perturbation $v=z^2/2$ (\ref{eq:KT_V}). Let us denote by $\tilde{\vartheta},\tilde{\varphi}$ the spherical angular coordinates measured with respect to the $y$-axis. For $\alpha=0$ the action-angle variables are $(j=\cos\tilde{\vartheta}=y,\theta=\tilde{\varphi})$ and we can use the preceding theory. The squeezing parameter for spin coherent states is (\ref{eq:SU2_coh})
\begin{equation}
\Lambda=1/\sin^2{\tilde{\vartheta}}=1/(1-y^2).
\label{eq:l_su2}
\end{equation}
To calculate the decay time $\tau_{\rm r}$ we need the derivative of the average perturbation. For $\alpha=0$ the evolution is very simple, just a rotation by $\pi/2$ around the $y$-axis, resulting in
\begin{equation}
\bar{v}=\frac{1}{4}(1-j^2),\qquad \bar{v}'=-\frac{j}{2}.
\end{equation}
Theoretical decay of fidelity is therefore 
\begin{equation}
\Fn=\exp{\left(-\delta^2 S t^2 y^2(1-y^2)/8 \right)},
\label{eq:reg_cohtheo}
\end{equation}
where we used $\hbar=1/S$. In Figure~\ref{fig:regcoh} we show the results of numerical calculation for a coherent packet placed at spherical coordinates (with respect to $z$-axis) $(\vartheta^*,\varphi^*)=\pi(1/\sqrt{3},1/\sqrt{2})$ resulting in action $y=0.77$ (note that we projected the initial state onto the invariant OE subspace (\ref{eq:KT_subspaces})). Spin number is chosen $S=100$. Excellent agreement with the theoretical Gaussian decay can be seen. In addition, we can observe a revival of the fidelity at $t_{\rm b}$. For our OE subspace the initial state is composed of two symmetrically positioned images, so that the beating time is $t_{\rm b}=2\pi/y\delta$, which nicely agrees with the numerics.

\subsubsection{Average Fidelity}
\begin{figure}[ht!]
\centerline{\includegraphics{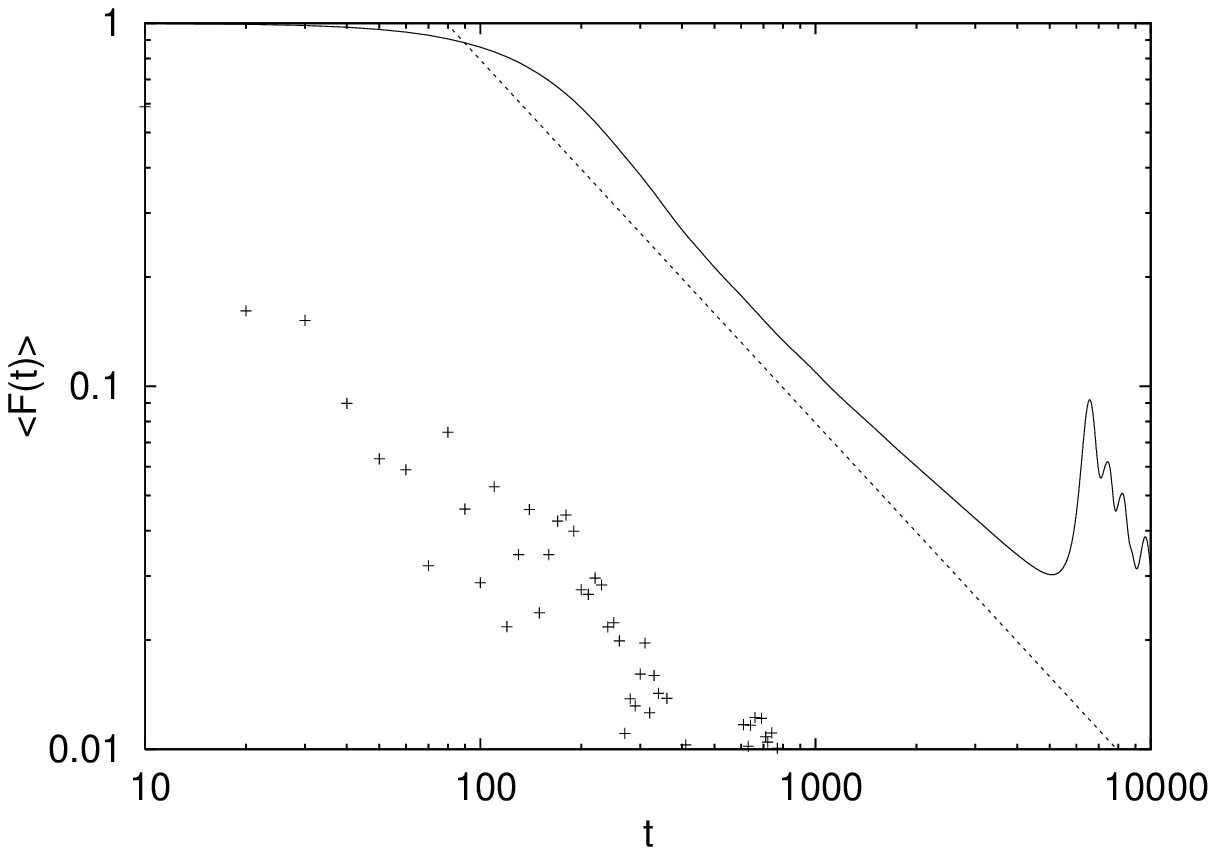}}
\caption{Average fidelity decay $\ave{F(t)}_\vec{j}$
  (\ref{eq:CIS_avg}) for a regular kicked top with $\alpha=0.1$,
  $\gamma=\pi/2$ and perturbation of strength $\delta=10^{-3}$ in the
  parameter $\alpha$. The averaging is done over $200$ coherent
  initial states randomly placed on a sphere and projected onto OE
  subspace (\ref{eq:KT_subspaces}). The Dotted line is the theoretical
  power law decay (\ref{eq:avgZ}). Pluses show the absolute value
  square of the average fidelity amplitude.}
\label{fig:avgZ}
\end{figure}
If $F(t,\vec{j})$ is Gaussian fidelity decay (\ref{eq:Fn_regcoh}) for a coherent initial state centred at $\vec{j}$, we can define the average fidelity
\begin{equation}
\ave{F(t)}_\vec{j}:=\frac{(2\pi)^d}{\cal V}\int{\!{\rm d}^d \vec{j} F(t,\vec{j})}.
\label{eq:CIS_avg}
\end{equation}
Packing all $\vec{j}$ dependent terms in a non-negative scalar function $g(\vec{j})$, the fidelity decay for a single coherent state can be written as $F(t,\vec{j}):=\exp{(-\delta^2 t^2 g(\vec{j})/\hbar)}$. For large $\delta^2 t^2/\hbar$ the main contribution to the average will come from regions of small $g(\vec{j})$ where the fidelity decay is slow. In general the function $g(\vec{j})$ will have zeros in action space, for simplicity let us assume there is a single zero at $\vec{j}^*$ of order $\eta$. The asymptotic decay can then be calculated and scales as
\begin{equation}
\ave{F(t)}_\vec{j} \asymp \left( \frac{\hbar}{\delta^2 t^2}\right)^{d/\eta},
\label{eq:F_asim}
\end{equation}
where the sign $\asymp$ will denote ``in the asymptotic limit'' troughout the
work. For infinite phase space the order of a zero $\eta$ can only be
an even number, whereas for a finite space $\eta$ can also be odd. 
\par
To illustrate the theory, we calculated the average decay for the kicked top and the perturbation in $\alpha$ used before. The function $g$ is $g(j)=j^2(1-j^2)/8$. We have three zeros, two of order $\eta=1$ at the boundary of the phase space $j=\pm 1$ and one of order $\eta=2$ at $j=0$. Asymptotically the zero with $\eta=2$ will dominate, giving the decay
\begin{equation}
\ave{F(t)}_{\vec{j}} \asymp \frac{\sqrt{2\pi}}{\delta t \sqrt{S}}.
\label{eq:avgZ}
\end{equation}
Figure~\ref{fig:avgZ} shows that we indeed get asymptotic $\sim
t^{-1}$ decay. If one averages the fidelity amplitude, the result is
close to zero (pluses in the figure) because the phase present in the
fidelity amplitude (\ref{eq:fn_phases}) averages to zero\footnote{One
  would have to have zeros of $\bar{v}$ and $\vec{\bar{v}}'$ at the
  same position $\vec{j}^*$ to get a nonzero average, which is
  generally not the case.}. 
\par
\begin{figure}[ht!]
\centerline{\includegraphics{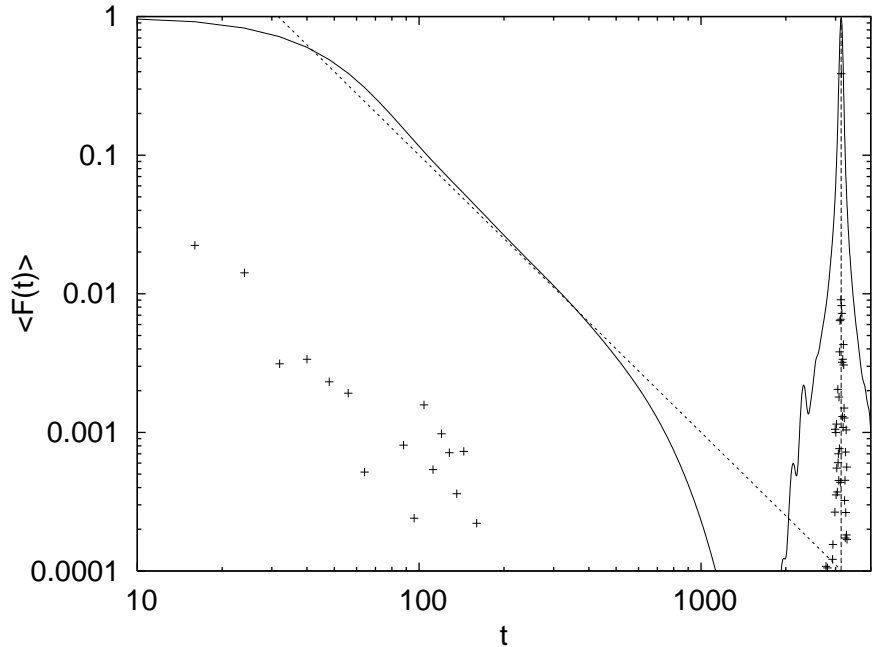}}
\caption{Average fidelity decay $\ave{F(t)}_\vec{j}$
  (\ref{eq:CIS_avg}) for a regular kicked top. Everything is the same
  as in Figure~\ref{fig:avgZ} apart from the perturbation being in the
  angle $\gamma$. Dotted line is the theoretical power law decay
  (\ref{eq:avgY}) without any fitting parameter. Note that we have different power law decay as in Figure~\ref{fig:avgZ}. The vertical dotted line shows the position of the beating time $t_{\rm b}=\pi/\delta$ (\ref{eq:tb}) which is in this case independent of $j$. If $t_{\rm b}$ depends on $j$ (as in Figure~\ref{fig:avgZ}) such recovery is absent in the average fidelity.}
\label{fig:avgY}
\end{figure}
As a second example, we take the same regular kicked top system and parameters, only the perturbation is now in rotation angle $\gamma$, i.e. $U_\delta:=U(\gamma+\delta,\alpha)$. The function $g$ is in this case $g(j)=(1-j^2)/2$. As opposed to the previous case, now we have only two zeros of order $\eta=1$ at $j=\pm 1$. The asymptotic decay is therefore
\begin{equation}
\ave{F(t)}_{\vec{j}} \asymp \frac{1}{\delta^2 t^2 S}.
\label{eq:avgY}
\end{equation}
Denominator is now quadratic in time, whereas for the perturbation
$S_{\rm z}^2$ it was linear. The theoretical decay law is again confirmed by the numerical experiment shown in Figure~\ref{fig:avgY}. The asymptotic power law decay of average fidelity is therefore {\em system dependent}.
\par
An entirely analogous method will be used in the next section to calculate the asymptotic decay for a random initial state. The only difference is that for {\em a random initial state} the zeros of $\bar{v}$ are the relevant quantity, whereas for an {\em average} fidelity decay of {\em a coherent state} the zeros of $\tau_{\rm r}$, i.e. zeros of $\vec{\bar{v}}'$ are important. Therefore, in a regular system the asymptotic decay for a random initial state is usually not the same as the average decay for a coherent state. 

  \subsection{Random Initial State}
\label{sec:RIS}
\begin{figure}[ht]
\centerline{\includegraphics{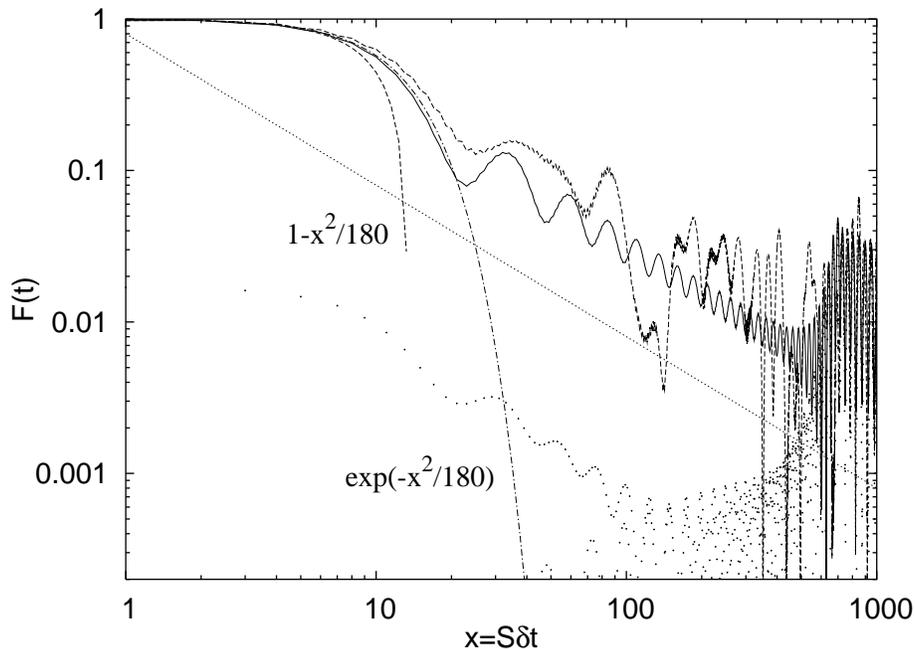}}
\caption{Fidelity decay for the regular kicked top, $\alpha=0.1$,
  $\gamma=\pi/2$, $\delta=0.01$, $S=100$. The solid curve gives the
  results of a numerical simulation for $\rho=\mathbbm{1}/{\cal N}$ on
  the OE subspace (\ref{eq:KT_subspaces}). Isolated dots denote the difference between numerical calculation and analytic formula (\ref{eq:Ferfi}) for $\alpha\to 0$, 
$|F_{\rm num.}(t)-F_{\rm anali.}(t)|$. The dotted line gives a predicted asymptotic decay $\propto t^{-1}$, and the dashed/chain curves are the 
predicted fidelity decays at small times, namely the second order expansion $F(t)=1-(St\delta)^2/180$, and 
'improved' one by an exponential. The wiggling dashed curve represents the numerics for a single random state.}
\label{fig:analit}
\end{figure}

In the previous section we have seen that for localized packets the
fidelity decay is dictated by the separation of two packets due to the
perturbation. The opposite possible choice of initial condition is to
take completely {\em delocalised} state, a uniform mixture of all
states $\rho=\mathbbm{1}/{\cal N}$ being the extreme case. For uniform
initial density matrix $\rho=\mathbbm{1}/{\cal N}$, the $D_\rho$ takes
the form $D_\rho(\vec{J}) = 1/{\cal N}$. The classical limit of
$D_\rho$ is obtained by calculating the number of states ${\cal N}$
using the Thomas-Fermi rule, resulting in the classical limit
$d_\rho(\vec{j}) = (2\pi\hbar)^d/{\cal V}$, where ${\cal V}$ is the
size of the classical phase space. While for small times fidelity will
decay quadratically (\ref{eq:reg_lr}), for large $\delta t/\hbar$ the integral in $\fn$ (\ref{eq:fn_ASI}) can again be calculated using the method of stationary phase. In contrast to a coherent state case, now one can have more than one stationary point. If we have $p$ stationary points, $\vec{j}_\eta, \eta=1,\ldots,p$, where the phase is stationary, $\partial\bar{v}(\vec{j}_\eta)/\partial\vec{j}=\vec{0}$, the integral results in
\begin{equation}
\fn = \frac{(2\pi)^{3d/2}}{\cal V}\left|\frac{\hbar}{t\delta}\right|^{d/2}\sum_{\eta=1}^p
\frac{\exp\{-\ii t\bar{v}(\vec{j}_\eta)\delta/\hbar - \ii \nu_\eta \}}{|\det \ma{\bar{V}}_\eta|^{1/2}},
\label{eq:reg_power}
\end{equation}
where $\{\ma{\bar{V}}_\eta\}_{kl} := \partial^2 \bar{v}(\vec{j}_\eta)/\partial j_k\partial j_l$ is a matrix of second
derivatives at the stationary point $\eta$, and $\nu_\eta = \pi(m_+ - m_-)/4$
where $m_{\pm}$ are the numbers of positive/negative eigenvalues of the matrix
$\ma{\bar{V}}_\eta$. We also assumed that the phase space is
infinite. In a finite phase space we will have diffractive oscillatory
corrections in the stationary phase formula, see
end of Section~\ref{sec:ran_freeze} or numerical results below. Note also
that for a large Hilbert space dimension ${\cal N}$ the fidelity for a {\em single initial random state} will also decay according to the asymptotic formula (\ref{eq:reg_power}). The decay time of the quantum fidelity for random initial states scales therefore as $\tau_{\rm r}\sim \hbar/\delta$ and is by a factor $\sqrt{\hbar}$ shorter than for coherent initial states. We repeat that the stationary phase formula (\ref{eq:reg_power}) is expected to be correct in the range 
${\rm const}\,\hbar/\delta < t < {\rm const}'/\delta$ so it will give
asymptotic decay of fidelity. Most interesting to note is the
power-law dependence on time and perturbation strength, $\Fn \asymp \left[\hbar/(t\delta)\right]^d$. With increasing dimensionality $d$ of a system the decay gets faster. This allows for a possible crossover to a 
Gaussian decay when approaching the thermodynamic limit $d\to\infty$, observed in a class of kicked spin chains by~\citet{Prosen:02}. Agreement beyond 
linear response with the Gaussian decay is frequently observed also
for a finite $d$, e.g. in a spin model of quantum
computation~\citep{Prosen:01} or even in a one dimensional kicked top seen in Figure~\ref{fig:analit}.
\par
For numerical demonstration we take the same kicked top system as for the case of coherent initial states, $\alpha=0.1$, $\gamma=\pi/2$ and $S=100$ and perturbation in parameter $\alpha$, giving $\bar{v}=(1-j^2)/4$. Rather than comparing the numerical results with the asymptotic formula (\ref{eq:reg_power}) we will try to directly calculate the fidelity decay in order to demonstrate oscillatory corrections due to a finite Hilbert space. We start with the quantum expression for the fidelity amplitude (\ref{eq:LDOS}), using the eigenvalues $(2m-1)^2$ of operator $S_{\rm y}^2$ on OE subspace,
\begin{equation}
\Fn=\left|\frac{2}{S} \sum_{m=1}^{S/2}{\exp{(-\ii \delta t (2m-1)^2/4S)}}\right|^2.
\label{eq:Fexact}
\end{equation}
For large $S$ we can replace the sum with an integral and get
\begin{equation}
\Fn=\frac{\pi}{\delta S t} \left|{\rm erfi}(\frac{1}{2} {\rm e}^{\ii \pi/4} \sqrt{\delta S t})\right|^2,
\label{eq:Ferfi}
\end{equation}
where ${\rm erfi}(z)=\frac{2}{\ii\sqrt{\pi}}\int_0^{\ii z}{{\rm e}^{-t^2} {\rm d}t}$ is a complex error function with the limit $\lim_{x \to \infty}{|{\rm erfi}(\frac{1}{2} {\rm e}^{\ii \pi/4} \sqrt{x})|}=1$ to which it approaches by oscillating 
around $1$. We therefore have an analytic expression for the fidelity (\ref{eq:Ferfi}) in the case of an uniform 
average over the whole Hilbert space or, equivalently, over one random initial state. Its asymptotic decay is proportional to $t^{-1}$ which 
agrees with the general semiclassical asymptotic (\ref{eq:reg_power})
but in addition we have an oscillatory ${\rm erfi}$ correction due to
a finite space. For small times we can get initial quadratic linear response decay by expanding the full theoretical formula (\ref{eq:Ferfi}) or more easily by simply calculating $\oC$ according to formula (\ref{eq:oC}). One can even use the classical calculation. For $\alpha=0$ we have $\bar{v}=(1-y^2)/4$, so the classical average correlation functions is
\begin{equation}
\bar{C}_{\rm cl}=\frac{1}{16}( \overline{y^4}-\overline{y^2}^2)=\frac{1}{180},
\end{equation}
where the averages are performed over a uniform distribution on a sphere. In Figure~\ref{fig:analit} we compare theoretical decay (\ref{eq:Ferfi}) with the numerical as well as with the linear response formula $\Fn=1-(\delta S)^2 \oC_{\rm cl} t^2$. Also, beyond the linear response there is a significant agreement with the Gaussian approximation $\Fn=\exp{(-\delta^2 S^2 \oC_{\rm cl} t^2)}$ obtained by exponentiating the linear response expression.

 \section{Time Scales}
\label{sec:time_scales}
Let us close the present chapter with an overview of time scales of different fidelity decays depending on parameters $\delta$ and $\hbar$ and on the initial state. 
\par
For regular dynamics we have only three relevant time scales, the
classical averaging time $t_{\rm ave}$ in which the average
perturbation operator converges to $\bar{V}$, the quantum fidelity
decay time $\tau_{\rm r}$ and the time $t_\infty$ when the fidelity
reaches a fluctuating plateau due to a finite dimension of Hilbert space. For times smaller than $t_{\rm ave}$ decay is system and state specific and can not be discussed in general. After $t_{\rm ave}$ the fidelity decay is first quadratic in time as dictated by the linear response formula (\ref{eq:reg_lr}). The decay time $\tau_{\rm r}$ scales as $\sim \sqrt{\hbar}/\delta$ (\ref{eq:Fn_regcoh}) for coherent initial states and as $\sim \hbar/\delta$ (\ref{eq:reg_power}) for random initial states. Beyond linear response the functional dependence of the decay is Gaussian for coherent initial states and power law for random initial states with the before mentioned decay time $\tau_{\rm r}$. This decay persists until the finite size plateau $\oFn$ (\ref{eq:Fn_avg}) is reached at time $t_\infty$. The time $t_\infty$ again depends on the initial state as well as on the Hilbert space dimension ${\cal N}$. For random initial states the power law decay gets faster with increasing dimensionality $d$ of the system, and is conjectured to approach a Gaussian decay in the thermodynamic limit.  
\par
\begin{figure}[ht]
\centerline{\includegraphics{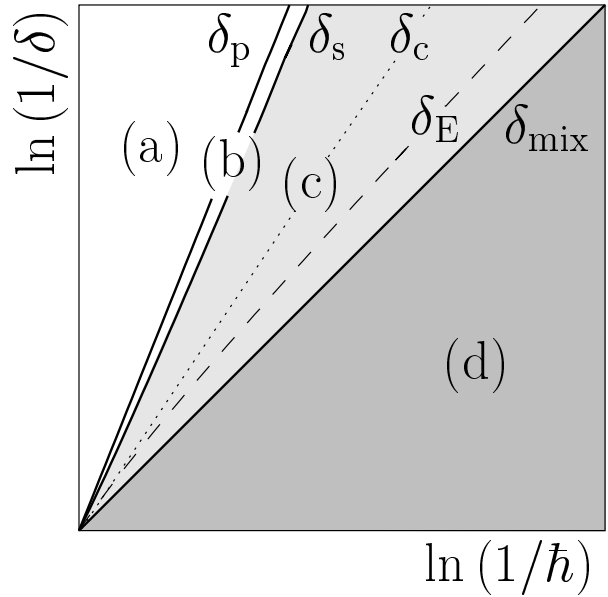}}
\caption{Schematic view of different fidelity decay regimes for mixing dynamics. For details see text.}
\label{fig:delta1}
\end{figure}

For mixing dynamics we have a more complicated situation. There are five relevant time scales (even six for coherent initial states): The classical mixing time $t_{\rm mix}$ on which correlation functions decay; the quantum decay time of the fidelity $\tau_{\rm m}$; the Heisenberg time $t_{\rm H}$ after which the system starts for ``feel'' finiteness of Hilbert space; the decay time $\tau_{\rm p}$ of perturbative Gaussian decay present after $t_{\rm H}$; the time $t_\infty$ when the fidelity reaches finite size plateau; for coherent initial states we have in addition the Ehrenfest time $t_{\rm E}$ up to which we have quantum-classical correspondence. Depending on the interrelation of these time scales, i.e. depending on the perturbation strength $\delta$, Planck's constant $\hbar$ and the dimensionality $d$, we will also have different decays of fidelity. All different regimes can be reached by fixing $\hbar$ and increasing $\delta$. Let us follow different decay regimes as we increase $\delta$ (shown in Figure~\ref{fig:delta1}):
\begin{enumerate}
\item[(a)] For $\delta<\delta_{\rm p}$ we will have $t_{\rm H} < \tau_{\rm m}$. This means that at the Heisenberg time, the fidelity due to exponential decay (\ref{eq:Fn_mixing}) will still be close to $1$, $F(t_{\rm H}) \approx 1$, and we will see only a Gaussian decay due to finite Hilbert space (\ref{eq:gaussian_mixing}). The critical $\delta_{\rm p}$ below which we will see this regime has already been calculated (\ref{eq:deltap}) and is
\begin{equation}
\delta_{\rm p}=\frac{\hbar}{\sqrt{\sigma_{\rm cl}{\cal N}}}=\hbar^{d/2+1}\,\frac{(2\pi)^{d/2}}{\sqrt{\sigma_{\rm cl}{\cal V}}}.
\label{eq:dp}
\end{equation}
For $\delta<\delta_{\rm p}$ the fidelity will have Gaussian decay with the decay time $\tau_{\rm p}$ (\ref{eq:gaussian_mixing})
\begin{equation}
\tau_{\rm p}=\frac{\hbar^{1-d/2}}{\delta}\sqrt{\frac{{\cal V}}{4\sigma_{\rm cl} (2\pi)^d}}.
\label{eq:taup}
\end{equation}
As we increase $\delta$, we will eventually reach a regime in which we will see initial exponential decrease and then at $t_{\rm H}$ crossover into a Gaussian decay until the plateau is reached at $t_\infty$.

\item[(b)] For $\delta_{\rm p}< \delta < \delta_{\rm s}$ we will have
  a crossover from the initial exponential decay (\ref{eq:Fn_mixing})
  to the asymptotic Gaussian decay (\ref{eq:gaussian_mixing}) at time $t_{\rm H}$. This regime will take place if $\tau_{\rm m} < t_{\rm H} < t_\infty$. With increasing perturbation, $t_\infty$ will decrease and the upper border $\delta_{\rm s}$ is determined by the condition $t_\infty=t_{\rm H}$. Denoting a finite size plateau by $\oFn\sim 1/{\cal N}^\mu$, with $\mu$ lying between $1$ and $2$, depending on the initial state (see Section~\ref{sec:time_averaged}), we have the condition $\exp{(-(t_{\rm H}/\tau_{\rm p})^2)}=\oFn$ which gives
\begin{equation}
\delta_{\rm s}=\frac{\hbar}{\sqrt{\sigma_{\rm cl}{\cal N}}} \sqrt{\mu \ln{{\cal N}}}=\delta_{\rm p} \sqrt{\mu \ln{{\cal N}}}.
\label{eq:ds}
\end{equation}
Further increasing the perturbation, we reach perhaps the most interesting regime, in which quantum fidelity can decay faster the more chaotic the systems is. In this regime the exponential decay persists until the plateau is reached.

\item[(c)] For $\delta_{\rm s} < \delta < \delta_{\rm mix}\, (\delta_{\rm E})$ we will have an exponential decay until $t_\infty$. The upper border $\delta_{\rm mix}$ is determined by the condition $\tau_{\rm m}=t_{\rm mix}$ which is a point where the argument leading to the factorisation of $n-$point correlation function breaks down. For random initial states $\delta_{\rm mix}$ does not depend on $\hbar$ and we get
\begin{equation}
\delta_{\rm mix}=\frac{\hbar}{\sqrt{2 \sigma_{\rm cl} t_{\rm mix}}}=\delta_{\rm p}\sqrt{\frac{{\cal N}}{2t_{\rm mix}}}.
\label{eq:dmix}
\end{equation}
Note that the relative size of this window $\delta_{\rm mix}/\delta_{\rm s}=\sqrt{{\cal N}/2\mu t_{\rm mix}\ln{{\cal N}}}$ increases both in the semiclassical $\hbar \to 0$ and in the thermodynamic $d \to \infty$ limit.
\par
For coherent initial states the quantum correlation function relaxes on a slightly longer time scale, namely on the Ehrenfest time $t_{\rm E}\sim -\ln{\hbar}/\lambda$. Until $t_{\rm E}$ quantum packet follows the classical trajectory and afterwards interferences start to build leading to the breakdown of quantum-classical correspondence. Equating $\tau_{\rm m}=t_{\rm E}$ gives the upper border for coherent states
\begin{equation}
\delta_{\rm E}=\frac{\hbar}{\sqrt{-\ln{\hbar}}}\, \sqrt{\frac{\lambda}{2\sigma_{\rm cl}}}= \delta_{\rm mix}\, \sqrt{\frac{\lambda t_{\rm mix}}{-\ln{\hbar}}}.
\label{eq:de}
\end{equation}

\item[(d)] For $\delta > \delta_{\rm mix}$ the perturbation is so strong that the quantum fidelity decays before $t_{\rm mix}$, i.e. perturbed and unperturbed dynamics are essentially unrelated and fidelity decays almost instantly.
\end{enumerate}
For coherent initial states the upper border of regime (c) is at $\delta_{\rm E}$ which is smaller than the lower border of regime (d) $\delta_{\rm mix}$ which opens up the possibility of another regime between (c) and (d), namely for $\delta_{\rm E}< \delta < \delta_{\rm mix}$ the fidelity will decay within the Ehrenfest time. In this regime the decay of quantum fidelity is the same as the decay of classical fidelity and can be explained in terms of classical Lyapunov exponents~\citep{Veble:04}. Note that the relative width of this regime $\delta_{\rm mix}/\delta_{\rm E}=\sqrt{\ln{(1/\hbar)}/\lambda t_{\rm mix}}$ grows only logarithmically in $1/\hbar$, i.e. much slower than the width of regime (c). 
\par
The fidelity decay time scales as $\sim 1/\delta$ for regular systems, while it is $\sim 1/\delta^2$ for mixing dynamics. This opens up an interesting possibility: is it possible that the fidelity would decay {\em faster} for regular systems than for chaotic ones? The answer is yes. Demanding $\tau_{\rm r}<\tau_{\rm m}$ we find that for sufficiently small $\delta$ one will indeed have faster fidelity decay in regular systems. This will happen for
\begin{equation}
\delta <
\left\{
\begin{array}{ll}
\delta_{\rm r}=\hbar\,\bar{C}^{1/2}/2\sigma_{\rm cl} \propto \hbar      & \hbox{random init.state} \\ 
\delta_{\rm c}=\hbar^{3/2}\sqrt{\bar{\vec{v}}'\cdot \Lambda^{-1} \bar{\vec{v}}'/8 \sigma^2_{\rm cl}}\propto \hbar^{3/2} & \hbox{coherent init.state}
\end{array}
\right. .
\label{eq:delta_con}
\end{equation}
We explicitly wrote the result for regular initial states $\delta_{\rm r}$ and coherent initial states $\delta_{\rm c}$ as the two have different scaling with $\hbar$. We can see that for random initial states $\delta_{\rm r}$ scales in the same way as $\delta_{\rm mix}$ and so one has faster decay of fidelity in regular systems just provided $\delta<\delta_{\rm r} \sim \delta_{\rm mix}$. For a coherent initial state this can be satisfied above the perturbative border $\delta > \delta_{\rm p}$ only in more than one dimension $d > 1$. In one dimensional systems $\delta_{\rm p}$ and $\delta_{\rm c}$ have the same scaling with $\hbar$ and whether we can observe faster decay of fidelity in regular systems than in chaotic ones depends on the values of $\sigma_{\rm cl}$ and $\bar{\vec{v}}'$. We stress that our result does not contradict any of the existing findings on quantum-classical correspondence. For example, a growth of quantum dynamical entropies~\citep{Alicki:96,Miller:99} persists only up to logarithmically short Ehrenfest time 
$t_{\rm E}$, which is also the upper bound for the so-called ``Lyapunov'' decay of the fidelity~\citep{Jalabert:01,Cucchietti:02} and within which one would always find $F^{\rm reg}(t) > F^{\rm mix}(t)$ (for coherent states) above the perturbative border $\delta > \delta_{\rm p}$, whereas our theory reveals new nontrivial 
quantum phenomena with a semiclassical prediction (but not correspondence!) much beyond that
time. If we let $\hbar\to 0$ first, and then $\delta \to 0$, i.e. we keep $\delta \gg \delta_{\rm r,\rm c}(\hbar)$,
then we recover the result supported by classical intuition, namely that the regular (non-ergodic) dynamics
is more stable than the chaotic (ergodic and mixing) dynamics. On the other hand, if
we let $\delta\to 0$ first, and only after that $\hbar\to 0$, i.e. satisfying (\ref{eq:delta_con}),
we find somewhat counterintuitive result saying that chaotic (mixing) dynamics is more stable than the regular one.
We can conclude the section by saying that we have {\em three non-commuting limits}, namely 
the {\em semiclassical limit} $\hbar \to 0$, 
the {\em perturbation strength} limit $\delta \to 0$, 
and the {\em thermodynamic limit} $d \to \infty$, 
such that no pair of these limits commutes.
\par
Similar regimes as for the fidelity decay were also obtained
by~\citet{Cohen:99,Cohen:00,Cohen:00a} when studying parametrically
driven systems and energy dissipation.

\subsection{Illustration with Wigner Functions}
\begin{figure}[htb!]
\centerline{\includegraphics[height=188mm]{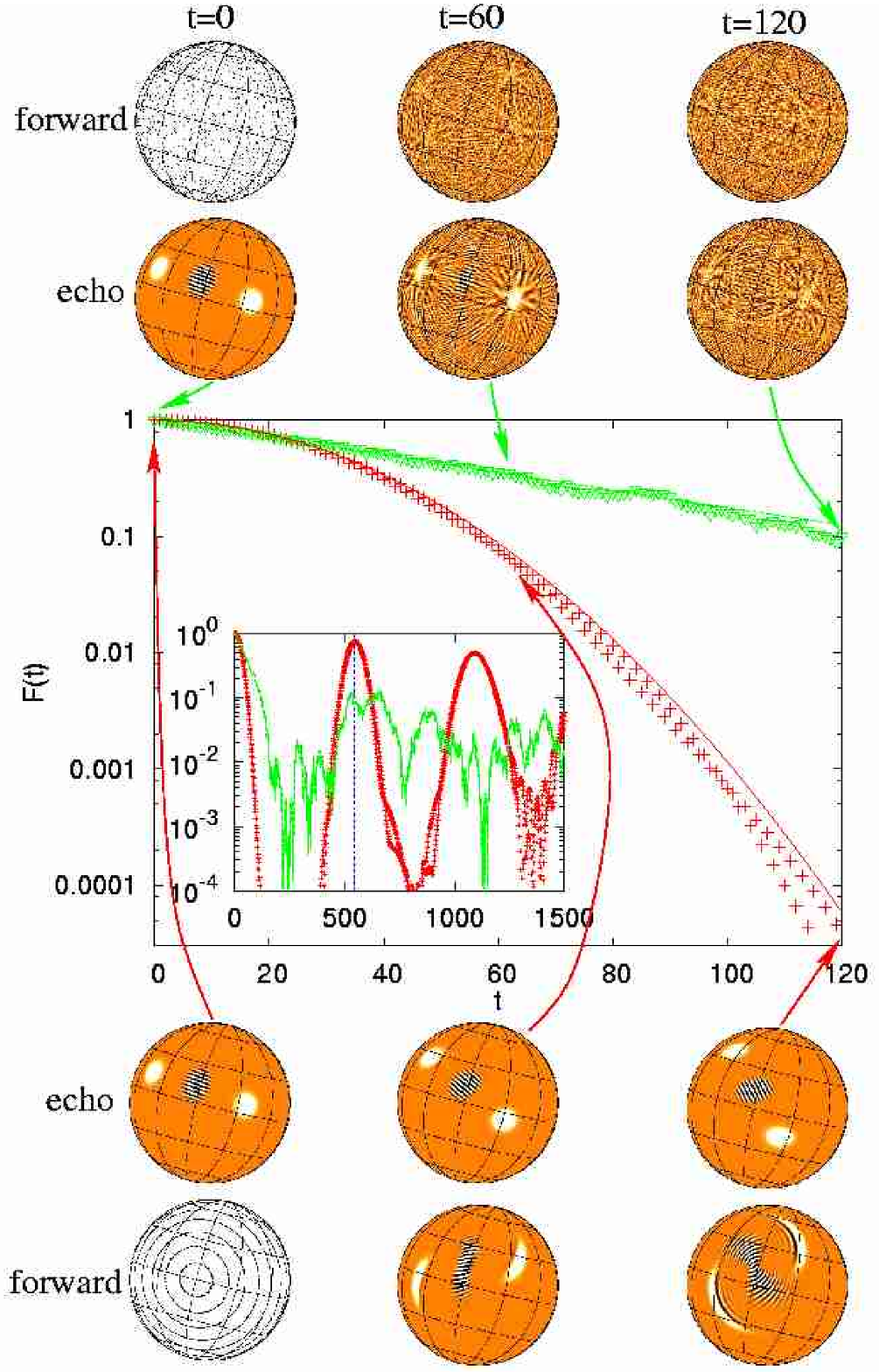}\includegraphics[angle=90,width=5mm]{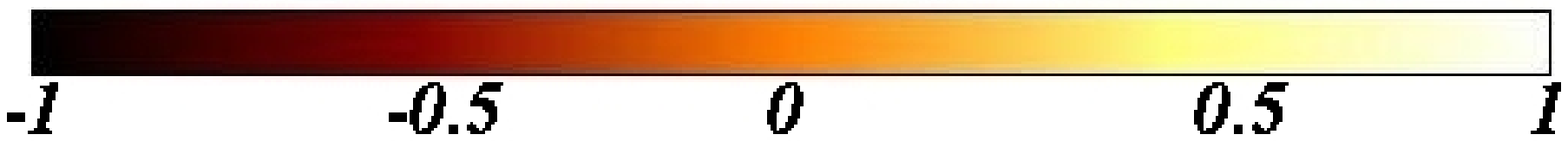}}
\caption{Fidelity decay for chaotic (top curve and pictures) and regular (bottom curve and pictures) kicked top. Initial conditions and the perturbation are the same in both cases (see text for details). Wigner functions after forward and echo evolution are shown.}
\label{fig:paradwig}
\end{figure}
In order to demonstrate faster decay of fidelity for a regular than
for a chaotic system we did a numerical simulation on the kicked top. For chaotic situation we choose $\alpha=30$, while for regular
one we take $\alpha=0.1$ as before. Other parameters are the same for the chaotic and the regular case, $\gamma=\pi/2$, $S=100$ and a coherent initial state  at $(\vartheta^*,\varphi^*)=\pi(1/\sqrt{3},1/\sqrt{2})$. Perturbation of strength $\delta=1.5\cdot 10^{-2}$ is in parameter $\alpha$. 
In Figure~\ref{fig:paradwig} we show the fidelity decay for both cases and an illustration of states in terms of Wigner functions. Spin Wigner function is a distribution on a sphere with a nice property that if we have two states $\rho_1$ and $\rho_2$ and their corresponding Wigner functions $W_{\rho_1}$ and $W_{\rho_2}$ the following equality holds,
\begin{equation}
\tr{(\rho_1 \rho_2)}=\int{\! W_{\rho_1}W_{\rho_2} {\rm d}\Omega}.
\end{equation}
For an exact definition of a spin Wigner function see
Appendix~\ref{app:wig}. Wigner functions enable us to represent
fidelity, which is an overlap of the initial state $\rho(0)$ with the
echo state $\rho^{\rm M}(t)$ (\ref{eq:rho_def}), as a phase space
overlap integral of two Wigner functions. Also, the classical quantity
analogous to Wigner function is just the classical density in phase
space. But note that Wigner function is not necessarily positive
whereas the classical density is, so the positivity of a Wigner
function is a necessary condition (but not sufficient) for the
classicality of a state. In Figure~\ref{fig:paradwig} we show for the
chaotic (triangles in the fidelity plot and pictures above) and the regular case (pluses in the fidelity plot and pictures below) two series of Wigner functions at
different times ($t=0,60,120$): the Wigner function after the unperturbed forward
evolution (row labeled ``forward'') and the Wigner function after the echo (row labeled ``echo''). We also show the
structure of the classical phase space. In the inset the same data for
the fidelity decay is shown on a longer time scale and the vertical
line shows the theoretical position of $t_{\rm b}$. The fidelity is
the overlap between the echo Wigner function and the initial Wigner
function. For our choice of coherent initial states, the initial
Wigner function is a Gaussian in the semiclassical limit
(\ref{eq:rho_clasSU2}). For chaotic dynamics the
forward Wigner function develops negative values around the Ehrenfest
time after which the quantum-classical correspondence is lost. For the
regular dynamics this correspondence persist much longer, namely until
the ``integrable'' Ehrenfest time $\sim \hbar^{-1/2}$ after which the
initial wave packet of size $\sim \hbar^{1/2}$ spreads over the phase
space. For a detailed study of Wigner functions in chaotic systems
see~\citep{Horvat:03,Lombardi:93} and references therein.
The echo Wigner function for regular dynamics moves ballistically from the initial position, causing the Gaussian decay of fidelity. But note that for regular dynamics the echo Wigner function does not have negative values even if they occur in the forward Wigner function. In our case the quantum fidelity agrees with the classical one for regular dynamics. In a chaotic case on the other hand, the echo image stays at the initial position and diffusively decays in amplitude, causing the fidelity to decay slower than in regular case. Classical fidelity follows quantum fidelity in the chaotic regime only up to the chaotic Ehrenfest time. 
\par
Finally, we would like to illustrate the dimensional dependence of the fidelity decay. We have seen that in $d>1$ border $\delta_{\rm c}$ is above the perturbative border and so one can get faster regular decay in a wider range of $\delta$. In addition, strong revivals of fidelity in one dimensional regular systems for coherent initial states should be absent for $d>1$.  

\subsection{Coupled Kicked Tops}
As already remarked, for a one dimensional system ($d=1$), 
the `surprising' behaviour of the regular decay time being smaller than the mixing one, $\tau_{\rm r} < \tau_{\rm m}$, is for coherent initial states possible only around the border (\ref{eq:dp}) $\delta_{\rm p}$ (unless $\sigma_{\rm cl}$ is very small) where the exponential decay in the mixing regime goes over to a Gaussian decay due to a finite ${\cal N}$. However, for more than one degree of freedom, such behaviour is generally possible well above the
finite size perturbative border $\delta_{\rm p}$. In order to illustrate this phenomenon we will now briefly consider a 
numerical example of a pair of coupled kicked tops where $d=2$. 
\begin{figure}[h!]
\centerline{\includegraphics{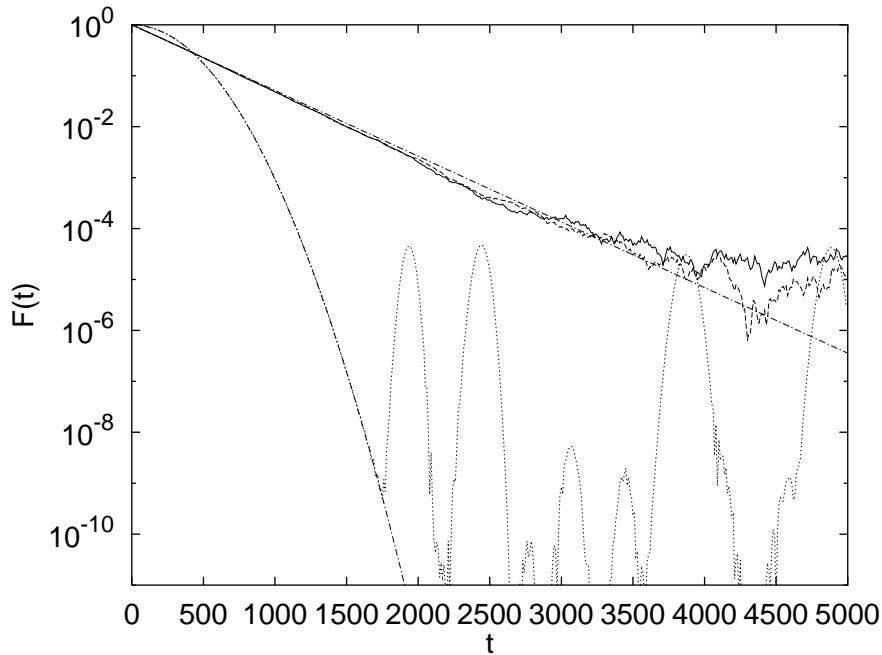}}
\caption{Fidelity decay for two coupled kicked tops, $\delta=8\cdot
  10^{-4}$ and $S=200$. The upper curves are for $\epsilon=20$ (mixing
  regime), the solid curve for a coherent initial state and the dashed
  curve for a random one. The lower dotted curve is for $\epsilon=1$ (regular regime) and a coherent initial state. The exponential and Gaussian chain curves give, respectively, the expected theoretical decays (\ref{eq:Fn_mixing}) and (\ref{eq:Fn_regcoh}), with the Gaussian decay time determined by the best fit.}
\label{fig:parad2}
\end{figure}   
\par 
We take a simplified version of coupled kicked tops (\ref{eq:2KT_def}) with a unitary propagator 
\begin{equation}
U(\epsilon)=
{\rm e}^{-\ii \frac{\pi}{2} S_{\rm 1y}} {\rm e}^{-\ii \frac{\pi}{2} S_{\rm 2y}} {\rm e}^{-\ii \epsilon S_{\rm 1z} S_{\rm 2z}/S}.
\label{eq:2KT}
\end{equation}
where $\vec{S_1}$ and $\vec{S_2}$ are two independent quantum angular momentum vectors. The perturbed propagator is obtained by perturbing the parameter $\epsilon$, so that $U_\delta=U(\epsilon+\delta)$. The perturbation generator is therefore
\begin{equation}
V=\frac{1}{S^2} S_{\rm 1z} S_{\rm 2z},
\label{2KTA}
\end{equation}
with $\hbar=1/S$. We have used the propagator (\ref{eq:2KT}) over 
the full $(2S+1)^2$ dimensional Hilbert space, without taking into account the symmetry classes of the double kicked top.
\par
The classical limit is obtained by $S\to \infty$ and writing the classical angular momentum vectors in terms of two unit 
vectors on the sphere $\vec{r}_{1,2}=\vec{S}_{1,2}/S$. In component notation we get the classical map
\begin{eqnarray}
x_{1,2}'&=& z_{1,2}\\
y_{1,2}'&=& y_{1,2} \cos(\epsilon z_{2,1}) + x_{1,2} \sin(\epsilon z_{2,1}) \nonumber \\
z_{1,2}'&=&-x_{1,2} \cos(\epsilon x_{2,1})+y_{1,2} \sin(\epsilon z_{2,1}) \nonumber 
\label{eq:KT2class}
\end{eqnarray}
We have chosen two regimes, non-ergodic (KAM) regime for $\epsilon=1$ where a vast majority of classical 
orbits are stable, and mixing regime for $\epsilon=20$ where no significant traces of stable classical orbits
were found and where the correlation sum was to a very good accuracy given by the first term only
\begin{equation}
\sigma \approx \frac{1}{2} C(0)=\frac{1}{2 S^4 {\cal N}} \tr{S_{\rm 1z}^2 S_{\rm 2z}^2}=\frac{1}{18}\left(1 + \frac{1}{S}\right)^2.
\label{eq:2KTsigma}
\end{equation}
The value of $\sigma$ if the whole correlation sum is calculated is $\sigma=0.058$, only slightly larger than $1/18$ given by $C(0)$. Our motivation here was to compare the regular and mixing fidelity decays for the coherent initial state which is here the product of spin coherent states (\ref{eq:SU2_coh}) for each top
\begin{equation}
\ket{\vartheta,\varphi}=\ket{\vartheta_2,\varphi_2} \otimes \ket{\vartheta_1,\varphi_1}.
\label{eq:ket}
\end{equation} 
\begin{figure}[ht!]
\centerline{\includegraphics{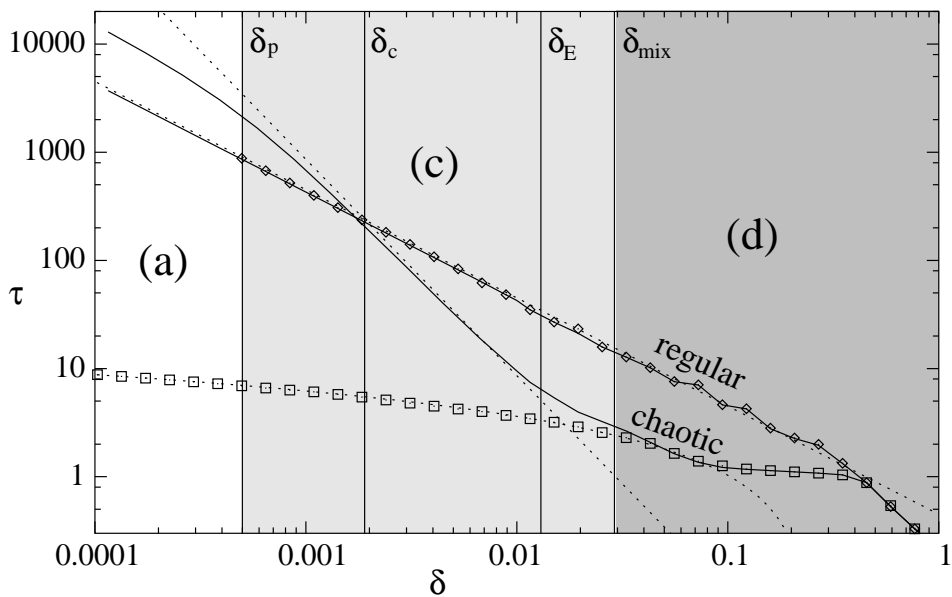}}
\caption{Numerically calculated decay time of the quantum and
  classical fidelity for the double kicked top. The two full curves show
  quantum fidelity, the one with a higher slope for the chaotic regime
  and the other for the regular regime. Symbols are decay times of
  classical fidelity, diamonds for the regular and squares for the
  chaotic regime. The two dotted straight lines are theoretical decay
  time predictions for the regular (\ref{eq:taur}) and chaotic
  (\ref{eq:taum}) regimes. The dotted curve (overlapping with squares)
  is the theoretical classical decay time $\tau_{\rm
  clas}=\ln{(0.25/\delta)}/\lambda$. Vertical lines show the
  theoretical position of perturbation borders (\ref{eq:borders}). The
  shading and the letters (a),(c) and (d) correspond to the regimes
  described in Figure~\ref{fig:delta1}. The Zeno regime corresponds to
  very short times $\tau<1$ (i.e. strong perturbations $\delta>0.4$).}
\label{fig:2ktopd}
\end{figure}
In Figure~\ref{fig:parad2} we show the fidelity decay at $S=200$ and $\delta=8\cdot 10^{-4}$ in regular and mixing cases starting from the
coherent state (\ref{eq:ket}) with $(\vartheta_1,\varphi_1)=(\vartheta_2,\varphi_2)=\pi(1/\sqrt{3},1/\sqrt{2})$. We find excellent agreement between the theoretical predictions (\ref{eq:Fn_mixing}) and (\ref{eq:Fn_regcoh}) and the
numerics. Note that we are here already in the regime $\delta < \delta_c$ where the fidelity decay in the mixing
regime is slower than in regular regime. In mixing regime ($\epsilon=20$) we show for comparison also the fidelity decay for a random initial state
for which the decay is (due to ergodicity) almost identical to the one for coherent initial state. Overall the fidelity decay here is similar as in a one-dimensional case,
however, the scaling of various time and perturbation scales on $\hbar=1/S$ is different as discussed in Section~\ref{sec:time_scales}. Observe also that in the regular regime the revivals of fidelity (quantum recurrences at $t_{\rm b}$) are much less pronounced in 
$d=2$ than in $d=1$ (e.g. in Figure~\ref{fig:regcoh}).
\par
To furthermore illustrate theoretical regimes of the fidelity decay as
explained in the previous section (Figure~\ref{fig:delta1}) we
numerically calculated the dependence of the time $\tau$ at which the
fidelity falls to value $0.37$ on the perturbation strength $\delta$ for the double kicked top model (\ref{eq:2KT}). The initial state is the same product coherent state as before. We also computed the decay time for the classical fidelity in order to compare the quantum and the classical fidelity. The results for $S=100$ are on Figure~\ref{fig:2ktopd}.  
For both the classical and the quantum fidelity we calculated two sets of data, one for regular regime at $\varepsilon=1$ and one for chaotic regime at $\varepsilon=5$. Numerical data is then compared with the theoretical predictions. For the chaotic decay time we use previously calculated $\sigma=0.058$ (\ref{eq:2KTsigma}) to get $\tau_{\rm m}$ (\ref{eq:Fn_mixing})
\begin{equation}
\tau_{\rm m}=\frac{8.6}{\delta^2 S^2}.
\label{eq:taum}
\end{equation}
In the regular case we used fitting of the decay in Figure~\ref{fig:parad2} to get the theoretical prediction (\ref{eq:Fn_regcoh}), i.e. to obtain $\sqrt{\bar{v}'\Lambda^{-1}\bar{v}'}$,
\begin{equation}
\tau_{\rm r}=\frac{4.5}{\delta\sqrt{S}},\qquad \sqrt{\bar{v}'\Lambda^{-1}\bar{v}'}=0.31. 
\label{eq:taur}
\end{equation}
\begin{figure}[b!]
\centerline{\includegraphics[width=95mm]{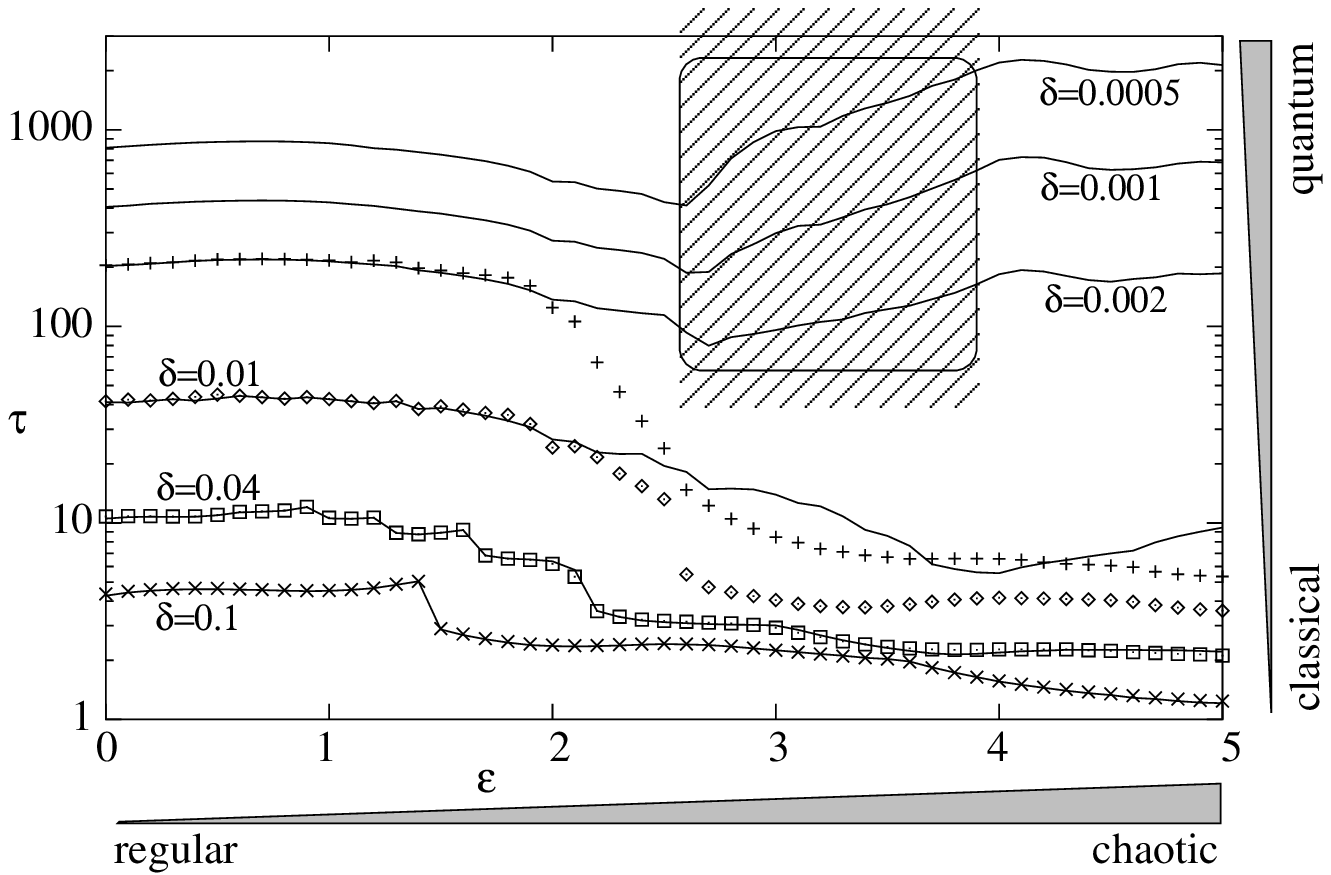}\includegraphics[width=67mm]{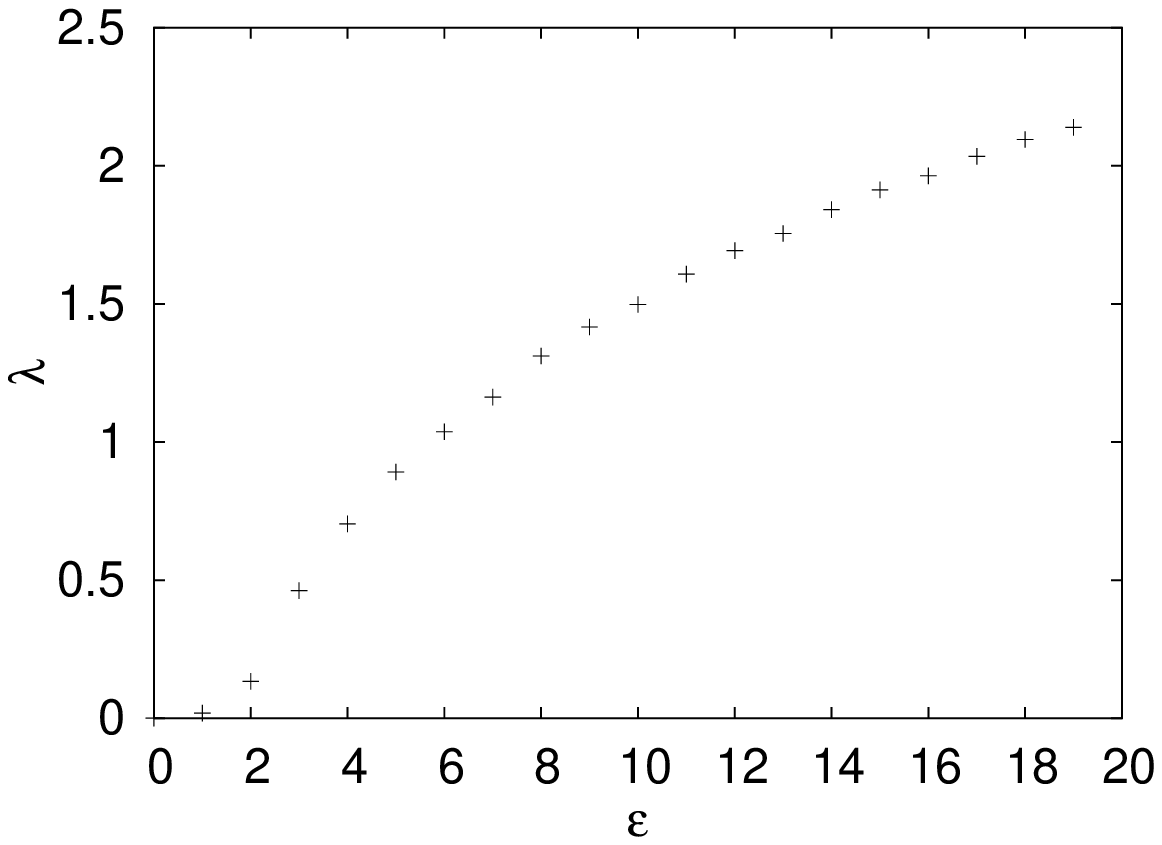}}
\caption{Numerically calculated dependence of quantum (solid lines) and classical (symbols) fidelity decay times on the parameter
  $\varepsilon$ for the double kicked top. Different curves are for
  different perturbation strengths $\delta$. By increasing
  $\varepsilon$ the classical dynamics goes from regular to mixing
  regime, see also right figure showing the dependence on
  $\varepsilon$ of numerically calculated Lyapunov exponents. By
  decreasing $\delta$ on the other hand, we go from the regime of
  quantum-classical correspondence for $\delta>\delta_{\rm E}$ towards
  a genuinely quantum regime, where in a chaotic regime we can {\em
    increase} the decay time by {\em increasing} chaoticity -- shaded
  region for the three smallest $\delta$.}
\label{fig:2ktop1}
\end{figure}
Note that the coefficient $\sqrt{\bar{v}'\Lambda^{-1}\bar{v}'}$ has been obtained by numerical fitting only for convenience. In principle it could be obtained from classical dynamics, but we would again have to resort to numerical calculations as the system at $\varepsilon=1$ is in a mixed KAM-like regime. The values of $\sigma$ and $\sqrt{\bar{v}'\Lambda^{-1}\bar{v}'}$ can then be used to calculate various perturbation borders as discussed in the beginning of Section~\ref{sec:time_scales}. For our choice of $S=100$ we get (\ref{eq:dp},\ref{eq:delta_con},\ref{eq:de},\ref{eq:dmix})
\begin{equation}
\delta_{\rm p}=0.0005,\quad \delta_{\rm c}=0.0019,\quad \delta_{\rm E}=0.013,\quad \delta_{\rm mix}=0.029.
\label{eq:borders}
\end{equation}
These theoretical borders are denoted with vertical lines in
Figure~\ref{fig:2ktopd}. We can see, that in regular regime the
quantum and the classical fidelity agree in the whole range of
$\delta$ and furthermore agree with the theoretical $\tau_{\rm r}$. In
the chaotic regime things are a bit more complicated. By decreasing
perturbation from $\delta=1$ we are at first in regime of very strong
perturbation where the fidelity decay happens faster than any
dynamical scale and it does not depend on whether we look at chaotic
or regular system or quantum or classical fidelity. This is the
so-called Zeno regime. For smaller $\delta$ the regular and chaotic
regimes start to differ but in the chaotic regime we still have
quantum-classical correspondence. This correspondence breaks down
around $\delta_{\rm mix}$ where the quantum fidelity starts to follow
the theoretical $\tau_{\rm m}$, while the classical fidelity decay is
$\tau_{\rm clas}=\log{(0.25/\delta)}/\lambda$, with $0.25$ being a
fitting parameter (depending on the width of the initial packet) and
$\lambda=0.89$ is the Lyapunov exponent, read from
Figure~\ref{fig:2ktop1}. For a brief explanation of this classical
decay see Section~\ref{sec:class_fid}. Incidentally, in our system at
$\varepsilon=5$ the classical mixing time is very short\footnote{This
  is the reason why $\sigma$ (\ref{eq:2KTsigma}) is almost entirely
  given by $C(0)$.}, $t_{\rm mix}\sim 1$, and we see that the
correspondence breaks already slightly before $\delta_{\rm E}$. The
quantum fidelity decay time $\tau_{\rm m}$ is valid until a
perturbative border $\delta_{\rm p}$ is reached, when a finite Hilbert
space dimension effects become important and the decay time becomes equal to $\tau_{\rm p}$. Note that for
$\delta<\delta_{\rm c}$ we indeed have a faster fidelity decay for a
{\em chaotic} dynamics than for a {\em regular} one.
\par
Another interesting aspect of our correlation function formalism is that the decay rate of the fidelity in a mixing situation is proportional to the integral of the correlation function $\sigma$. 
As stronger chaoticity will usually result in a {\em faster decay} of $C(t)$ and therefore in smaller $\sigma$, this means that {\em increasing} chaoticity (of the classical system) will {\em increase} quantum fidelity, i.e. stabilise quantum dynamics. Of course, for this to be observable we have to be out of the regime of quantum-classical correspondence. All this is illustrated in Figure~\ref{fig:2ktop1}, where we show similar decay times as in Figure~\ref{fig:2ktopd}, i.e. the same system, initial condition and $S=100$, but depending on the parameter $\varepsilon$ for six different perturbation strengths $\delta$. Parameter $\varepsilon$ controls the chaoticity of the classical dynamics. At $\varepsilon=1$ we are in the regular regime and for larger $\varepsilon$ we get into the chaotic regime, also seen from the dependence of the Lyapunov exponent. We can see that in the regular regime ($\varepsilon < 2$) the classical fidelity agrees with the quantum one regardless of $\delta$. In the chaotic regime though, the agreement is present only for the two largest $\delta$ shown, where we have $\delta>\delta_{\rm mix}$ (\ref{eq:borders}). For $\delta<\delta_{\rm c}$ and for chaotic dynamics (three smallest $\delta$) we get into the non-intuitive regime (shaded region in Figure~\ref{fig:2ktop1}) where the quantum fidelity will increase if we increase chaoticity. Note that this growth of the decay time stops at around $\varepsilon \sim 4$ because the classical mixing time $t_{\rm mix}$ gets so small that the transport coefficient is given by its time independent term $\sigma=C(0)/2$ alone.


\chapter{Special Case: Zero Time Averaged Perturbation}
\label{ch:freeze}
\begin{flushright}
\baselineskip=13pt
\parbox{70mm}{\baselineskip=13pt
\sf One should always keep an open mind, but not so open that one's brains
fall out.}\medskip\\
---{\sf \itshape Bertrand Russell}\\\vspace{20pt}
\end{flushright}

In the previous chapter we considered general perturbations, for which the double correlation sum was growing with time either linearly for mixing dynamics or quadratically for regular systems. What about the third possibility, namely if it {\em does not} grow with time. This situation will be the subject of the present chapter. We will demand that $\ave{\Sigma^2(t)} \sim t^0$ for any initial state, i.e. all matrix elements of $\Sigma^2$ must be ${\cal O}(t^0)$,
\begin{equation}
\Sigma^2_{jk}(t) \propto t^0,\qquad j,k=1,\ldots,{\cal N}.
\label{eq:freeze_cond} 
\end{equation}
We will write the perturbation $V$ as the sum of its time average $\oV$ and the rest, called the {\em residual} part $\Vres$
\begin{equation}
V=:\oV+\Vres.
\label{eq:Vres_def}
\end{equation}
For a nondegenerate\footnote{For degenerate spectra we have
$\oV=\sum_{k,l}{\delta(\phi_k-\phi_l)V_{kl}\ket{\phi_k}\bra{\phi_l}}$.}
spectrum of the unperturbed propagator $U_0$ we have seen that the
time averaged perturbation equals to the diagonal part of $V$
(\ref{eq:Vbar_matrix}) therefore, the residual $\Vres$ is just the
off-diagonal part of $V$ and has zeros on the diagonal,
$(\Vres)_{kk}\equiv 0$. The operator $\Sigma(t)$ can then be written
as $\Sigma(V,t)=\oV t +{\cal O}(t^0)$, where the second part depends
just on $\Vres$ and does not grow with time. One can conclude that to
satisfy condition (\ref{eq:freeze_cond}) the time averaged
perturbation must be zero, $\oV \equiv 0$. The subject of the present
chapter are perturbations with $\oV=0$, also called {\em residual}
perturbations because $V=\Vres$. For non-residual perturbations all
essential physics was contained in the operator $\Sigma(t)$. For residual perturbations on the other hand, the second term involving $\Gamma(t)$ in the BCH form (\ref{eq:Md_exp}) of the echo operator will also be important. Matrix elements of $\Gamma(t)$ in the eigenbasis of $U_0$ are
\begin{eqnarray}
\!\!\!\!\!\!\!\!\!\frac{\bra{\phi_j}\Gamma(t)\ket{\phi_j}}{t} &=& \frac{1}{\hbar} \sum_{k\neq j}
|V_{jk}|^2 \cot[{\textstyle\frac{1}{2}}(\phi_k-\phi_j)] + 
{\cal O}(t^{-1}),\label{eq:Gamma_matrix}\\
\!\!\!\!\!\!\!\!\!\frac{\bra{\phi_j}\Gamma(t)\ket{\phi_k}}{t} &=& \frac{1}{\hbar}(V_{jj}\!-\!V_{kk})V_{jk}
\frac{e^{-\ii\frac{1}{2}(\phi_j\!-\!\phi_k)}\!+\!e^{-\ii(\phi_j\!-\!\phi_k)(\frac{1}{2}-t)}}
{2\sin[\frac{1}{2}(\phi_k-\phi_j)]} + {\cal O}(t^{-1}), \quad j\neq k.\nonumber
\end{eqnarray}
Note that the matrix elements of $\Gamma(t)$ can not grow faster than linearly with $t$, despite the double sum over time in the definition of $\Gamma$. We see that, provided the perturbation is residual,
the limit of {\em time-averaged} $\Gamma(t)$ defined as
\begin{equation}
\oG = 
\lim_{t\to\infty}\frac{\Gamma(t)}{t}
= \frac{\ii}{\hbar}\lim_{t\to\infty}\frac{1}{t}\sum_{t'=0}^{t-1}\sum_{t''=t'}^{t-1} 
[V(t'),V(t'')]
\label{eq:oG_def}
\end{equation}
exists and is {\em diagonal} in the eigenbasis of $U_0$:
\begin{equation}
\oG =\sum_j \oG_{jj} \ket{\phi_j}\bra{\phi_j},\qquad
\oG_{jj} = 
\frac{1}{\hbar}\sum_{k\neq j}
|V_{jk}|^2 \cot[{\textstyle\frac{1}{2}}(\phi_k-\phi_j)].
\end{equation}
The operator $\oG$ is again a constant of motion, $[U_0,\oG]=0$.
\par
Any residual perturbation can be defined in terms of another operator $W$ by the following prescription
\begin{equation}
V=W(1)-W(0),\qquad W(t):=U_0^{-t} W U_0^t.
\label{eq:W_def}
\end{equation}
In the continuous time case we would have definition $V=:(d/dt)W=\frac{\ii}{\hbar}[H_0,W]$. Indeed, given a residual perturbation one easily determines the matrix elements of $W$ as
\begin{equation}
W_{jk}:=\frac{V_{jk}}{\exp{(\ii (\phi_j-\phi_k))}-1}.
\label{eq:W_matrix}
\end{equation}
Note that instead of the unperturbed evolution operator $U_0$ one could use any other unitary operator that has the same degeneracy and eigenvectors as $U_0$ in the definition of $V$ in terms of $W$ (\ref{eq:W_def}). With the newly defined operator $W$, the expression for $\Sigma(t)$ is extremely simple, 
\begin{equation}
\Sigma(t)=W(t)-W(0).
\label{eq:S_W}
\end{equation}
Similarly, the expression for $\Gamma(t)$ is also considerably simplified,
\begin{equation}
\Gamma(t)=\Sigma_{\rm R}(t)-\frac{\ii}{\hbar}[W(0),W(t)],\qquad R:=\frac{\ii}{\hbar}[W(0),W(1)],
\label{eq:G_R}
\end{equation}
and 
\begin{equation}
\Sigma_{\rm R}(t):=\sum_{t'=0}^{t-1}{R(t')},\qquad R(t):=U_0^{-t} R U_0^t.
\label{eq:SR_def}
\end{equation}
The operator $\Gamma(t)$ is, apart from the term $\frac{\ii}{\hbar}[W,W(1)]$, similarly as $\Sigma(t)$ the sum of $R(t)$. In the continuous time case we have $R:=\frac{\ii}{\hbar}[W,({\rm d}/{\rm d}t)W]=\hbar^{-2}[W,[W,H_0]]$ and $\Gamma(t)=\int_0^t{{\rm d}t' R(t')}-\frac{\ii}{\hbar}[W,W(t)]$. The fidelity decay will be given by the echo operator
\begin{equation}
\Md=\exp{\left\{-\frac{\ii}{\hbar}\left( \Sigma(t) \delta+\frac{1}{2}\Gamma(t) \delta^2 + \cdots \right) \right\}},
\label{eq:Md_res}
\end{equation}
with $\Sigma(t)$ (\ref{eq:S_W}) and $\Gamma(t)$ (\ref{eq:SR_def}) expressed in terms of the operator $W$. We see that for small times $t<t_2$, with $t_2\sim 1/\delta$, the second term involving $\Gamma(t)$ can be neglected. Therefore, for $t<t_2$ the fidelity amplitude is simply
\begin{equation}
\fn=\ave{\exp{(-\ii \delta (W(t)-W(0))/\hbar)}}.
\end{equation}
Expanding $\fn$ to the second order in $\delta$, we find
$\Fn = 1 - \frac{\delta^2}{\hbar^2}(\kappa_0^2 + \kappa_t^2 - C(t) - C(t)^*)$ where
$\kappa_k^2:=\ave{W^2(k)}-\ave{W(k)}^2$, $C(t):=\ave{W(t) W(0)}-\ave{W(t)}\ave{W(0)}$.
Using the Cauchy-Schwartz inequality $|C(t)| \le \kappa_0 \kappa_t$ and the fact that for a bounded operator $W$ the sequence $\kappa_t$ is bounded, say by $r$, we find a {\em freeze of fidelity}
\begin{equation}
1-\Fn \le 4\frac{\delta^2}{\hbar^2}r^2,\qquad t < t_2
\end{equation}
for {\em arbitrary} quantum dynamics, irrespective of the existence
and the nature of the classical limit. For residual perturbation the
fidelity therefore stays high up to a classically long time $t_2$ and only then starts to decay again. After $t_2$ the second term $\Gamma(t)$ will become important and this will cause the fidelity to decay. We have to stress that the freeze of fidelity is of purely quantum origin. The classical fidelity does not exhibit such behaviour. 
\par
Although residual perturbations might seem artificial at first sight,
there are two possibilities how they can arise. First, the average
perturbation $\oV$ commutes with the Hamiltonian $H_0$ generating
$U_0$ and can sometimes be put together with the original $H_0$ into
the unperturbed Hamiltonian, thereby resulting in a residual
perturbation. This moving of $\oV$ into $H_0$ just changes the
eigenenergies and is often done in various mean field
approaches. Another possibility to have $\oV=0$ is due to
symmetries. For instance, having a unitary symmetry $P$ commuting with
the unperturbed evolution, $[U_0,P]=0$, and perturbation $V$
anticommuting with the symmetry, $V P=-P\,V$, will result in a
residual perturbation. But note that in order to have a well defined
operator $W$ (\ref{eq:W_matrix}) with no singularities near the
diagonal, so that our theory can be applied, the
matrix elements of $V$ must increase smoothly away from the diagonal.
\par
In the next two sections we will discuss two examples for which one can use the echo operator (\ref{eq:Md_res}) to calculate fidelity decay to all orders in $\delta$. This can be done for completely mixing dynamics and for the opposite extreme of regular dynamics. In both cases we will also assume the operators $V$, $W$, and therefore also $R$, to have well defined classical limits, so we will be able to use semiclassical arguments. For small times $t<t_2$ the fidelity will freeze to a constant value - the plateau. The value of the plateau will be determined by the non-increasing $\Sigma(t)$,
\begin{equation}
F_{\rm plat}(t)=\left|\ave{\exp{\left(-\ii \frac{\delta}{\hbar} \left\{W(t)-W(0)\right\} \right)}}\right|^2.
\label{eq:Fn_plateau}
\end{equation}
For large times $t>t_2$ instead, the fidelity decay will be determined by the operator $R$ in $\Gamma(t)$.

 \section{Mixing Dynamics}
\subsection{The Plateau}
We begin with the linear response evaluation of the fidelity plateau
(\ref{eq:Fn_plateau}). For classically mixing dynamics we can assume
that for times larger than some mixing time $t_1$ the time
correlations vanish in leading semiclassical order, $C(t) \to {\cal
O}(\hbar)$. Also, quantum expectation values become time independent
and equal to their classical values in leading order. We will denote
by $\ave{A}_{\rm cl}:=\int{{\rm d}\mu A_{\rm cl}}$ the classical
average value of observable $A$. Therefore, for times $t_1 < t < t_2$ the linear response expression for fidelity plateau, i.e. its time-independent value is
\begin{equation}
F_{\rm plat}\cong 1-\frac{\delta^2}{\hbar^2}(\kappa_0^2+\kappa_{\rm
cl}^2),\qquad \kappa_t^2:=\< W^2(t) \>-\ave{W(t)}^2,
\label{eq:Fplat_kappa}
\end{equation}
where $\kappa_{\rm cl}^2:=\ave{w^2}_{\rm cl}-\ave{w}^2_{\rm cl}$ is
time-independent classical limit of $\kappa^2_t$ with the classical
limit $w(\vec{q},\vec{p})$ of operator $W$. We will consider two kinds
of initial states, namely coherent states and random states. For coherent initial states (CIS) the mixing time $t_1$ is equal to the Ehrenfest time, $t_1 \sim -\ln{\hbar}/\lambda$. As the initial state variance of $W$ is proportional to the spread of the CIS, $\kappa_0^2 \propto \hbar$, it can be neglected with respect to $\kappa_{\rm cl}^2$. The fidelity plateau for CIS is therefore $F_{\rm plat}^{\rm CIS}\cong 1-(\delta/\hbar)^2 \kappa^2_{\rm cl}$. For random initial states (RIS), the mixing time is $\hbar$ independent, $t_1 \propto \hbar^0$. Also initial state average for RIS is equal to ergodic average and the fidelity plateau is in this case $F_{\rm plat}^{\rm RIS}\cong 1-2(\delta/\hbar)^2 \kappa^2_{\rm cl}$. Summarising, the linear response value of the fidelity plateau is for CIS and RIS
\begin{equation}
1-F_{\rm plat}^{\rm CIS}\cong \frac{\delta^2}{\hbar^2}\kappa^2_{\rm cl},\qquad 1-F_{\rm plat}^{\rm RIS}\cong 2\frac{\delta^2}{\hbar^2}\kappa^2_{\rm cl},
\label{eq:plat_LR}
\end{equation}
i.e. it is twice as large for RIS than for CIS.
\par
One can go beyond linear response in approximating (\ref{eq:Fn_plateau}) using a simple fact that in the leading order in $\hbar$ quantum observables commute, and as before, that for $t > t_1$ the 
time correlations vanish, namely $\<\exp(-\ii\frac{\delta}{\hbar}(W(t)-W))\> \cong
\<\exp(-\ii\frac{\delta}{\hbar}W(t))\> \<\exp(\ii\frac{\delta}{\hbar} W)\>$. For $t>t_1$ expectation values become time independent and so
\begin{equation}
F_{\rm plat} \cong 
\left|\ave{\exp{(-\ii w \delta/\hbar)}}_{\rm cl}\ave{\exp{(\ii W\delta/\hbar)}}\right|^2.
\label{eq:NLRP}
\end{equation}
Note that the the right average is an average over initial state
whereas the left average is the classical average over the invariant
measure. The fidelity plateau can be compactly expressed in terms of a generating function $G(\delta/\hbar)$
\begin{equation}
G(z):=\ave{\exp(-\ii z \,w)}_{\rm cl}=\frac{1}{\cal V}\int{\! {\rm d}^d \vec{q}\, {\rm d}^d \vec{p} \,\exp{\left\{-\ii z \,w(\vec{q},\vec{p}) \right\}} }.
\label{eq:G_def}
\end{equation}
For CIS one can neglect the initial state average over a localized packet, i.e. the second term, and gets
\begin{equation}
F_{\rm plat}^{\rm CIS}\cong |G(\delta/\hbar)|^2.
\label{eq:plat_CIS}
\end{equation}
For RIS on the other hand, initial state average is equal to the classical average in the leading order and we have
\begin{equation}
F^{\rm RIS}_{\rm plat}\cong |G(\delta/\hbar)|^4.
\label{eq:plat_RIS}
\end{equation}
We have a universal relation between the plateau for CIS and RIS, namely $F^{\rm RIS}_{\rm plat}\cong (F^{\rm CIS}_{\rm plat})^2$. If the argument $z=\delta/\hbar$ of the generating function is large, the analytic function $G(z)$ can be 
calculated generally by the method of stationary phase.
In the simplest case of a single isolated stationary point $\vec{x}^*$ in $N$ dimensions we obtain
\begin{equation}
|G(z)| \asymp \left|\frac{\pi}{2z}\right|^{N/2}\left|{\rm det\,}\partial_{x_j} \partial_{x_k} 
w(\vec{x}^*)\right|^{-1/2}.
\label{eq:Gasym}
\end{equation}
This expression gives an asymptotic power law decay of the plateau height 
independent of the perturbation details. Note that for a finite phase space we will have 
oscillatory {\em diffraction corrections} to Eq.~(\ref{eq:Gasym}) due to a finite range of 
integration $\int{\!{\rm d}\mu}$ which in turn causes an interesting situation for 
specific values of $z$, namely that by 
increasing the perturbation strength $\delta$ we can actually increase the value of the plateau. 
\begin{figure}[ht]
\centerline{\includegraphics[angle=-90,width=140mm]{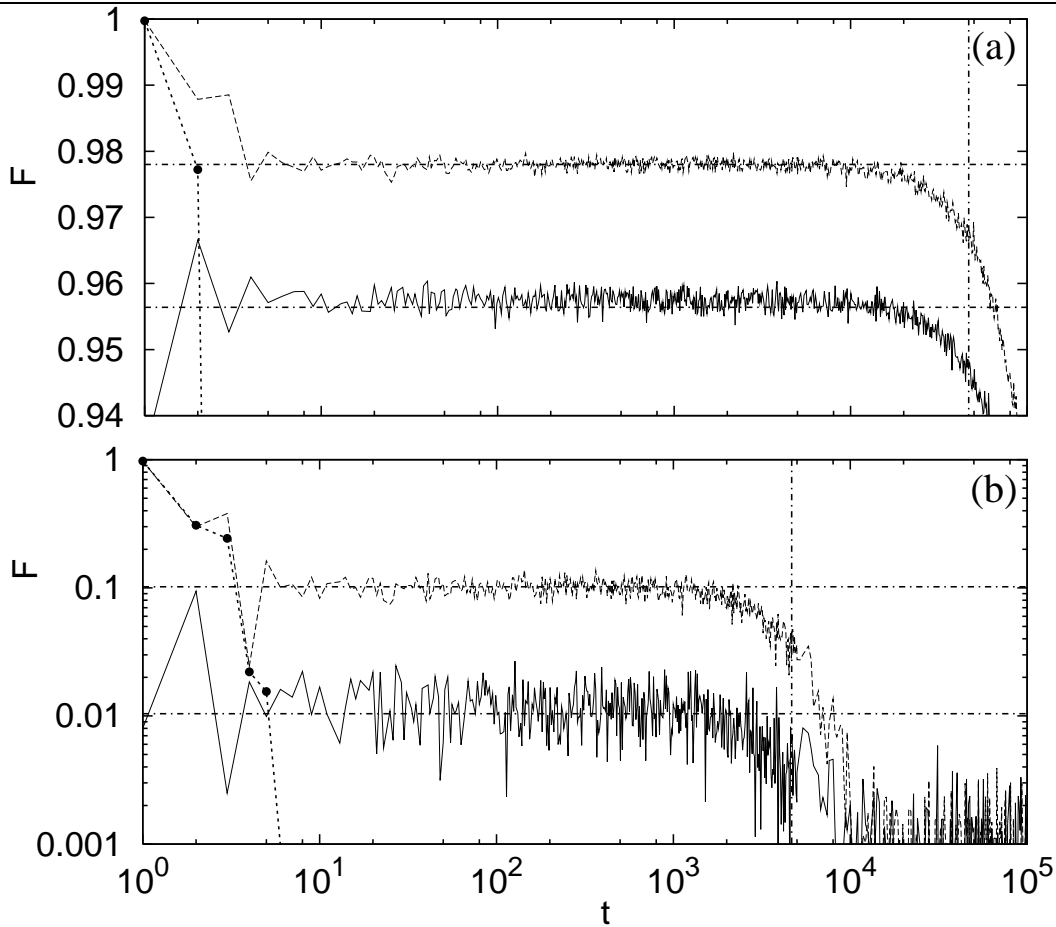}}
\caption{$\Fn$ for the kicked top with perturbations $\delta=10^{-3}$ (a), and $\delta=10^{-2}$ (b). 
In each plot the upper curve is for CIS and the lower for
RIS. Horizontal chain lines are theoretical plateau values, linear
response (\ref{eq:plat_LR}) in (a) and full
(\ref{eq:plat_RIS},\ref{eq:plat_CIS}) in (b). Vertical chain lines are theoretical values of $t_2$ (\ref{eq:t2}). The full circles represent calculation of
the corresponding classical fidelity for the CIS which follows quantum fidelity 
up to the Ehrenfest ($\log\hbar$) barrier and exhibits no freezing.}
\label{fig:1ktop}
\end{figure}
\par
We tested the above theory by numerical experiments. For the system we choose the kicked top, with slightly different order of factors as before (\ref{eq:KT_def}). One step propagator is
\begin{equation}
U_0:=\exp{\left(-\ii \alpha \frac{S_{\rm z}^2}{2S}\right)} \exp{\left(-\ii \frac{\pi}{2}S_{\rm y}\right)}.
\label{eq:U0_def}
\end{equation}
We take $\alpha=30$ ensuring fully chaotic classical dynamics whereas spin size is taken $S=1000$ giving effective Planck's constant $\hbar=1/S=1\cdot 10^{-3}$. Again we choose two initial states, a random one and a coherent one centred at $(\vartheta^*,\varphi^*)=(1,1)$ and both are projected onto OE subspace (\ref{eq:KT_subspaces}). To get a residual perturbation we take $W:=S_{\rm z}^2/2S^2$ so that $W(1)=S_{\rm x}^2/2S^2$ giving the perturbation generator
\begin{equation}
V:=\frac{S_{\rm x}^2-S_{\rm z}^2}{2S^2}.
\label{eq:Vres}
\end{equation}
The perturbed propagator is as always $U_\delta=U_0 \exp{(-\ii \delta V/\hbar)}$. We choose two perturbation strengths. A weak $\delta=10^{-3}$ to check the linear response expressions (\ref{eq:plat_LR}). The results are seen in Figure~\ref{fig:1ktop}a. The classical value of $\kappa_{\rm cl}^2$ used in theoretical formulas is easily calculated for $w=z^2/2$ and one gets $\kappa_{\rm cl}^2=1/45$. In Figure~\ref{fig:1ktop}b we show numerical results for larger $\delta=10^{-2}$ where the plateau is very low and full formulas must be used. Generating function (\ref{eq:G_def}) can be calculated exactly without stationary phase approximation, resulting in
\begin{equation}
G(\delta S)=\sqrt{\frac{\pi}{2\delta S}}\, {\rm erfi}\left({\rm e}^{\ii \pi/4}\sqrt{\delta S/2}\right).
\label{eq:Gz}
\end{equation}
We have an asymptotic power law decay as predicted by the general
stationary phase formula\footnote{Note that $w=z^2/2$ has one
stationary point in $N=1$ dimension, despite the phase space being
two-dimensional.} (\ref{eq:Gasym}). In addition, due to a finite phase
space we have a diffractive oscillatory ${\rm erfi}$
correction. Theoretical prediction for the plateau using $G(\delta S)$ (\ref{eq:Gz})
agrees well with the numerical results shown in the figure. Small quantum fluctuations around the theoretical plateau values lie beyond the leading semiclassical approximation used in our theoretical derivations. Note also that the quantum fidelity and its plateau values have been
expressed (in the leading order in $\hbar$) in terms of classical quantities only. Yet, the freezing of fidelity is a purely quantum phenomenon as one can also see in Figure~\ref{fig:1ktop} where the classical fidelity does not exhibit freezing. 

\subsection{Beyond the Plateau}
Next we shall consider the regime of long times $t > t_2$. 
Then the second term in the exponential of the echo operator
(\ref{eq:Md_res}) dominates the first one, however even the first term
may not be negligible for large $\delta$. Up to terms of order ${\cal
O}(t \delta^3)$ we can factorize Eq.~(\ref{eq:Md_res}) as $\Md \approx
\exp(-\ii\frac{\delta}{\hbar}(W(t)-W))\exp(-\ii\frac{\delta^2}{2\hbar}\Gamma(t))$.
When computing the expectation value we again use the fact that in
leading semiclassical order operator ordering is irrelevant and that,
since $t \gg t_1$, any time-correlation can be factorized thus the
second term $\frac{\ii}{\hbar}[W,W(t)]$ of $\Gamma(t)$ (\ref{eq:G_R})
vanishes and we have
\begin{equation}
\Fn \cong F_{\rm plat} \left|\ave{\exp\left(-\ii\frac{\delta^2}{2\hbar}\Sigma_{\rm R}(t)\right)}\right|^2,\qquad t > t_2.
\label{eq:Fn_fact}
\end{equation}
This result is quite intriguing. It tells us that apart from a pre-factor $F_{\rm plat}$, the decay of the fidelity due to a residual perturbation is formally the same (in the leading semiclassical order when time ordering is not important) as the fidelity decay with a generic non-residual perturbation, eq.~(\ref{eq:Md_prod}), when one substitutes the 
operator $V$ with $R$ and the perturbation strength $\delta$ with $\delta_R=\delta^2/2$. Thus we can directly apply the semiclassical theory of fidelity decay for general perturbations explained in Section~\ref{sec:mixing},
using a renormalised perturbation $R$ of renormalised strength $\delta_R$. Here we simply rewrite the key results in the 'non-Lyapunov' perturbation-dependent 
regime, $\delta_R < \hbar$. Using the classical transport rate $\sigma_{\rm R}$, \begin{equation}
\sigma_{\rm R}:=\lim_{t \to \infty}{\frac{\ave{\Sigma_{\rm R}^2(t)}_{\rm cl}-\ave{\Sigma_{\rm R}}_{\rm cl}^2}{2t}},
\label{eq:sigmaR}
\end{equation}
we have either an exponential decay 
\begin{equation}
\Fn \cong F_{\rm plat}
\exp{\left(-\frac{\delta^4}{2\hbar^2} \sigma_{\rm R} t \right)},\quad
t < t_{\rm H},
\label{eq:Fexp}
\end{equation}
or a (perturbative) Gaussian decay 
\begin{equation}
\Fn \cong F_{\rm plat}
\exp{\left(-\frac{\delta^4}{2\hbar^2} \sigma_{\rm R} \frac{t^2}{t_{\rm H}}\right)},\quad
t > t_{\rm H}.
\label{eq:Fgau}
\end{equation}
\begin{figure}[ht]
\centerline{\includegraphics[angle=-90,width=140mm]{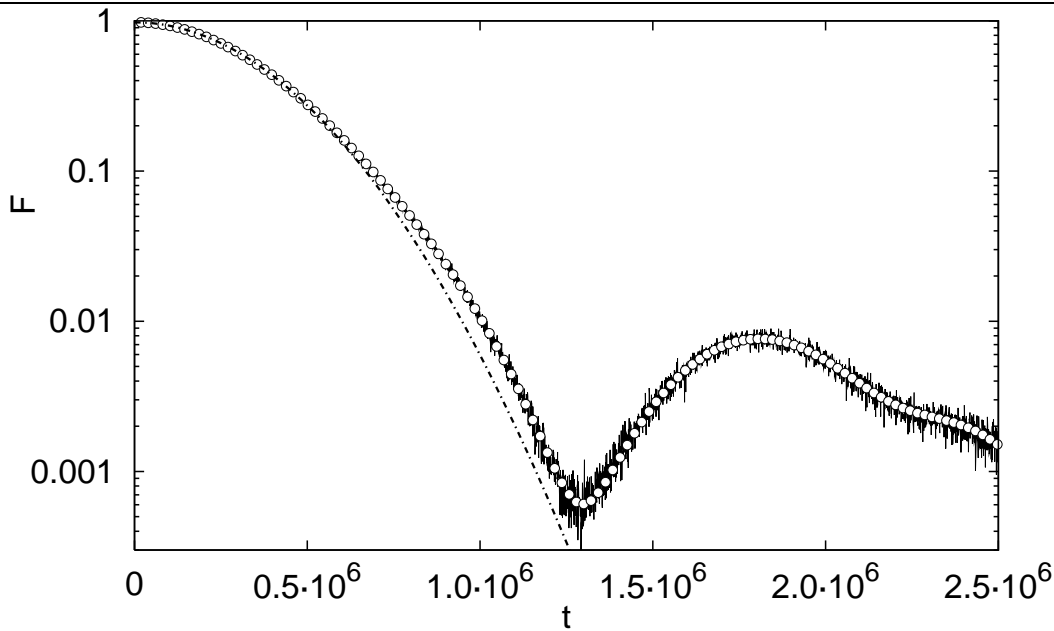}}
\caption{Long time Gaussian decay (\ref{eq:Fgau}) for the same CIS and
parameters as in Figure~\ref{fig:1ktop}a. The full curve is a direct numerical simulation, empty circles are numerical calculation using a renormalised perturbation $R$ of strength $\delta_{\rm R}$ (\ref{eq:R}), while the chain curve gives the theoretical decay.}
\label{fig:1ktopL}
\end{figure}
The crossover between the two decay regimes happens at the Heisenberg
time $t_{\rm H}={\cal N}/2$ (\ref{eq:np_def}). Of course, the same
consideration regarding the asymptotic saturation value of the
fidelity $\oFn$ due to a finite ${\cal N}$ applies here as well (see
discussion of time scales in Section~\ref{sec:time_scales}). The
prefactor $F_{\rm plat}$ can be calculated as described in the
previous section, Eq.~(\ref{eq:NLRP}), and {\em depends} on the
initial state. The exponential terms of (\ref{eq:Fexp},\ref{eq:Fgau})
on the other hand {\em do not} depend on the initial state. 
\par
From the two possible regimes of long-time fidelity decay we can also more precisely specify time $t_2$ when the plateau ends. Comparing the two factors in (\ref{eq:Fexp},\ref{eq:Fgau}) with the plateau value (Eq.~\ref{eq:plat_LR} for $\delta < \hbar$), we obtain a semiclassical estimate of $t_2$
\begin{equation}
t_2 \approx {\rm min}\left\{\sqrt{\frac{t_H \kappa_{\rm cl}^2}{\sigma_{\rm R}}}\frac{1}{\delta},
\frac{\kappa^2_{\rm cl}}{\sigma_{\rm R} \delta^2}\right\}.
\label{eq:t2}
\end{equation}
In Figure~\ref{fig:1ktop} we can see nice agreement of theoretical
$t_2$ with the results of the simulation.
\par
Interestingly though, as we have another time scale $t_2$, the
duration of the plateau, not present for a general case of
non-residual perturbation, the exponential regime (\ref{eq:Fexp}) can
only take place if $t_2 < t_{\rm H}$, otherwise we immediatelly get a
Gaussian decay after the plateau. If one wants to keep $F_{\rm plat}
\sim 1$ (i.e. high) and have an exponential decay in the full range
until the asymptotic $\oFn \sim 1/{\cal N}$, the condition on
dimensionality is imposed. Namely, demanding high plateau $\delta/\hbar<1$ for
$t<t_2$ and low fidelity in the limit $\hbar \to 0$ at the Heisenberg time, $F(t_{\rm H})=\exp{(-(\delta/\hbar)^4 \sigma_{\rm
R}/4\hbar^{d-2})}$, gives condition on dimensionality $d \ge 2$. In
one dimensional systems and in the semiclassical limit the exponential decay (\ref{eq:Fexp})
therefore can not be seen.
\par
We again compared theory with numerics. The system is the one
dimensional kicked top already used in the previous section describing
the plateau. The perturbation generator $R$ for our choice of $W$ is
\begin{equation}
R=-\frac{1}{2S^3}(S_{\rm x}S_{\rm y}S_{\rm z}+S_{\rm z}S_{\rm y}S_{\rm x}),
\label{eq:R}
\end{equation}
having a classical limit $R_{\rm cl}=r=-xyz$. We numerically calculated the integral of the classical correlation function (\ref{eq:sigmaR}) giving the transport coefficient $\sigma_{\rm R}=5.1\cdot 10^{-3}$. In Figure~\ref{fig:1ktopL} we show three different sets of data. Direct numerical calculation, ``a renormalised calculation'' obtained by taking perturbed dynamics as $U_\delta=U_0 \exp{(-\ii \delta_{\rm R} R/\hbar)}$, i.e. perturbation generator $R$ with the strength $\delta_{\rm R}=\delta^2/2$ and theoretical Gaussian decay (\ref{eq:Fgau}) where all parameters have been calculated classically. Apart from the plateau prefactor $F_{\rm plat}$ (which is close to $1$) the decay does not depend on the initial state.

\subsubsection{Double Kicked Top}
\begin{figure}[ht!]
\centerline{\includegraphics[angle=-90,width=140mm]{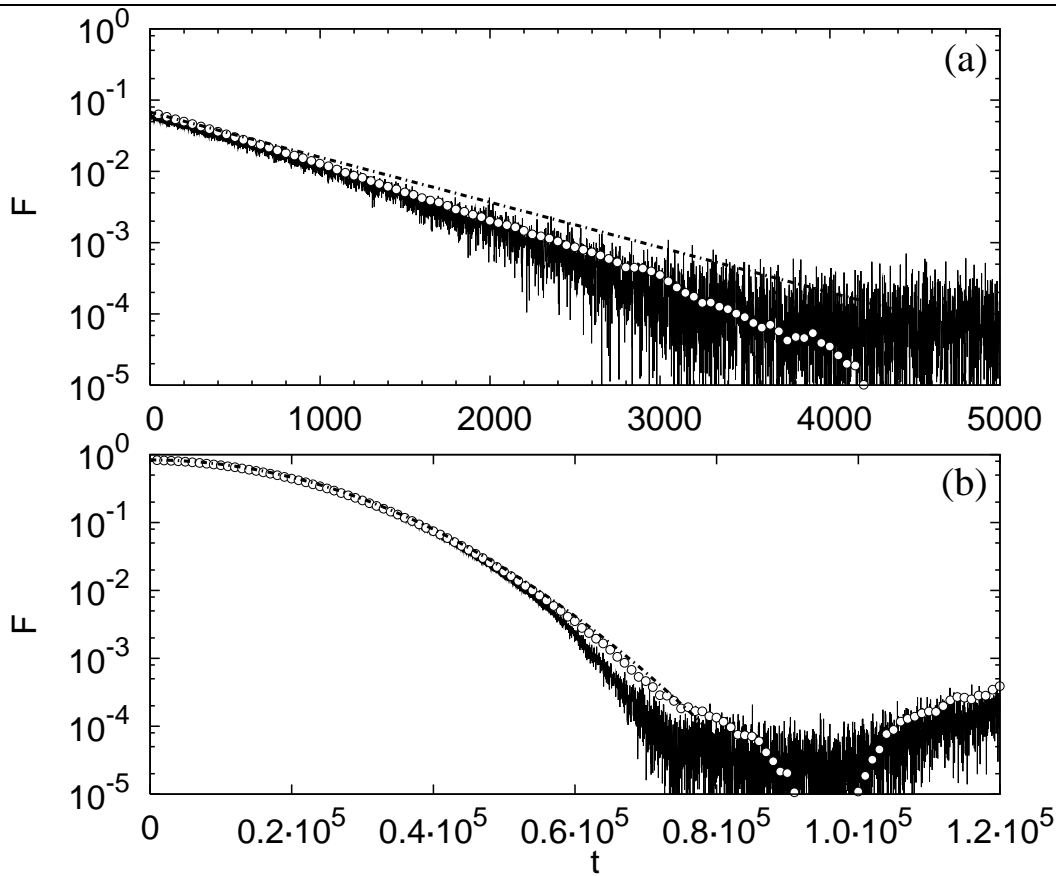}}
\caption{Long time fidelity decay for two coupled kicked tops. For strong perturbation $\delta=7.5\cdot10^{-2}$ in (a) we obtain an exponential decay (\ref{eq:Fexp}), and for smaller $\delta=2\cdot 10^{-2}$ in (b) we get a Gaussian decay (\ref{eq:Fgau}). The three curves have the same meaning as in Figure~\ref{fig:1ktopL}.}
\label{fig:2ktop}
\end{figure}
To demonstrate that for $d>1$ we can also get an exponential decay of the fidelity, we also consider a two dimensional system. We look at two ($d=2$)
coupled tops $\vec{S}_1$ and $\vec{S}_2$ described by a propagator
\begin{equation}
U_0=\exp{(-\ii \varepsilon S_{\rm z1}S_{\rm z2})} \exp{(-\ii \pi S_{\rm y1}/2)}\exp{(-\ii \pi S_{\rm y2}/2)},
\end{equation}
with a perturbation generated by 
\begin{equation}
W=A_1 \otimes \mathbbm{1} + \mathbbm{1}\otimes A_2,\qquad A=S_{\rm z}^2/2S^2.
\end{equation}
We set $S=1/\hbar=100$, and $\varepsilon=20$ in order to be in a 
fully chaotic regime. The initial state is 
always a direct product of spin coherent states centred at 
$(\varphi_{1,2},\vartheta_{1,2})=(1,1)$ which is subsequently projected on an invariant subspace of dimension ${\cal N}=S(S+1)$ spanned by 
$\{ {\cal H}_{\rm OE} \otimes {\cal H}_{\rm r} \}_{\rm sym}$, where 
${\cal H}_{\rm r}={\cal H} \setminus {\cal H}_{\rm OE}$ and $\{\cdot\}_{\rm sym}$ is a subspace symmetric with respect to the exchange of the two tops. The results of numerical simulation are shown in Figure~\ref{fig:2ktop}. We show only the long-time decay, as the situation in the 
plateau is qualitatively the same as for $d=1$. For sufficiently large perturbation one 
obtains an exponential decay shown in Figure~\ref{fig:2ktop}a, while for smaller 
perturbation we have a Gaussian decay shown in Figure~\ref{fig:2ktop}b. 
Numerical data have been successfully compared with the theory
(\ref{eq:Fexp},\ref{eq:Fgau}) using a classically calculated
$\sigma_{\rm R}=9.2 \cdot 10^{-3}$ together with theoretical $F_{\rm plat}$, and with the ``renormalised'' numerics using the operator $R$, similarly as in Figure~\ref{fig:1ktopL} for a one dimensional system. 

 \section{Regular Dynamics}
\label{sec:denominator}
The procedure of calculating the fidelity decay for the case of
regular dynamics will be similar as for general perturbations in
Section~\ref{sec:regular}. We will use the classical action-angle
variables (for definition see Section~\ref{sec:actions}) and
semiclassical methods to calculate the fidelity plateau as well as its
long time decay. The only difference will be that in contrast to the
case of a general perturbation, where only $\Sigma(t)$ was important
in the BCH form of the echo operator, here the second term involving
$\Gamma(t)$ will also be relevant for long times. The same form of the
echo operator (\ref{eq:Md_res}) has already been used for the case of
a residual perturbation in mixing systems as described in the
previous section. The existence of action-angle variables enables us
to expand the classical limit $v$ of the quantum perturbation generator $V$ into Fourier series
\begin{equation}
v(\vec{j},\vec{\theta}) := \sumn{v_{\vec{m}}(\vec{j}) 
{\rm e}^{\ii\vec{m}\cdot\vec{\theta}}}.
\label{eq:fourierV}
\end{equation}
The fact that the perturbation is residual is reflected in the zeroth
Fourier coefficient which is zero, $v_\vec{0}\equiv 0$, so we
explicitly excluded this term from the summation. Classically this
means that the perturbation only changes the shapes of tori as the
average change of the frequency along the unperturbed tori is
zero. For the explanation of the decay of classical fidelity in
case of a residual perturbation see~\citep{Benenti:03}. Similarly, we
can expand the classical limit $w(\vec{j},\vec{\theta})$ of a quantum observable $W$,
\begin{equation}
w(\vec{j},\vec{\theta}) := \sumn w_{\vec{m}}(\vec{j}) 
{\rm e}^{\ii\vec{m}\cdot\vec{\theta}}.
\label{eq:fourierW}
\end{equation}
The zeroth Fourier mode $w_\vec{0}$ can be set to zero as it cancels in the definition of $v$ in terms of $w$. 
Using these expansions we can easily calculate the leading semiclassical forms of $\Sigma(t)$ and $\Gamma(t)$. For $\Sigma(t)$ (\ref{eq:S_W}) we have
\begin{equation}
\Sigma(t)\cong  w(\vec{J},\vec{\Theta}+\vec{\omega}(\vec{J})t)-w(\vec{J},\vec{\Theta}) \cong \sumn{w_\vec{m}(\vec{J}) ({\rm e}^{\ii \vec{m}\cdot \vec{\omega}(\vec{J})t}-1){\rm e}^{\ii \vec{m}\cdot \vec{\Theta}}}.
\label{eq:S_w}
\end{equation}
Note that this is still an operator (capital $\vec{J}$ and $\vec{\Theta}$) and is correct in the leading semiclassical order (the sign $\cong$). Coefficients $w_\vec{m}$ can be also expressed in terms of $v_\vec{m}$ as
\begin{equation}
w_\vec{m}(\vec{j})=-\ii {\rm e}^{-\ii \vec{m}\cdot\vec{\omega}(\vec{j})t/2}\frac{v_\vec{m}(\vec{j})}{2\sin{(\frac{1}{2}\vec{m}\cdot\vec{\omega}(\vec{j}))}}.
\label{eq:w_v}
\end{equation}
In the continuous time case we have $w_\vec{m}=-\ii v_\vec{m}/\vec{m}\cdot\vec{\omega}$. 
\par
The operator $\Sigma(t)$ will determine the plateau. Long time decay
on the other hand will be dictated by the operator $\Gamma(t)$. For
long times when this decay will take place we can define an average
$\oG$ (\ref{eq:oG_def}) and approximate $\Gamma(t) \approx \oG
t$. This approximation is justifiable as we will see, because the
fidelity decay will happen on a long time scale $\sim 1/\delta^2$,
whereas the average of $\Gamma(t)$ converges in a much shorter classical averaging time $t_{\rm ave}$. As one can see from the definition of $\Gamma(t)$ in Eq.~(\ref{eq:G_R}), the average $\oG$ is nothing but the time averaged operator $R$. For $R$ the semiclassical limit $r$ can be calculated using the Poisson brackets instead of commutators, $r=-\{ w,w(1)\}$. When we average $r$ over time, only the zeroth Fourier mode survives resulting in
\begin{equation}
\oG(\vec{J}) \cong \bar{r}(\vec{J}),\qquad \bar{r}(\vec{j})=-\sumn{\vec{m}\cdot\partial_\vec{j}\left\{\, |w_\vec{m}(\vec{j})|^2\sin{\left[\vec{m}\cdot\vec{\omega}(\vec{j})\right]}\right\}\,}.
\label{eq:oG_w}
\end{equation}
In the continuous time case we would have $\bar{r}= -\sumn{\vec{m}\cdot\partial_\vec{j}\left\{|w_\vec{m}|^2 \, \vec{m}\cdot\vec{\omega}\right\}}$.
\par
From the equation for $w_\vec{m}$ (\ref{eq:w_v}) in terms of
$v_\vec{m}$ we see that the denominator
$\sin{(\vec{m}\cdot\vec{\omega}/2)}$ could cause problems at points of
vanishing $\vec{m}\cdot\vec{\omega}(\vec{j})=\vec{0}$. This divergence
carries over to all operators like $\oG$ or $\Sigma(t)$. In classical
perturbation theory this so-called ``small denominator'' problem is
well known and cannot be avoided. For our quantum mechanical
calculation though, there is an easy remedy. It is sufficient to
remember that we are dealing with a residual perturbation, i.e. the
one for which all matrix elements $V_{jk}$ are zero if the
corresponding eigenphases are equal, $\phi_j=\phi_k$. But the term
$\vec{m}\cdot\vec{\omega}$ is nothing else than the semiclassical
approximation for $\phi_j-\phi_k$. Therefore, one can see that
actually all the diverging terms have to be excluded when evaluating
expectation values. This is furthermore confirmed if we compare the
semiclassical expression for $\oG$ (\ref{eq:oG_w}) with the matrix
elements of operator $\Gamma(t)$ (\ref{eq:Gamma_matrix}). An example
of such a calculation will be presented when describing the decay for random initial states. 
\par
Using semiclassical expressions for $\Sigma(t)$ and $\oG$ we can write the echo operator as
\begin{equation}
\Md=\exp{\left\{-\frac{\ii}{\hbar}\left( \Sigma(t) \delta+\frac{1}{2}\oG t \delta^2 \right) \right\}},
\label{eq:Md_resreg}
\end{equation}
with $\Gamma(t)$ and $\oG$ given in
Eqs.~(\ref{eq:S_w},\ref{eq:oG_w}). The third order term in the BCH
form of the echo operator (\ref{eq:Md_resreg}) can be shown to grow no
faster than $\delta^3 t$ and can therefore be neglected. For times
smaller than $t_2$ (specified later) the term involving $\oG$ can be
neglected and the fidelity will exhibit freeze. The fidelity amplitude
of the plateau is $\fn_{\rm plat}\cong \ave{\exp{(-\ii \delta
\Sigma(t)/\hbar)}}$ with semiclassical $\Sigma(t)$. Furthermore, for
$t>t_1$ the fidelity plateau is time independent and can be calculated
by averaging the fidelity amplitude over time
\begin{equation}
f_{\rm plat} \cong \lim_{t \to \infty}{\frac{1}{t}\sum_{t'=0}^t{f(t')}}.
\end{equation}
Time averaging the classical version of
$\Sigma(t)=\Sigma(\vec{j},\vec{\theta}(t))$ (\ref{eq:S_w}) is
equivalent to averaging over the angle $\vec{\theta}$, resulting in the fidelity plateau
\begin{equation}
f_{\rm plat} \cong \ave{\exp{\left( \ii \frac{\delta}{\hbar} w(\vec{J},\vec{\Theta}) \right)} \int{\frac{{\rm d}^d\vec{x}}{(2\pi)^d} \exp{\left( -\ii\frac{\delta}{\hbar} w(\vec{J},\vec{x}) \right)} } },
\label{eq:fplat_reg}
\end{equation}
where we assumed ergodicity of $\vec{m}\cdot\vec{\omega}t$ so we could replace a sum over $t$ with an integral over angles. 
\par
For long times $t>t_2$ the second term in the echo operator becomes dominant. If the plateau is small ($\delta < \hbar$), the first term with $\Sigma(t)$ can be neglected and the fidelity is
\begin{equation}
\Fn=\left|\ave{\exp{\left\{-\ii \frac{\delta^2 t}{2\hbar} \oG(\vec{J}) \right\} }} \right|^2.
\label{eq:F_long}
\end{equation}
For strong perturbations, when the plateau is not negligible, both terms should be retained. But as opposed to the mixing situation where we could factorize the two contributions, here one can not make any general statements. In the following theoretical calculations we will focus on the case when the plateau can be neglected in comparison with the long time decay. The formula (\ref{eq:F_long}) is completely analogous to the long time fidelity decay for a non-residual perturbation and we can use the same ASI representation in terms of an integral over classical actions. The role of $\oV$ is played by $\oG$ and the perturbation is here $\delta^2/2$. The fidelity amplitude in the ASI formulation is therefore 
\begin{equation}
\fn \cong \hbar^{-d}\int{{\rm d}^d \vec{j} \exp{\left(-\ii \frac{\delta^2 t}{2\hbar} \bar{r}(\vec{j})\right)} \,d_\rho(\vec{j})},
\label{eq:res_ASI}
\end{equation}
with $d_\rho(\vec{j})$ being the classical limit of $D_\rho(\hbar \vec{n})=\bracket{\vec{n}}{\rho}{\vec{n}}$ and recall that $\bar{r}$ is the classical limit of $\oG$ (\ref{eq:oG_w}). The ASI representation is valid in a time range $t_1<t<t_{\rm a}$, where the upper limit $t_{\rm a}$ is determined by the variation of the phase over one Planck's cell,
\begin{equation}
t_{\rm a}=\frac{2}{|\partial_j \bar{r}|_{\rm ef}}\frac{1}{\delta^2}\sim \hbar^0 \delta^{-2}.
\label{eq:res_t*}
\end{equation}
\par
Before going on with the evaluation of the plateau (\ref{eq:fplat_reg}) or of the long time decay (\ref{eq:res_ASI}) for random and coherent initial states, let us justify why we were allowed to time average the plateau.   

\subsubsection{Justification of Time Average Plateau}
\label{sec:justify}
By expanding the echo operator into power series over $\delta$, the calculation of the plateau is turned into calculation of expectation values of moments $\Sigma^k(t)$,
\begin{equation}
\Sigma^k(t)\cong \sum_{\vec{m}_1,\ldots,\vec{m}_k \neq \vec{0}}{\prod_{l=1}^k{ w_{\vec{m}_l}(\vec{J}) ({\rm e}^{\ii \vec{m}_l\cdot\vec{\omega}(\vec{J})t}-1){\rm e}^{\ii \vec{m}_l \cdot \vec{\Theta}} }}.
\label{eq:Sk}
\end{equation}
We can average over the initial density matrix $\rho$ with matrix elements $\rho_{\vec{n},\vec{n}'}$ in the action eigenbasis. Taking into account that the exponential of operator $\vec{\Theta}$ acts like a shift operator (\ref{eq:shift}) and that eigenvalues of $\vec{J}$ are $\hbar(\vec{n}+\vec{\alpha})$, we see that the expectation value will be a sum of terms of the form
\begin{equation}
\sum_{\vec{m}_1,\ldots,\vec{m}_k \neq \vec{0}} \sum_{\vec{n}}{ g(\hbar \vec{n}) {\rm e}^{\ii \vec{m}\cdot\vec{\omega}(\hbar \vec{n})t} \rho_{\vec{n},\vec{n}+\vec{m}'} },
\label{eq:sumsum}
\end{equation}
where the function $g$ depends on indices $\vec{m}_l$, index
$\vec{m}':=\sum_l^k{\vec{m}_l}$ and $\vec{m}:=\sum_l^k{ s_l
\vec{m}_l}$ with $s_l=0$ or $1$ depending on which terms we take from
the product in Eq.~(\ref{eq:Sk}). We also neglected Maslov indices as
they are negligible in the leading semiclassical order. For our
derivation only the inner sum over $\vec{n}$ is important. Let us consider two cases of initial states $\rho$, random initial state and coherent initial state.
\par
For a random initial state we average over a random ensemble,
resulting in $\rho_{\vec{n},\vec{m}} \to
\delta_{\vec{n},\vec{m}}/{\cal N}$. From this we immediately conclude
that index $\vec{m}'$ in Eq.~(\ref{eq:sumsum}) must be zero, and as a
consequence also $\vec{m}= \vec{0}$ is zero. Therefore, for random
initial states only terms with all $\vec{m}_l \equiv \vec{0}$ survive
the averaging over random ensemble (which is the same as if we would
average over time instead of over random ensemble). The time scale
$t_1$ after which approximation with a time average is permissible is
determined just by frequencies, $t_1 \sim 1/|\vec{m}\cdot
\vec{\omega}|$ and does not depend on $\hbar$ or perturbation strength
$\delta$, i.e. for random initial states we have
\begin{equation}
t_1 \sim \delta^0 \hbar^0.
\label{eq:t1_RIS}
\end{equation}  
\par
For localized initial states a little more work is needed to show that
the expectation value equals the time average. For definiteness we
will consider coherent initial states, for which the matrix elements
have a Gaussian distribution (\ref{eq:coh}). Furthermore, we will assume that
the number of relevant Fourier modes is smaller than the width (in
$\vec{n}$) of the initial packet. As the width of the packet grows as $\hbar^{-1/2}$ this is justifiable in the semiclassical limit provided $w(\vec{j},\vec{\theta})$ is sufficiently smooth, so that its Fourier coefficients decrease sufficiently fast. In this approximation we have
\begin{equation}
\rho_{\vec{n},\vec{n}+\vec{m}'} \cong d_\rho(\hbar \vec{n}) {\rm e}^{\ii \vec{m}'\cdot\vec{\theta}^*},
\end{equation}
with $d_\rho$ a classical limit of the quantum structure function (\ref{eq:drho_coh}). The inner summation over quantum actions $\hbar\vec{n}$ in Eq.~(\ref{eq:sumsum}) can now be replaced with an integral over the classical action $\vec{j}$ and the method of stationary phase can be used to calculate the resulting integral. Expanding frequencies around the centre of the packet $\vec{j}^*$, $\vec{\omega}(\vec{j}^*+\vec{x})=\vec{\omega}(\vec{j}^*)+\Omega\vec{x}+\cdots$, where 
\begin{equation}
\Omega_{jk}:=\frac{\partial \omega_j(\vec{j}^*)}{\partial j_k},
\end{equation}
is a matrix of frequency derivatives, we can  calculate the sum
\begin{eqnarray}
&&\sum_{\vec{n}} g(\hbar\vec{n}) {\rm e}^{\ii\vec{m}\cdot\vec{\omega}(\hbar\vec{n})t} d_\rho(\hbar\vec{n})
\cong \hbar^{-d}\int {\rm d}^d\vec{j} g(\vec{j}) {\rm e}^{\ii\vec{m}\cdot\vec{\omega}(\vec{j})t} d_\rho(\vec{j})\nonumber\\
&&\cong g(\vec{j}^*){\rm e}^{\ii\vec{m}\cdot\vec{\omega}(\vec{j}^*)t} \left(\frac{\hbar}{\pi}\right)^{d/2}
\!\!\!\left|\det\Lambda\right|^{1/2}
\int {\rm d}^d \vec{x} \exp\left(-\frac{1}{\hbar} \vec{x}\cdot \Lambda \vec{x} + \ii t \vec{m}\cdot\Omega\vec{x}\right)
\nonumber\\
&&= g(\vec{j}^*){\rm e}^{\ii\vec{m}\cdot\vec{\omega}(\vec{j}^*)t}
\exp\left(-\frac{\hbar t^2}{4}\vec{m}\cdot \Omega \Lambda^{-1} \Omega^{\rm T}\vec{m}\right).
\label{eq:gauss}
\end{eqnarray}
We see that all terms with $\vec{m}\neq \vec{0}$ decay to zero. The
longest decay time scale of Gaussian envelopes is estimated as
\begin{equation}
t_1 = \left( \frac{\hbar}{4} \min_{\vec{m}\neq\vec{0}}\left(\vec{m}\cdot \Omega \Lambda^{-1}\Omega^{\rm T}\vec{m}\right)\right)^{-1/2}
\propto \hbar^{-1/2}.
\label{eq:t1}
\end{equation}
This means that for times longer than $t_1$ only term with zero $\vec{m}$ will survive which is equivalent to time averaging the classical expression. 
Note that the Gaussian decay (\ref{eq:gauss}) is absent if $\Omega=0$,
{\em e.g.} in the case of a $d$-dimensional harmonic oscillator. Fidelity decay for a residual perturbation in a harmonic oscillator will be discussed for a Jaynes-Cummings model in Section~\ref{sec:har_freeze}. There may also be a general problem 
with the formal existence of the scale $t_1$ (\ref{eq:t1}) if the derivative matrix $\Omega$ is singular, but this may not actually affect the fidelity for 
sufficiently fast converging or finite Fourier series
(\ref{eq:fourier}). Time $t_1$ can be interpreted as the {\em
integrable Ehrenfest time}, up to which quantum-classical
correspondence will hold. After $t_1$ a quantum wave packet of size
$\sim\sqrt{\hbar}$ will spread over a classically large region of size
$\sim \hbar^0$ and interference phenomena will become important. This
will also be reflected in the fidelity. As the perturbation $w_\vec{m}$ will couple different tori, the echo packet will also start to exhibit interferences after $t_1$ and therefore we can expect agreement between quantum and classical fidelity only up to time $t_1$. This must be contrasted with the case of a general non-residual perturbation, where there were no upper limits on the correspondence between quantum and classical fidelity for localized initial packets under certain conditions.

\subsection{The Numerical Model}
\label{sec:freeze_regular}
For the sake of numerical demonstration in the present section of residual perturbations in a regular dynamics we take an integrable kicked top model with a slightly different unperturbed one step propagator
\begin{equation}
U_0=\exp{\left( -\ii \frac{\alpha}{2} S \left\{\frac{S_{\rm z}}{S}-\beta \right\}^2 \right) },
\label{eq:KT_reg}
\end{equation}
where we introduced a second parameter $\beta$. The classical map corresponding to $U_0$ is simply a twist around ${\rm z}$-axis
\begin{eqnarray}
x_{t+1}&=&x_t \cos{(\alpha (z_t-\beta))}-y_t\sin{(\alpha (z_t-\beta))} \nonumber \\
y_{t+1}&=&y_t \cos{(\alpha (z_t-\beta))}+x_t\sin{(\alpha (z_t-\beta))} \\
z_{t+1}&=&z_t. \nonumber
\label{eq:Cmap}
\end{eqnarray}
Unperturbed evolution represents a continuous time system with the
classical Hamiltonian $h_0$ generating a frequency field $\omega(j)$
\begin{equation}
h_0(j) = \frac{1}{2}\alpha (j-\beta)^2,\qquad \omega(j) = \alpha (j-\beta).
\label{eq:h0}
\end{equation}
Here we used a canonical transformation from a unit-sphere to an action-angle pair 
$(j,\theta)\in [-1,1]\times [0,2\pi)$, namely
\begin{equation}
x = \sqrt{1-j^2}\cos\theta,
\qquad
y = \sqrt{1-j^2}\sin\theta,
\qquad
z = j.
\qquad
\label{eq:can}
\end{equation}
\par
We perturb the Hamiltonian by periodic kicking with a transverse pulsed field in $x$ direction,
\begin{equation}
h_\delta(j,\theta,\tau) = \frac{1}{2}\alpha (j-\beta)^2 + 
\delta \sqrt{1-j^2}\cos\theta \sum_{k=-\infty}^\infty \delta(\tau-k).
\end{equation}
Perturbed quantum evolution is given by a product of two unitary propagators
\begin{equation}
U_\delta = U_0\exp{(-\ii \delta S_{\rm x})},
\label{eq:Jx}
\end{equation}  
so the perturbation generator is 
\begin{equation}
V = S_{\rm x}/S.
\label{eq:V}
\end{equation}
The classical perturbation has only one Fourier component, namely
\begin{equation}
v(j,\theta) = \sqrt{1-j^2}\cos\theta,\quad v_{\pm 1}(j) = \frac{1}{2}\sqrt{1-j^2},
\label{eq:VFour}
\end{equation}
whereas zeroth Fourier component is zero, $v_0\equiv 0$, indicating that we have a residual perturbation, $\bar{v}=0$ and $\bar{V}=0$. The classical limit $w$ of $W$ is also readily calculated, giving
\begin{equation}
w(j,\theta)=\frac{1}{2}\sqrt{1-j^2}\,\frac{\sin{(\theta-\omega/2)}}{\sin{(\omega/2)}}, 
\label{eq:w_clas}
\end{equation}
with $\omega=\alpha(j-\beta)$. The reason we introduced parameter $\beta$ is to be able to control a possible singularity in $w(j,\theta)$ at points where $\omega=0$.

   \subsection{Coherent Initial State}
\label{sec:CIS_V0}
\subsubsection{The Plateau}
For times $t<t_1$ quantum fidelity will follow the classical fidelity
and will be system specific. For times $t_1<t<t_2$ quantum fidelity
will exhibit the plateau while the classical fidelity will continue to
decay. To calculate the quantum plateau we have to evaluation the time-average formula for the plateau (\ref{eq:fplat_reg}). We shall make use 
of the fact that for coherent states we have the expectation value
\begin{equation}
\ave{\exp(-(\ii\delta/\hbar)g(\vec{J},\vec{\Theta}))} \cong
\exp(-(\ii\delta/\hbar)g(\vec{j}^*,\vec{\theta}^*)).
\end{equation}
for some smooth function $g$, provided that the size of the wave-packet
$\sim\sqrt{\hbar}$ is smaller than the oscillation scale of the
exponential $\sim \hbar/\delta$, {\em i.e.} provided $\delta \ll \hbar^{1/2}$.
Then the squared modulus of $f_{\rm plat}$ (\ref{eq:fplat_reg}) reads
\begin{equation}
F_{\rm plat}^{\rm CIS} \cong \frac{1}{(2\pi)^{2d}}
\left|\int {\rm d}^d\vec{\theta} \exp\left(-\frac{\ii\delta}{\hbar}w(\vec{j}^*,\vec{\theta})\right)\right|^2.
\label{eq:plateauCS}
\end{equation}
It is interesting to note that the angle $\vec{\theta}^*$ 
does not affect the probability $F_{\rm plat}$ as it only rotates the
phase of the amplitude $f_{\rm plat}$. Note the similarity of this
result with the plateau for coherent initial states and mixing
dynamics, Eq.~(\ref{eq:plat_CIS}). In the mixing case we had an
equilibrium average of $w$ in $2d$ dimensional phase space, whereas
for regular dynamics we have an average over only $d$ dimensional
angle-space of $\vec{\theta}$. For weak perturbation $\delta<\hbar$
only the linear response expression for the plateau is needed. Expanding general $F_{\rm plat}$ we obtain
\begin{equation}
1-F_{\rm plat}^{\rm CIS}=\frac{\delta^2}{\hbar^2}\nu_{\rm CIS},\qquad \nu_{\rm CIS}= \sumn{|w_\vec{m}(\vec{j}^*)|^2}.
\label{eq:nuCIS}
\end{equation}
\par
\begin{figure}[h!]
\centerline{\includegraphics{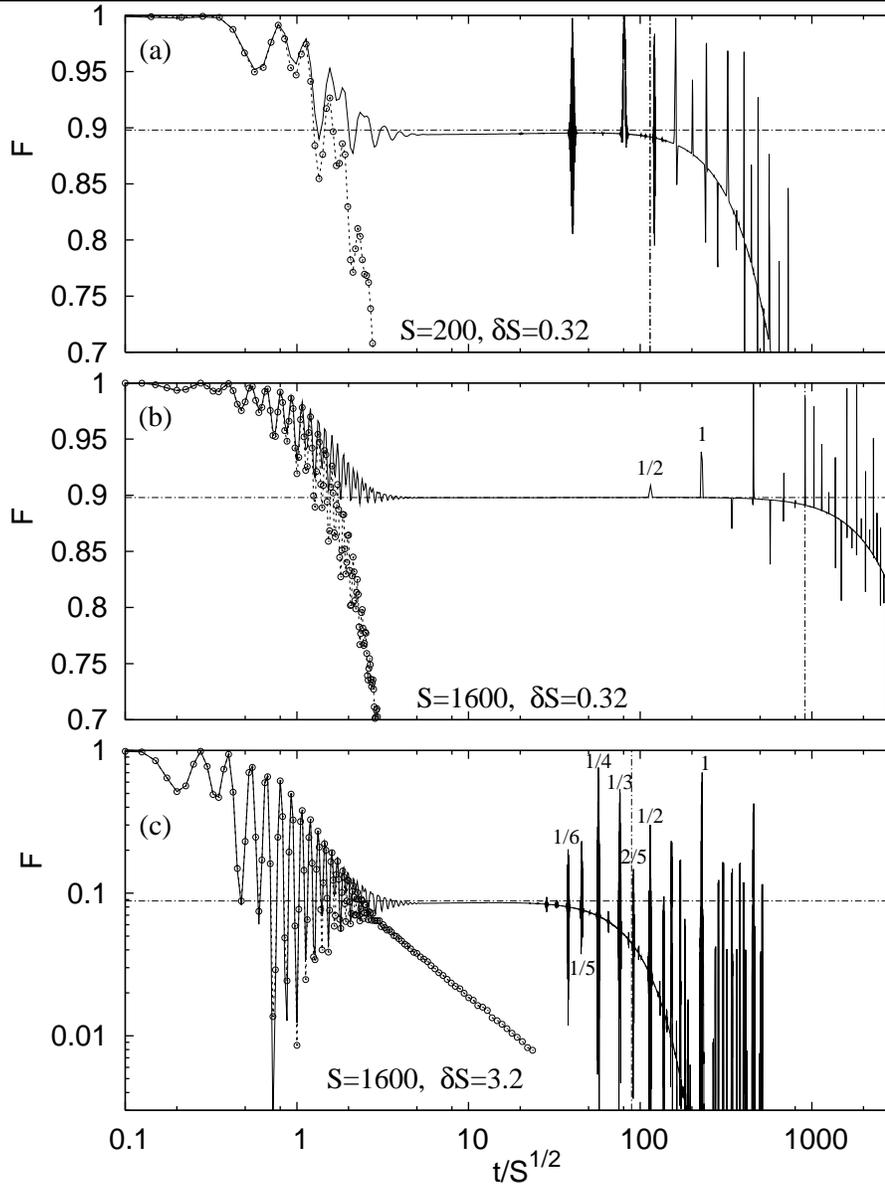}}
\caption{Short time decay of the fidelity for a quantized top is shown for the coherent initial state,
for $S=200$ (a), and $S=1600$ (b), with a fixed product $\delta
S=0.32$, where the plateau is well described by linear response. In
(c) we show $S=1600$ and a stronger perturbation with 
$\delta S = 3.2$.
Note that the time axis is rescaled as $t/t_1$.
Symbols connected with dashed lines denote the corresponding classical fidelity. The horizontal chain line 
denotes the theoretical value of the plateau (\ref{eq:bessel}), while the vertical chain line denotes the estimated 
theoretical value for $t_2$ (\ref{eq:t2coh}). In (b,c) we also indicate fractional $2\pi k/p$ resonances with 
$k/p$ marked on the figure (see text for details).} 
\label{fig:shortCoh}
\end{figure}   
Results of the numerical simulation for $\beta=0$, $\alpha=1.1$ are shown in Figure~\ref{fig:shortCoh}. For our choice of the perturbation (\ref{eq:w_clas}) the integral in the semiclassical expression of the plateau (\ref{eq:plateauCS}) is readily calculated. Actually, the calculation of $F_{\rm plat}$ can be analytically carried out for any perturbation with a single nonzero Fourier mode $\pm \vec{m}_0$, with the result
\begin{equation}
F_{\rm plat}^{\rm CIS}=J_0^2\left(2\frac{\delta}{\hbar} |w_{\vec{m}_0}(\vec{j}^*)|\right),
\end{equation}
were $J_0$ is the zero order Bessel function. For a more general multi-mode perturbations we have to evaluate the integral 
(\ref{eq:plateauCS}) numerically. For our single mode perturbation $w$ (\ref{eq:w_clas}) we have
\begin{equation}
F_{\rm plat}^{\rm CIS}=J_0^2\left(\delta S \frac{\sqrt{1-j^{*2}}}{2\sin{(\alpha j^*/2)}} \right).
\label{eq:bessel}
\end{equation}
The linear response expansion of this general result reads
\begin{equation}
1-F_{\rm plat}^{\rm CIS}=(\delta S)^2 \nu_{\rm CIS},\qquad \nu_{\rm CIS}=\frac{1-j^{*2}}{8\sin^2{(\alpha j^*/2)}}.
\label{eq:plat_LRcoh}
\end{equation}
\par
For numerical illustration in Figure~\ref{fig:shortCoh} the initial coherent packet has been placed at
$(\vartheta^*,\varphi^*)=(1,1)$. We can see that until time $t_1$
(\ref{eq:t1}) quantum fidelity follows the classical one (circles in Figure~\ref{fig:shortCoh}). After $t_1$ quantum fidelity freezes at the plateau, whose value nicely agrees with the linear response formula (\ref{eq:plat_LRcoh}) or with the full expression (\ref{eq:bessel}) for strong perturbation. Vertical chain lines show theoretical values of $t_2$, which is the time when the plateau ends. Also, at certain times the quantum fidelity exhibits resonances, i.e. strong revivals of fidelity. These ``spikes'' occurring at regular intervals are prominent also in a long time decay of fidelity in Figure~\ref{fig:longCoh}. These will be called the {\em echo resonances} and are 
particular to one-dimensional systems.

\subsubsection{Long Time Decay}
\begin{figure}[h!]
\centerline{\includegraphics{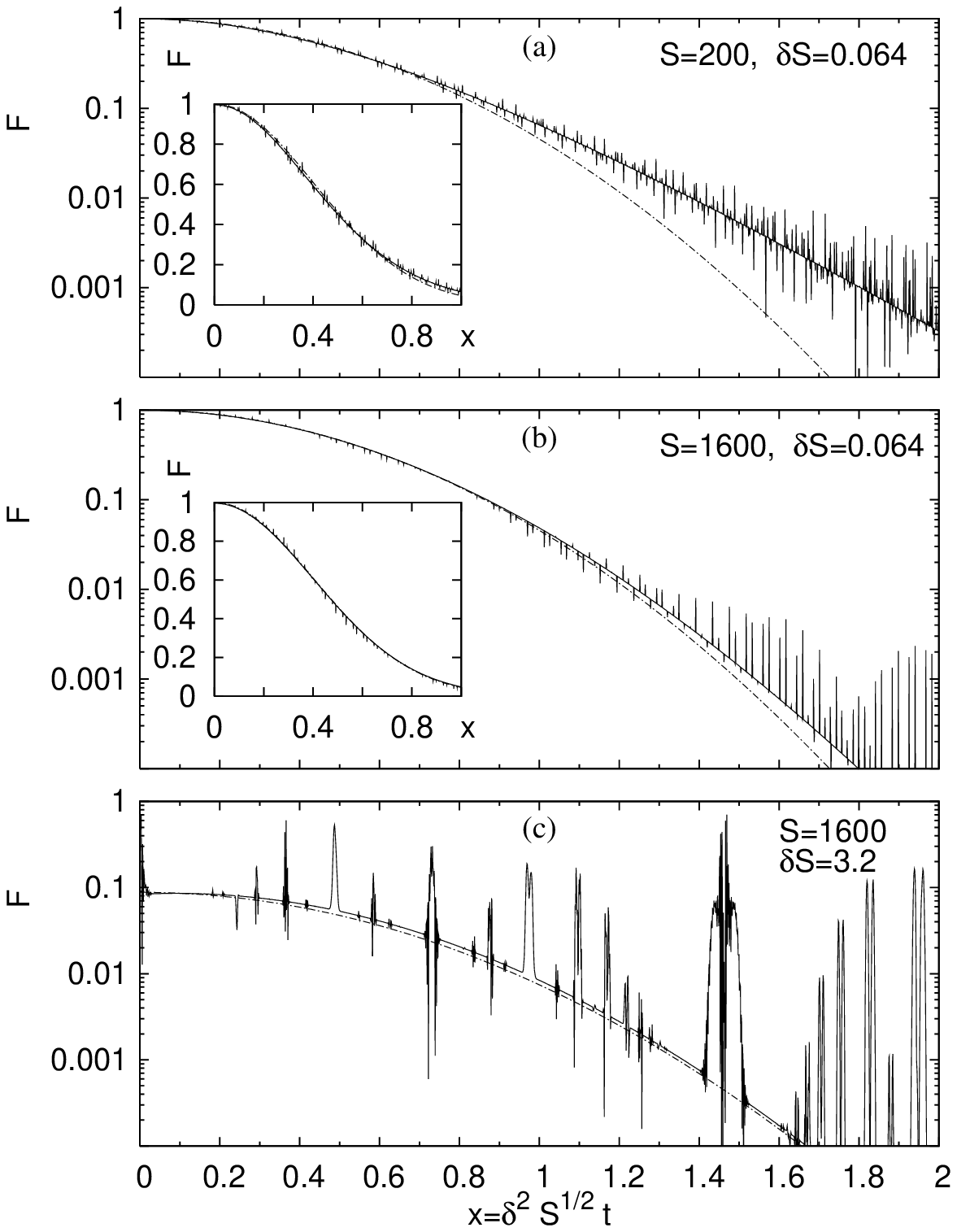}}
\caption{Long time ballistic decay of the fidelity for the kicked top with coherent initial state is shown 
for cases $S=200$ (a), and $S=1600$ (b), of weak perturbation $\delta S=0.064$,
and for strong perturbation $S=1600, \delta S=3.2$ (c).
Chain curves indicate a theoretical Gaussian (\ref{eq:res_long}) with analytically computed coefficients,
except in case (c), where we multiply the theoretical Gaussian decay by a prefactor $0.088$ being equal to 
the theoretical value of the plateau (\ref{eq:bessel}), and re-scale the exponent of the Gaussian by a factor $0.8$
taking into account the effect of a non-small first term in the exponent of (\ref{eq:Md_resreg}). 
Note that in the limit $S \to \infty$ the agreement with semiclassical theory improves and that the size of 
the resonant spikes is of the same order as the drop in the linear response plateau. The insets show the
data and the theory on the normal scale.}
\label{fig:longCoh}
\end{figure}
To obtain long time decay of quantum fidelity one has to evaluate the ASI in Eq.(\ref{eq:res_ASI}). Everything is analogous to the case of general non-residual perturbation described in Section~\ref{sec:regular}. In the formula for $F(t)=\exp{(-(t/\tau_{\rm r})^2)}$ (\ref{eq:Fn_regcoh}) we only have to replace the perturbation $\delta$ with $\delta^2/2$ and $\bar{v}$ with $\bar{r}$, so that the fidelity decay is
\begin{equation}
\Fn\cong \exp{\left\{ -(t/\tau_{\rm rr})^2\right\}} ,\qquad \tau_{\rm rr}=\frac{1}{\delta^2}\sqrt{\frac{8\hbar}{\vec{\bar{r}}'\cdot\Lambda^{-1}\vec{\bar{r}}'}},
\label{eq:res_long}
\end{equation}
with the semiclassical $\bar{r}$ given in Eq.~(\ref{eq:oG_w}) and its
derivative $\vec{\bar{r}}':=\partial_\vec{j}\bar{r}$. The decay time
scales as $\tau_{\rm rr}\sim \hbar^{1/2} \delta^{-2}$ and is thus
smaller than the upper limit $t_{\rm a}\sim \hbar^0 \delta^{-2}$ of
the validity of the stationary phase approximation used in deriving the Gaussian decay. Remember that the above formula is valid only if the plateau is small and the $\Sigma(t)$ term can be neglected, i.e. for $\delta < \nu_{\rm CIS}^{-1/2}\hbar$. For such small $\delta$ the crossover time $t_2$ from the plateau to the long time decay can be estimated by comparing the linear response plateau (\ref{eq:plat_LRcoh}) with the long time decay (\ref{eq:res_long}),  $(\delta/\hbar)^2 \nu_{\rm CIS} \approx (t_2/\tau_{\rm rr})^2$, resulting in $t_2\approx \tau_{\rm rr} \delta \sqrt{\nu_{\rm CIS}}/\hbar\sim\hbar^{-1/2}\delta^{-1}$. For stronger perturbations, namely up to $\delta \sim \sqrt{\hbar}$ time $t_2$ can be simply estimated by $\tau_{\rm rr}$. We therefore have
\begin{equation}
t_2 = \min\{1,\frac{\delta}{\hbar}\nu_{\rm CIS}^{1/2}\} \tau_{\rm rr} = 
\min\{{\rm const}\ \hbar^{1/2}\delta^{-2},{\rm const}\ \hbar^{-1/2}\delta^{-1}\}.
\label{eq:t2coh}
\end{equation}
Time scale $t_2$ can be seen in Figure~\ref{fig:shortCoh} as the point of departure of 
fidelity from the plateau value. Using our model with $\alpha=1.1$ and
$\beta=0$ and the position of the initial coherent state at $(\vartheta^*,\varphi^*)=(1,1)$ 
this can be calculated to be $t_2= \min\{0.57 \sqrt{S}/\delta, 0.57/(\delta^2 \sqrt{S}) \}$ (similarity of numerical prefactors is just a coincidence). These theoretical positions of $t_2$ are shown with vertical chain lines in Figure~\ref{fig:shortCoh} and are given by $\tau_{\rm rr}\nu_{\rm CIS}^{1/2}\delta/\hbar$ in Figures~\ref{fig:longCoh}a,b while it is $\tau_{\rm rr}$ for a strong perturbation $\delta S=3.2$ in Figure~\ref{fig:shortCoh}c. Long time decay of fidelity is shown in 
figure \ref{fig:longCoh}. Theoretical Gaussian decay (\ref{eq:res_long}), shown with a chain curve, 
is again confirmed by numerical results. The average $\oG$ given by classical $\bar{r}$ is (\ref{eq:oG_w})
\begin{equation}
\bar{r}=\frac{1}{8\sin^2{(\alpha j/2)}}\left\{\alpha(1-j^2)+2j \sin{(\alpha j)} \right\}.
\end{equation}
The derivative $\bar{r}'$ is
\begin{equation}
\bar{r}'=\frac{1}{8}\left\{ 4\cot{(\alpha j/2)}-\alpha \left[ 4j+\alpha(1-j^2)\cot{(\alpha j/2)} \right]/\sin^2{(\alpha j/2)} \right\},
\label{eq:r'}
\end{equation}
which gives using $\Lambda=1/(1-j^2)$ for spin coherent states the decay time (\ref{eq:res_long}) $\tau_{\rm rr}=0.57 \delta^{-2} \hbar^{1/2}$. Note that we do not have any fitting parameters, except in the
case of a strong perturbation ($\delta S \gg 1$, Figure~\ref{fig:longCoh}c) where the prefactor and the exponent of a Gaussian had to be slightly adjusted due to
the non-negligible effect of the first term in (\ref{eq:Md_resreg}) [see caption for details].
\subsubsection{Average fidelity}
\begin{figure}[ht]
\centerline{\includegraphics{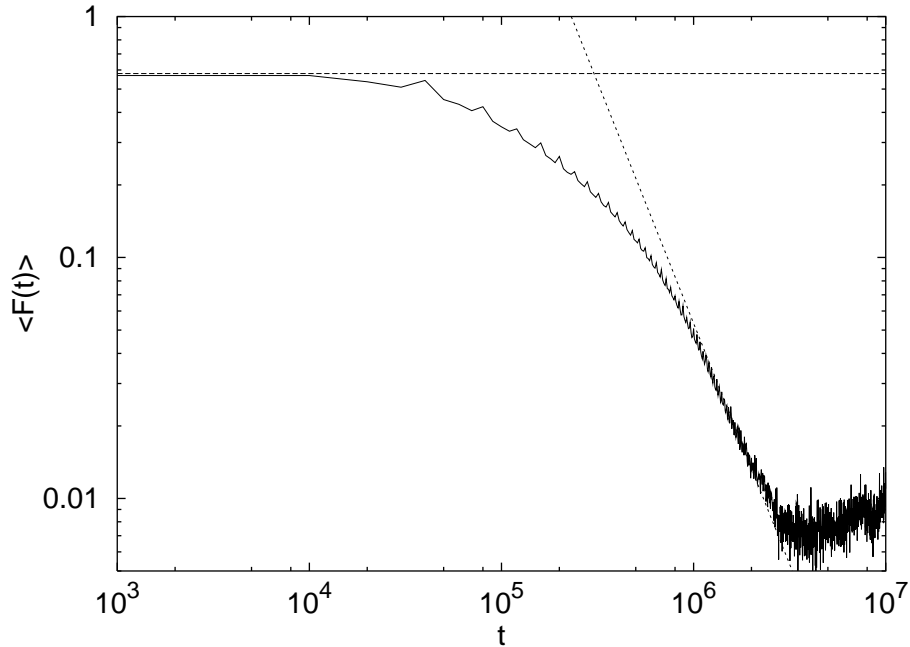}}
\caption{Average fidelity decay $\ave{F(t)}_\vec{j}$ (\ref{eq:avgX})
for the regular kicked top with freezing (no fitting parameters!). The dotted line is the theoretical power law decay (\ref{eq:avgX}). The horizontal line gives the theoretical value of the plateau (\ref{eq:RIS_sing}).}
\label{fig:avgX}
\end{figure}

We should remark that, although we obtain asymptotically Gaussian decay of fidelity for a 
single coherent initial state, one may be interested in an {\em effective fidelity} 
averaged with respect to phase-space positions of the initial coherent state, similarly as for the case of general perturbations (\ref{eq:CIS_avg}). For times $t<t_2$ when we have a plateau, the average fidelity $\ave{\Fn}_{\vec{j}}$ will also have a plateau of the same height as the plateau for a random initial state calculated in Section~\ref{sec:ran_freeze}, Equation~(\ref{eq:RIS_plat}). In the linear response regime this plateau is just twice as large as for a single coherent state. Long time decay of the average fidelity $\ave{\Fn}_{\vec{j}}$ will be given by the phase space average of the regular decay time $\tau_{\rm rr}$. The calculation is very similar to the one for a general perturbation described at the end of Section~\ref{sec:CIS_gen} so we wont repeat is here. Asymptotically we will have a power law decay with the power determined by the order $\eta$ of zeros of $\vec{\bar{r}}'\cdot \Lambda^{-1}\vec{\bar{r}}'$ and dimensionality $d$ (see Eq.~\ref{eq:F_asim}) which is not universal as claimed by some authors~\citep{Jacquod:03}. To demonstrate asymptotic power law decay (\ref{eq:F_asim}) we take the same regular kicked top model (\ref{eq:KT_reg}) as used throughout this section with parameter $\alpha=1.1$, $\beta=0$ and spin size $S=1000$. We take perturbation (\ref{eq:V}) of strength $\delta=5\cdot 10^{-4}$ and average fidelity over $200$ coherent initial states placed randomly over sphere. The results are in Figure~\ref{fig:avgX}.
The $\tau_{\rm rr}^2$, i.e. $\vec{\bar{r}}'\cdot
\Lambda^{-1}\vec{\bar{r}}'$ has for our $\vec{\bar{r}}'$ (\ref{eq:r'})
two zeros of the first order, $\eta=1$, at $j=\pm 1$. The derivative
(i.e. the first nonzero term in the expansion around zero) of the term   $\vec{\bar{r}}'\cdot \Lambda^{-1}\vec{\bar{r}}'$ evaluated at $j=-1$ is
\begin{equation}
c=\frac{2\alpha-\sin{\alpha}}{2\sin^2{(\alpha/2)}}.
\end{equation}
The asymptotic theoretical decay (\ref{eq:F_asim}) should then be
\begin{equation}
\ave{\Fn}_{\vec{j}}\asymp \frac{8}{\delta^4 S t^2 c},
\label{eq:avgX}
\end{equation}
where for $\alpha=1.1$ the coefficient is $c=2.4$. This theoretical decay agrees with numerics in Figure~\ref{fig:avgX}. The theoretical value of the plateau according to the Equation~(\ref{eq:RIS_sing}) for a random initial state should be $0.58$ which also agrees with numerics.

\subsubsection{Illustration in Terms of Wigner Function}


\begin{figure}[h!]
\centerline{\includegraphics[width=110mm]{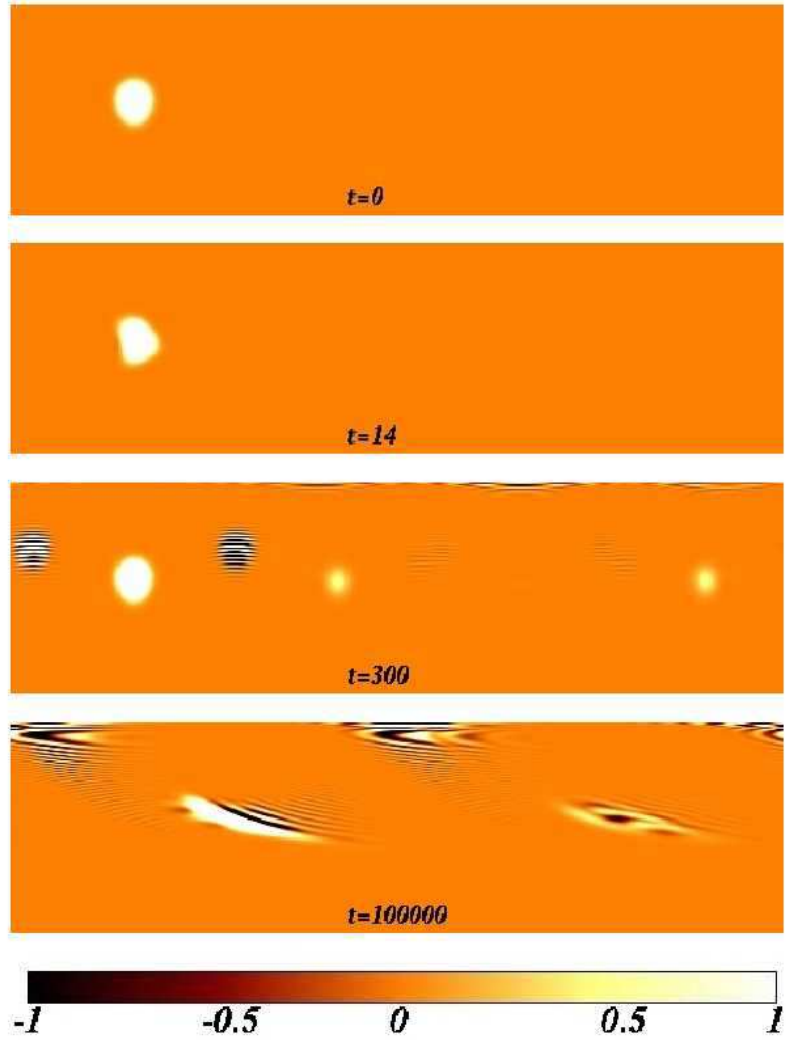}}
\caption{
[Movies online~\citep{movie:reg}]
Snapshots of the Wigner function of echo-dynamics for a quantized top 
$\alpha=1.1,\beta=0,\gamma=0$ with $S=200$ and for $\delta=1.6\cdot 10^{-3}$ 
(same as in Figures~\ref{fig:shortCoh}a,\ref{fig:resonances}a).
The upper hemisphere is shown with $j=\cos{\vartheta}\in[0,1]$ on the
vertical axis and $\theta=\varphi\in [0,2\pi]$ on the horizontal axis.
From top to bottom we show: the initial state at $t=0$, the state at 
$t=14\approx t_1$ when we are around the regular Ehrenfest time, 
at $t=300$ in the middle of the plateau, and at $t=100000$ in the ballistic regime.}
\label{fig:movies}
\end{figure}

All the phenomena described theoretically in the preceding subsections
can be nicely illustrated in terms of the {\em echoed Wigner function} --- 
the Wigner function $W_{\! \rho^{\rm M}}(\cos{\vartheta},\varphi)$ 
of the echo-dynamics. For details regarding the spin Wigner functions see Appendix~\ref{app:wig} and the discussion at the end of Section~\ref{sec:time_scales}. The fidelity $F(t)$ is given simply by the overlap of the echoed Wigner function and the Wigner function of the initial state.
Therefore, the phase-space chart of the echoed Wigner function
contains the most detailed information on echo-dynamics 
and illustrates the essential differences between different regimes of
fidelity decay.
This is shown in Figure~\ref{fig:movies} for the quantized top, see also online movies~\citep{movie:reg}. In the initial {\em classical regime}, $t < t_1$, the echoed Wigner function
has not yet developed negative values and is in point-wise agreement with 
the Liouville density of the classical echo-dynamics.
In the {\em plateau regime}, $t_1 < t < t_2$, the echoed Wigner function
decomposes into several pieces, one of which freezes at the position of the 
initial packet providing significant and constant overlap --- the plateau.
At very particular values of time, namely at the echo resonances, 
different pieces of the echoed Wigner function somehow magically recombine
back into the initial state.
In the asymptotic, {\em ballistic regime}, $t > t_2$, even the frozen piece
starts to drift ballisticly away from the position of the initial packet,
thus explaining a fast Gaussian decay of fidelity.

\subsection{Echo Resonances for Coherent Initial States}

\begin{figure}[h!]
\centerline{\includegraphics{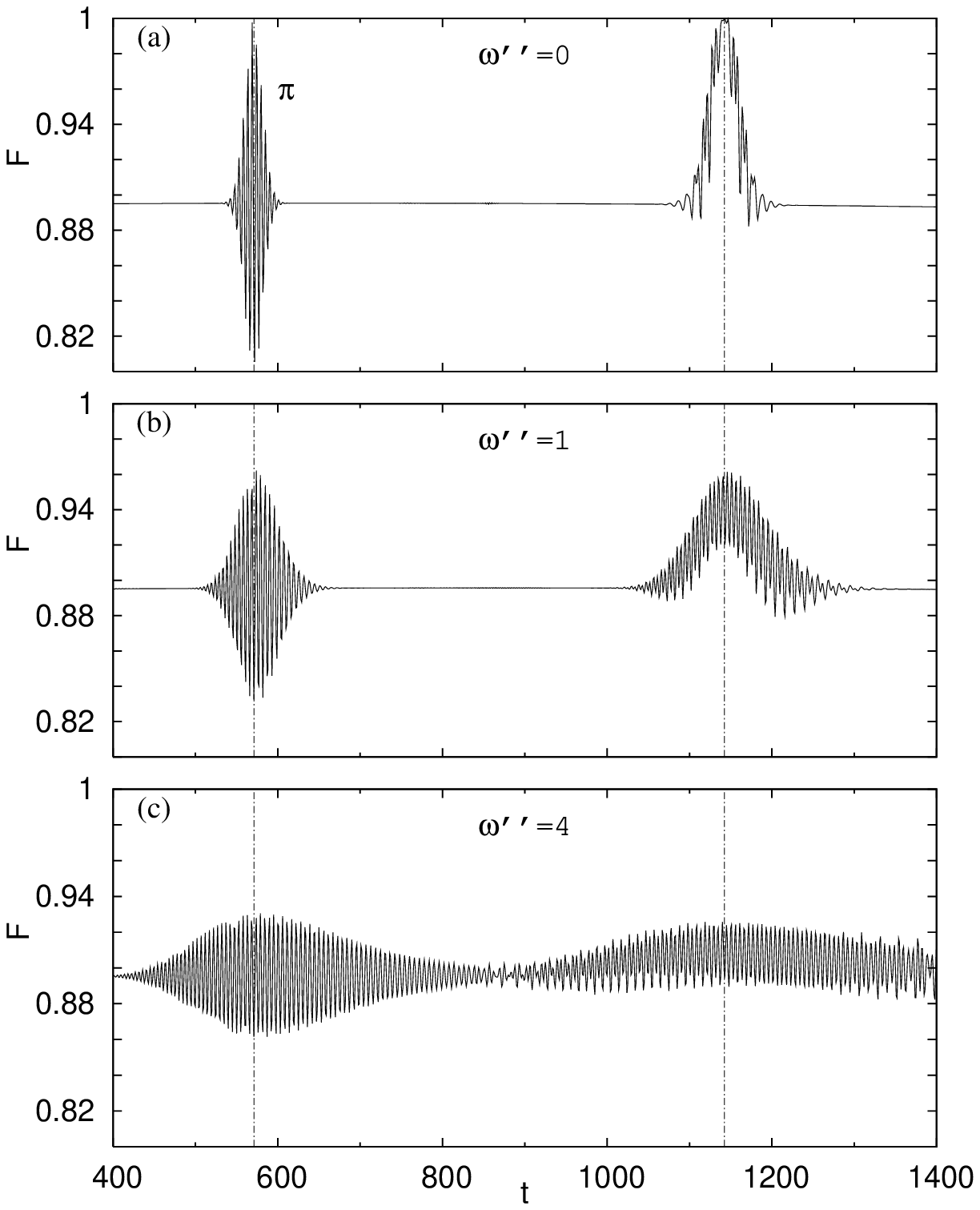}}
\caption{Structure of echo resonances for coherent initial states of the modified quantum top 
$U_0=\exp{(-\ii S [\alpha (S_z/S)^2/2 + \gamma (S_z/S - j^*)^3/6])}$, 
$S=200, \delta=1.6 \cdot 10^{-3}$, for increasing values of $\omega''=\gamma=0$ (a),
$\omega''=1$ (b), and $\omega''=4$ (c), which weakens and broadens the resonances.
Note that in (a), $\omega''=0$ we have the same data as 
in Figure~\ref{fig:shortCoh}a. 
Vertical chain lines show theoretical times $t_{\rm r}/2$ for $\pi$, and 
$t_{\rm r}$ for $2\pi$  resonances.}
\label{fig:resonances}
\end{figure}   

Let us now discuss the behaviour of the fidelity for initial wave-packets 
in the regime of linear response approximation in some more detail. 
We shall consider possible deviations from the time average plateau (\ref{eq:fplat_reg}). Specifically we will explain the resonances observed {\em e.g.} in Figure~\ref{fig:shortCoh}.
\par
For such a resonance to occur the phases of terms $\< \Sigma^k(t)\>$ (\ref{eq:Sk}) have to build up in a
constructive way and this is clearly impossible in a generic case, unless:
\begin{itemize}
\item[(i)] We have one 
dimension $d=1$, so we sum up over a one-dimensional array of integers $n$ in action space\footnote{
In more than one dimension we would clearly need a strong condition on 
commensurability of frequency derivatives over the entire region of the action lattice 
where the initial state is supported.}.
\item[(ii)] The wave-packet is localized over a classically small region of the action 
space/lattice such that a variation of the frequency derivative 
${\rm d}\omega(j)/{\rm d} j$ over this region is sufficiently small. 
\end{itemize}
In this subsection we thus consider a one-dimensional case, $d=1$.  
For simplicity we will focus on a linear response regime, so we will
study time-dependent terms $\ave{\Sigma^2(t)}$ and $\ave{\Sigma(t)}$
with $\Sigma(t)$ given in Eq.~(\ref{eq:S_w}). In the justification of
the time average plateau, Section~\ref{sec:justify}, we replaced a sum
over $n$ with an integral over action space. Here the time is not
small enough and furthermore the very graininess of quantum actions is
the source of resonances and we will have to explicitly consider sums
over $n$. We seek a condition, such that consecutive phases in the 
exponential build up an interference pattern.

\subsubsection{$2\pi$-resonance:}

Let us expand the frequency around the centre of the packet
\begin{equation}
\omega(j) = \omega^* + (j-j^*)\omega' + \frac{1}{2}(j-j^*)^2\omega'' + \ldots
\end{equation}
where $\omega^* = \omega(j^*)$, $\omega' = {\rm d}\omega(j^*)/{\rm d} j$ and $\omega'' = {\rm d}^2\omega(j^*)/{\rm d} j^2$. The phases in the sums of the form (\ref{eq:gauss}) with factors ${\rm e}^{\ii m \omega(\hbar n)}$ come into resonance, 
for $m = 1$ and hence for any higher $m\ge 1$, when they change by $2\pi$ per
quantum number, which happens at time $t_{\rm r}$, 
\begin{equation}
\hbar\omega' t_{\rm r} = 2\pi,\quad t_{\rm r} = \frac{2\pi}{\hbar\omega'},
\label{eq:resonance}
\end{equation}
and its integer multiples. It is interesting to note that
these resonant times correspond precisely to the condition for 
{\em revivals} of the wave-packet in the {\em forward} evolution (apart 
from a phase-space translation) studied in~\citep{Braun:96} and~\citep{Leichtle:96}. In addition, we need that the coherence of linearly increasing 
phases is not lost along the size of the wave-packet, {\em i.e.}
\begin{equation}
\zeta := m\omega'' t \Delta_j^2 = \frac{\hbar m \omega'' t}{2\Lambda} < 2\pi,
\label{eq:ResCond}
\end{equation}
where we denoted by $\Delta_j := \ave{(j-j^*)^2}^{1/2} = \sqrt{\frac{\hbar}{2\Lambda}}$ the action-width of the initial wave-packet. Therefore, if $\zeta \ll 2\pi$ we will observe echo resonances at integer multiples of $2\pi$-resonance time $t_{\rm r}$.
\par 
The shape of the echo resonance can also be derived. Let time $t$ be {\em close} to $k t_{\rm r}, k\in\Z$, and write $t = k t_{\rm r} + t'$ where 
$t' \ll t_{\rm r}$, so that $\hbar \omega' t' \ll 2\pi$. We can estimate the general time dependent sum over $n$ in entirely analogous fashion to Eq.~(\ref{eq:gauss}) by: (i) shifting the time variable
to $t'$, (ii) incorporating the resonance condition, and (iii) 
approximating the resulting sum by an integral, due to the smallness of $t'$,
\begin{eqnarray}
&& \sum_n g(\hbar n) {\rm e}^{\ii m \omega(\hbar n)t} d_\rho(\hbar n) \cong \sum_n g(\hbar n){\rm e}^{\ii m \left(\omega^* + (\hbar n - j^*)\omega' + 
\frac{1}{2}(\hbar n -j^*)^2\omega''\right)(t_{\rm r} + t')}d_\rho(\hbar n) 
\nonumber \\
&=& {\rm e}^{\ii m \omega^* t}\sum_n g(\hbar n) {\rm e}^{\ii m \left((\hbar n - j^*)\omega' t' + \frac{1}{2}(\hbar n - j^*)^2\omega'' t\right)} d_\rho(\hbar n) \nonumber \\
&\cong& 
{\rm e}^{\ii m \omega^* t} g(j^*) \sqrt{\frac{\Lambda}{\pi\hbar}} \int\!{\rm d} j\,{\rm e}^{\ii m \left((j-j^*)\omega' t' + \frac{1}{2}(j-j^*)^2\omega'' t\right) - \frac{\Lambda}{\hbar}(j-j^*)^2} \nonumber \\
&=& 
{\rm e}^{\ii m \omega^* t} g(j^*) 
\frac{1}{\sqrt{1-\ii \zeta}}
\exp\left(-\frac{\hbar m^2\omega'^2 t'^2}{4\Lambda} \frac{1 + \ii \zeta}{1 + \zeta^2}\right).
\label{eq:resatom}
\end{eqnarray}
We see that the resonance has a Gaussian envelope, modulated with an
oscillation frequency $\approx \omega^*$. At the resonance centre
$t'=0$, and assuming $\zeta \ll 1$, we can easily calculate the linear
response terms in the fidelity, getting
$\ave{\Sigma^2(t)}=\ave{\Sigma(t)}^2$. Therefore, for small $\zeta$
the fidelity at a $2\pi$ echo resonance is $1$. We get a perfect
revival of fidelity. For non-negligible $\zeta$, the resonance height
dies as $(1+\zeta^2)^{-1/2}$. As $\zeta$ depends on time $t$, the
resonances will decrease in height with the increasing order $k$ of
the resonance. That is, for the $k$-th resonance occurring at $t=k t_{\rm r}$ we have
\begin{equation}
\zeta=k \frac{\pi m}{\Lambda} \frac{\omega''}{\omega'},
\label{eq:ResCond1}
\end{equation}
and the magnitude of the resonances therefore falls as $\sim 1/k$. We will get strong and numerous resonances, i.e. small $\zeta$, provided that either the second derivative $\omega''$ is small, or the initial state is squeezed such that $\Lambda \gg 1$. For example, if the second derivative vanishes everywhere, $\omega''\equiv 0$, 
then the resonances may appear even for extended states. This is the case for our numerical model, where resonances can be seen also for a random state in Figure~\ref{fig:shortRan}. The Gaussian envelope of the resonance has an effective width
\begin{equation}
\Delta_{\rm r} = \frac{m \omega'}{\Delta_j} \sqrt{1 + \zeta^2} \sim k/\sqrt{\hbar}.
\end{equation}
In the semiclassical limit, the resonance positions scale as 
$t_{\rm r}\propto \hbar^{-1}$, while their widths grow only as $\Delta_{\rm r}\propto \hbar^{-1/2}$, so they are well separated. On the other hand, with increasing order $k$ the resonance width grows, and they eventually start to overlap at $k\sim 1/\sqrt{\hbar}$. This overlapping takes place at time $\sim \hbar^{-3/2}$ and is smaller than $t_2$ provided $\delta < \nu_{\rm CIS}^{-1/2} \hbar$. 
\par
The structure of $2\pi-$resonance is nicely illustrated in a numerical example in 
Figure~\ref{fig:resonances}, where we consider a slightly modified model with
\begin{equation}
U_0=\exp{(-\ii S [\alpha (S_z/S)^2/2 + \gamma (S_z/S - j^*)^3/6])},
\end{equation}
corresponding to $h_0(j) = \alpha j^2/2 + \gamma (j-j^*)^3/6$ and a
coherent initial state at $(\vartheta,\varphi)=(1,1)$. For such a system $\omega'' = \gamma $ may not be
identically vanishing and $\omega'=\alpha$.
 
\subsubsection{$\pi$- and $2\pi/m$-resonances:}
\begin{figure}[h]
\vspace{-1cm}
\centerline{\includegraphics[angle=-90,width=140mm]{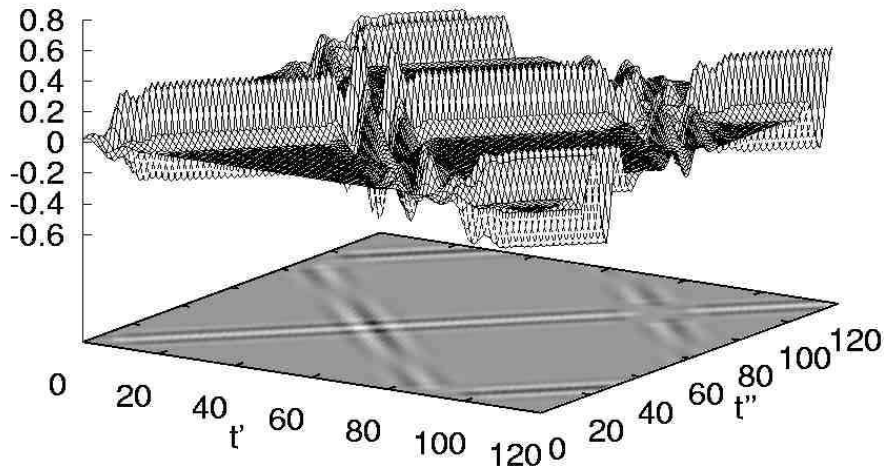}}
\vspace{-1cm}
\caption{Two-time correlation function $C(t',t'')$ (\ref{eq:Cjk}) for the quantized top with 
$\alpha=1.1, \gamma=0, S=16$, and coherent initial state at $(\vartheta,\varphi)=(1,1)$. The structures giving rise to a $\pi$ and $2\pi$ resonance can be seen.}
\label{fig:Cor}
\end{figure}   
We note that one may obtain a resonance condition for a single
time-dependent term (\ref{eq:resatom}) with fixed $m$ even at a shorter time, namely for $t = t_{\rm r}/m$. This is trivially the
case for perturbations with many, or at least more than one Fourier components with $|m|> 1$.
However, in such cases only selected time-dependent terms of the moment will be affected, so the fidelity will generically {\em not come back to} $1$, even in the linear 
response regime and in the strongly resonant case $\zeta \ll 1$. 
Such (incomplete) resonances at fractional times 
$(k/m)t_{\rm r}$ will be called $2\pi/m$-resonances.
\par
However, we may obtain a resonant condition at $t=t_{\rm r}/2$ even for the first Fourier 
component $m=1$ of the perturbation, as taking the square of the operator 
$\Sigma(t)$ produces Fourier components $m+m'=\pm 2$. Such a resonant behaviour at times $(k + \frac{1}{2})t_{\rm r}$ will 
be called a $\pi-$resonance.
\par
So for perturbations with a {\em single} Fourier mode $m=\pm 1$, or more generally with only
odd-numbered Fourier modes $m=2l+1$, the $\pi-$resonance can affect only the term having ${\rm e}^{\ii (m+m')\omega(\hbar n)t}$ in 
the expression for the second moment. All other terms (having a single
$m$ in the phase) result in their time-averaged values. To see this,
we observe that the time dependent parts of the form $g(\hbar n) {\rm
e}^{\ii m \omega t}$ are proportional to $\sum_n{d_\rho(\hbar n)
g(\hbar n) (-1)^{mn}}$. As $m$ is an odd number and $d_\rho(\hbar n)
g(\hbar n)$ is a smooth function of $n$, this sum averages to
zero. This allows us to again compute explicitly 
the fidelity in linear response close to the peak in a strongly
resonant case. We find 
\begin{eqnarray}
\ave{\Sigma(t)} &\cong& -\sum_{m=2l+1} w_m(j^*) e^{\ii m \theta^*},\quad{\rm for}\quad 
|t-(k+\frac{1}{2})t_{\rm r}| \ll 
\Delta_t,\;\;{\rm and}\;\;\zeta \ll 1,\nonumber \\
\<\Sigma^2(t)\> &\cong& \Bigl(\sum_{m=2l+1} w_m(j^*) e^{\ii m \theta^*}\Bigr)^2 
+ \Bigl( \sum_{m=2l+1} w_m(j^*) e^{\ii m (\theta^*+\omega^* t)}\Bigr)^2, \\
1-F(t) &\cong& \frac{\delta^2}{\hbar^2}\left(\sum_{m=2l+1} |w_m(j^*)|\cos(m\omega^* t + \beta_m)\right)^2, \nonumber
\end{eqnarray}
where $\beta_m$ are phases of complex numbers $w_m(j^*){\rm e}^{\ii m\theta^*}$.
So we have learned that the fidelity at the peak of a $\pi$-resonance displays 
an oscillatory pattern, oscillating precisely around the plateau value 
$F_{\rm plat}^{\rm CIS}$ (\ref{eq:nuCIS}) with an amplitude of oscillations equal to 
$1-F_{\rm plat}^{\rm CIS}$ so that the fidelity comes back to $1$ close to the peak
of the resonance.
\par
Again, our numerical example illustrates such an oscillatory structure of $\pi-$resonance in 
Figure~\ref{fig:resonances}. The resonances can also be nicely seen in `short-time' Figure~\ref{fig:shortCoh}, and because $\zeta=0$ also in the 'long-time' Figure~\ref{fig:longCoh}. In Figure~\ref{fig:Cor} we depict the structure of $\pi-$ and $2\pi-$ resonance as reflected
in the two-time correlation function $C(t',t'')$. Note that the first intersection of the
soliton-like-trains for $t'-t''={\rm const}$ and $t+t'={\rm const}$ happens at $t_{\rm r}/2$ and
produces a $\pi-$resonance, while the second intersection at $t_{\rm r}$ produces a $2\pi$-resonance.
\par
In analogy to the emergence of a $\pi-$resonance as a consequence of the contribution from the 
{\em second} moment of $\Sigma(t)$, even for the first Fourier mode $m=1$, we shall eventually 
obtain also fractional $2\pi/p$-resonances
at times $(k/p)t_{\rm r}$ in the non-linear-response regimes where higher moments 
$\ave{\Sigma^p(t)}$ contribute to $F(t) \sim 
\ave{\exp(-\ii\Sigma(t) \delta/\hbar)}$. This is illustrated numerically in Figure~\ref{fig:shortCoh}c showing the case of strong 
perturbation $\delta S=3.2$ so that higher orders are important. One indeed obtains 
fractional resonances, some of which have been marked on the figure.

   \subsection{Random Initial States}
\label{sec:ran_freeze}
The second specific case of interest is that of random initial states.
Here we shall assume that our Hilbert space has a finite dimension 
${\cal N}$, like e.g. in the case of the kicked top or a general quantum map
with a finite classical phase space, or it is determined by some
large classically invariant region of phase space, e.g. we may consider
all states $\ket{\vec{n}}$ between two energy
surfaces $E_1 < h_0(\hbar\vec{n}) < E_2$ of an autonomous system. 
In any case we have the scaling
\begin{equation}
{\cal N}\cong \frac{{\cal V}}{(2\pi \hbar)^d},
\end{equation}
where ${\cal V}$ is the volume of the relevant classical phase space.
Throughout this section we will assume that the Hilbert space size
${\cal N}$ is sufficiently large, so that the difference between the
expectation value in a single random state and an average over the whole space can be neglected. Also, as discussed in Section~\ref{sec:state_avg}, the difference between the average fidelity amplitude squared and the average fidelity is semiclassically small. 

\subsubsection{The Plateau}
\begin{figure}[h!]
\centerline{\includegraphics{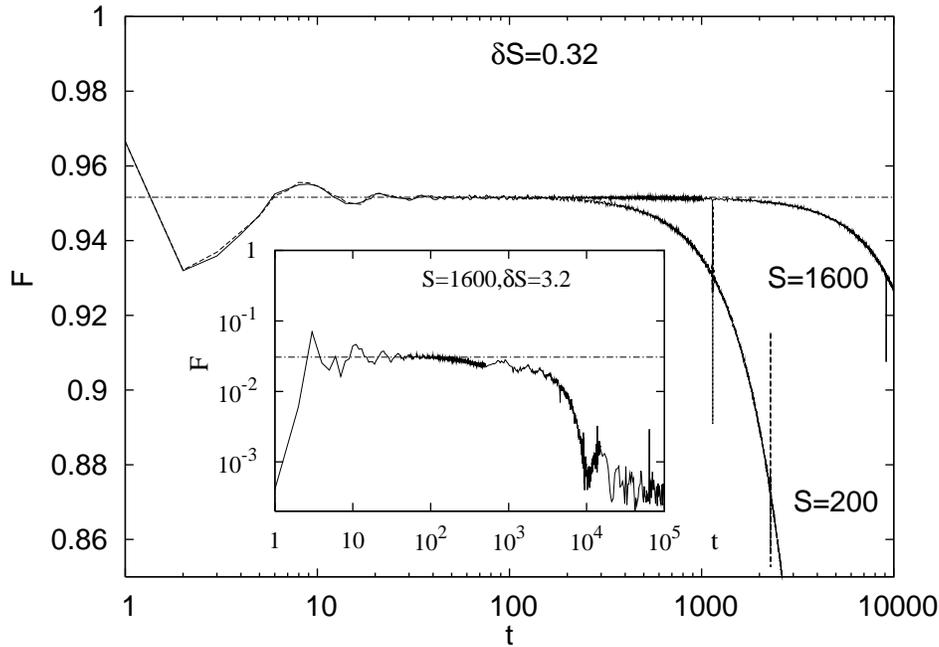}}
\caption{Short time fidelity for a quantized top with $\beta=1.4$
(having no singularities) and a random initial state. 
In order to reduce statistical fluctuations, averaging over $20$ random initial states is 
performed for $S=1600$, and over $100$ initial states for $S=200$. The horizontal chain line shows the
semiclassical theory (\ref{eq:RIS_plat}). Echo resonances (two spikes for $S=200$) are here present due to the special
property $\omega''(j)=0$ and will be absent for a more generic
unperturbed system. The main figure shows the case of the weak
perturbation $\delta S=0.32$, whereas the inset is for the strong perturbation $\delta S=3.2$.} 
\label{fig:shortRan}
\end{figure}   
\begin{figure}[h!]
\centerline{\includegraphics{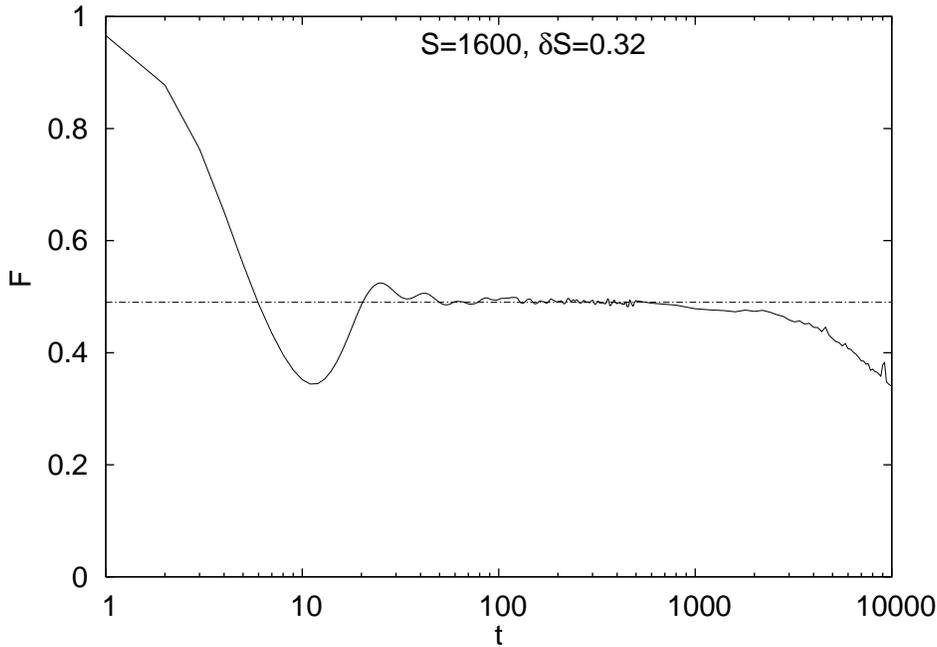}}
\caption{Short time fidelity decay for $\beta=0$ (we have
singularities) and a random initial state. The chain line shows the theoretical value of the plateau as computed from Eq.~(\ref{eq:RIS_sing}).
}
\label{fig:shortRanSing}
\end{figure}   
To calculate the plateau we replace the quantum expectation value $\<
\bullet \>$ with a classical phase space average, resulting in a fidelity amplitude
\begin{equation}
f_{\rm plat}^{\rm RIS}\cong\frac{(2\pi)^d}{\cal V}\int{\!{\rm d}^d \vec{j} \left| \int\!\frac{{\rm d}^d\vec{\theta}}{(2\pi)^d} 
\exp\left(-\frac{\ii\delta}{\hbar}w(\vec{j},\vec{\theta})\right) \right|^2}.
\end{equation}
Interestingly, the plateau for a random initial state is just the {\em
average} plateau for a coherent state squared, where averaging is done
over the position of the initial coherent state. If we denote the plateau for a coherent initial state centred at $\vec{j}^*$ by $F_{\rm plat}^{\rm CIS}(\vec{j}^*)$ (Eq.~\ref{eq:plateauCS}), then the plateau for a random initial state  $F_{\rm plat}^{\rm RIS}$ is simply
\begin{equation}
F_{\rm plat}^{\rm RIS}\cong\left| \frac{(2\pi)^d}{\cal V}\int{\!{\rm d}^d \vec{j} F_{\rm plat}^{\rm CIS}(\vec{j})}\, \right|^2.
\label{eq:RIS_plat}
\end{equation}
Similarly as in mixing dynamics there is a square relationship between
the plateau for RIS and CIS. In the mixing case the plateau for CIS
does not depend on the position of the initial state due to ergodicity,
here though the plateau for CIS does depend on $\vec{j}^*$ and one has
to take a square of the average plateau for CIS. In the linear
response approximation one has a simple formula for the plateau
\begin{equation}
1-F_{\rm plat}^{\rm RIS}\cong\frac{\delta^2}{\hbar^2}\nu_{\rm RIS},\qquad \nu_{\rm RIS}=2\frac{(2\pi)^d}{\cal V}\int{\!{\rm d}^d\vec{j} \sumn{|w_\vec{m}(\vec{j})|^2}}.
\end{equation}
\par
In Section~\ref{sec:denominator} we already discussed possible
divergence problems in $w_\vec{m}$. If we express $w_\vec{m}$ in terms
of $v_\vec{m}$, Eq.~(\ref{eq:w_v}), we have denominators
$\sin{(\vec{m}\cdot\vec{\omega}(\vec{j})/2)}$. We therefore have
divergences at points in phase space where
$\vec{m}\cdot\vec{\omega}(\vec{j})=2\pi k$ with $k$ an integer. For a
coherent initial state this was not a real problem as it would occur
only if we placed initial packet at such a point. For random initial
state though, there is an average over the whole action space in the
plateau formula and if there is a single diverging point somewhere in
the phase space it will cause divergence. The solution is very simple as
explained in Section~\ref{sec:denominator}. The integral is actually
an approximation for a sum over $\hbar\vec{n}$ and so we have to
retain the original sum over the eigenvalues of the action operator and exclude possible diverging terms. The formula for the plateau in the case of such divergences is 
\begin{equation}
F_{\rm plat}^{\rm RIS}\cong \left| \frac{1}{\cal N}\sum_{\vec{n}}^{\vec{m}\cdot\vec{\omega}(\hbar\vec{n})\neq 2\pi k}{F_{\rm plat}^{\rm CIS}(\hbar\vec{n})}\, \right|^2,
\label{eq:RIS_sing}
\end{equation}
where in the summation over $\vec{n}$ we exclude all terms for which for any constituent Fourier mode $\vec{m}$ we would have $\vec{m}\cdot\vec{\omega}(\hbar\vec{n})= 2\pi k$.
\par
Again we find an excellent confirmation of our theoretical predictions in the numerical experiment for the same system as for a coherent initial state (\ref{eq:KT_reg}). In the first calculation we choose the shift $\beta = 1.4$ so that we have no singular frequency in the action space. 
In Figure~\ref{fig:shortRan} we demonstrate the plateau, which in the case of random states starts earlier than for coherent states, namely at $t_1 \propto \hbar^0 \delta^0$ (\ref{eq:t1_RIS}). The value of the plateau can be immediately written in terms of the result for coherent states,
\begin{equation}
F_{\rm plat}^{\rm RIS} \cong \left[ \frac{1}{2}\int_{-1}^1{\rm d} j J_0^2\left(\delta S\frac{\sqrt{1-j^2}}{2\sin{\left\{ \alpha (j-\beta)/2\right\}}}\right) \right]^2.
\label{eq:BessR}
\end{equation}
The integral has to be calculated numerically.
 Horizontal chain lines in Figure~\ref{fig:shortRan} correspond to this theoretical values and agree with 
numerics, both for the weak perturbation $\delta S = 0.32$ and the strong perturbation $\delta S = 3.2$ (inset). Small echo resonances visible in the figures are due to the fact that the Hamiltonian is a quadratic function of the action and therefore $\omega'' \equiv 0$, so that the 
resonance condition (\ref{eq:resonance}) is satisfied also for extended states (\ref{eq:ResCond}). 
For a more generic Hamiltonian these narrow resonant spikes will be absent. 
In Figure~\ref{fig:shortRanSing} we also demonstrate the plateau in the fidelity for the
zero-shift case $\beta=0$ with a singular-frequency, $\omega(j=0) = 0$, where we again find an excellent
agreement with the theoretical prediction (\ref{eq:RIS_sing}). In this case the theoretical value has been 
obtained by replacing an integral in (\ref{eq:BessR}) with a sum as implied by Eq.~(\ref{eq:RIS_sing}) and summing over all quantum numbers except $n=0$. Observe that the value of the plateau is much lower than in the case of 
a non-zero shift $\beta=1.4$ in Figure~\ref{fig:shortRan} as quantum numbers around $n=0$ will be close-to resonant.

\subsubsection{Long Time Decay}

\begin{figure}[h]
\centerline{\includegraphics{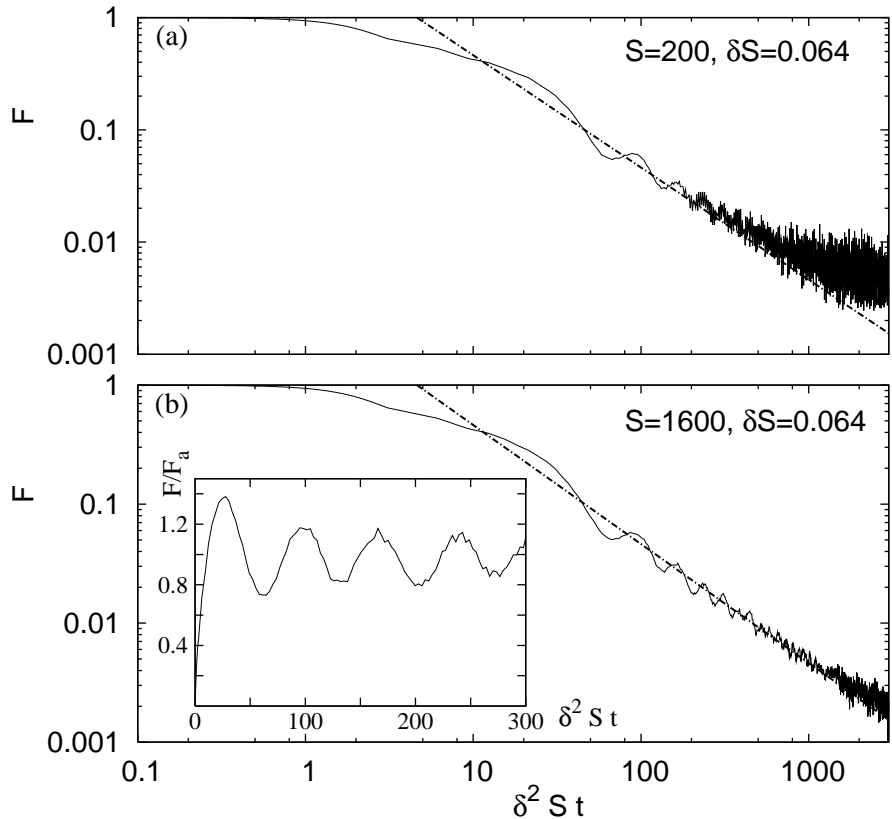}}
\caption{Long time power law fidelity decay for random states in the kicked top with $\beta=0$ and $\alpha=1.1$, for
$S=200$ (a), and $S=1600$ (b). 
Here averaging for $S=200$ is performed over $1000$ 
initial random states, otherwise all is the same as for Figure~\ref{fig:shortRan}. 
The heavy chain line shows the theoretical asymptotic decay
(\ref{eq:RIS_asymp}) with an analytically computed prefactor
(no fitting parameters). The inset in the bottom figure shows the diffractive quotient between the
numerical fidelity and the asymptotic formula (\ref{eq:RIS_long}) (chain line in the main figure).}
\label{fig:longRan}
\end{figure}
After a sufficiently long time $t>t_2$ fidelity will start to decay. To calculate this decay we have to evaluate the ASI formula (\ref{eq:res_ASI}). For a uniform average over Hilbert space we have $D_\rho=1/{\cal N}$ with the classical limit $d_\rho(\vec{j})=(2\pi \hbar)^d/{\cal V}$. The fidelity amplitude is therefore
\begin{equation}
\fn \cong \frac{(2\pi)^d}{\cal V}\int{{\rm d}^d \vec{j} \exp{\left(-\ii \frac{\delta^2 t}{2\hbar} \bar{r}(\vec{j})\right)}}.
\label{eq:RIS_long}
\end{equation}
The stationary phase procedure is completely analogous to the situation for the case of non-residual perturbation, described in Section~\ref{sec:RIS}, Eq.~(\ref{eq:reg_power}). We will only write the asymptotic result,
\begin{equation}
F(t)\asymp \left(\frac{t_{\rm ran}}{t} \right)^d,\qquad t_{\rm ran} = {\rm const}\times \frac{\hbar}{\delta^2}.
\label{eq:RIS_asymp}
\end{equation}
Here we should remember that the asymptotic formula
(\ref{eq:RIS_asymp}) has been obtained as a stationary phase
approximation of an integral in the limit of an infinite action
space. If we have a finite region of the action space, as is the case
for the kicked top, the stationary phase approximation gives an
additional oscillating prefactor, whose amplitude is damped as $(\hbar/t)^{1/2}$ for $\hbar\to 0$ and/or $t\to\infty$, and which can be interpreted as a {\em diffraction}. 
This oscillating prefactor can be seen in numerical data for the
fidelity in the inset of Figure~\ref{fig:longRan}. For random states
the time $t_2$ when the plateau ends is independent of $\hbar$
and is determined by the ratio of two competing terms in the BCH form or the echo operator, $t_2 \sim |\delta \Sigma(t)/\hbar|/|\delta^2 \oG/2\hbar |$, therefore we have
\begin{equation}
t_2 \sim 1/\delta.
\end{equation}
This agrees with the numerical results shown in Figures~\ref{fig:shortRan} and~\ref{fig:shortRanSing}.


\chapter{Coupling with the Environment}
\label{ch:coupling}
\begin{flushright}
\baselineskip=13pt
\parbox{85mm}{\baselineskip=13pt
\sf When a man tells you that he got rich through hard work, ask him: 'Whose?'
}\medskip\\
---{\sf \itshape Don Marquis}\\\vspace{20pt}
\end{flushright}
  
The fidelity might not always be the relevant measure of
 stability. Coupling with the environment is usually unavoidable so
 that the evolution of our system is no longer Hamiltonian. To preserve the Hamiltonian formulation we have to include the environment in our description. We therefore have a ``central system'', denoted by a subscript ``${\rm c}$'', and an environment, denoted by subscript ``${\rm e}$''. The names {\em central system} and {\em environment} will be used just to denote two pieces of a composite system, without any connotation on their properties, dimensionality etc. The central system will be that part which is of interest and the environment the rest. The Hilbert space is a tensor product ${\cal H}={\cal H}_{\rm c} \otimes {\cal H}_{\rm e}$ and the evolution of a whole system is determined by a Hamiltonian or a propagator on the whole Hilbert space ${\cal H}$ of dimension ${\cal N}={\cal N}_{\rm c} {\cal N}_{\rm e}$. The unperturbed state $\ket{\psi(t)}$ and the perturbed one $\ket{\psi_\delta(t)}$ are obtained with propagators $U(t)$ and $U_\delta(t)$ (\ref{eq:psit}). Fidelity would in this case be the overlap of two wave functions on the {\em whole space}. But if we are not interested in the environment, this is clearly not the relevant quantity. Namely, the fidelity will be low even if the two wave functions are the same on the subspace of the central system and differ only on the environmental part.      
 \section{Reduced Fidelity, Purity Fidelity}
We can define a quantity analogous to the fidelity, but which will measure the overlap just on the subspace of interest, i.e. on the subspace of the central system. Let us define a reduced density matrix of a central subsystem
\begin{equation}
\rho_{\rm c}(t):=\tre{\rho(t)},\qquad \rho_{\rm c}^{\rm M}(t):=\tre{\rho^{\rm M}(t)},
\label{eq:reduced_rho}
\end{equation}
where $\tre{\bullet}$ denotes a trace over the environment and $\rho^{\rm M}(t)=\Md \rho(0)\Md^\dagger$ is the so-called echo density matrix. Throughout this chapter we will assume that the initial state is a pure product state, i.e. a direct product,
\begin{equation}
\ket{\psi(0)}=\ket{\psi_{\rm c}(0)}\otimes\ket{\psi_{\rm e}(0)}=:\ket{\psi_{\rm c}(0);\psi_{\rm e}(0)},
\label{eq:psi_prod}
\end{equation}
where we also introduced a short notation $\ket{\psi_{\rm c};\psi_{\rm e}}$ for pure product states. The resulting initial density matrix $\rho(0)=\ket{\psi(0)}\bra{\psi(0)}$ is of course also pure. The fidelity was defined as $F(t)=\tr{\lbrack \rho(0)\rho^{\rm M}(t)\rbrack}$ (\ref{eq:Fn_trace}) and in a similar fashion we will define a {\em reduced fidelity}~\citep{Znidaric:03} denoted by $\Fr$,
\begin{equation}
\Fr:=\trc{\rho_{\rm c}(0) \rho_{\rm c}^{\rm M}(t)}.
\label{eq:Fr_def}
\end{equation}
The reduced fidelity measures the distance between the initial reduced
density matrix and the reduced density matrix after the echo. Note
that our definition of the reduced fidelity agrees with the
information-theoretic fidelity (\ref{eq:Uhl_fid}) on a central
subspace ${\cal H}_{\rm c}$ only if the initial state is a pure product state, so that $\rho_{\rm c}(0)$ is also a pure state.  
\par
One of the weirdest features of quantum mechanics is entanglement. Of
some interest is therefore also whether the coupling with the
environment will produce entanglement and how fast. Due to the
coupling between the central system and the environment the initial
product state will evolve after an echo into the pure entangled state $\Md \ket{\psi(0)}$ and therefore the reduced density matrix $\rho^{\rm M}_{\rm c}(t)$ will be a mixed one. For a pure state $\ket{\psi(t)}$ the criterion for entanglement is very simple. It is quantified by a {\em purity} $I(t)$, defined as
\begin{equation}
I(t):=\trc{\rho_{\rm c}^2(t)},\qquad \rho_{\rm c}(t):=\tre{\ket{\psi(t)}\bra{\psi(t)}}.
\label{eq:I}
\end{equation}
Purity, or equivalently von Neumann entropy $\tr{(\rho\ln{\rho})}$, is a standard quantity used in decoherence studies~\citep{Zurek:91}. Iff the purity is less than one, $I<1$, then the state $\ket{\psi}$ is entangled (between the environment and the central system), otherwise it is a product state. Similarly, one can define a purity after an echo, called {\em purity fidelity}~\citep{Prosen:02spin} $\Fp$,
\begin{equation}
\Fp:=\trc{\{ \rho^{\rm M}_{\rm c}(t) \}^2}.
\label{eq:Fp_def}
\end{equation}
All three quantities, the fidelity $\Fn$, the reduced fidelity $\Fr$ and the purity fidelity $\Fp$ measure stability to perturbations. If the perturbed evolution is the same as the unperturbed one, they are all equal to one, otherwise they are less than one. The fidelity $\Fn$ measures the stability of a 
whole state, the reduced fidelity gives the stability on ${\cal H}_{\rm c}$ and the purity fidelity measures separability of $\rho^{\rm M}(t)$. One expects that fidelity 
is the most restrictive quantity of the three - $\rho(0)$ and $\rho^{\rm M}(t)$ must 
be similar for $\Fn$ to be high. For $\Fr$ to be high, only the reduced density matrices 
$\rho_{\rm c}(0)$ and $\rho_{{\rm c}}^{\rm M}(t)$ 
must be similar, and finally, the purity fidelity $\Fp$ is high if only $\rho^{\rm M}(t)$ factorizes. It looks though as fidelity is the strongest criterion for stability. 

\subsection{Inequality Between Fidelity, Reduced Fidelity and Purity Fidelity}
Actually, one can prove the following inequality for an arbitrary {\em pure} state $\ket{\psi}$ and an arbitrary {\em pure product} state $\ket{\phi_{\rm c};\phi_{\rm e}}$~\citep{Znidaric:03,Prosen:03corr},
\begin{equation}
\left|\braket{\phi_{\rm c};\phi_{\rm e}}{\psi} \right|^4 \le \left| \bracket{\phi_{\rm c}}{\rho_{\rm c}}{\phi_{\rm c}} \right|^2 \le \trc{\rho_{\rm c}^2},
\label{eq:ineq_gen}
\end{equation}
where $\rho_{\rm c}:=\tre{\ket{\psi}\bra{\psi}}$.
\par
{\bf Proof.} Uhlmann's theorem~\citep{Uhlmann:76}, i.e. noncontractivity of the fidelity, states that tracing over an arbitrary subsystem can not decrease the fidelity,
\begin{equation}
\tr{\lbrack \ket{\phi_{\rm c};\phi_{\rm e}}\bra{\phi_{\rm c};\phi_{\rm e}} \ket{\psi}\bra{\psi} \rbrack} \le \tr{\lbrack \ket{\phi_{\rm c}}\bra{\phi_{\rm c}} \rho_{\rm c} \rbrack}. 
\end{equation}
Then, squaring and applying the Cauchy-Schwartz inequality 
$|\tr{\lbrack A^\dagger B \rbrack}|^2 \le \tr{\lbrack AA^\dagger\rbrack}\tr{\lbrack BB^\dagger\rbrack}$ we immediately obtain the wanted inequality (\ref{eq:ineq_gen}).
\par
The rightmost quantity in the inequality $I=\tr{\lbrack \rho_{\rm c}^2
\rbrack}$ is nothing but the purity of state $\ket{\psi}$ and so does
not depend on $\ket{\phi_{\rm c};\phi_{\rm e}}$. One can think of
inequality (\ref{eq:ineq_gen}) as giving us a {\em lower bound} on
purity. An interesting question for instance is, which state
$\ket{\phi_{\rm c};\phi_{\rm e}}$ optimises this bound for a given
$\ket{\psi}$, i.e. what is the maximal attainable overlap
$\left|\braket{\phi_{\rm c};\phi_{\rm e}}{\psi} \right|^4$ (fidelity)
for a given purity. The rightmost inequality is optimised if we choose
$\ket{\phi_{\rm c}}$ to be the eigenstate of the reduced density
matrix $\rho_{\rm c}$ corresponding to its largest eigenvalue
$\lambda_1$, $\rho_{\rm c}\ket{\phi_{\rm c}}=\lambda_1 \ket{\phi_{\rm
c}}$. To optimise the left part of the inequality, we have to choose
$\ket{\phi_{\rm e}}$ to be the eigenstate of $\rho_{\rm
e}:=\trc{\rho}$ corresponding to the same largest eigenvalue
$\lambda_1$, $\rho_{\rm e}\ket{\phi_{\rm e}}=\lambda_1 \ket{\phi_{\rm
e}}$. The two reduced matrices $\rho_{\rm e}$ and $\rho_{\rm c}$ have
the same eigenvalues~\citep{Araki:70}, $\lambda_1 \ge \lambda_2 \ge \ldots
\ge\lambda_{{\cal N}_{\rm c}}$. For such choice of
$\ket{\phi_{\rm c};\phi_{\rm e}}$ the left inequality is actually an equality, $\left|\braket{\phi_{\rm c};\phi_{\rm e}}{\psi} \right|^4 = \left| \bracket{\phi_{\rm c}}{\rho_{\rm c}}{\phi_{\rm c}} \right|^2=\lambda_1^2$ and the right inequality is
\begin{equation}
\lambda_1^2 \le \tr{\lbrack \rho_{\rm c}^2 \rbrack}=\sum_{j=1}^{{\cal N}_{\rm c}}{\lambda_j^2},
\end{equation}
with equality iff $\lambda_1=1$. In the case when the largest eigenvalue is close to one, $\lambda_1=1-\epsilon$, the purity will be $I =(1-\epsilon)^2+{\cal O}(\epsilon^2) \sim 1-2\epsilon$ and the difference between the purity and the overlap will be of the {\em second order} in $\epsilon$, $I-\left|\braket{\phi_{\rm c};\phi_{\rm e}}{\psi} \right|^4\sim \epsilon^2$. Therefore, for high purity the optimal choice of $\ket{\phi_{\rm c};\psi_{\rm s}}$ gives a sharp lower bound, i.e. its deviation from $I$ is of second order in the deviation of $I$ from unity.
\par
For our purpose of studying stability to perturbations, a special case of the general inequality (\ref{eq:ineq_gen}) is especially interesting. Namely, taking for $\ket{\psi}$ the state after the echo evolution $\Md \ket{\psi(0)}$ and for a product state $\ket{\phi_{\rm c};\phi_{\rm e}}$ the initial state $\ket{\psi(0)}$ (\ref{eq:psi_prod}), we obtain  
\begin{equation}
F^2(t)\le F_{\rm R}^2(t) \le F_{\rm P}(t).
\label{eq:ineq}
\end{equation}
Immediate consequence of this inequality is that if the fidelity is high, the reduced fidelity and the purity fidelity will also be high. In the case of perturbations with a zero time average in Chapter~\ref{ch:freeze}, the fidelity freezes at the plateau and from the inequality we immediately know that the same phenomenon will be present for the reduced fidelity and the purity fidelity.

 \subsection{Uncoupled Unperturbed Dynamics}
Special, but very important case is when the unperturbed dynamics $U_0$ represents two uncoupled systems, so that we have 
\begin{equation}
U_0=U_{\rm c}\otimes U_{\rm e}.
\label{eq:U0_uncoupled}
\end{equation}
This is a frequent situation if the coupling with the environment is ``unwanted'', so that our ideal evolution $U_0$ would be an uncoupled one. The reduced fidelity $\Fr$ and the purity fidelity $\Fp$ have especially nice forms in such case. 
\par
The reduced fidelity (\ref{eq:Fr_def}) can be rewritten as
\begin{equation}
\Fr=\trc{\rho_{\rm c}(0) \rho_{\rm c}^{\rm M}(t)}=\trc{\rho_{\rm c}(t) \rho^\delta_{\rm c}(t)},
\end{equation}
where $\rho_{\rm c}(t)$ is the unperturbed state of the central system and $\rho^\delta_{\rm
c}(t):=\tre{U_{\delta}(t)\rho(0)U^\dagger_{\delta}(t)}$ the corresponding state obtained by perturbed evolution. Whereas for a general unperturbed evolution the reduced fidelity was an overlap of the initial state with an echo state, for a factorized unperturbed evolution it can also be interpreted as the overlap of the (reduced) unperturbed state at time $t$ with a perturbed state at time $t$, similarly as the fidelity.
\par
The purity fidelity can also be simplified for uncoupled unperturbed evolution. As the $U_0$ is in factorized form, we can bring it out of the innermost trace in the definition of the purity fidelity and use the cyclic property of the trace, finally arriving at
\begin{equation}
\Fp=\trc{\{ \rho^{\rm M}_{\rm c}(t) \}^2}=\trc{\{ \rho^\delta_{\rm c}(t)\}^2}=I(t).
\end{equation}
The purity fidelity is therefore equal to the purity of the forward evolution. The general inequality gives in this case 
\begin{equation}
F^2(t) \le F^2_{\rm R}(t) \le I(t),
\end{equation}
and so the fidelity and the reduced fidelity give a lower bound on the decay of purity. Because the purity is frequently used in studies of decoherence this connection is especially appealing.
\par
In all our theoretical derivations regarding the purity fidelity we will assume a general unperturbed evolution, but one should keep in mind that the results immediately carry over to purity in the case of uncoupled unperturbed dynamics. Also a large part of our numerical demonstration in next two sections will be done on systems with an uncoupled unperturbed dynamics as this is usually the more interesting case.

\subsection{Linear Response Expansion}
\label{sec:LRE}
We proceed with the linear response expansion of the reduced fidelity (\ref{eq:Fr_def}) and the purity fidelity (\ref{eq:Fp_def}). We will use notation $\rho_{\rm c}:=\ket{\psi_{\rm c}(0)}\bra{\psi_{\rm c}(0)}$ for initial pure density matrix on a central system and $\rho_{\rm e}:=\ket{\psi_{\rm e}(0)}\bra{\psi_{\rm e}(0)}$ for the environment. The perturbed propagator is defined in exactly the same way as for the fidelity (\ref{eq:Ud_def}) in terms of the perturbation generator $V$. To order ${\cal O}(\delta^4)$ we get,
\begin{eqnarray}
1-\Fn&=&\left( \frac{\delta}{\hbar} \right)^2 \<\Sigma(t)(\mathbbm{1}\otimes\mathbbm{1}-\rho_{\rm c}\otimes\rho_{\rm e})\Sigma(t)\> \nonumber\\
1-\Fr&=&\left( \frac{\delta}{\hbar} \right)^2 \<\Sigma(t)(\mathbbm{1}-\rho_{\rm c})\otimes\mathbbm{1}\Sigma(t)\> \nonumber\\
1-\Fp&=&2\left( \frac{\delta}{\hbar} \right)^2 \<\Sigma(t)(\mathbbm{1}-\rho_{\rm c})\otimes(\mathbbm{1}-\rho_{\rm e})\Sigma(t)\>,
\label{eq:LR_echo}
\end{eqnarray}
where $\ave{\bullet}=\tr{\lbrack (\rho_{\rm c}\otimes \rho_{\rm
e})\bullet\rbrack}$ denotes the quantum expectation value in the
initial product state and $\Sigma(t)$ is the sum of $V(t)$
(\ref{eq:Sigma_def}). If the expectation values are written explicitly
in terms of expectations in the base states $\ket{j;\nu}$, $j=1,\ldots,{\cal N}_{\rm c}$, $\nu=1,\ldots,{\cal N}_{\rm e}$, with the convention that the first base state $\ket{1;1}:=\ket{\psi_{\rm c};\psi_{\rm e}}$ is the initial state, we have
\begin{eqnarray}
1-\Fn&=&\left( \frac{\delta}{\hbar} \right)^2 \left\{ \bracket{1;1}{\Sigma^2(t)}{1;1}-\bracket{1;1}{\Sigma(t)}{1;1}^2 \right\} \nonumber\\
1-\Fr&=&\left( \frac{\delta}{\hbar} \right)^2 \left\{ \bracket{1;1}{\Sigma^2(t)}{1;1}-\sum_{\nu=1}^{{\cal N}_{\rm e}}{ \left| \bracket{1;\nu}{\Sigma(t)}{1;1}\right|^2} \right\} \\
1-\Fp&=&2\left( \frac{\delta}{\hbar} \right)^2 \left\{ \bracket{1;1}{\Sigma^2(t)}{1;1}-\sum_{\nu=1}^{{\cal N}_{\rm e}}{ \left| \bracket{1;\nu}{\Sigma(t)}{1;1}\right|^2} -\sum_{j=2}^{{\cal N}_{\rm c}}{ \left| \bracket{j;1}{\Sigma(t)}{1;1}\right|^2} \right\} \nonumber.
\label{eq:LR_echo11}
\end{eqnarray}
As one can see, the linear response expansion of course also satisfies
the general inequality (\ref{eq:ineq}). The difference between $\Fr$
and $\Fn$ as well as between $\Fp$ and $\Fn$ is in {\em off-diagonal}
matrix elements of operator $\Sigma(t)$. Somehow reminiscent
perturbative expansion, although without time dependence, has been
obtained in studying the eigenvalues of the reduced density
matrix~\citep{Kubler:73}. Depending on the growth of linear response
terms with time we will again have two general categories, that of
mixing dynamics and that of regular dynamics.

\subsubsection{Mixing Dynamics}
For mixing dynamics the correlations decay and the linear response term will grow linearly with time. For large times one can argue that $\Sigma(t)$ should look like a random matrix and the terms giving the difference between the fidelity and the purity fidelity and the reduced fidelity can be estimated as
\begin{equation}
\frac{\sum_{j=2}^{{\cal N}_{\rm c}}{ \left| \bracket{j;1}{\Sigma(t)}{1;1}\right|^2}}{\bracket{1;1}{\Sigma^2(t)}{1;1}}\sim \frac{\sum_{j}{|[\Sigma(t)]_{(j;1),(1;1)}|^2}}{\sum_{j,\nu}{|[\Sigma(t)]_{(1;1),(j,\nu)}|^2}}\sim \frac{1}{{\cal N}_{\rm e}},
\label{eq:rm}
\end{equation}
because there are more terms in the sum for fidelity. Therefore we can
estimate the difference $\Fp-F^2(t)\sim 1/{{\cal N}_{\rm c}}+1/{{\cal
N}_{\rm e}}$ and $\Fr-\Fn \sim 1/{{\cal N}_{\rm c}}$. Provided both
dimensions ${\cal N}_{\rm c,e}$ are large and for sufficiently long
times, so the ``memory'' of the initial state is lost and the
correlations decay, we can expect the decay of all three quantities to be the same.

\subsubsection{Regular Dynamics}
For regular dynamics on the other hand, $\Sigma(t)$ will not approach
a random matrix but will grow with time with a well defined long time
limit $\Sigma(t) \to \oV t$. This will happen for times larger than
the averaging time $t_{\rm ave}$ (\ref{eq:Vbar_def}). Expectation value of $\Sigma^2(t)$ will then grow quadratically with time. In a similar way as we defined the average correlation function $\oC$ (\ref{eq:Cbar_reg}), we can also define a time average of the correlation functions occurring in the linear response expressions for $\Fr$ and $\Fp$,
\begin{eqnarray}
\oCr &:=& \lim_{t \to \infty}{\frac{\<\Sigma(t)(\mathbbm{1}-\rho_{\rm c})\otimes\mathbbm{1}\Sigma(t)\>}{t^2}}=\<\oV(\mathbbm{1}-\rho_{\rm c})\otimes\mathbbm{1}\oV\> \nonumber\\ 
\oCp &:=& \lim_{t \to \infty}{\frac{\<\Sigma(t)(\mathbbm{1}-\rho_{\rm
c})\otimes(\mathbbm{1}-\rho_{\rm e})\Sigma(t)\>}{t^2}}=\< \oV(\mathbbm{1}-\rho_{\rm c})\otimes(\mathbbm{1}-\rho_{\rm e})\oV\>.
\label{eq:Cbars}
\end{eqnarray}
In the linear response regime purity fidelity and reduced fidelity will therefore decay quadratically with time, if $\oCr$ and $\oCp$ are nonzero.
\par
For coherent initial states the average correlation function $\oC$ is
proportional to $\hbar$ (\ref{eq:Fn_regcoh}) provided
$\bar{v}(\vec{j})$ is sufficiently smooth. In the semiclassical limit
we can make an expansion around the centre of the wave packet
$\bar{v}(\vec{j})\approx
\bar{v}(\vec{j}^*)+\vec{\bar{v}}'(\vec{j}^*)\{\vec{j}-\vec{j}^*\}+\cdots$.
The second moment $\< \bar{v}^2\>_{\rm cl}-\ave{\bar{v}}_{\rm cl}^2$
is then proportional to the dispersion in $\vec{j}$ of the initial
packet (\ref{eq:drho_coh}), i.e. to $\hbar$. As we take a product initial coherent packet we have a $d_{\rm c}+d_{\rm e}$ dimensional squeezing matrix $\Lambda$
\begin{equation}
\Lambda=\pmatrix{ \Lambda_{\rm c} & 0 \cr
  0 & \Lambda_{\rm e}},
\label{eq:Lambda_def}
\end{equation}
and we can write derivatives of classical $\bar{v}$ as
\begin{equation}
\vec{\bar{v}}'=:(\vec{\bar{v}}_{\rm c}',\vec{\bar{v}}_{\rm e}'),\qquad \vec{\bar{v}}_{\rm c}':=\frac{\partial \vec{\bar{v}}(\vec{j}^*)}{\partial \vec{j}_{\rm c}},\qquad \vec{\bar{v}}_{\rm e}':=\frac{\partial \vec{\bar{v}}(\vec{j}^*)}{\partial \vec{j}_{\rm e}},
\end{equation}
where we split actions $\vec{j}$ into two components $\vec{j}=:(\vec{j}_{\rm c},\vec{j}_{\rm e})$. The formula for $\oC$ (\ref{eq:Fn_regcoh}) can then be written as
\begin{equation}
\oC=\frac{1}{2}\hbar \left( \vec{\bar{v}}_{\rm c}'\Lambda^{-1}_{\rm c}\vec{\bar{v}}'_{\rm c}+ \vec{\bar{v}}_{\rm e}'\Lambda^{-1}_{\rm e}\vec{\bar{v}}'_{\rm e} \right),
\label{eq:oC'}
\end{equation}
By a similar method one can calculate also $\oCr$ (\ref{eq:Cbars}) and gets
\begin{equation}
\oCr=\frac{1}{2}\hbar \left( \vec{\bar{v}}_{\rm c}'\Lambda^{-1}_{\rm c}\vec{\bar{v}}'_{\rm c} \right). 
\label{eq:oCr}
\end{equation}
For $\oCp$ on the other hand, the lowest order expansion of $\vec{\bar{v}}'$ used for $\oC$ and $\oCr$ gives zero. To get $\oCp$ we must expand $\bar{v}$ to the {\em second} order in $(\vec{j}-\vec{j}^*)$
\begin{equation}
\bar{v}(\vec{j})\approx \bar{v}(\vec{j}^*)+\vec{\bar{v}}'(\vec{j}^*)\{\vec{j}-\vec{j}^*\}+ \frac{1}{2}\{\vec{j}-\vec{j}^*\}\cdot \bar{v}''(\vec{j}^*)\{\vec{j}-\vec{j}^*\}+\cdots,
\label{eq:ov_series}
\end{equation}
where $\bar{v}''(\vec{j}^*)$ is a symmetric matrix of second derivatives evaluated at the position of initial packet $\vec{j}^*$,
\begin{equation}
\bar{v}''=\pmatrix{\bar{v}''_{\rm cc} & \bar{v}''_{\rm ce}\cr \bar{v}''_{\rm ec} & \bar{v}''_{\rm ee}},\qquad \bar{v}''_{lk}(\vec{j}^*):=\frac{\partial^2 \bar{v}(\vec{j}^*)}{\partial j_l \partial j_k}.
\label{eq:v''}
\end{equation}
The only nonzero contribution to $\oCp$ comes from off-diagonal terms $\bar{v}''_{\rm ce}$ and $\bar{v}''_{\rm ec}=(\bar{v}''_{\rm ce})^T$, with the final result being
\begin{equation}
\oCp=\left(\frac{1}{2}\hbar\right)^2 \tr{\lbrack u \rbrack},\qquad u:=\Lambda_{\rm c}^{-1}\bar{v}''_{\rm ce}\Lambda_{\rm e}^{-1}\bar{v}''_{\rm ec}.
\label{eq:oCp}
\end{equation}
The result for $\oCp$ is very interesting as it means that the purity fidelity in regular systems will decay as $\Fp =2(\delta/\hbar)^2 \oCp t^2$ on {\em $\hbar$-independent} time scale because of $\oCp \propto \hbar^2$. Note that to reach this conclusion of $\hbar$-independent decay, we need only the existence of a smooth classical limit of $\Sigma(t)$, therefore $\Fp$ will decay on $\hbar$-independent time scale even before $\oV$ converges, i.e. for $t<t_{\rm avg}$. 
\par
A simple special case of time averaged perturbation $\oV$ is the
tensor product form $\oV=\bar{V}_{\rm c}\otimes\bar{V}_{\rm e}$. Then
we have $\bar{v}''_{\rm ce}=\vec{\bar{v}}'_{\rm
c}\otimes\vec{\bar{v}}'_{\rm e}/\bar{v}$ (note that in order to comply with our previous notation we have $\vec{\bar{v}}'_{\rm c}=(\vec{\partial}_{\rm c} \bar{v}_{\rm c}) \bar{v}_{\rm e}$). If in addition $\Lambda_{\rm c,e}$ are diagonal, then the average $\oCp$ is
\begin{equation}
\oCp=\left(\frac{1}{2\bar{v}}\hbar\right)^2 \left( \vec{\bar{v}}_{\rm c}'\Lambda^{-1}_{\rm c}\vec{\bar{v}}'_{\rm c} \right) \left( \vec{\bar{v}}_{\rm e}'\Lambda^{-1}_{\rm e}\vec{\bar{v}}'_{\rm e} \right).
\end{equation}
As we can see, for a tensor product forms of the time averaged perturbation there is a relation between $\oC$, $\oCr$ and $\oCp$,
\begin{equation}
\bar{v}^2 \oCp\equiv \oCr (\oC-\oCr).
\label{eq:uni_rel}
\end{equation}
The purity fidelity will be high, i.e. decoherence will be slow, if we
either make $\oCr$ small or make $\oCr$ close to $\oC$. Note that if
$\oCr$ is small, then the inequality (\ref{eq:ineq}) already tells
that $\oCp$ will be also small, but here we have a stronger result,
that $\oCp$ is of second order in $\hbar$ whereas $\oCr$ is only of first order. 

\subsection{Numerical Illustration}
Here we would like to briefly demonstrate quadratic decay of the
purity fidelity for regular systems and linear decay for chaotic
systems. Furthermore, we will show that one can also have a situation
where the purity fidelity $\Fp$ decays slower for chaotic systems
than for regular. We take a Jaynes-Cummings model
(Section~\ref{sec:Jaynes}) which is a two degrees of freedom model of
a harmonic oscillator coupled to a spin. As initial state we
always take a product state of two coherent states, for the spin the
coherent state (\ref{eq:SU2_coh}) is centred at
$(\vartheta^*,\varphi^*)=(1,1)$ and for the boson coherent state
(\ref{eq:boson_coh}) we take $\alpha=1.15$. We take two different sets
of parameters. For the chaotic regime we choose $\omega=\epsilon=0.3$
and $G=G'=1$, for which the energy of the initial state is $E=1.0$ and
the classical Poincar\' e section shows a single ergodic
component. The second set of parameters is $\omega=\epsilon=0.3$ and
$G=1$, while $G'=0$, for which the systems is integrable and the
energy is $E=0.63$. The perturbation is chosen to be in the spin
energy $\epsilon$, i.e. a detuning, corresponding to the perturbation generator $V$
\begin{equation}
V=\hbar S_{\rm z}.
\end{equation}
Because the simulation is very time consuming for the chaotic case, we
choose a small spin size $S=4$, resulting in an effective Planck's constant $\hbar=1/4$. 
\par
\begin{figure}[ht]
\centerline{\includegraphics{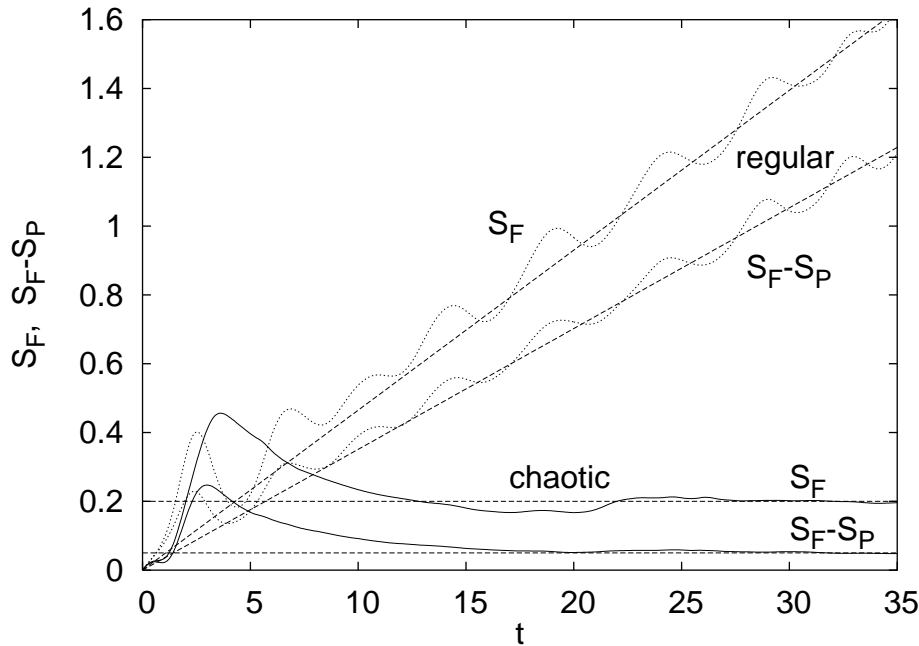}}
\caption{The correlation integral $S_{\rm F}(t)$ and $S_{\rm
F}(t)-S_{\rm P}(t)$ (\ref{eq:St}) in the Jaynes-Cummings model for
regular dynamics (upper curves) and chaotic dynamics (lower
curves). The horizontal dashed lines (for the chaotic case) are at
$2\sigma=0.2$ and $0.2/4$. The two linearly increasing dashed lines for regular dynamics have slopes $\oC=0.046$ and $\oC(1-0.98/4)$.}
\label{fig:corr}
\end{figure}

Note that, as opposed to previous kicked top model, here the system is conservative, and the time index $t$ is continuous. First we want to check the growth of linear response terms with time. In Figure~\ref{fig:corr} we show the correlation integral $S_{\rm F}(t)$ for fidelity and the difference $S_{\rm F}(t)-S_{\rm P}(t)$ between the correlation integrals for fidelity $S_{\rm F}(t)$ and for purity fidelity $S_{\rm P}(t)$,
\begin{eqnarray}
S_{\rm F}(t)&:=&\frac{1}{t}\<\Sigma(t)(\mathbbm{1}\otimes\mathbbm{1}-\rho_{\rm c}\otimes\rho_{\rm e})\Sigma(t)\>\nonumber\\
S_{\rm P}(t)&:=&\frac{1}{t}\<\Sigma(t)(\mathbbm{1}-\rho_{\rm c})\otimes(\mathbbm{1}-\rho_{\rm e})\Sigma(t)\>.
\label{eq:St}
\end{eqnarray}
Note that for chaotic systems $S_{\rm F}(t)$ should converge to $2\sigma$ (\ref{eq:sigma_def}), where $\sigma$ is the transport coefficient, i.e. the integral of the correlation function. The difference $S_{\rm F}(t)-S_{\rm P}(t)$ should on the other hand be of the order $\sim 2 \sigma (1/{\cal N}_{\rm c}+1/{\cal N}_{\rm e}) \sim \sigma/2$ (due to $1/{\cal N}_{\rm c}+1/{\cal N}_{\rm e}\approx 1/S$). These two predictions are nicely confirmed by the lower two curves in Figure~\ref{fig:corr}. 
We can see that the correction to the purity fidelity is really of
order $1/S$ in the chaotic case and would therefore vanish in the semiclassical limit. For regular dynamics we expect $S_{\rm F}(t)$ to grow linearly with time, with the slope given by $\oC$, $S_{\rm F}(t) \to \oC t$, with $\oC \propto 1/S$ (\ref{eq:oC'}). The purity fidelity integral should grow as $S_{\rm P}(t) \to \oCp t$ with $\oCp \propto 1/S^2$ (\ref{eq:oCp}). The difference should therefore be $S_{\rm F}(t)-S_{\rm P}(t) \sim S_{\rm F}(1-{\rm const}/S)$. This is again confirmed in Figure~\ref{fig:corr}. We checked that $S_{\rm P} \propto 1/S^2$ also for larger $S$, up to $S=24$.
\par
\begin{figure}[h!]
\centerline{\includegraphics[width=160mm]{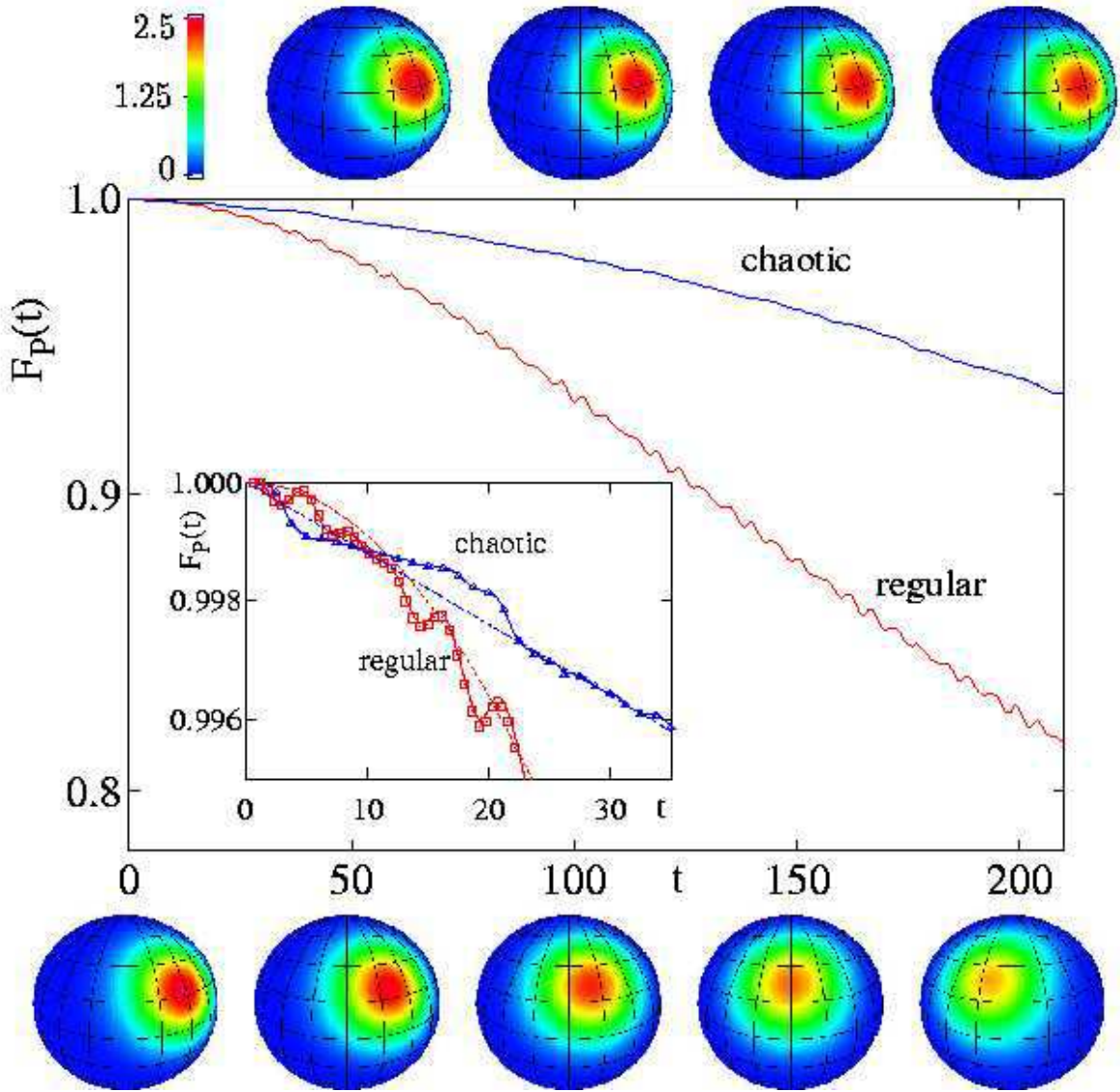}}
\caption{Echo dynamics in the Jaynes-Cummings model.
 The square of the Wigner function for chaotic dynamics (top diagrams)
 and integrable dynamics (bottom diagrams) is shown as a function of 
time at times corresponding to the axis. The purity fidelity is shown in the main frame on the same time scale and for short times in the inset. 
Red curves give the integrable and blue curves the chaotic evolution. In the inset full curves show the complete numerics, 
symbols the evaluation starting from the numerical correlation integrals of Figure~\ref{fig:corr} and dashed curves the linear or quadratic approximation using $\sigma$ and $\oCp$.}
\label{fig:bigwig}
\end{figure}
Next, we would like to demonstrate a faster decay of the purity
fidelity in the regular regime than in the chaotic one. We set the perturbation strength to $\delta=0.005$ for which we expect the purity fidelity to become lower in the regular regime at the time determined by $1-\Fp=2\delta^2S^2 \oCp t^2 \approx 1-\Fn=\delta^2S^2 2 \sigma t$, i.e. at $t \approx 9$, where we used $\sigma=0.1$ and $\oCp=0.011$ as given in Figure~\ref{fig:corr}. In Figure~\ref{fig:bigwig} we show the purity-fidelity for chaotic and regular regimes. We can see the crossover at the predicted time $t\approx 9$. We also illustrate the evolution of the purity fidelity with the square of the Wigner function corresponding to the reduced density matrix $\rho_{\rm c}^{\rm M}(t)$ on the central spin subspace. If $W_{\!\rho_{\rm c}^M}(\vartheta,\varphi)$ is the Wigner function, purity fidelity can be written as a phase space integral
\begin{equation}
\Fp=\int{\!{\rm d}\Omega\, W_{\rho_{\rm c}^M}^2(\vartheta,\varphi)}.
\end{equation}
For details about the spin Wigner function see Appendix~\ref{app:wig}. Near the top and bottom of Figure~\ref{fig:bigwig} we see this evolution for the chaotic and the integrable Hamiltonian respectively. In the centre of the figure we plot the purity fidelity on the same time scale as the Wigner 
functions in the main frame and an amplification of short times in the inset.
We observe detailed agreement of numerics with the results obtained from the numerical values of
the correlation integrals $S_{\rm F}(t)$ and $S_{\rm P}(t)$, reproducing the 
oscillatory structure of the decay. From the same correlation integrals we obtained the coefficients 
for the linear and quadratic decay, which agree well if we discard the oscillations. It is important to remember that the integral over the square of the Wigner function gives the purity fidelity
and therefore the fading of the picture will be indicative of the
purity fidelity decay. On the other hand the movement of the centre is
an indication of the rapid decay of fidelity (not shown in the figure).

\section{Mixing Dynamics}
Discussing linear response results (Section~\ref{sec:LRE}) in the case
of mixing dynamics we have shown that the linear decay is the same for all three quantities. Similar random matrix arguments as for the linear response can be used also for higher order terms and therefore one expects that in the semiclassical limit of small $1/{{\cal N}_{\rm c}}+1/{{\cal N}_{\rm e}}$ we will have the same exponential decay (\ref{eq:Fn_mixing})
\begin{equation}
 F_{\rm P}(t) \approx F_{\rm R}^2(t) \approx F^2(t)=\exp{(-2 t/\tau_{\rm m})},
\label{eq:three}
\end{equation}
with the decay time $\tau_{\rm m}=\hbar^2/2\delta^2\sigma_{\rm cl}$
(\ref{eq:Fn_mixing}) independent of the initial state. This result is
expected to hold when $\Sigma(t)$ can be approximated with a random
matrix for large times (\ref{eq:rm}) and $V$ does not contain terms
acting on only one subspace. Such terms could cause fidelity to
decay while having no influence on purity.
\par
\begin{figure}[h]
\centerline{\includegraphics{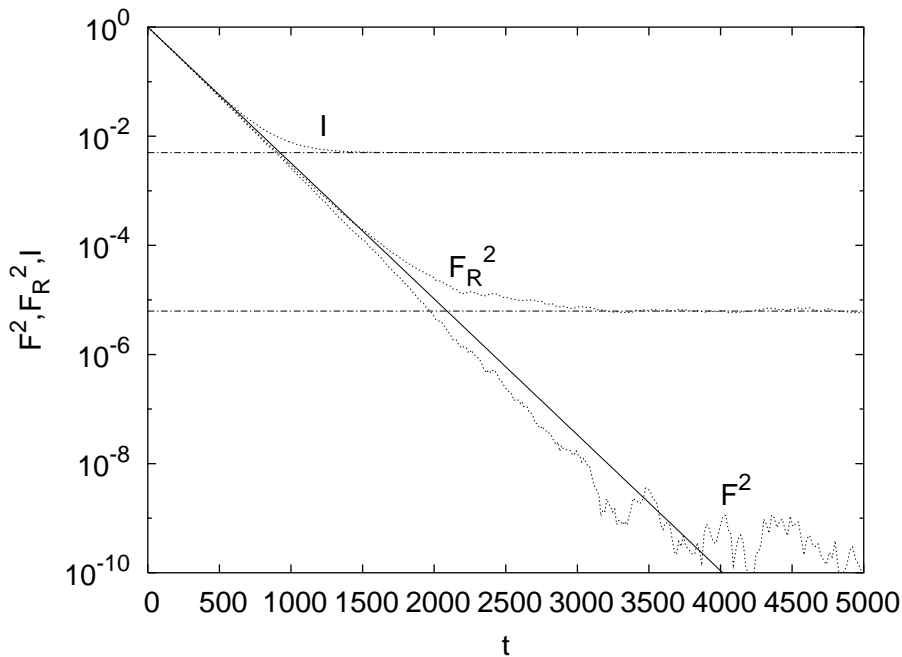}}
\caption{Decay of $F^2(t),F^2_{\rm R}(t)$ and $I(t)$ (dotted curves) in the mixing regime of the double kicked top. The solid line gives the theoretical exponential 
decay (\ref{eq:three}) with $\tau_{\rm m}$ calculated from the classical $\sigma_{\rm cl}=0.056$. Horizontal chain lines give the saturation values of the purity and the reduced fidelity, $1/200$ and $1/400^2$, respectively.}
\label{fig:fid3030}
\end{figure}
For numerical verification of this result we chose a double kicked top
system (\ref{eq:2KT_def}) with $\varepsilon=0$, so that the
unperturbed evolution is uncoupled. For the double kicked top model we
will always have an uncoupled unperturbed evolution in all our
numerical demonstrations in this chapter. Other parameters are
$\gamma_{\rm c,e}=\pi/2.1$ and $\alpha_{\rm c,e}=30$, ensuring chaotic
classical dynamics. The spin size is chosen to be $S=200$, so that we
have ${\cal N}_{\rm c,e}=2S+1$. The perturbation is
\begin{equation}
V=\frac{1}{S^2} S_{\rm z}\otimes S_{\rm z},
\label{eq:V_SzSz}
\end{equation}
with the strength $\delta=8 \cdot 10^{-4}$. The coherent product initial state is placed at $\vartheta_{\rm c,e}^*=\pi/\sqrt{3}$, $\varphi_{\rm c,e}^*=\pi/\sqrt{2}$. We show in Figure~\ref{fig:fid3030} the decay of the fidelity $F(t)$, the reduced fidelity 
$\Fr$ and the purity $I(t)$. Clean exponential decay is observed in all three cases, on a time scale $\tau_{\rm m}$ (\ref{eq:three}) given by the classical transport coefficient $\sigma_{\rm cl}$. We numerically calculated the classical correlation function
\begin{equation}
C_{\rm cl}(t)=\lbrack \ave{z_{\rm c}(t)z_{\rm c}(0)}_{\rm cl} \rbrack^2,
\end{equation}
where we took into consideration that the unperturbed dynamics is uncoupled and is the same for both subsystems and that $\ave{z}_{\rm cl}=0$. Taking only the first term $C_{\rm cl}(0)=1/9$ would give $\sigma_{\rm cl}=1/18$ (\ref{eq:sigma_def}) while the full sum of $C_{\rm cl}(t)$ gives a slightly larger value $\sigma_{\rm cl}=0.056$. Exponential decay, of course, persists only up to the saturation value determined by a finite Hilbert space size (see Section~\ref{sec:time_averaged}).

\section{Regular Dynamics}

\subsection{Beyond Linear Response}
\label{sec:I}
\subsubsection{Purity Fidelity}
For regular dynamics and coherent initial states one can calculate
purity fidelity to all orders in the semiclassical limit and not just
the linear response expansion (\ref{eq:oCp}). For times larger than
the averaging time $t_{\rm ave}$ the echo operator goes towards $\Md
\to \exp{(-\ii \delta \oV t/\hbar )}$. In the semiclassical limit we
can use the classical limit $\bar{v}$ instead of $\oV$. Classical
$\bar{v}(\vec{J})$ is a function of action operators only and so
similarly to the evaluation of fidelity one can use the ASI
(\ref{eq:fn_ASI}) for the evaluation of purity fidelity. A partial trace over the environment gives an integral over $\vec{j}_{\rm e}$ and due to a square in the definition of purity fidelity we end up with an integral over $2(d_{\rm c}+d_{\rm e})=2d$ dimensions, if $d_{\rm c}$ is the dimension of $\vec{j}_{\rm c}$ and $d_{\rm e}$ of $\vec{j}_{\rm e}$,
\begin{equation}
\Fp \cong \hbar^{-2d} \int{\!{\rm d}\Gamma \exp{\left[ -\ii \frac{\delta}{\hbar} t \left\{ \bar{v}(\vec{j}_{\rm c},\vec{j}_{\rm e})-\bar{v}(\vec{j}_{\rm c}',\vec{j}_{\rm e})+\bar{v}(\vec{j}_{\rm c}',\vec{j}_{\rm e}')-\bar{v}(\vec{j}_{\rm c},\vec{j}_{\rm e}') \right\} \right] } d_\rho(\vec{j}) d_\rho(\vec{j}') },
\label{eq:Fp_ASI}
\end{equation} 
where ${\rm d}\Gamma={\rm d}^{d}\vec{j}{\rm d}^{d}\vec{j}'$,
$\vec{j}=(\vec{j}_{\rm c},\vec{j}_{\rm e})$, $\vec{j}'=(\vec{j}_{\rm
c}',\vec{j}_{\rm e}')$ and $d_\rho(\vec{j})$ is the classical limit of
$\bracket{\vec{n}}{\rho}{\vec{n}}$ (\ref{eq:fn_ASI}). Next we expand $\bar{v}(\vec{j})$ around the position of the initial packet (\ref{eq:ov_series}). The constant term $\bar{v}(\vec{j}^*)$ and the linear terms cancel exactly as well as the diagonal quadratic terms, regardless of the position of the initial packet. The argument of the exponential function which remains is then
\begin{equation}
-\ii \frac{\delta t}{\hbar} \left[ (\vec{j}_{\rm c}-\vec{j}_{\rm
 c}')\cdot\bar{v}_{\rm ce}''(\vec{j}^*)(\vec{j}_{\rm e}-\vec{j}_{\rm e}') \right],
\end{equation}
with the matrix $\bar{v}''$ of second derivatives given in
Equation~(\ref{eq:v''}). The resulting expansion can be used in $\Fp$
(\ref{eq:Fp_ASI}) to calculate the purity fidelity for initial states
having well defined classical $d_\rho(\vec{j})$. For coherent initial
states $d_\rho(\vec{j})$ are Gaussian (\ref{eq:drho_coh}) and so the whole integral is
also a Gaussian, giving 
\begin{equation}
\Fp=\frac{1}{\sqrt{\det{(\mathbbm{1}-\ii \delta t L^{-1} \tilde{V}'')} }},
\end{equation}
where $L$ and $\tilde{V}''$ are $2d$ dimensional matrices
\begin{equation}
L:=\pmatrix{ \Lambda_{\rm c} & 0 & 0 & 0 \cr 0 &\Lambda_{\rm c} & 0 &
0 \cr 0 & 0 & \Lambda_{\rm e} & 0 \cr 0 & 0 & 0 & \Lambda_{\rm e}},\qquad \tilde{V}'':=\frac{1}{2}\pmatrix{ 0 & 0 & \bar{v}''_{\rm ce} & -\bar{v}''_{\rm ce} \cr
0 & 0 & -\bar{v}''_{\rm ce} & \bar{v}''_{\rm ce} \cr
\bar{v}''_{\rm ec} & -\bar{v}''_{\rm ec} & 0 & 0\cr
-\bar{v}''_{\rm ec} & \bar{v}''_{\rm ec} & 0 & 0},
\end{equation}
with matrices of squeezing parameters $\Lambda_{\rm c}$ and
$\Lambda_{\rm e}$ (\ref{eq:Lambda_def}) of dimensions $d_{\rm c}$ and
$d_{\rm e}$, respectively, and $d_{\rm c}\times
d_{\rm e}$ dimensional matrix $\bar{v}''_{\rm ce}$
(\ref{eq:v''}). Note that the determinant is real despite the
imaginary unit. The determinant of the matrix $\mathbbm{1}-\ii \delta
t L^{-1} \tilde{V}''$ can be
simplified using the following identity for block matrices
\begin{equation}
\det{\pmatrix{A & B \cr C & D}}=\det{(A)}\det{(D-CA^{-1}B)}=\det{(D)}\det{(A-BD^{-1}C)},
\end{equation}
with $m \times n$ dimensional matrices $B$ and $C^T$. Noting that the
matrix $\mathbbm{1}-\ii \delta
t L^{-1} \tilde{V}''$ has $\mathbbm{1}$ on the diagonal, we obtain 
\begin{equation}
\det{(\mathbbm{1}-\ii \delta t L^{-1} \tilde{V}'')}=\det{(\mathbbm{1}+(\delta t)^2 Z ) },
\end{equation}
with $Z$ being the $2d_{\rm c}\times 2d_{\rm c}$ dimensional matrix
\begin{equation}
Z:=\frac{1}{2}\pmatrix{\Lambda_{\rm c}^{-1} \bar{v}''_{\rm ce} \Lambda_{\rm e}^{-1}\bar{v}''_{\rm ec} & -\Lambda_{\rm c}^{-1} \bar{v}''_{\rm ce} \Lambda_{\rm e}^{-1}\bar{v}''_{\rm ec} \cr
 -\Lambda_{\rm c}^{-1} \bar{v}''_{\rm ce} \Lambda_{\rm e}^{-1}\bar{v}''_{\rm ec} & \Lambda_{\rm c}^{-1} \bar{v}''_{\rm ce} \Lambda_{\rm e}^{-1}\bar{v}''_{\rm ec}}.
\end{equation}
To simplify the determinant of $\mathbbm{1}+(\delta t)^2 Z$ we use the following identity
\begin{equation}
\det{\pmatrix{A & B \cr C & D}}=\det{(DA-BC)},
\end{equation}
for square matrices $A,B,C$ and $D$ and commuting $B$ and $D$,
 $[B,D]=0$. Using this finally gives
\begin{equation}
\Fp=\frac{1}{\sqrt{\det{\{ \mathbbm{1}+(\delta t)^2 \,u \}}}},\qquad u:=\Lambda_{\rm c}^{-1} \bar{v}''_{\rm ce} \Lambda_{\rm e}^{-1}\bar{v}''_{\rm ec}.
\label{eq:Fp_reg}
\end{equation}
Note that $u$ is the same $d_{\rm c}\times d_{\rm c}$ matrix we had in
the expression for $\oCp$ (\ref{eq:oCp}). This very simple, yet
important result deserves a little discussion. It gives the purity
fidelity (and thereby as a special case also the purity) decay for
regular systems and coherent initial states to all orders in
$\delta$. Its validity is limited to times larger than the averaging
time $t_{\rm ave}$ in which $\oV$ converges and for sufficiently small
$\hbar$. Planck's constant must be small to allow the replacement of
$\oV$ with its classical limit $\bar{v}$ and $\bar{v}$ must be smooth
on the scale of the wave packet size $\sqrt{\hbar}$. Furthermore, we
replaced a sum over the quantum numbers with an integral over the
actions, the validity of this being given by the condition $\delta t
|\bar{v}''_{\rm ce}| \hbar^2/\hbar \ll 1$, i.e. $\delta t_{\rm
u}<1/\hbar$, if $t_{\rm u}$ is the upper time limit of validity of the ASI. As we will see, the asymptotic decay of purity fidelity as
given by Equation~(\ref{eq:Fp_reg}) is between $1/(\delta t)$ and
$1/(\delta t)^{d_{\rm c}}$, depending on the matrix $u$. The $\Fp$ at
the upper border $t_{\rm u}$ is therefore between $\hbar$ and
$\hbar^{d_{\rm c}}$. The purity fidelity will for long times saturate
at the plateau given by the finite Hilbert space size, $\bar{F}_{\rm
P}\approx 1/{\cal N}_{\rm c}\sim \hbar^{d_{\rm c}}$. Comparing this,
we see that the upper border $t_{\rm u}$ coincides with the point
where the asymptotic saturation $\bar{F}_{\rm P}$ is reached for one
degree of freedom systems. For $d_{\rm c}>1$ the ASI break time $t_{\rm u}$
is only by a constant factor smaller than the time when we reach $\bar{F}_{\rm P}$. Summarising, the purity fidelity decay (\ref{eq:Fp_reg}) is valid from $t_{\rm ave}$ all the way to the asymptotic plateau $\bar{F}_{\rm P}$ provided $\hbar$ is sufficiently small, without any bound on $\delta$.
\par
Let us now explore the asymptotic decay of $\Fp$. To simplify
theoretical arguments, we assume that the squeezing parameters are all
equal to one\footnote{Usually this can be achieved by the right choice
of actions.}, $L\equiv \mathbbm{1}$ so that we have
$u=\bar{v}''_{\rm ce} \bar{v}''^T_{\rm ce}$. In $1+ d_{\rm e}$ degrees
of freedom systems ($d_{\rm c}=1$), the matrix $u$ is just a number, and the purity fidelity decays as
\begin{equation}
\Fp=\frac{1}{\sqrt{1+u\,(\delta t)^2}},\qquad d_{\rm c}=1.
\label{eq:Fp_1d}
\end{equation}
A single parameter $u$ is already fixed by the linear response, i.e. by the value of $\oCp$ (\ref{eq:oCp}). Asymptotically we get $\Fp \asymp 1/(\delta t)$ decay {\em regardless} of the second dimension $d_{\rm e}$. For general systems with $d_{\rm c}>1$ one can see that the determinant $\det{\{ \mathbbm{1}+(\delta t)^2 \,u \}}$ is a polynomial of order at most $d_{\rm c}$ in $(\delta t)^2$. Furthermore, as the matrix $u$ is symmetric and positive definite, its eigenvalues are positive, meaning that all the coefficients of the polynomial are positive and the $\Fp$ is always less than one. In a special case, when the matrix $u$ can be written as a tensor product of two vectors, like $u=\bar{v}''_{\rm ce} \bar{v}''^T_{\rm ce}=:\vec{x}\otimes\vec{y}$, the determinant is again simple regardless of the dimensions involved and we get
\begin{equation}
\Fp=\frac{1}{\sqrt{1+(\delta t)^2 \vec{x}\cdot\vec{y}}},\qquad u=\vec{x}\otimes\vec{y}.
\end{equation}
In such case we again get the asymptotic $\Fp\asymp 1/(\delta t)$
decay, but here {\em regardless} of {\em both} dimensions. Matrix $u$
has such a form for instance if all the matrix elements of
$\bar{v}''_{\rm ce}$ are the same (e.g. for the perturbation
$\bar{v}=\sum_{k,l}{ (\vec{j}_{\rm c}\otimes \vec{j}_{\rm e}
)_{kl}}$). Our result of course applies also to the purity decay in
weakly coupled systems and does not agree with the result of~\citet{Jacquod:04}.

\subsubsection{Reduced Fidelity}
For the reduced fidelity a similar procedure based on the ASI can be used. In the expansion of $\bar{v}(\vec{j})$ around $\vec{j}^*$ (\ref{eq:ov_series}) the first non vanishing term is a linear one. The resulting Gaussian integrals are analogous to the ones in the calculation of the fidelity decay for coherent initial states (but there are twice as many), the final result being a Gaussian decay
\begin{equation}
\Fr=\exp{\left(-\frac{\delta^2}{\hbar^2} \oCr t^2 \right)},
\label{eq:Fr_exact}
\end{equation} 
with a single parameter $\oCr=\frac{1}{2}\hbar \left( \vec{\bar{v}}_{\rm c}'\Lambda^{-1}_{\rm c}\vec{\bar{v}}'_{\rm c} \right)$ given by the linear response alone (\ref{eq:oCr}).

\subsubsection{Numerical Illustration}
\begin{figure}[h]
\centerline{\includegraphics{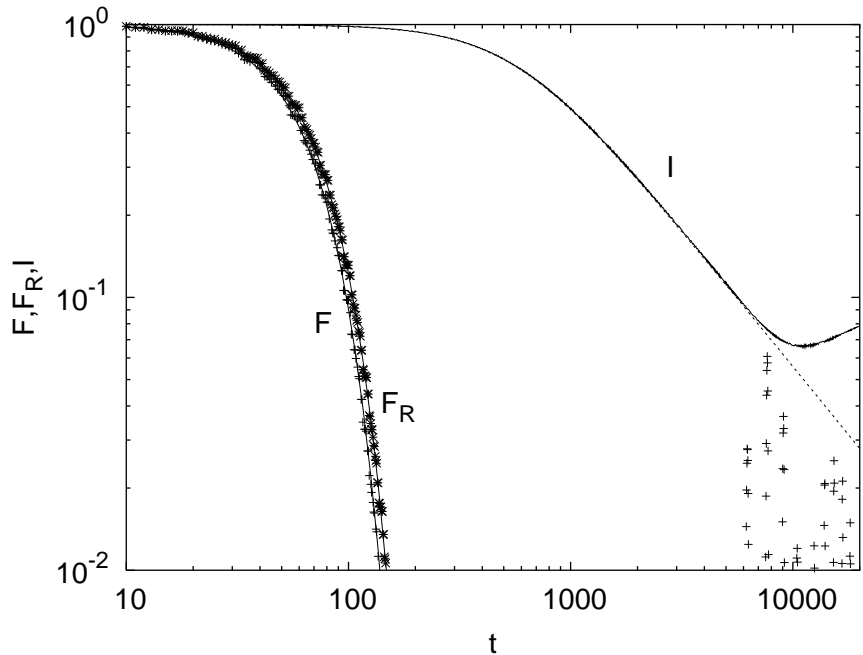}}
\caption{Decay of $F(t),F_{\rm R}(t)$ and $I(t)$ in the regular regime
of the double kicked top. For parameters see text. The theoretical
Gaussian decay for the fidelity (\ref{eq:Fn_regcoh}) and the reduced
fidelity (\ref{eq:Fr_exact}), with the theoretical value of $\oC$
(\ref{eq:oc_th}) and $\oCr$ (\ref{eq:ocr_th}), overlap with the
numerics (symbols) within the line width. Theoretical purity decay
(\ref{eq:Fp_1d}) (dotted curve) as determined by the theoretical $u$
(\ref{eq:u_th}) also nicely agrees with the numerics (full curve) up to the plateau.}
\label{fig:fidJ100_0_0}
\end{figure}
To numerically confirm the above formulas for the decay of purity
fidelity and reduced fidelity we take a double kicked top model
(\ref{eq:2KT_def}). To be in a regular regime we take $\alpha_{\rm
c,e}=0$ and $\gamma_{\rm c}=\pi/2.1$ and $\gamma_{\rm
e}=\pi/\sqrt{7}$. Different unperturbed frequencies $\gamma_{\rm
c}\neq \gamma_{\rm e}$ are chosen in order to have a general
situation, i.e. that the subsystems are not in resonance. Unperturbed
dynamics is again uncoupled, $\varepsilon=0$, so that the purity
fidelity equals purity. The spin size is chosen $S=100$ and the initial
product coherent state is placed at $(\vartheta^*,\varphi^*)_{\rm
c}=\pi(1/\sqrt{3},1/\sqrt{2})$ and $(\vartheta^*,\varphi^*)_{\rm
e}=\pi(1/\sqrt{3},3/\sqrt{7})$. The perturbation with strength
$\delta=0.01$ is of the form
\begin{equation}
V=\frac{1}{S^4} S_{\rm z}^2 \otimes S_{\rm z}^2.
\end{equation}
Unperturbed classical evolution is very simple, namely rotation around
$y$-axes for angles $\gamma_{\rm c}$ and $\gamma_{\rm e}$ for the
central system and the environment, respectively. The classical limit
of $\oV$ is readily calculated and expressed in terms of the actions $j_{\rm c}=y_{\rm c}$ and $j_{\rm e}=y_{\rm e}$,
\begin{equation}
\bar{v}=\frac{1}{4} (1-j_{\rm c}^2)(1- j_{\rm e}^2).
\end{equation}
The derivatives of $\bar{v}$ are $\bar{v}'_{\rm c}=-j_{\rm c}(1-j_{\rm
e}^2)/2$ and similarly for $\bar{v}'_{\rm e}$. The squeezing parameter
for the spin coherent states is $\Lambda=1/(1-j^2)$ and the average correlation function $\oC$ (\ref{eq:oC'}) is
\begin{equation}
\oC=\frac{1}{S}(1-j_{\rm c}^2)(1-j_{\rm e}^2)\{j_{\rm c}^2(1-j_{\rm e}^2)+j_{\rm e}^2(1-j_{\rm c}^2) \},
\label{eq:oc_th}
\end{equation}
which for our choice of the initial coherent packet gives $\oC=0.024/S$. For the reduced fidelity we get
\begin{equation}
\oCr=\frac{1}{S}j_{\rm c}^2(1-j_{\rm c}^2)(1-j_{\rm e}^2),
\label{eq:ocr_th}
\end{equation}
which evaluates to $\oCr=0.021/S$. The above two parameters $\oC$ and
$\oCr$ completely determine the Gaussian fidelity (\ref{eq:Fn_regcoh})
and reduced fidelity (\ref{eq:Fr_exact}) decays. As we have $1+ 1$
degrees of freedom system, the purity is determined by a single parameter $u$ (\ref{eq:Fp_1d}). For our perturbation we get
\begin{equation}
u=j_{\rm c}^2 j_{\rm e}^2 (1-j_{\rm c}^2)(1-j_{\rm e}^2),
\label{eq:u_th}
\end{equation}
giving $u=0.032$ for our initial product coherent state. As our perturbation $\oV$ is of the product form, we could as well use a universal relation between $\oCp$, $\oCr$ and $\oC$ (\ref{eq:uni_rel}) to calculate $\oCp$ (or equivalently $u$). The numerical results are shown in Figure~\ref{fig:fidJ100_0_0}, together with the theory. Agreement is excellent.

\subsection{The Jaynes-Cummings Model}
\label{sec:jaynes_reg}
In this section we will consider in detail the stability of a
Jaynes-Cummings model, described in Section~\ref{sec:Jaynes}, under
various perturbations. The Jaynes-Cummings system can be
experimentally realized and so it is a possible model on which one could
experimentally study the quantum stability. In experiments one usually has only a
co-rotating term, i.e. $G'$ is zero, therefore we will focus on a
situation when we have $G'=0$ in the unperturbed dynamics and so the
classical limit of the unperturbed system is integrable. For the
initial state we will always choose a product of coherent states for
a harmonic oscillator, given by a real parameter $\alpha$ (\ref{eq:boson_coh}), and
a spin, given by the initial position $(\vartheta,\varphi)$
(\ref{eq:SU2_coh}). In studies of the reduced fidelity and purity
fidelity we will consider the spin as a central system, and the
harmonic oscillator  as an environment. For regular dynamics with coherent initial states the decay of fidelity and reduced fidelity is Gaussian with the decay time determined by a single parameter $\oC$ and $\oCr$, while the decay of the purity fidelity is $\Fp=1/\sqrt{1+u\,(\delta t)^2}$ (\ref{eq:Fp_1d}) and is again determined by a single parameter $u$. With this in view, it is sufficient to determine only the linear response parameters $\oC$ (\ref{eq:oC'}), $\oCr$ (\ref{eq:oCr}) and $\oCp$ (\ref{eq:oCp}), to get the decay to all orders in $\delta$. Therefore, we will focus only on the calculation of these coefficients. The calculation of the expectation values in the coherent initial state is described in Appendix~\ref{app:expect}. 
\par
Our unperturbed system will be
\begin{equation}
H_0=\hbar \omega a^+ a+\hbar \varepsilon S_{\rm z}+G \frac{\hbar}{\sqrt{2S}} (a S_+ + a^+ S_-).
\label{eq:JC_H0}
\end{equation}
For the perturbation we will look at four different situations:
\begin{itemize}
\item variation of $\omega$, corresponding to $V=\hbar a^+ a=:[\delta \omega]$
\item variation of $\varepsilon$, corresponding to $V=\hbar S_{\rm z}=:[\delta \varepsilon]$
\item variation of $G$, corresponding to $V=\frac{\hbar}{\sqrt{2S}}(a S_+ + a^+ S_-)=:[\delta G]$
\item variation of $G'$, corresponding to $V=\frac{\hbar}{\sqrt{2S}}(a S_- + a^+ S_+)=:[\delta G']$,
\end{itemize}
where we introduced short notation $V=[\delta \bullet]$ symbolically
denoting the perturbation in the parameter $\bullet$. In the last case
of $V=[\delta G']$ and if $G=0$ all correlations average to zero,
corresponding to the residual perturbation (i.e. freezing), and this
case will be considered separately in Section~\ref{sec:har_freeze}. 
\par
As an additional simplifying assumption, we will assume that we have
$\omega=\varepsilon$, i.e. spin and an oscillator are in
resonance. For $G\neq 0$ we have numerically checked that the average
correlation functions are quite insensitive to the resonance
condition. The only exception being the $V=[\delta G]$ perturbation,
for which $\oC$, $\oCr$ and $\oCp$ vanish if we are out
of resonance and have $G=0$. Because this freezing case will be studied separately, we do not loose generality by taking $\omega=\varepsilon$ but gain in the simpler theoretical calculations. Namely, in resonance the $G$ term commutes with the unperturbed Hamiltonian, $[a S_+ + a^+ S_-,H_0]=0$. In addition, in experimental implementations we are usually close to the resonance condition. 

\begin{figure}[h!]
\centerline{\includegraphics[width=160mm]{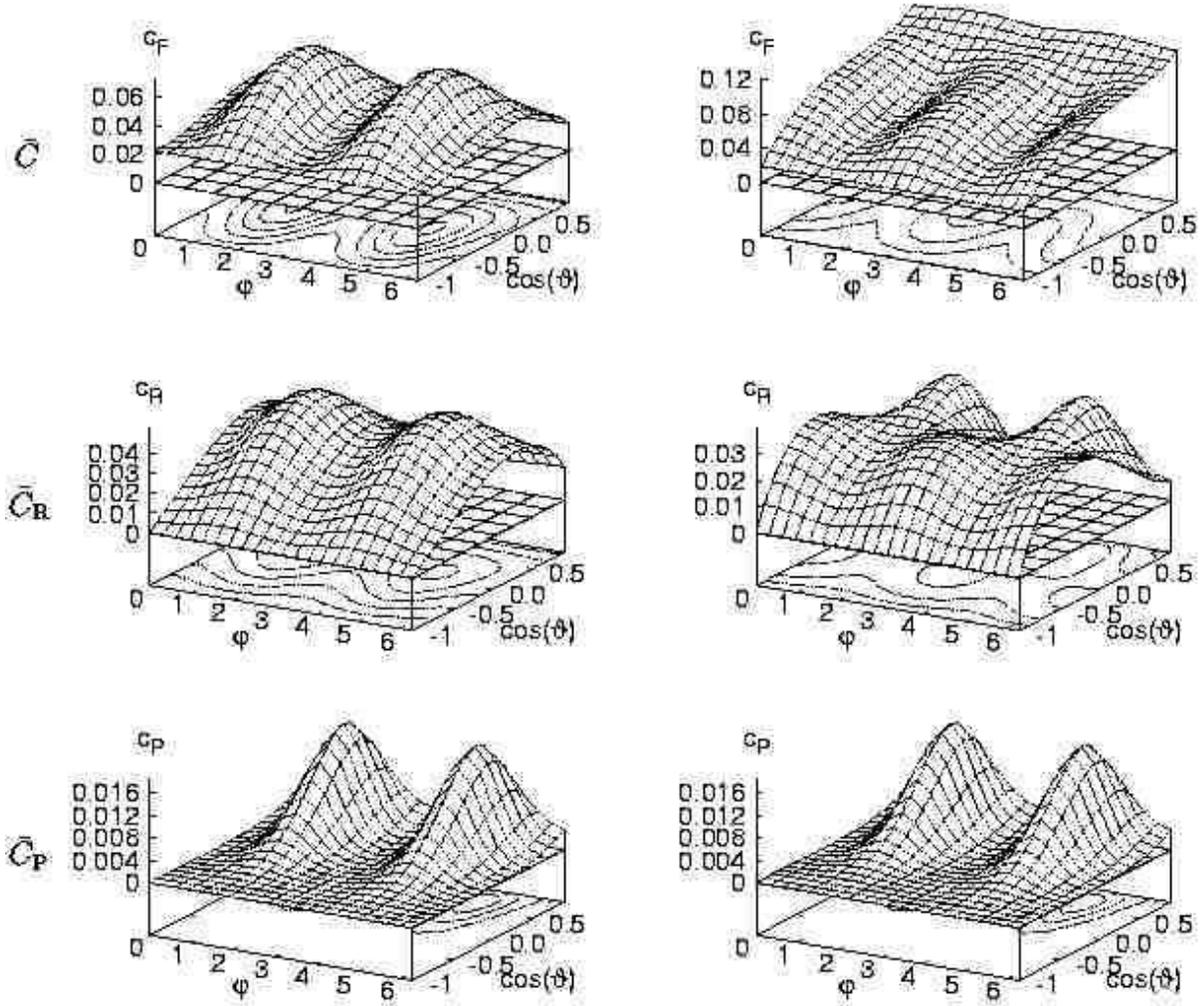}}
\caption{Numerically calculated dependence of $\bar{C}$'s on the
position of the spin coherent state $\varphi$ and $\cos{\vartheta}$
for $\omega=\varepsilon=0.3$, $G=1$ (everything is independent of $G$
provided it is nonzero), $G'=0$, $\alpha=1.15$ and $J=4$. Left column
shows the results for $\varepsilon$ perturbation and the right for the
perturbation in $\omega$.}
\label{fig:cbar}
\end{figure}
\subsubsection{Perturbation of the frequency $\omega$}
We will first analyse the case when the perturbation consists of the variation of $\omega$, so that the perturbation generator is
\begin{equation}
V=\hbar a^+ a.
\label{eq:varomega}
\end{equation}
Due to the resonance condition $\omega=\varepsilon$ the unperturbed
propagator $U_0$ factorises into a part proportional to the $G$ term and the rest. The perturbation in the Heisenberg picture is therefore 
\begin{equation}
V(t)={\rm e}^{\ii G (a^+ S_- + a S_+) t /\sqrt{2S}} (\hbar a^+ a) {\rm e}^{-\ii G (a^+ S_- + a S_+) t /\sqrt{2S}}.
\label{eq:omegaV}
\end{equation}
To calculate the average correlation function one needs an integral
of the perturbation $\Sigma(t)=\int_0^t{V(\tau){\rm d}\tau}$ in the
limit $t \to \infty$. As the constant $G$ just re-scales time in the perturbation $V(t)$ (\ref{eq:omegaV}), the $\bar{C}$'s which are an infinite time limit property of $\Sigma(t)$ will not depend on $G$, except possibly at $G=0$ where the symmetry of $H_0$ changes and we might have an effect due to degeneracies. Direct calculation in this simple case of $G=0$ gives 
\begin{equation}
\oC=\frac{1}{S}\frac{\alpha^2}{S},\qquad \oCr=\oCp=0,\qquad \hbox{(if $G=0$)},
\end{equation}
for the initial product coherent state (\ref{eq:SU2_coh},\ref{eq:boson_coh}). Note that $\alpha^2/S$ has a
well defined classical limit, namely the energy of the oscillator. The
last two results are expected. The $\oCp=0$ because the perturbed
dynamics is also uncoupled and $\oCr=0$ because the perturbation is
only in the oscillator (``environment'') part of $H_0$. For $G>0$ we
get an additional angle dependent term of order $\hbar^2=1/S^2$ which
can be seen in numerically calculated $\bar{C}$'s in
Figure~\ref{fig:cbar}. Therefore, there is a discontinuous jump in all
three $\bar{C}$'s at $G=0$, the discontinuity being proportional to $\hbar^2$ and thus of higher order for $\oC$ while it is the leading order term for $\oCr$ and $\oCp$. Otherwise the values of the plateaus $\bar{C}$ are independent of the coupling $G$, but of course the time scale on which we get the convergence of $\bar{C}$'s scales as $1/G$. For small $G$ the discontinuity will happen at large times.  

\subsubsection{Perturbation of the spin energy $\varepsilon$}
In this case the perturbation generator is
\begin{equation}
V=\hbar S_{\rm z}.
\label{eq:epsV}
\end{equation}
Everything is analogous to the previous case and in the simple uncoupled case of $G=0$ (and $\omega=\varepsilon$) we get 
\begin{equation}
\oC=\oCr=\frac{1}{S}\frac{\sin^2{\vartheta}}{2},\qquad \oCp=0,\qquad  \hbox{(if $G=0$)}.
\end{equation}
The $\oC$ now does not depend on the oscillator coherent state
parameter $\alpha$ but instead depends on the position of the spin
coherent state. Also, the $\oCr$ is now nonzero as we make the
perturbation in our central system. At $G=0$ there is again a
discontinuity of order $\hbar^2$ in all three $\bar{C}$'s, otherwise
they are independent of $G$. For $\oC$ and $\oCr$ this discontinuity
is of higher order in $\hbar$ and can be neglected in the
semiclassical limit. Dependence of numerically calculated $\bar{C}$'s
on the initial position of the spin coherent packet is shown in Figure~\ref{fig:cbar}. One can see, that $\oCp$ is equal for $[\delta\omega]$ and $[\delta\varepsilon]$ perturbations. Therefore the decoherence (purity fidelity) is insensitive to which frequency we detune, whereas fidelity and reduced fidelity are not.   

\begin{figure}[h!]
\centerline{\includegraphics{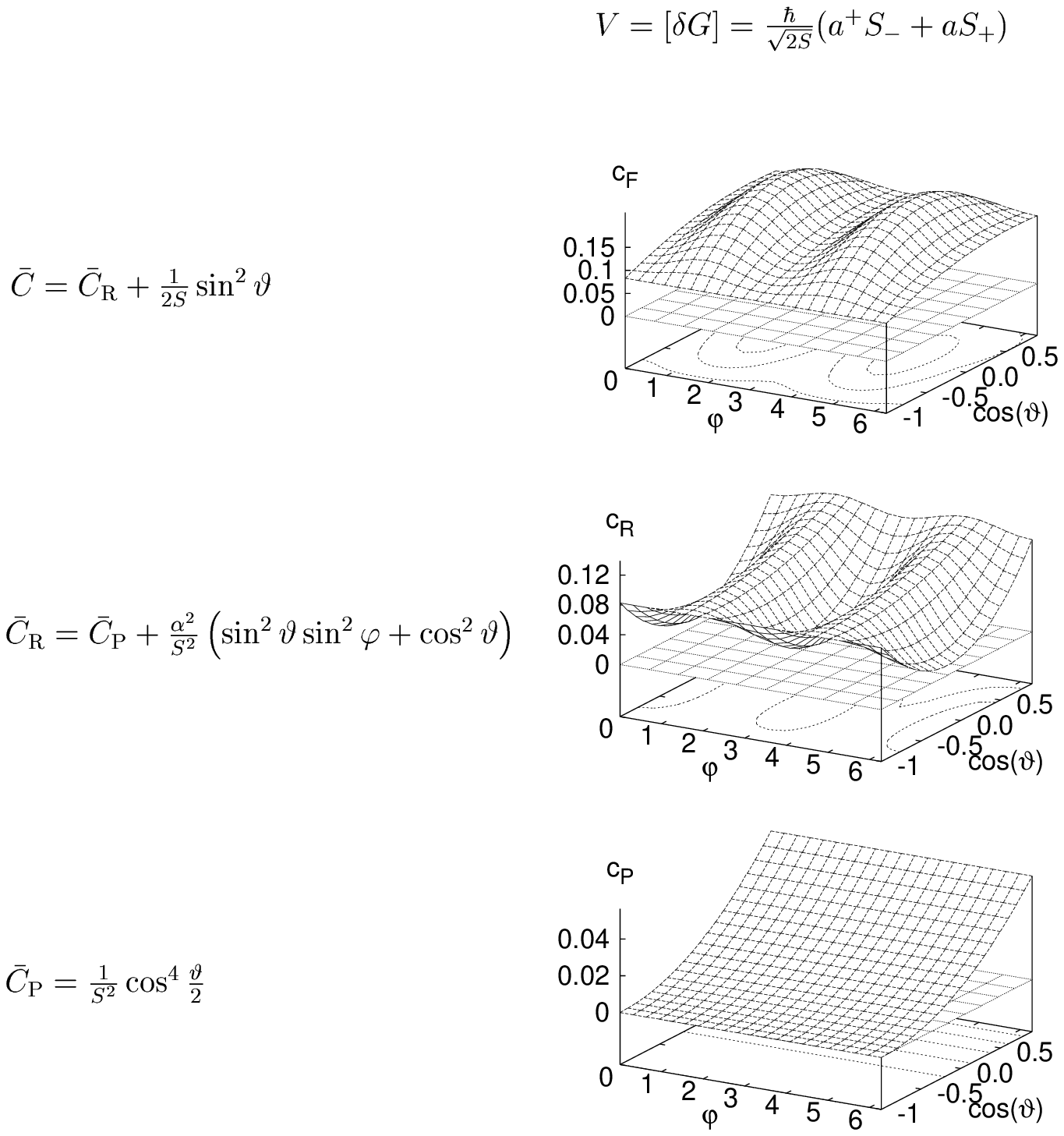}}
\caption{Dependence of $\bar{C}$'s on the position of the spin
coherent initial state ($\varphi$ and $\cos{\vartheta}$) for
$\omega=\varepsilon=0.3$, $G=1$ (everything is independent of $G$),
$G'=0$, $\alpha=1.15$ and $J=4$. The left column shows exact
theoretical results whereas the right one shows numerical plots of this dependence. All is for the perturbation in the coupling constant $G$.}
\label{fig:cbar_G}
\end{figure}
\subsubsection{Perturbation in the coupling $G$}
The perturbation generator is 
\begin{equation}
V=\frac{\hbar}{\sqrt{2S}} (a^+ S_- + a S_+).
\label{eq:GV}
\end{equation}
We again take the resonant condition $\omega=\varepsilon$. Now the
perturbation is constant in time $V(t)=V(0)$ regardless of the value
of $G$ and we can calculate all three $\bar{C}$'s for an arbitrary
$G$. These are written in Figure~\ref{fig:cbar_G} next to each plot
showing its dependence on the initial spin packet position. There is
no dependence on $G$ and also no discontinuity at $G=0$, contrary to
previous two examples. Also the $\oC$ and $\oCp$ are strictly larger
than in the case of $\omega$ or $\varepsilon$ perturbation whereas the
$\oCr$ is of order $\hbar$ and is smaller than for the $\varepsilon$
perturbation. Now $\oCp \propto \hbar^2 $ is of course nonzero and
gives us the decay time scale of purity fidelity (or purity in the case of $G=0$).

\section{Separation of Time Scales}
Until now we have discussed two broad categories, mixing systems in which the whole correlation function decays and regular systems where we had a plateau in the correlation function. In this section we will consider situation where the time scale of the environment is much smaller than that of the central system and the correlation function can be considerably simplified. We will furthermore consider perturbations of the product form
\begin{equation}
V=V_{\rm c}\otimes V_{\rm e}.
\end{equation}
For numerical illustrations we will use a coupled double kicked top
with $\varepsilon=0$, i.e. an uncoupled unperturbed system. To simplify notation we will use $\ave{A_{\rm c}}$ to denote the average of operator $A_{\rm c}$ acting only on the {\em central system} in the initial state of the {\em central system}, $\ave{A_{\rm c}}=\trc{A_{\rm c}\rho_{\rm c}(0)}$, and similarly for the environment. If the operator acts on the whole system, then $\ave{A}$ denotes the expectation in the {\em whole} initial state, as before. When the time scale of the environmental correlations $\ave{V_{{\rm e}}(t) V_{{\rm e}}(t')}$ 
is much smaller than the time scale of the central systems' correlations 
$\ave{V_{{\rm c}}(t) V_{{\rm c}}(t')}$, time averaging over the fast
environmental part of the perturbation can be performed. Regarding the
environmental correlation function two extreme situations are
possible. If the correlations of 
the environment decay, we will call such a case ``fast mixing
environment'', and we have a finite integral of the environmental 
correlation function. If the correlations of the environment do not decay, we will call it a ``fast regular environment'', and we 
have generically a non vanishing average correlation function of the environment.  

\subsection{Fast Mixing Environment}
\label{sec:fast_cha}
The situation, when the time scale $t_{{\rm e}}$ on which the correlation function for the environment decays
is much smaller than the time scale $t_{{\rm c}}$ of the central system, is of considerable physical interest. 
This includes various ``brownian'' like baths, where the correlation times are smaller than the dynamical times 
of the central system in question. The correlation sums $S_{\rm F}(t)$, $S_{\rm P}(t)$ (\ref{eq:St}) and $S_{\rm R}(t)$,
\begin{equation}
S_{\rm R}(t):=\frac{1}{t}\<\Sigma(t)(\mathbbm{1}-\rho_{\rm c})\otimes\mathbbm{1}\Sigma(t)\>,
\label{eq:Str}
\end{equation}
giving the linear response decay of the fidelity $\Fn=1-(\delta/\hbar)^2 t S_{\rm F}(t)$, the reduced fidelity $\Fr=1-(\delta/\hbar)^2 t S_{\rm R}(t)$ and of the purity (or generally purity fidelity) $I(t)=1-2(\delta/\hbar)^2 t S_{\rm P}(t)$ can be significantly simplified in such situation. We will furthermore assume $\overline{\ave{V_{{\rm e}}}}=0$, with $\overline{\ave{A}}=\lim_{t \to \infty}{t^{-1}\int_0^t{\ave{A(\xi)} d\xi}}$ denoting a time average. This assumption corresponds to an equilibrium situation where the average ``force'' $V_{{\rm e}}$ vanishes. The integration over the fast 
variable $V_{{\rm e}}$ can be carried out and we get for $t\gg t_{\rm c} \gg t_{\rm e}$
\begin{eqnarray}
S_{\rm F}(t)&=& 2 \sigma_{{\rm e}} \overline{\ave{V_{{\rm c}}^2}} \nonumber \\
S_{\rm R}(t)&=& 2 \sigma_{{\rm e}} \left\{ \overline{\ave{V_{{\rm c}}^2}}-\overline{\ave{V_{{\rm c}}}^2} \right\}  \nonumber \\
S_{\rm P}(t)&=& 2 \sigma_{{\rm e}} \left\{ \overline{\ave{V_{{\rm c}}^2}}-\overline{\ave{V_{{\rm c}}}^2} \right\},
\label{eq:fastCDE}
\end{eqnarray} 
with 
\begin{equation}
\sigma_{{\rm e}}:=\lim_{t \to \infty}{\< \Sigma_{{\rm e}}^2(t)\>/2t}, \qquad{\rm with}\qquad 
\Sigma_{{\rm e}}(t)=\int_0^t{\! V_{{\rm e}}(\xi) d\xi},
\label{eq:sigmae}
\end{equation}
being the integral of the autocorrelation function for the
environmental part of the perturbation $V_{{\rm e}}$ alone. The result does not depend on the initial state of the environment. Some interesting conclusions can be drawn from these linear response results (\ref{eq:fastCDE}).
\par
We can see that the decay time scale depends only on the time average diagonal correlations of the central system 
$\ave{V_{{\rm c}}^2(t)}$ and not on the full correlation function. This is a simple consequence of the separation of 
time scales and means that the decay of all three stability measures 
does not depend on the dynamics of the central system (e.g. being mixing or regular). Furthermore, reduced fidelity $\Fr$ and the purity $I(t)$ will decay on the same time scale (\ref{eq:fastCDE}), 
meaning that the decay of the reduced fidelity is predominantly caused by the loss of coherence, i.e. entanglement between the two factor spaces.
This in turn means that the reduced fidelity, which is a property of
echo dynamics, i.e. of comparison of two slightly 
different Hamiltonian evolutions, is equivalent to the decay of the purity.
\par
If the initial state of the central system $\rho_{{\rm c}}(0)$ is a Gaussian wave packet (coherent state) then the dispersion $\overline{\ave{V_{{\rm c}}^2}}-\overline{\ave{V_{{\rm c}}}^2}$ is by factor of order 
$\hbar$ {\em smaller} than $\overline{\ave{V_{{\rm c}}^2}}$. Thus for coherent initial states of the central system, irrespective of the 
initial state of the environment, the $\Fr$ and $I(t)$ are going to decay on a $1/\hbar$ times {\em longer} time 
scale than $\Fn$. We have therefore reached a general conclusion based on the assumption of chaotic fast
environment (which is often the case), namely that the 
coherent states are most robust against decoherence (decay of purity), provided $t_{{\rm e}} \ll t_{{\rm c}}$ and decoherence time is longer than the correlation time of the environment, $t_{{\rm e}} \ll t_{\rm dec}$. If 
decoherence is even faster than the time scale of the environment, as is the case for macroscopic superpositions, 
then formulas (\ref{eq:fastCDE}) are not valid any more as one is effectively in a regular regime with $S_{\rm P}(t) \propto t^2$. Decoherence time is then independent not just of systems dynamics but also of environmental dynamics characterised by $\sigma_{{\rm e}}$~\citep{Braun:01,Strunz:03,Strunz:03a}.
\par
\begin{figure}[h]
\centerline{\includegraphics{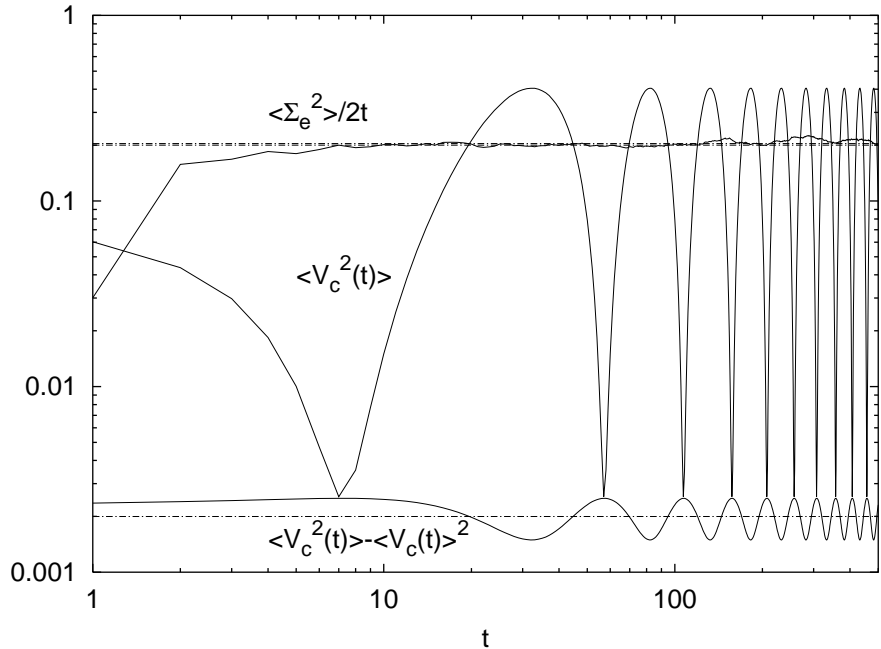}}
\caption{Various correlation sums from formulas (\ref{eq:fastCDE}) in a fast chaotic regime (solid curves, as indicated in the figure). 
Chain lines indicate corresponding theoretical time averages. For details see text.}
\label{fig:cor030}
\end{figure}
In the regime of fast chaotic environment one can derive a master equation for a reduced density 
matrix of the central system~\citep{Kolovsky:94,Qoptics}. We take a partial trace over the environment of the echo density matrix $\rho^{\rm M}(t)$ and write it for a small time step $\Delta t$. This time step 
$\Delta t$ must be larger than the correlation time $t_{{\rm e}}$ of the environment and at the same time smaller than 
the correlation time $t_{{\rm c}}$ of the central system. For the environmental part of the correlation function we assume 
fast exponential decay (particular exponential form is not essential) which is independent of the state $\rho(0)$
\begin{equation}
\tre{ V_{{\rm e}}(t) V_{{\rm e}}(t') \rho } \longrightarrow \frac{\sigma_{{\rm e}}}{t_{{\rm e}}} 
\exp{\{-|t-t'|/t_{{\rm e}}\}} \tre{\rho}.
\label{eq:expcorr}
\end{equation}
Assuming the perturbation to be a product $V(t)=V_{{\rm c}}(t) \otimes V_{{\rm e}}(t)$ and the average ``force'' 
$\tre{V_{{\rm e}}(t) \rho }$ to vanish together with the exponential decay of environmental correlations of the 
form (\ref{eq:expcorr}) for an arbitrary state, yields a master equation for the reduced density matrix 
$\rho^{\rm M}_{\rm c}(t):=\tre{\rho^{\rm M}(t)}$, 
\begin{equation}
\dot{\rho}^{\rm M}_{\rm c}(t)=-\frac{\delta^2 }{\hbar^2}\sigma_{{\rm e}} [V_{{\rm c}}(t),[V_{{\rm c}}(t),\rho^{\rm M}_{\rm c}(t)]].
\label{eq:master}
\end{equation}
Master equation is strictly valid in the Markovian limit of the
correlation function being a delta function in time or if considered on time
scales larger than the correlation time.  
\par
\begin{figure}[h]
\centerline{\includegraphics{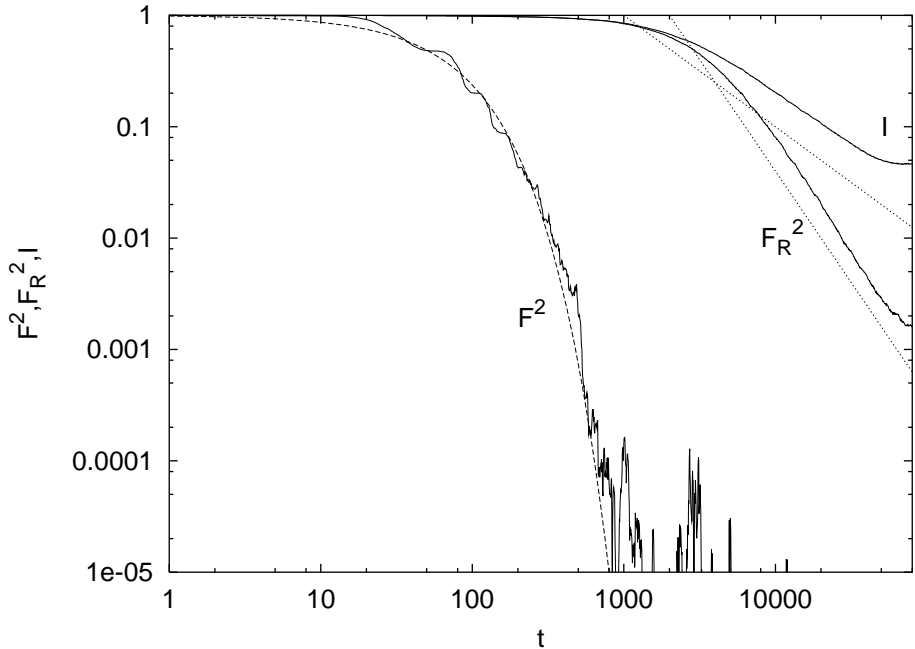}}
\caption{Decay of $F^2(t),F_{\rm R}^2(t)$ 
and $I(t)$ for fast chaotic environment. The dashed line is an exponential function with the 
exponent given by the values of $\sigma_{{\rm e}}$ and
$\overline{\ave{V_{{\rm c}}^2}}$ (\ref{eq:fastCDE}) while the two dotted lines have slopes $-2$ and $-1$. For parameters see text.}
\label{fig:fid030}
\end{figure}
\begin{figure}[h]
\centerline{\includegraphics{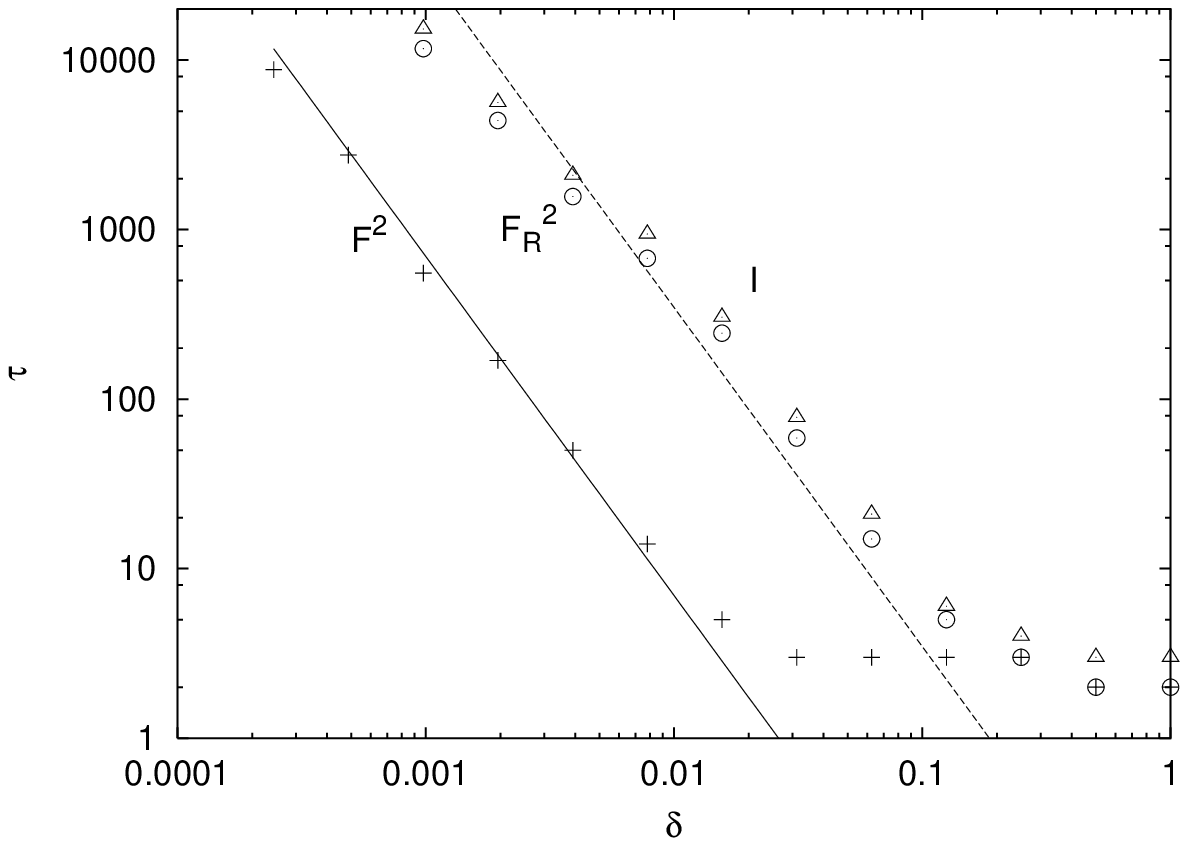}}
\caption{Times $\tau$ at which $F^2(t),F_{\rm R}^2(t),I(t)$
fall to level $0.37$ for different $\delta$ and fast chaotic 
environment. Symbols give the numerics and lines give the theoretical dependence of $\tau$. All is for $S=100$. For other parameters see text.}
\label{fig:presek030}
\end{figure}
For numerical demonstration we use a double kicked top with: $V_{\rm c,e}=S_{\rm z}/S$, $S=200$, $\delta=1.5 \cdot 10^{-3}$, coherent initial 
state at $(\vartheta,\varphi)_{\rm c,e}=(\pi/\sqrt{3},\pi/\sqrt{2})$ and parameters $\alpha_{{\rm c}}=0$, 
$\gamma_{{\rm c}}=\pi/50$ for the central system and $\alpha_{{\rm e}}=30$, $\gamma_{{\rm e}}=\pi/2.1$ for the environment. 
Actually, we could take any value of $\alpha_{{\rm c}}$ and would get qualitatively similar results. The only advantage 
of using regular central dynamics $\alpha_{{\rm c}}=0$ is that it is then possible to explicitly calculate averages 
$\overline{\ave{V_{{\rm c}}^2}}$ 
and $\overline{\ave{V_{{\rm c}}^2}}-\overline{\ave{V_{{\rm c}}}^2}$. Namely, if $\alpha_{{\rm c}}=0$ and 
$\gamma_{{\rm c}} \ll 1$ we get
\begin{eqnarray}
\overline{\ave{V_{{\rm c}}^2}} &=&\frac{1}{2}(1-y_{{\rm c}}^2)+\frac{1}{4S} (1+y_{{\rm c}}^2) \nonumber \\
\overline{\ave{V_{{\rm c}}^2}}-\overline{\ave{V_{{\rm c}}}^2} &=& \frac{1}{4S} (1+y_{{\rm c}}^2).
\label{eq:Vth}
\end{eqnarray}
The values of these two quantities for our initial condition are $\overline{\ave{V_{{\rm c}}^2}}=0.202$ and $\overline{\ave{V_{{\rm c}}^2}}-\overline{\ave{V_{{\rm c}}}^2}=0.399/S$, and are shown in Figure~\ref{fig:cor030} with two dotted lines (by pure coincidence we have $\sigma_{\rm e}=0.20 \approx \overline{\ave{V_{{\rm c}}^2}}$), together with numerically calculated time dependent (not yet averaged) $\ave{V_{{\rm c}}^2(t)}$ and $\ave{V_{{\rm c}}^2(t)}-\ave{V_{{\rm c}}(t)}^2$. This time dependent 
values oscillate on a time scale $\approx 50$, which is much longer than the time $\approx 10$ in which 
$\sigma_{{\rm e}}$ converges and so the assumption $t_{{\rm e}} \ll t_{{\rm c}}$ is justified. The values 
of all three quantities are then used in the linear response formulas (\ref{eq:fastCDE}) to give us the time scales 
on which $F,F_R$ and $I$ decay. The results are shown in Figure~\ref{fig:fid030}. We can see that the fidelity again 
decays exponentially as predicted, but the reduced fidelity and the
purity have a power-law tails. The decay time can be estimated by the lowest order expansions (\ref{eq:fastCDE}). Using the values of $\sigma_{{\rm e}}$ and $\overline{\ave{V_{{\rm c}}^2}}-\overline{\ave{V_{{\rm c}}}^2}$ (theoretical 
expression (\ref{eq:Vth})), we get $\tau_{\rm F}\approx 1/(0.16 \delta^2 S^2)$ and $\tau_{\rm R,P}\approx 1/(0.32 \delta^2 S)$ ($\tau_{\rm F}$ and $\tau_{\rm R}$ are for $F^2$ and $F_{\rm R}^2$) which agrees with numerics in Figure~\ref{fig:presek030}. The same general conclussion again holds: the more chaotic the environment is (smaller $\sigma_{\rm e}$), the slower the decay of all three quantities. Purity and reduced fidelity both decay on a $1/\hbar$ longer time scale than the fidelity in accordance with expressions (\ref{eq:Vth}) for coherent initial states. 

\subsection{Fast Regular Environment}
\label{sec:fast_reg}
\begin{figure}[h]
\centerline{\includegraphics{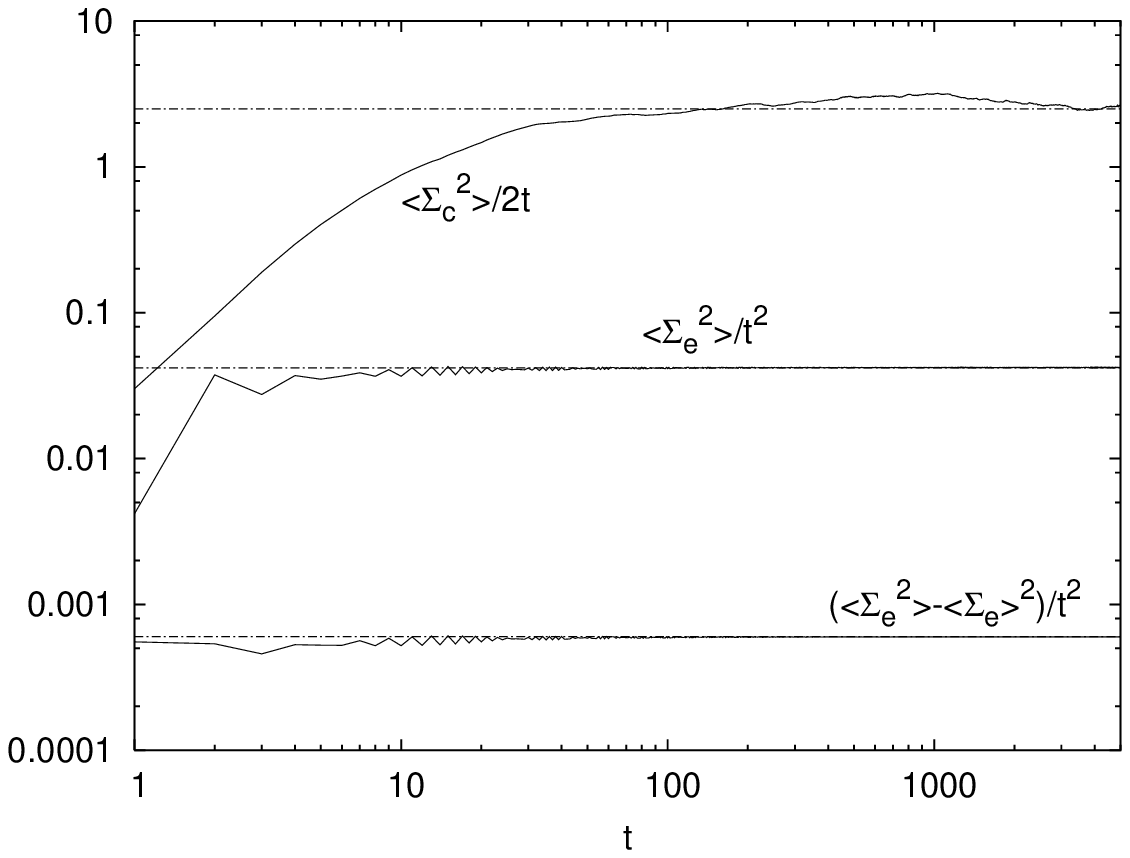}}
\caption{Correlation sums occurring in (\ref{eq:fastRegCDE}) (solid
curves) for $S=200$. The top chain line gives the best fit for
$\sigma_{{\rm c}}$ and the two lower chain 
lines give theoretical time averaged correlation functions for the
environment (\ref{eq:Cth}). All is for a fast regular environment. See text for details.}
\label{fig:cor300}
\end{figure}
Here we will explore perhaps a less physical situation of a regular environmental dynamics, i.e. one with non-decaying correlation function. The double integral of environmental correlations grows as $\propto t^2$ and we can define the average correlation function
\begin{equation}
\bar{C}_{{\rm e}}:=\lim_{t \to \infty}{\< \Sigma_{{\rm e}}^2(t)\>/t^2}.
\label{eq:avgce}
\end{equation}
If in addition the correlations of the system also do not decay then the correlation sum of the total system 
will grow as $\propto t^2$ which is just the regular regime already discussed before. Here, we will focus on a different situation where the integral of systems correlation function converges, i.e. the dynamics of the central system is mixing.
We will additionally assume the average ``position'' $V_{{\rm c}}$ to be zero $\overline{\ave{V_{{\rm c}}}}=0$. The 
transport coefficient of the central system $\sigma_{{\rm c}}$ is then
\begin{equation}
\sigma_{{\rm c}}:=\lim_{t \to \infty}{\<\Sigma_{{\rm c}}^2(t)\>/2t}, \qquad \Sigma_{{\rm c}}(t)=\int_0^t{\! V_{{\rm c}}(\xi) d\xi}.
\label{eq:sigmas}
\end{equation}
The expressions for $S_{\rm F}(t)$, $S_{\rm P}(t)$ (\ref{eq:St}) and
$S_{\rm R}(t)$ (\ref{eq:Str}) are independent of time and can be simplified to
\begin{eqnarray}
S_{\rm F}&=& 2 \sigma_{{\rm c}} \bar{C}_{{\rm e}}  \nonumber \\
S_{\rm R}&=& 2 \sigma_{{\rm c}} \bar{C}_{{\rm e}}   \nonumber \\
S_{\rm P}&=& 2 \sigma_{{\rm c}} \left\{ \bar{C}_{{\rm e}}-\overline{\ave{V_{{\rm e}}}}^2  \right\}.
\label{eq:fastRegCDE}
\end{eqnarray}
Note that now the reduced fidelity $\Fr$ decays on the same time scale as the 
fidelity $F(t)$. This must be contrasted to the case of a fast mixing environment (\ref{eq:fastCDE}), where 
$\Fr$ decayed on the same time scale as the purity. If the initial state of the environment $\rho_{{\rm e}}(0)$ is a 
coherent state, then purity will decay on a $1/\hbar$ times longer time scale than fidelity or reduced fidelity.
On the other hand, for a random initial state of the environment, the average force $\ave{V_{{\rm e}}}=0$ vanishes, and all three quantities decay on the same time scale.
\par
\begin{figure}[h]
\centerline{\includegraphics{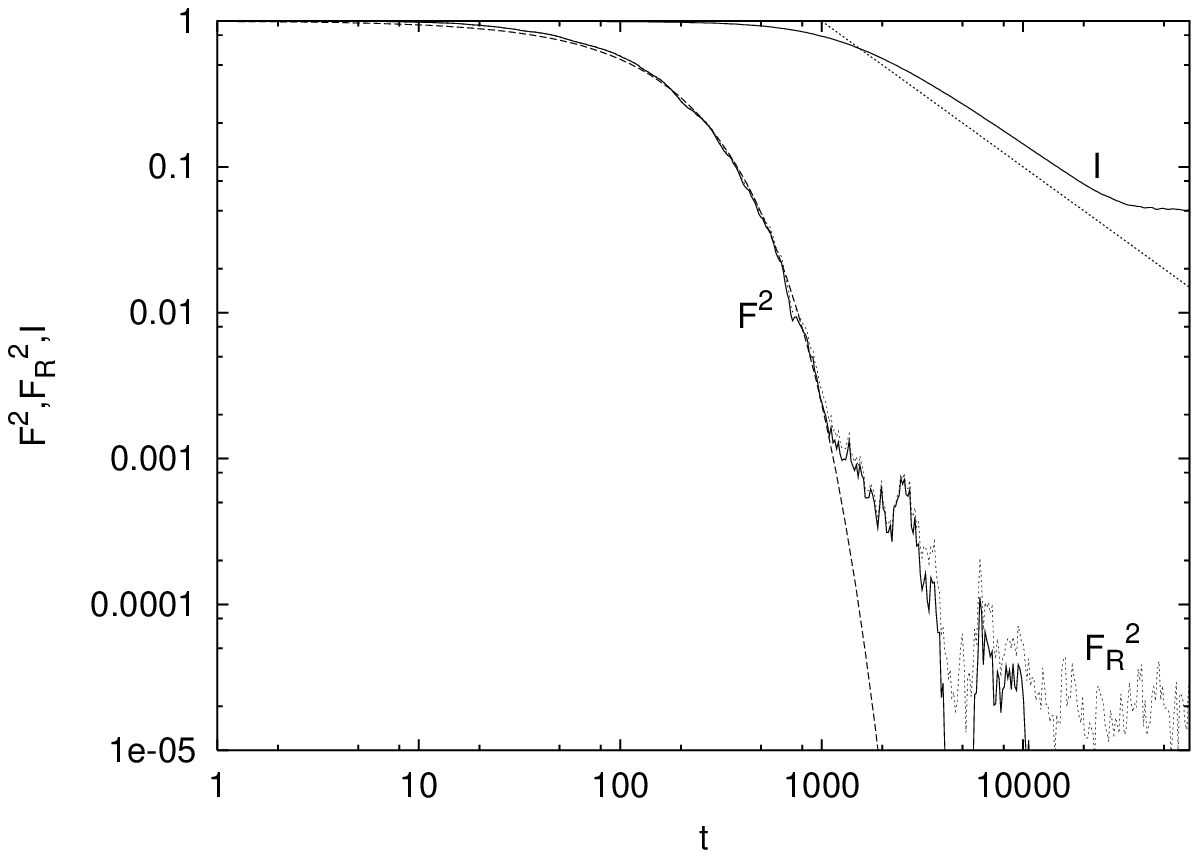}}
\caption{Decay of $F^2(t),F^2_{\rm R}(t)$ and $I(t)$ for fast regular
environment. The dashed line is the exponential function with the exponent given by a product 
of $\sigma_{{\rm c}}$ and $\bar{C}_{\rm e}$ (\ref{eq:fastRegCDE}). The
straight dotted line has slope $-1$. See text for details.}
\label{fig:fid300}
\end{figure}
\begin{figure}[h]
\centerline{\includegraphics{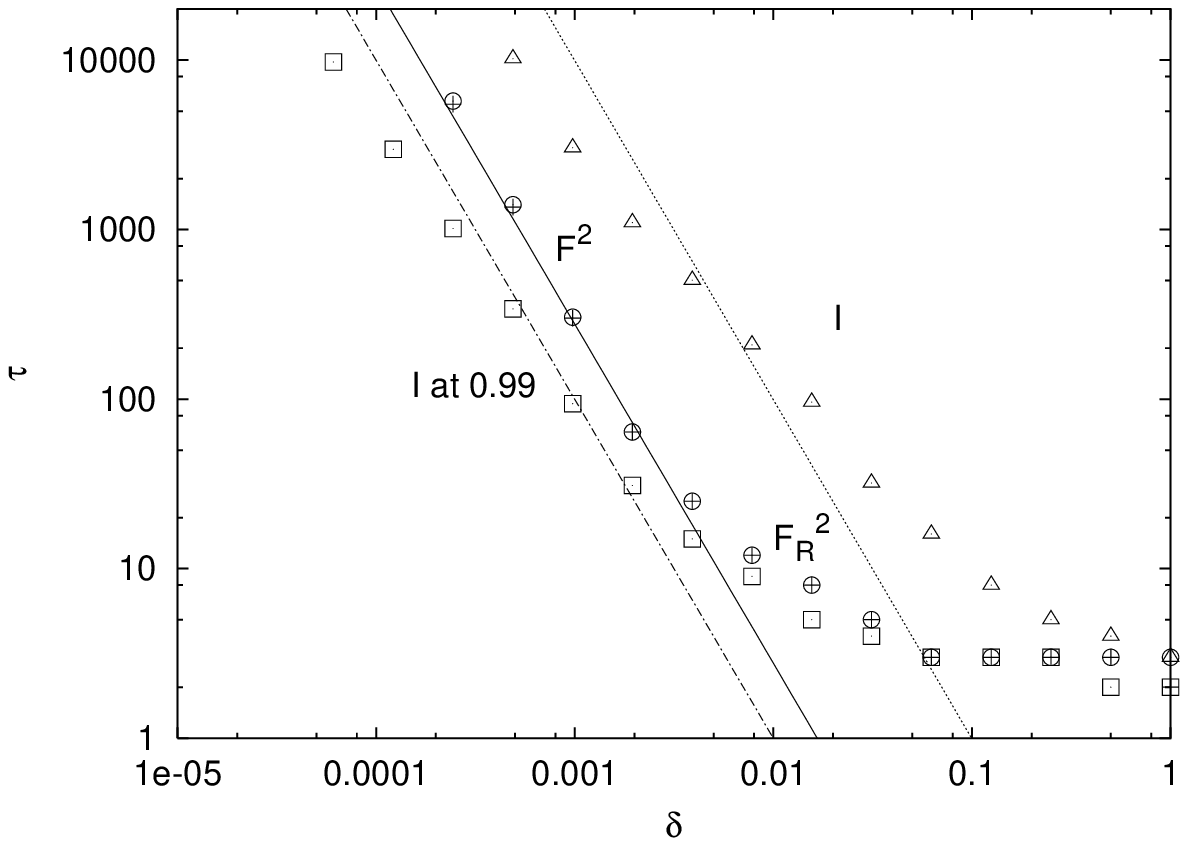}}
\caption{Times $\tau$ at which $F^2(t)$ (pluses), $F^2_{\rm R}(t)$ (circles) and $I(t)$ (triangles) fall to level $0.37$, 
and times $\tau$ when $I(t)$ falls to $0.99$ (squares) for varying $\delta$ and fast regular environment. Symbols are the numerics and lines 
give theoretical dependences of $\tau$. Everything is for $S=100$.}
\label{fig:presek300}
\end{figure}
For the purpose of numerical experiment we chose $V_{{\rm c}}=S_{\rm z}/S$, and $V_{{\rm e}}=S_{\rm z}^2/S^2$ in order to have
a less trivial situation of non-vanishing average force $\ave{V_{{\rm e}}}$. The initial condition is again $(\vartheta,\varphi)_{\rm c,e}=(\pi/\sqrt{3},\pi/\sqrt{2})$ and the parameters are 
$J=200$, $\alpha_{{\rm c}}=30$, $\gamma_{{\rm c}}=\pi/7$ and $\alpha_{{\rm e}}=0$, $\gamma_{{\rm e}}=\pi/2.1$ and the perturbation 
strength $\delta=6 \cdot 10^{-4}$. By choosing the explicitly solvable case $\alpha_{{\rm e}}=0$ we can calculate $\bar{C}_{{\rm e}}$ and $\overline{\ave{V_{{\rm e}}}}^2$, say for 
the simple case of a $\pi/2$ rotation, $\gamma_{{\rm e}}=\pi/2$,
\begin{eqnarray}
\bar{C}_{{\rm e}} &=& \frac{1}{4} (1-y_{{\rm e}}^2)^2+\frac{1}{4S}(-3 y_{{\rm e}}^4+2 y_{{\rm e}}^2 +1) + {\cal O}(1/S^2) \nonumber \\
\bar{C}_{{\rm e}}-\overline{\ave{V_{{\rm e}}}}^2 &=& \frac{1}{2S} y_{{\rm e}}^2(1-y_{{\rm e}}^2)+\frac{1}{16S^2}(11 y_{{\rm e}}^4-11 y_{{\rm e}}^2 +2) + {\cal O}(1/S^3).
\label{eq:Cth}
\end{eqnarray}
For our parameters we have $y_{\rm c,e}=0.772$ giving $\bar{C}_{\rm e}=0.0407$ and $\bar{C}_{{\rm e}}-\overline{\ave{V_{{\rm e}}}}^2=0.120/S$. The values of these coefficients are shown in Figure~\ref{fig:cor300} (lower two dotted lines) and nicely agree with the numerics. In Figure~\ref{fig:fid300} we can observe the exponential decay of fidelity and reduced fidelity on the same time scale (both curves almost overlap) and the decay of the purity on a $1/\hbar$ longer time scale. For longer times purity decay is again algebraic. In Figure~\ref{fig:presek300} we show dependence of the decay times on $\delta$. The dependence for purity is quite interesting. 
If one looks at the time the purity falls to $0.99$ one has agreement 
with linear response (by definition). But if one looks at the purity level $0.37$, they don't agree as well, meaning
that the shape of purity decay may change (not only the scale) as one varies $\delta$ or $\hbar$.
On the other hand, this may also be simply a finite size effect.

\section{Freeze in a Harmonic Oscillator}
\label{sec:har_freeze}
Previously we have discussed the so called freeze of fidelity in regular systems having a nonsingular derivative of the unperturbed frequencies, $\Omega=\partial\vec{\omega}/\partial\vec{j}\neq 0$. In the present section we will consider the case of a harmonic oscillator, for which $\Omega=0$ and the theory explained in Section~\ref{sec:denominator} can not be used. For the sake of numerical demonstration we will use a Jaynes-Cummings model (see Section~\ref{sec:Jaynes}) with the perturbation in the $G'$ parameter, i.e. the only case not explained in Section~\ref{sec:jaynes_reg}. The initial state will be a product of coherent state for a spin and an oscillator. The perturbation will be
\begin{equation}
V=\frac{\hbar}{\sqrt{2S}} (a^+ S_+ + a S_-)=[\delta G'].
\label{eq:GcV}
\end{equation}
Provided we have $G'=0$ for the unperturbed system the correlation
function has a zero time average regardless of other parameters. This
is a simple consequence of the symmetries as the perturbation is a
``counter-rotating'' term, while the unperturbed Hamiltonian has only
a ``co-rotating'' term. Due to the symmetry the perturbation is
residual. The fact that the perturbation is residual has nothing to do with the perturbation $G'$ breaking integrability. For instance, if we make a perturbation in $G'$ and any other parameter at the same time, we will have a perturbation that breaks integrability but is not residual.
\par
Because we want to study the case of $\Omega=0$, we choose $G=0$ so the unperturbed Hamiltonian,
\begin{equation}
H_0=\hbar \omega a^+ a+\hbar \varepsilon S_{\rm z},
\end{equation}
is uncoupled. As a consequence purity fidelity equals purity. Residual perturbation can be written as a time derivative of another operator $W$ (\ref{eq:W_def}), which is in our case 
\begin{equation}
V:=\frac{\ii}{\hbar}[H_0,W]=\frac{dW}{dt}, \qquad W=\frac{\ii \hbar}{\sqrt{2S}(\omega+\varepsilon)}\left( a^+ S_+ - a S_- \right).
\end{equation}
The echo operator can than be written as
\begin{equation}
\Md = \exp\left\{ -\frac{\ii}{\hbar}\left( \{ W(t)-W(0) \} \delta 
+ \frac{1}{2} \Gamma(t) \delta^2 + \ldots\right)\right\},
\label{eq:echo}
\end{equation}
with $\Gamma(t)$ given by
\begin{eqnarray}
\Gamma(t)&:=&\int_0^t{R(\tau)d\tau}-\frac{\ii}{\hbar}[W(0),W(t)],\nonumber \\
R&:=&\frac{\ii}{\hbar}[W,\frac{dW}{dt}]=\frac{\hbar^2}{(\omega+\varepsilon)}\left( \mathbf{S}^2-S_{\rm z}^2-S_{\rm z}-2 a^+ a S_{\rm z} \right).
\label{eq:Rr}
\end{eqnarray}
For the semiclassical calculations we will also need the classical
limits of $W$ and $R$. Introducing canonical action angle variables
for the spin (i.e. ``central'') system,
\begin{equation}
\hbar S_{\rm z}\to j_{\rm c}=\cos{\vartheta},\qquad \hbar S_{\pm} \to \sqrt{1-j_{\rm c}^2} \,{\rm e}^{\pm \ii \theta_{\rm c}},\quad \theta_{\rm c}=\varphi,
\end{equation}
and for the oscillator (``environment''),
\begin{equation}
a \to \frac{\ii}{\sqrt{2\hbar}}\sqrt{2j_{\rm e}}\, {\rm e}^{-\ii \theta_{\rm e}},\qquad j_{\rm e}=|\alpha|^2/S,
\end{equation}
the classical Hamiltonian reads
\begin{equation}
h_0=\vec{\omega}\cdot\vec{j},\qquad \vec{\omega}=(\varepsilon,\omega),\quad \vec{j}=(j_{\rm c},j_{\rm e}).
\end{equation}
The classical limits $w$ of $W$ and $r$ of $R$ are
\begin{eqnarray}
w&=&-\frac{\sqrt{2 j_{\rm e}(1-j_{\rm c}^2)}}{\omega+\varepsilon} \cos{(\theta_{\rm c}+\theta_{\rm e})}\nonumber \\
r&=&\frac{1}{\omega+\varepsilon} \left(1-j_{\rm c}^2-2j_{\rm c}j_{\rm e} \right).
\end{eqnarray}

\subsection{The Plateau}
\begin{figure}[h]
\centerline{\includegraphics{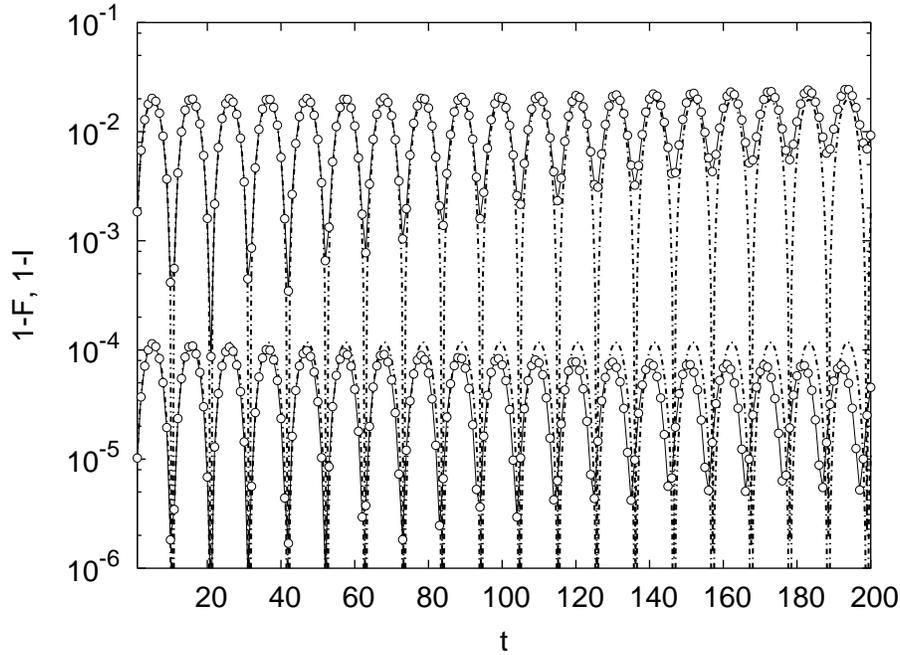}}
\caption{The plateau $1-F(t)$ (upper symbols) and $1-I(t)$ (lower symbols) for $S=50$ and coherent initial state. Perturbation of strength $\delta=0.01$ is in $G'$, $\omega=\varepsilon=0.3,G=0$, $\alpha=1.15$. Chain lines represent the theory (\ref{eq:har_plat_th},\ref{eq:har_i_th}) and circles full numerical simulations.}
\label{fig:j50g0_dGcS}
\end{figure}
The height of the plateau is given by the expectation value of the echo operator
\begin{equation}
F_{\rm plat}=|\ave{\exp{(-\ii \delta \{W(t)-W(0)\}/\hbar)}}|^2.
\end{equation}
For a harmonic oscillator one has to evaluate this expression explicitly and can not use time averaging as has been done for a general unperturbed dynamics in Section~\ref{sec:denominator}. We immediately see, that the leading semiclassical order will vanish for coherent initial states. We would get $|\exp{(-\ii \delta s(\vec{j}^*,\vec{\theta}^*,t)/\hbar)}|^2 = 1$, where we denoted
\begin{equation}
s(\vec{j},\vec{\theta},t):=w(\vec{j},\vec{\theta}+\vec{\omega}t)-w(\vec{j},\vec{\theta}).
\label{eq:s_def}
\end{equation}
Recall that in case of $\Omega \neq 0$ this leading order already gave a nonzero contribution. Here though, we have to calculate the next order. For coherent initial states this is easily done using the stationary phase method, i.e. expanding the phase around the position of the packet to lowest order. The calculation is actually very similar to the calculation of the fidelity decay for coherent initial state and non-residual perturbations. Expansion of the phase gives
\begin{equation}
s(\vec{j},\vec{\theta},t)=s(\vec{j}^*,\vec{\theta}^*,t)+\vec{s}'\cdot (\vec{j}-\vec{j}^*,\vec{\theta}-\vec{\theta}^*)+\cdots,
\end{equation}
with
\begin{equation}
\vec{s}':=(\frac{\partial s(\vec{j}^*,\vec{\theta}^*,t)}{\partial \vec{j}},\frac{\partial s(\vec{j}^*,\vec{\theta}^*,t)}{\partial \vec{\theta}}).
\end{equation}
Note that now $s(\vec{j},\vec{\theta},t)$ depends also on the angles and
therefore the ASI method cannot be used directly. Instead, we will use
the {\em classical} averaging over the initial Gaussian
distribution in a {\em phase space}. We will replace the quantum
expectation value with the classical average. Using the compact notation $\vec{x}:=(\vec{j},\vec{\theta})=(\vec{j}_{\rm c},\vec{j}_{\rm e},\vec{\theta}_{\rm c},\vec{\theta}_{\rm e})$, we can write the classical density corresponding to the coherent initial state as
\begin{equation}
\rho(\vec{x};\vec{x}^*)=\left( \frac{2}{\pi \hbar}\right)^{d} \sqrt{{\rm det}D}\, \exp{\left(-(\vec{x}-\vec{x}^*)\cdot D(\vec{x}-\vec{x}^*)/\hbar\right)},
\label{rhocl}
\end{equation}
with $\vec{x}^*$ being the position of the initial packet and a matrix $D$ (of size $2d\times 2d$) determining the squeezing of the initial packet in $d=d_{\rm c}+d_{\rm e}$ degrees of freedom system. The above classical density is normalised as $\int{\! \rho\, {\rm d}\vec{x}}=1$. For our choice of the initial state being a product of a spin coherent state (\ref{eq:rho_clasSU2}) and an oscillator coherent state (\ref{eq:rho_classbos}) the matrix $D$ is diagonal with elements
\begin{equation}
D_{11}=\frac{1}{1-j_{\rm c}^2},\quad D_{22}=\frac{1}{2j_{\rm e}},\quad D_{33}=1-j_{\rm c}^2,\quad D_{44}=2j_{\rm e}.
\label{eq:D}
\end{equation}
The fidelity plateau is now calculated as
\begin{equation}
F_{\rm plat}=\left| \int{\!{\rm d}\vec{x} \rho(\vec{x};0) \exp{(-\ii \delta \vec{s}'\cdot\vec{x}/\hbar )}} \right|^2,
\end{equation}
resulting in a Gaussian function
\begin{equation}
F_{\rm plat}^{\rm CIS}=\exp{\left(-\frac{\delta^2}{\hbar} \nu_{\rm har}  \right)},\qquad \nu_{\rm har}:=\frac{1}{2}\vec{s}'\cdot D^{-1}\vec{s}'.
\label{eq:har_plat}
\end{equation}
Note that the result is formally very similar to the expression for
the fidelity decay for a non-residual perturbation
(\ref{eq:Fn_regcoh}). Linear response expansion of the plateau is of
course $F_{\rm plat}^{\rm CIS}=1-\frac{\delta^2}{\hbar} \nu_{\rm har}$
and so the plateau for a harmonic oscillator is by factor $1/\hbar$
{\em higher} than for general systems (\ref{eq:nuCIS}). Beyond linear
response though, the plateau will decay faster (i.e. as a Gaussian)
for a harmonic oscillator than for systems with $\Omega \neq 0$ where the asymptotic decay of the plateau was a power law (e.g. $1/(\delta S)$ for the numerical model used in Section~\ref{sec:denominator}). 
\par
For our Jaynes-Cummings model we have
\begin{equation}
s(\vec{j},\vec{\theta},t)=2\sin{\left(\frac{\omega+\varepsilon}{2}t \right)}\frac{\sqrt{2j_{\rm e}(1-j_{\rm c}^2)}}{\omega+\varepsilon} \sin{\left(\theta_{\rm c}+\theta_{\rm e}+ t\frac{\omega+\varepsilon}{2}\right)}.
\end{equation}
The coefficient $\nu_{\rm har}$ is then
\begin{equation}
\nu_{\rm har}=\frac{2}{(\omega+\varepsilon)^2}\sin^2{\left(\frac{\omega+\varepsilon}{2}t \right)} \left\{ (1-j_{\rm c}^2)+2j_{\rm e}\left[ j_{\rm c}^2+(1-j_{\rm c}^2)\cos^2{\left( \theta_{\rm c}+\theta_{\rm e}+t \frac{\omega+\varepsilon}{2} \right)}\right] \right\}.
\end{equation}
The same $\nu_{\rm har}$ would be obtained from the quantum calculation of the second moment of $\Sigma(t)$.
\par
Results of the numerical calculation are shown in Figure~\ref{fig:j50g0_dGcS}. The initial packet was at $(\vartheta,\varphi)=(1,1)$ and $\alpha=1.15$, giving actions $j_{\rm c}=0.54$ and $j_{\rm e}=0.026$ and angles $\theta_{\rm c}=1$ and $\theta_{\rm e}=\pi/2$. Because $j_{\rm e} \ll j_{\rm c}$ we can neglect the second term in $\nu_{\rm har}$ and get
\begin{equation}
1-F_{\rm plat}^{\rm CIS}=3.93\, \delta^2 S\sin^2{(0.3 t)}.
\label{eq:har_plat_th}
\end{equation}
The agreement of this theoretical prediction with numerics can be seen in Figure~\ref{fig:j50g0_dGcS}.
\par
For the reduced fidelity and the purity one can go trough similar calculations. We will just list the linear response result obtained from the quantum expectation values,
\begin{eqnarray}
1-F_{\rm R}&=&\delta^2 S \frac{2}{(\omega+\varepsilon)^2}\sin^2{\left(\frac{\omega+\varepsilon}{2}t \right)} 2j_{\rm e}\left[ j_{\rm c}^2+(1-j_{\rm c}^2)\cos^2{\left( \theta_{\rm c}+\theta_{\rm e}+t \frac{\omega+\varepsilon}{2} \right)}\right] \nonumber \\
1-I&=& \delta^2 \frac{2}{(\omega+\varepsilon)^2}\sin^2{\left(\frac{\omega+\varepsilon}{2}t \right)} (1-j_{\rm c})^2.
\label{eq:har_i_th}
\end{eqnarray}
Theory for the purity shown in Figure~\ref{fig:j50g0_dGcS} also agrees well with numerics. The purity plateau for coherent initial states is $\hbar$-independent, just as the whole decay of purity.
 
\subsection{Beyond the Plateau}
After a sufficiently long time, when the second term $\Gamma(t)$
becomes important in the echo operator (\ref{eq:echo}), the decay will
be determined by the operator $R$. As our $R$ is non-residual, we can
use the theory for general perturbations, just using a
``renormalised'' perturbation strength $\delta^2/2$. The fidelity and
the reduced fidelity will decay as Gaussians, with the decay times
given by $\oC$ (\ref{eq:oC'}) and $\oCr$ (\ref{eq:oCr}), respectively,
while the decay of the purity in our $1+ 1$ degrees of freedom system will be $I(t)=1/\sqrt{1+u\, (\delta t)^2}$ (\ref{eq:Fp_1d}), with $u=4 \oCp/\hbar^2$ (\ref{eq:oCp}). For our simple example, the average $\bar{r}$ is equal to $r$ and the derivatives occurring in the expressions for $\oC$'s are
\begin{equation}
\frac{\partial\bar{r}}{\partial j_{\rm c}}=-\frac{2}{\omega+\varepsilon}(j_{\rm c}+j_{\rm e}),\qquad \frac{\partial\bar{r}}{\partial j_{\rm e}}=-\frac{2}{\omega+\varepsilon} j_{\rm c},\qquad \frac{\partial^2 \bar{r}}{\partial j_{\rm c}\partial j_{\rm e}}=-\frac{2}{\omega+\varepsilon}.
\end{equation}
Taking into account that the squeezing parameters for coherent initial states are $\Lambda_{\rm c}=1/(1-j_{\rm c}^2)$ and for an oscillator $\Lambda_{\rm e}=1/(2 j_{\rm e})$, we obtain
\begin{eqnarray}
\oC&=&\hbar \frac{2}{(\omega+\varepsilon)^2}\left\{ (j_{\rm c}+j_{\rm e})^2(1-j_{\rm c}^2)+2 j_{\rm c}^2 j_{\rm e}\right\} \nonumber \\
\oCr&=&\hbar \frac{2}{(\omega+\varepsilon)^2}(j_{\rm c}+j_{\rm e})^2(1-j_{\rm c}^2) \nonumber \\
\oCp &=& \hbar^2 \frac{2}{(\omega+\varepsilon)^2} (1-j_{\rm c}^2)j_{\rm e}.
\label{eq:avgCF}
\end{eqnarray}
This then immediately gives the fidelity decay (\ref{eq:res_long}), the reduced fidelity (\ref{eq:Fr_exact}) and the purity (\ref{eq:Fp_1d}),
\begin{equation}
\Fn=\exp{(-\delta^4 S^2 \oC t^2/4)},\qquad \Fr=\exp{(-\delta^4 S^2 \oCr t^2/4)},\qquad I(t)=\frac{1}{\sqrt{1+ \oCp S^2 \delta^4 t^2}}.
\label{eq:teorija}
\end{equation}
Note the $\hbar$-independent decay of purity.
\begin{figure}[h]
\centerline{\includegraphics{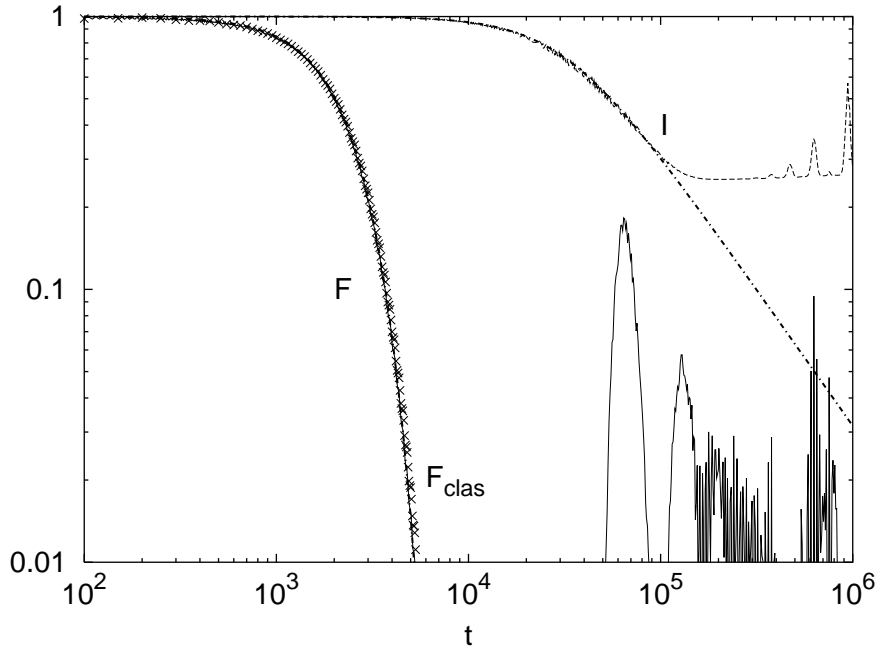}}
\caption{Long time fidelity and purity decay for the same data as in
Figure~\ref{fig:j50g0_dGcS}. The theoretical decay for purity (chain curve)(\ref{eq:teorija}), agrees with the numerical $I(t)$ until the asymptotic plateau is reached. Similarly, theory for the fidelity overlaps with the numerics. Symbols show the numerical result for classical fidelity which in this case agrees with the quantum fidelity.}
\label{fig:j50g0_dGc}
\end{figure}
\par
In Figure~\ref{fig:j50g0_dGc} we show numerical results for the
Jaynes-Cummings model with spin size $S=1/\hbar=50$ and a coherent
initial state placed at $(\vartheta,\varphi)=(1,1)$ and $\alpha=1.15$,
giving actions $j_{\rm c}=0.54$ and $j_{\rm e}=0.026$. The
perturbation is $V=[\delta G']$ of strength $\delta=0.01$, while the
parameters of the unperturbed Hamiltonian are
$\omega=\varepsilon=0.3$, $G=0$. The average correlation functions
(\ref{eq:avgCF}) are $\oC S=1.35$, $\oCr S=1.26$ and $\oCp
S^2=0.10$. For clarity we show in the figure only the fidelity and the purity as the reduced fidelity would almost overlap with the
fidelity. This theoretical values are then used to compare with
numerics. Agreement is excellent. We also show a numerical calculation of the classical fidelity (symbols). One can see that for a {\em harmonic oscillator} and a {\em residual} perturbation the classical fidelity {\em agrees} with the quantum one. Recall, that for a {\em residual} perturbation and a general unperturbed system, i.e. having $\Omega\neq 0$, the classical fidelity followed the quantum only up to $t_1\propto \sqrt{1/\hbar}$ time. For harmonic oscillators though, the quantum fidelity is equal to the classical despite the perturbation being residual. Purity can also be calculated purely classically. The classical purity can be defined as
\begin{equation}
I_{\rm cl}:=\int{\!{\rm d}\vec{x}_{\rm c} \rho_{\rm c}^2(t)},\qquad \rho_{\rm c}(t):=\int{\!{\rm d}\vec{x}_{\rm e} \rho(t)},
\label{eq:Iclas}
\end{equation}
with $\rho(t)$ being the classical density at time $t$,
i.e. $\rho(t)=\rho(\vec{\phi}^{-t}(\vec{x});\vec{x}^*)$ with $\vec{\phi}(\vec{x})$
being the Hamiltonian flow in phase space\footnote{Here the density
has to be square normalised, $\int{\!{\rm d}\vec{x} \rho^2}=1$.}. Such
classical purity (linear entropy) has been used
before~\citep{Wehrl:78,Angelo:04}. For our system and perturbation
the integrals in (\ref{eq:Iclas}) are Gaussian and the result is {\em
the same} as the one obtained from the quantum definition of purity (\ref{eq:teorija}).

\section{Decoherence for Cat States}
In this section we want to study decoherence for macroscopic
superpositions of states, the so-called Schr\" odinger cat states. We would like to demonstrate the accelerated decoherence for macroscopic superpositions, without resorting to any effective master equation description. The goal is to show that if we start with the initial state of the central system in the superposition of two macroscopically separated states $\ket{\tau_1}_{\rm }$ and $\ket{\tau_2}_{\rm }$, the decoherence (decay of purity) is much faster than if we start with a single state $\ket{\tau_1}_{\rm }$ of the central system.
\par
We can explain the accelerated purity decay for cat states using our
results for the reduced fidelity decay for coherent initial
states. Decoherence of a cat state will cause the reduced density
matrix to evolve from a coherent superposition of two packets at the
beginning to a mixture of two packets after the decoherence time $t_{\rm dec}$,
\begin{equation}
\rho_{\rm c}(0) \sim \frac{1}{2} \left( \ket{\tau_1}\bra{\tau_1}+\ket{\tau_2}\bra{\tau_2}+\ket{\tau_1}\bra{\tau_2}+\ket{\tau_2}\bra{\tau_1} \right) \stackrel{t_{\rm dec}}{\longrightarrow} \frac{1}{2} \left( \ket{\tau_1}\bra{\tau_1}+\ket{\tau_2}\bra{\tau_2} \right),
\label{eq:dec}
\end{equation}
The purity $I(t)$ of the initial coherent state is $I(0)=1$ while the
purity of the final state is $I(t_{\rm dec})=1/2$, if the states
$\ket{\tau_1}$ and $\ket{\tau_2}$ are orthogonal. 
\par
Let us suppose that the initial state is a product state of the
coherent state of the environment $\ket{\alpha}_{\rm }$ and a cat
state of the central system, i.e. a superposition of two coherent states $\ket{\tau_1}$ and $\ket{\tau_2}$,
\begin{equation}
\ket{\psi(0)}=\frac{1}{\sqrt{2}} \left( \ket{\tau_1}+\ket{\tau_2} \right)\otimes\ket{\alpha}.
\label{eq:cat}
\end{equation}
We will assume that we are in a regular regime, for which the decay of
purity of individual product coherent states has been derived before
(\ref{eq:Fp_reg}) and was seen do decay on a $\hbar$-independent time
scale $t_{\rm d}\sim 1/\delta$, where $\delta$ is a
coupling strength between the central system and the environment. For times smaller than the decoherence time of individual coherent states $t_{\rm d}$, the initial cat state (\ref{eq:cat}) will evolve into a superposition of product states,
\begin{equation}
\ket{\psi(t)} \approx \ket{\chi_1} \otimes \ket{\beta_1} +\ket{\chi_2}\otimes \ket{\beta_2},
\label{eq:cohcat}
\end{equation}
where we used the notation $\ket{\chi_1}\otimes \ket{\beta_1} \approx
U(t) \ket{\tau_1}\otimes \ket{\alpha}$ and $\ket{\chi_2}\otimes
\ket{\beta_2} \approx U(t) R_{\rm c} \ket{\tau_1}\otimes
\ket{\alpha}$, where for the second state we have written
$\ket{\tau_2}=:R_{\rm c} \ket{\tau_1}$ with some unitary matrix
$R_{\rm c}$. Note that the propagator $U(t)$ has a
coupling of strength $\delta$ between the two subsystems. The second state can therefore be thought of to be
obtained from the same initial state $\ket{\tau_1}\otimes\ket{\alpha}$
under a slightly different, perturbed evolution $U_\delta(t):=U(t)
R_{\rm c}$. The coherences in the reduced matrix (off diagonal
elements in $\rho_{\rm c}$) of the evolved state (\ref{eq:cohcat}) are
proportional to the overlap $\braket{ \beta_1}{\beta_2}$. The square
of the overlap $\braket{ \beta_1}{\beta_2}$ is in fact the overlap of
two reduced density matrices, one obtained with the unperturbed
evolution $U(t)$ and the other with the perturbed one
$U_\delta(t)$. Because the final states $\ket{\beta_{1,2}}$ as well as
the perturbed evolution $U_\delta(t)$ depend on the initial states
of the central system $\ket{\tau_{1,2}}$, the square of the overlap is nothing but the reduced fidelity~\footnote{Actually we used a slightly different definition
of the reduced fidelity as an overlap of the initial and the echo
state.} of the
environment~\footnote{In general the states $\ket{\beta_{1,2}}$ can
not be obtained by a Hamiltonian evolution on the environmental
subspace alone.}. The decay of coherences and therefore of the purity will be
given by the reduced fidelity of a product coherent initial state
$\ket{\tau_1}\otimes\ket{\alpha}$. As we have seen before
(\ref{eq:Fr_exact}), the $\oCr$ for a coherent state in a regular
regime is $\oCr= \frac{1}{2}\hbar (\vec{\bar{v}}'_{\rm e}\cdot
\Lambda^{-1}_{\rm e}\vec{\bar{v}}'_{\rm e})$ giving the reduced fidelity
decay $\exp{(-l^2 \oCr t^2/\hbar^2)}$, which in turn determines
the decay of coherences and therefore also purity decay. Here the
perturbation strength $l$ depends on the unitary matrix
$R_{\rm c}$, i.e. on the ``distance'' between the initial states
$\ket{\tau_{1,2}}$ of the central system, as well as on $U(t)$. This perturbation strength
$l$ is of course different than the coupling $\delta$ of the
evolution $U(t)$. For instance, in the simple
case of $\ket{\tau_2}$ being a space shifted packet $\ket{\tau_1}$,
$\ket{\tau_2}={\rm e}^{-\ii x\, p/\hbar}\ket{\tau_1}$, the perturbation
strength is proportional to the distance $x$ between the two
states. As the coupling $\delta$ causes decoherence of individual
packets on a time scale $t_{\rm d}\sim 1/\delta$ while the
decoherence of the cat state happens on the decay time scale of the reduced
fidelity $t_{\rm dec}\sim \sqrt{\hbar}/l$, it is sufficient to have
$l>\delta \sqrt{\hbar}$ in order to obtain faster
decoherence for cat states. The separation $l$ of the two packets
constituing the cat state has to scale only as $l \sim \sqrt{\hbar}$ and
therefore does not have to be macroscopically large in the semiclasical limit. For an environment with many degrees of freedom $d_{\rm e}$, and if
derivatives $\vec{\bar{v}}'_{\rm e}$ are all nonzero, $\oCr$ will be
proportional to $d_{\rm e}^2$. The reduced fidelity for the central
system on the other hand will have $\oCr$ proportional to $d_{\rm
c}^2$ and so if $d_{\rm e}\gg d_{\rm c}$, the reduced fidelity of the
environment will decay faster than the reduced fidelity of the central
system. Therefore, at the decoherence time $t_{\rm dec}$ we will still have
$\ket{\chi_{1,2}} \approx \ket{\tau_{1,2}}$, justifying the
decoherence scenario (\ref{eq:dec}). For cat states we therefore have
$\oCp \propto \hbar$ instead of $\oCp \propto \hbar^2$ as for a single coherent state, resulting in an accelerated decoherence. 
\par
Of course, the above argumentation is by no means a strict proof but is just a simple illustration of how an accelerated decoherence arises. An argumentation using the fidelity has been used by~\citet{Karkuszewski:02}, but without realizing the key role played by the time scales of purity and reduced fidelity decays. To repeat, several ingredients were needed: (i) the dynamics was assumed to be regular, (ii) for the reduced fidelity of product coherent states we have $\oCr \propto \hbar$, (iii) the purity of constituent product coherent states decays on much longer $\hbar$-independent time scale (i.e. $\oCp \propto \hbar^2$) and approximation (\ref{eq:cohcat}) was possible, (iv) furthermore, if $d_{\rm e}\gg d_{\rm c}$ the resulting decohered states $\ket{\chi}$ are still approximately equal to the initial states $\ket{\tau}$ of the central system as the reduced fidelity decay is faster for the environment than for the central system. All this together enabled us to ``derive'' a faster decoherence decay for macroscopic cat states. 
\par
For chaotic dynamics and times longer than the Ehrenfest time $\sim
\ln{\hbar}$ we have seen that fidelity, reduced fidelity and purity
all decay on the same time scale irrespective of the initial state. A
natural question arises: How does decoherence of macroscopic
superposition in chaotic system behave? Actually, decoherence for macroscopic superpositions is so fast, that it happens on time scales {\em shorter} than any dynamical scale of the system~\citep{Braun:01,Strunz:03,Strunz:03a}. It therefore happens on an ``instantaneous'' time scale on which {\em every} system behaves as a regular one (i.e. no decay of correlation functions). The chaoticity therefore has no influence on decoherence of macroscopic superpositions.     

\subsubsection{Numerical Example}
\begin{figure}[h]
\centerline{\includegraphics[angle=-90,width=80mm]{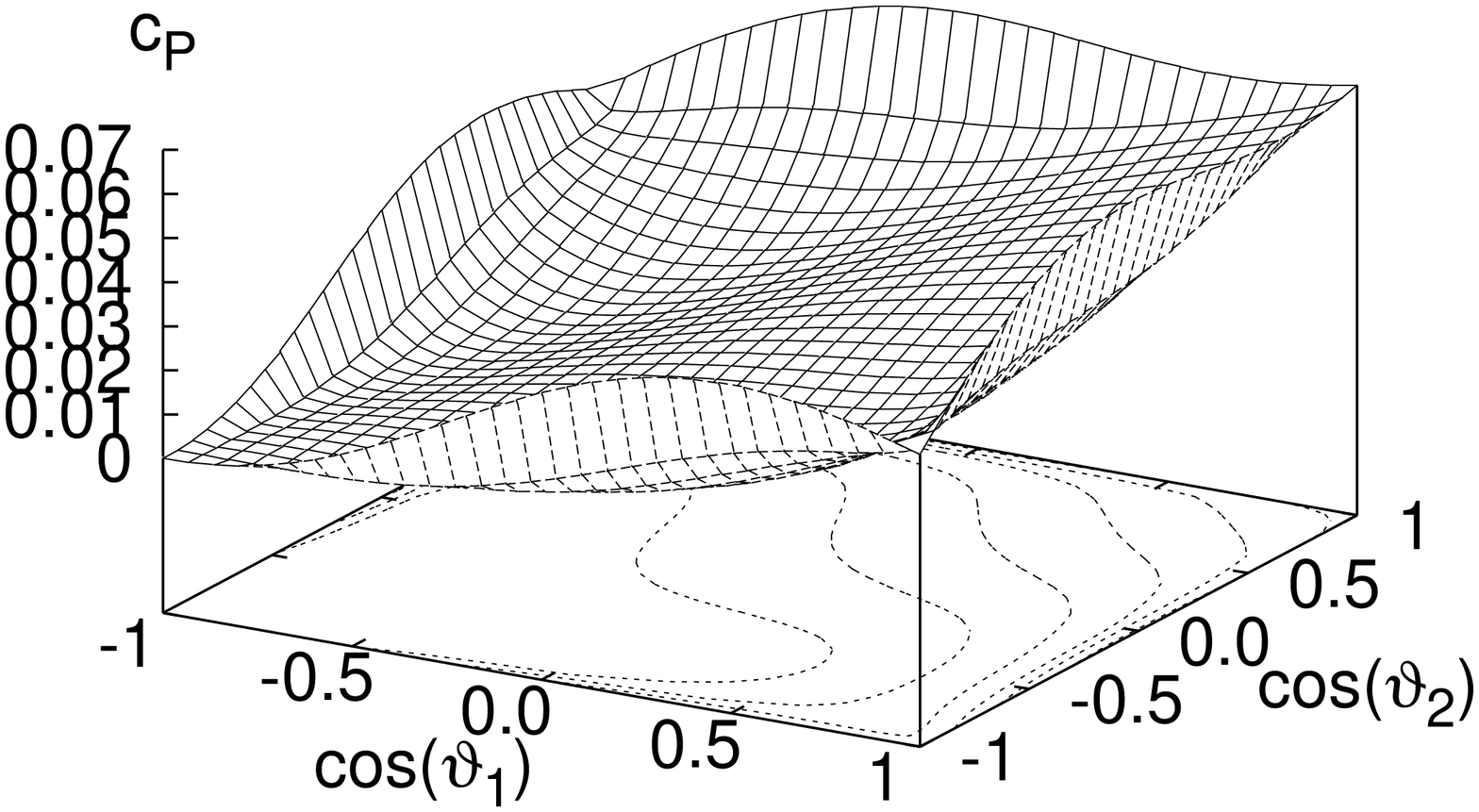}\includegraphics[angle=-90,width=80mm]{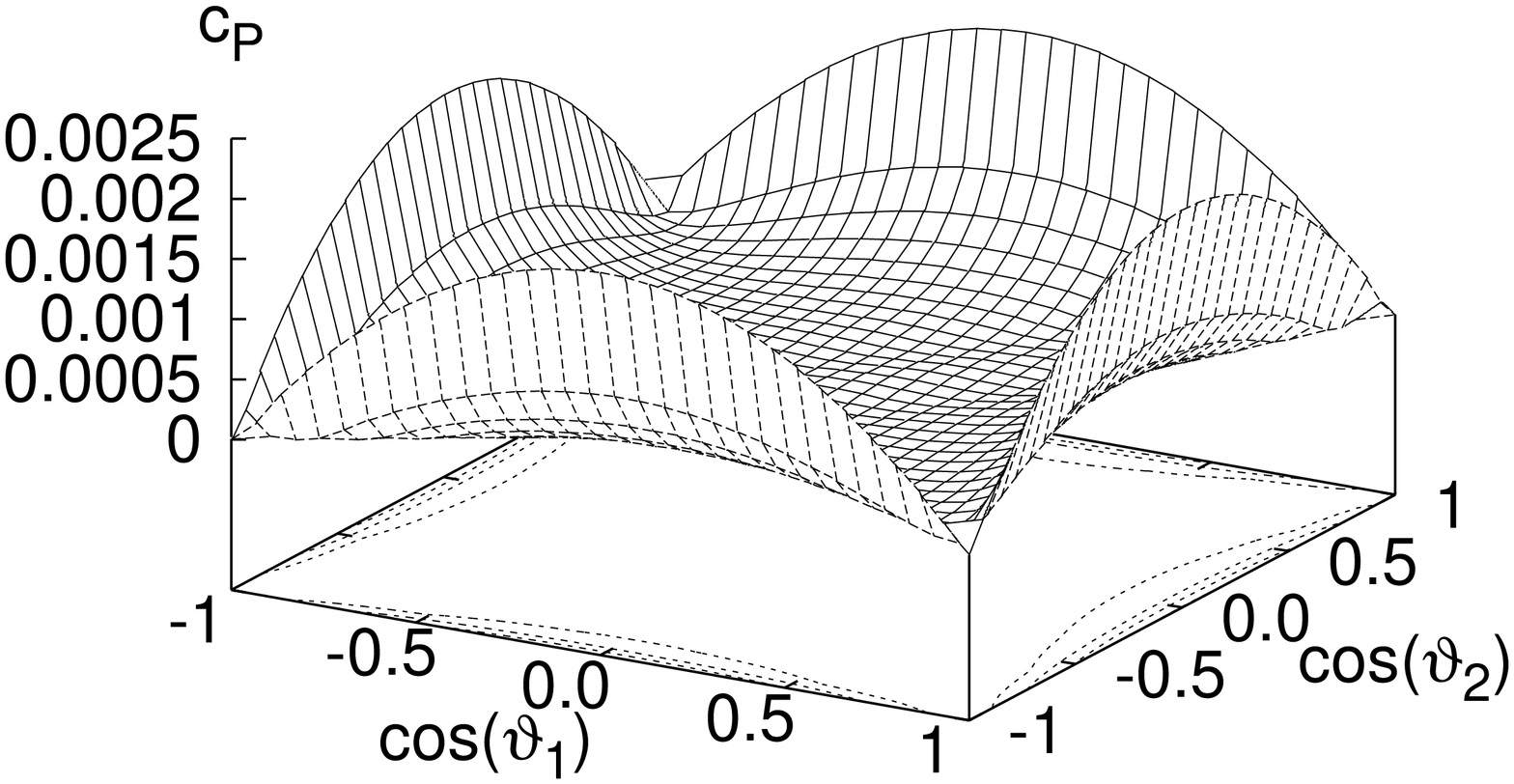}}
\caption{Dependence of $\oCp$ (Eq.~\ref{eq:l}, exact expression) for the cat state (\ref{eq:cat_SU2}) on the position angles of two constituent coherent packets. We have take $\varphi_1=\varphi_2=1$ and $S=4$ (left figure) and $S=64$ (right figure).
}
\label{fig:cat}
\end{figure}
We take a Jaynes-Cummings model with the unperturbed Hamiltonian
$H_0=\hbar \omega a^\dagger a + \varepsilon \hbar S_{\rm z}$ and the
perturbation in the coupling $G$ of strenght $\delta$, $V=[\delta G]$. Purity fidelity is then equal to the purity which will be used as a measure of decoherence. The initial state will be a product state of a coherent state $\ket{\alpha}$ for the oscillator acting as an environment and the superposition of two coherent packets for the spin,
\begin{equation} 
\ket{\psi(0)}_{\rm c} \propto \ket{\tau_1}_{\rm }+\ket{\tau_2}_{\rm }.
\label{eq:cat_SU2}
\end{equation}
\begin{figure}[t!]
\centerline{\includegraphics[angle=-90,width=140mm]{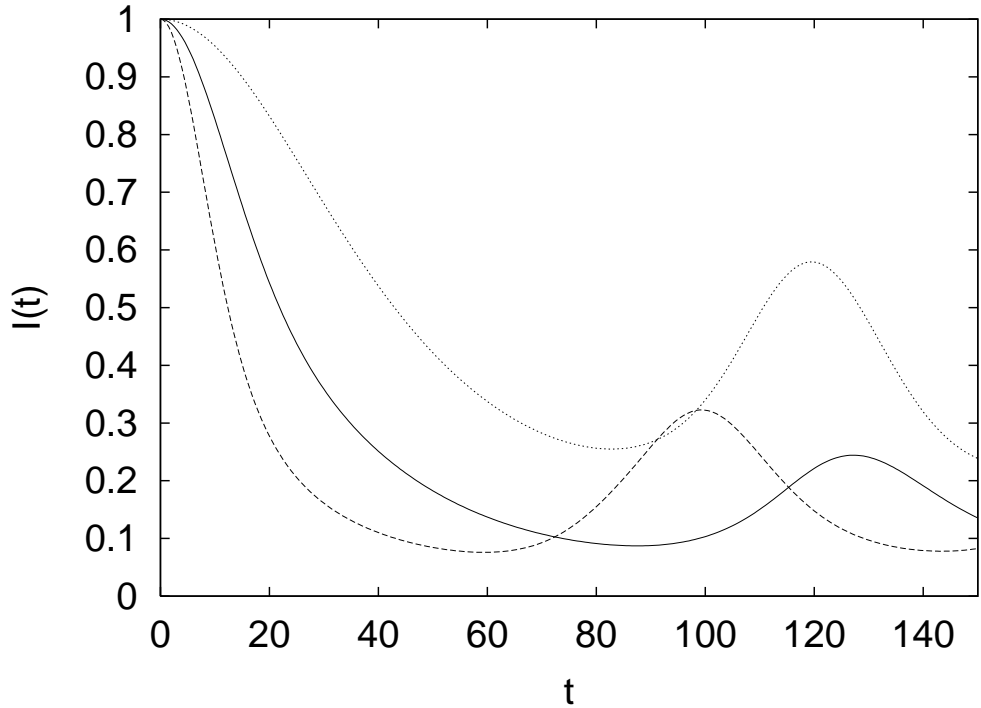}}
\caption{The purity decay for a random spin initial state (fastest
decaying  curve), Schr\" odinger cat state (middle curve) and coherent
state (slowest decaying curve). The oscillator initial state is always
a coherent state with $\alpha=1.15$. The parameters of the
Jaynes-Cummings model are $\omega=\varepsilon=0.3$, $G=G'=0$ and the
perturbation is in $G$ (i.e. in the coupling) with $\delta=0.02$,
$S=20$. The cat state consists of two coherent SU(2) packets at
$(\vartheta,\varphi)=(1,1)$ and $(0,1)$. The theoretical decay times
$\tau=\hbar/\delta\sqrt{2 \oCp}$ are $\tau=46$, $26$ and $13$ for the
coherent state, the cat state and the random state,
respectively. $\oCp$ for the coherent state has been calculated
according to the theoretical formula in Figure~\ref{fig:cbar_G}, for
the cat state according to (\ref{eq:l}) and for a random state using (\ref{eq:cpran}).}
\label{fig:pur}
\end{figure}
The Heisenberg evolution is simple, so that the integral of perturbation $\Sigma(t)$ is
\begin{equation}
\Sigma(t)=\frac{2\hbar}{\Delta \sqrt{2S}} \sin{\left( \frac{t \Delta}{2} \right)} \left\{ {\rm e}^{-\ii t\Delta  /2} a^+ S_- + {\rm e}^{\ii t\Delta  /2} a S_+ \right\},
\label{eq:Sigma} 
\end{equation}
where $\Delta:=\varepsilon -\omega$ is a detuning. From now on we set $\Delta=0$ in order to have a non vanishing $\oCp$. As we take a coherent oscillator initial state, the expectations over oscillator space in correlation function sum (\ref{eq:Cbars}) can be explicitly made and one ends up with the very neat result
\begin{figure}[t!]
\centerline{\includegraphics[width=140mm]{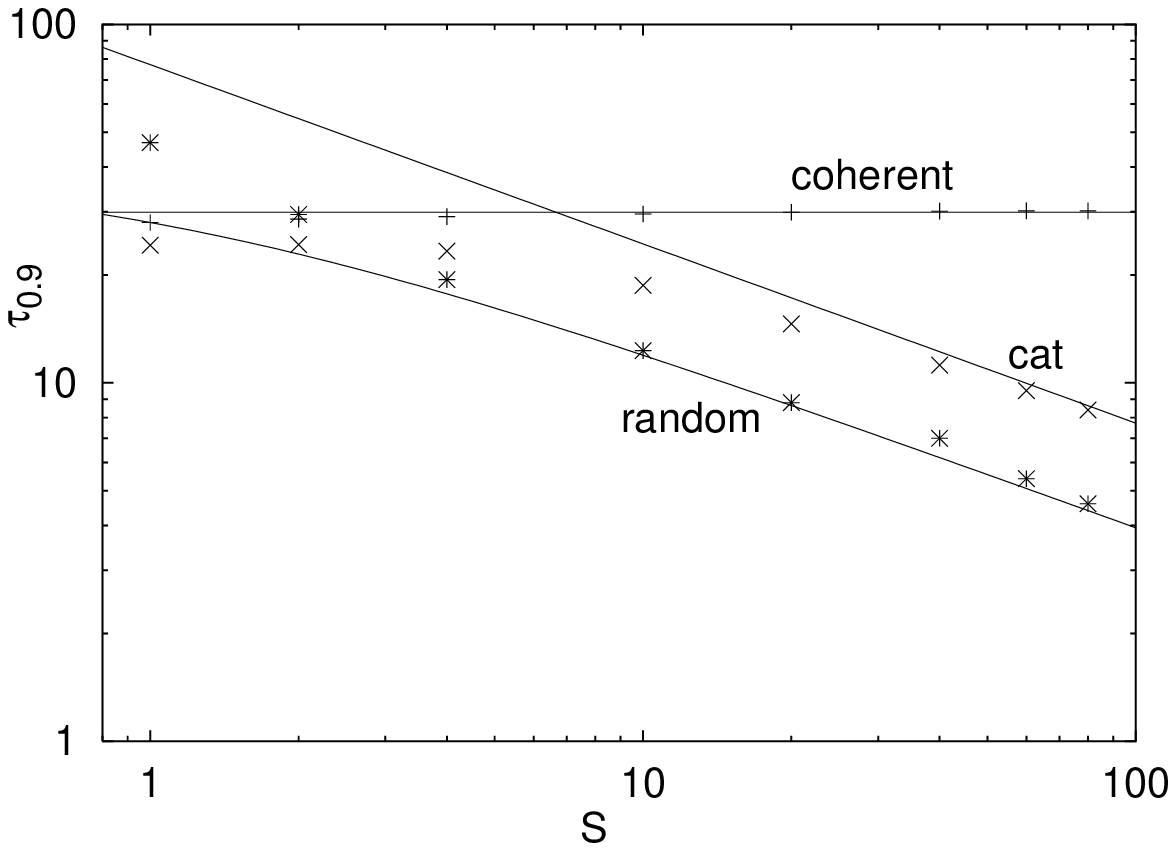}}
\caption{Dependence of numerically determined decay times of purity on
$S$ ($=1/\hbar$) for a random spin initial state, a Schr\" odinger cat
state and a coherent state. The perturbation strength is $\delta=0.01$ while all other parameters are the same as for Figure~\ref{fig:pur}. Theoretical decay times $\tau=\hbar/\delta\sqrt{2 \oCp}$ are shown with full lines.}
\label{fig:decoh}
\end{figure} 
\begin{equation}
\oCp=\frac{1}{2S^3} \left\{ \langle S_+ S_-\rangle - \langle S_+ \rangle \langle S_- \rangle \right\},
\label{eq:catcp}
\end{equation} 
where now the expectation values are only over a {\em spin state},
e.g. the cat state (\ref{eq:cat_SU2}) or any other. A similar result
has been obtained before trough a master equation approach to
decoherence, see for instance~\citep{Benedict:99,Foldi:01}. Our result
though has quite different perspective as it was derived directly from
the Hamiltonian dynamics without any resort to a Markovian description
of the reduced dynamics or to dissipation. The calculation of $\oCp$
(\ref{eq:catcp}) for a cat state (\ref{eq:cat_SU2}) using formulas
from Appendix~\ref{app:expect} is straightforward although tedious,
with the exact but considerably involved result. Rather than showing
the whole result, we concentrate on the leading order in $1/S$ (i.e. $\hbar$) which will dictate the decay in the semiclassical limit and therefore will also govern the decoherence of macroscopic superpositions. The result is
\begin{equation}
\oCp=\frac{l^2}{8S}+ {\cal O}(\hbar^2), \qquad l^2:=\sin^2{\vartheta_1}+\sin^2{\vartheta_2}-2\sin{\vartheta_1}\sin{\vartheta_2} \cos{(\varphi_1-\varphi_2)}.
\label{eq:l} 
\end{equation}  
The $l$ in the above expression has an interesting geometrical
interpretation: it is a distance between the projections of positions
of the two coherent states to the {\tt x-y} plane. There is indeed a
faster decoherence for cat states than for a single coherent
state. For a cat state $\oCp \propto \hbar=1/S$ whereas we had $\oCp
\propto \hbar^2$ for a single coherent packet. The exact dependence of
$\oCp$ on the position of both packets in a cat state can be seen in
Figure~\ref{fig:cat}. The
accelerated decoherence can be seen as the increasing of $\oCp$ away
from the diagonal ($\vartheta_1\neq \vartheta_2$) for $S=64$, where
the exact result for $\oCp$ can be approximated by the leading order
expression (\ref{eq:l}). Note that due to symmetry there is a slow
decoherence for states with $\vartheta_1+\vartheta_2=\pi$ despite of the
fact that this leads to a macroscopic superpositions. Linear response
decay time of purity for macroscopic superpositions is therefore
(\ref{eq:l}) $t_{\rm dec}=2\sqrt{\hbar}/(\delta l)$, while the decay
time for individual coherent packets is (using $\oCp$ from
Figure~\ref{fig:cbar_G}) $t_{\rm
d}=1/(\sqrt{2}\cos^2{(\vartheta/2)\delta})$. We get an accelerated
decoherence, $t_{\rm dec} \ll t_{\rm d}$, for $l \gg \sqrt{8\hbar} \cos^2{(\vartheta/2)}$.
\par
For random initial state of a spin and a coherent packet for an
oscillator we can calculate $\oCp$ using the expression (\ref{eq:catcp}) and get
\begin{equation}
\oCp=\frac{S+1}{3S^2},\qquad \hbox{random spin i.c.}.
\label{eq:cpran}
\end{equation}
Again we have $\oCp \propto \hbar$, just like for cat states, but with a smaller prefactor. If we had a random state on the whole Hilbert space, i.e. also for an oscillator, we would have $\oCp \propto \hbar^0$. Therefore, the decoherence rate for a cat state is {\em between} the decoherence rate of a single packet and the decoherence rate for a random state.    
\par
The accelerated decoherence can be seen in Figure~\ref{fig:pur} where
we show numerically calculated purity decays for all three different
initial spin states. Revivals of purity on a classical time scale $\sim 1/\delta$ due to integrability of unperturbed Hamiltonian can also be seen. This revivals will be absent in a general non-integrable Hamiltonian. To further illustrate the scaling of decay times with $S$ we show in Figure~\ref{fig:decoh} numerically calculated decay times in which the purity decays to level $0.9$ for all three different initial states, coherent, cat state and random state of a spin, while the oscillator is initially always in the same coherent state.


\chapter{Application: Quantum Computation}
\begin{flushright}
\baselineskip=13pt
\parbox{85mm}{\baselineskip=13pt
\sf In theory, there is no difference between theory and practice. In practice, there is no relationship between theory and practice.
}\medskip\\
---{\sf \itshape Anonymous}\\\vspace{20pt}
\end{flushright}

Quantum information theory is relatively recent, for a review
see~\citep{Nielsen:01,Steane:98,Ekert:96}. Its beginnings go back to '80 and in recent years theoretical concepts have been demonstrated in experiments. While quantum cryptography, a method of provably secure communication, is already commercially available, quantum computation on the other hand is still limited to small laboratory experiments.  
\par
In order to perform quantum computation you obviously need a quantum
computer. A quantum computer can be considered as a many-body system of $n$ elementary 
two-level quantum systems --- called {\em qubits}. The union of all
$n$ qubits is called a quantum register $\ket{\rm r}$. The size of the
Hilbert space ${\cal N}$ and therefore the number of different states of a
register grows with the number of qubits as ${\cal N}=2^n$. Quantum computation then consists of the following steps:
\begin{itemize}
\item Load the data for the quantum computation in the initial state
of the quantum register $\ket{\rm r}$, resulting in a general superposition of ${\cal N}$ basis states.
\item Then perform the actual computation, represented by a unitary
transformation $U$. As $U$ acts on an exponentially large space it is usually decomposed into simpler units $U=U_T\cdots U_2U_1$. Such sequence of $T$ elementary 
one-qubit and two-qubit {\em quantum gates} $U_t$, $t=1,2,\ldots,T$ is called a {\em quantum algorithm}.
\item Finally, we read out the result of the computation by performing measurements on the qubits of the final register state $U\ket{{\rm r}}$. 
\end{itemize}
A quantum algorithm is called {\em efficient} if the number of 
needed elementary gates $T$ grows with at most {\em polynomial} rate in $n = \log_2 {\cal N}$, and only in this case it can generally be expected to outperform the best
classical algorithm. At present only few efficient quantum algorithms 
are known, perhaps the most generally useful being the Quantum Fourier 
Transformation (QFT)~\citep{Shor:94,Coppersmith:94} which will also be the subject of our study.
\par
There are two major obstacles for performing practical quantum computation. First, there is the  
problem of {\em decoherence}~\citep{Chuang:95} resulting from an
unavoidable generally {\em time-dependent} perturbation due to the coupling between the qubits and the environment. If the perturbation couples only a small number of qubits at a time then such
errors can be eliminated at the expense of extra qubits by 
{\em quantum error correcting codes}~\citep{Steane:96,Calderbank:96,Steane:96a}, for another approach see~\citep{Tian:00}. Second, even if one knows an efficient error correcting code 
or assumes that quantum computer is ideally decoupled from the environment, 
there will typically exist small {\em unknown} or {\em uncontrollable} 
residual interaction among the qubits which one may describe by a general {\em static} perturbation. Therefore, understanding the {\em stability} of quantum algorithms with respect to various types 
of perturbations is an important problem, for some results on this topic see~\citep{Miquel:96,Miquel:97,Banacloche:98,Banacloche:99,Banacloche:00,Song:01,Georgeot:00,Berman:01,Berman:02b,Berman:02,Celardo:03}.
\section{General Framework}
Let us perturbe the $t$-th quantum gate by a perturbation\footnote{It is expected that the computation is stable
only when the evolution is close to unitary.} generated by $V_t$,
\begin{equation}
U^\delta_t:=\exp{(-\ii \delta V_t )}U_t.
\label{eq:pert}
\end{equation}
We set $\hbar=1$ and use the superscript $\delta$ to denote a
perturbed gate and the subscript $t$ to denote a discrete time index,
i.e. a gate number. We allow for different perturbations $V_t$ at different
gates. The perturbed algorithm is simply a product of perturbed gates,
$U^\delta=U^\delta_T\cdots U_1^\delta$. Fidelity will again serve as a
measure of stability and we have
\begin{equation}
F(T)=|\ave{\Md}|^2,\qquad M_\delta:=U(-T) U^\delta(T),
\label{eq:deffid}
\end{equation}
with $U(t):=U_t\cdots U_1$ and similarly for $U^\delta(t)$. For our
generally time dependent perturbation the echo operator equals to
\begin{equation}
M_\delta={\rm e}^{-\ii \delta V_T(T)}\cdots{\rm e}^{-\ii \delta V_2(2)}{\rm e}^{-\ii \delta V_1(1)},
\label{eq:prodfid}
\end{equation}
where $V_t(t):=U^\dagger(t)V_t U(t)$ is the perturbation of $t$-th
gate $V_t$ propagated with the unperturbed gates $U(t)=U_t\cdots
U_1$. Beware that $V_t(t)$ is time dependent due to two reasons, one
is due to the interaction picture (time in parentheses) and the second
is that the perturbation itself is explicitly time dependent, i.e. different
perturbation for different gate (time as a subscript). In quantum
computation one is usually interested in initial states containing a maximum amount of information, thereby being close to random states. With this in view, we take the initial state average to be a trace over the whole Hilbert space, 
\begin{equation}
\ave{\bullet}:=\frac{1}{\cal N}\tr{(\bullet)}.
\end{equation}
By this prescription we study the average fidelity over the whole
Hilbert space. Without sacrificing generality we furthermore assume
the average perturbation to be traceless,
$\sum_{t=1}^T{\tr{V_t}}=0$. Trace of the average perturbation only
changes the phase of the fidelity amplitude and has therefore no
influence on the fidelity (i.e. probability). The linear response
expansion of the fidelity then reads
\begin{equation}
F(T)=1-\delta^2 \sum_{t,t'=1}^T{C(t,t')},
\label{eq:Fcor}
\end{equation}
with the correlation function
\begin{equation}
C(t,t'):=\frac{1}{\cal N} \tr{\lbrack V_t(t) V_{t'}(t')\rbrack}.
\label{eq:corr}
\end{equation}
Here we used the fact that the average trace vanishes so the term
$\ave{\Sigma(t)}$ is zero. Decomposition of a given quantum algorithm
$U$ into quantum gates is by no means unique. An interesting question then is, given perturbations $V_t$, which form of the algorithm has the highest fidelity, i.e. is the most stable? We will see that the standard QFT algorithm can be rewritten in a non trivial way so that it becomes more stable against static perturbations. The guiding principle in the construction of this new algorithm will be to study the correlation function (\ref{eq:Fcor}) and trying to minimise its sum. 

\subsection{Time Dependent Perturbations}
If our perturbation $V_t$ is time dependent, i.e. we have different perturbations on different gates, then the decay of correlation function will not only depend on the unperturbed dynamics, but also on how strong these perturbations are correlated at different gates. Let us have a closer look at one extreme example. If the perturbation $V_t$ is an {\em uncorrelated noise}, as it would be in the case of 
coupling to an {\em ideal heath bath}, then the matrix elements of $V_t$ may be assumed to be 
{\em Gaussian random} variables which are uncorrelated in time,
\begin{equation}
\ave{V_{jk}(t) V_{lm}(t')}_{\rm noise} = \frac{1}{\cal N}\delta_{jm}\delta_{kl}\delta_{tt'}.
\end{equation} 
Hence one finds $\ave{C(t,t')}_{\rm noise} = \delta_{t t'}$, where we
have averaged over noise. In fact the average of the product in
$M_\delta$ (\ref{eq:prodfid}) equals to the product of the average and yields the noise-averaged fidelity
\begin{equation} 
\ave{F(T)}_{\rm noise} = \exp(-\delta^2 T),
\end{equation} 
which is {\em independent} of the quantum algorithm $U$. This result is completely general provided that the correlation time of the perturbation is {\em smaller} than the duration of a single gate. On the other hand, 
for a {\em static} perturbation $V_t\equiv V$ one may 
expect slower correlation decay, depending on the 'regularity' of the 
evolution operator $U$, and hence faster decay of fidelity. 
Importantly, note that in a physical situation, where perturbation is expected to 
be a combination $V_t = V_{\rm static} + V_{\rm noise}(t)$, the 
fidelity drop due to a static component
is expected to {\em dominate} long-time quantum computation $T\to\infty$ (i.e. large number of qubits $n$) over the
noise component, as soon as the quantum algorithm exhibits 
long time correlations of the operator $V_{\rm static}$. If $V_{\rm
static}=0$ is zero, the quantum computation can be stabilised by
making ``adiabatically'' slow evolution of gates. In the following we
will focus exclusively on the static perturbations being the worst ones.

 \section{Quantum Fourier Transformation}
   \label{sec:QFT}
We will consider Quantum Fourier Transformation algorithm (QFT) and
   will consider its stability against static random
   perturbations. The perturbation $V_t\equiv V$ will be a random
   hermitian matrix from a Gaussian unitary ensemble (GUE). Gaussian unitary
   ensamble is invariant under unitary transformations and the matrix
   elements in an arbitrary basis are independent random Gaussian
   variables~\citep{Mehta:91}. Due to hermitian symmetry they are real
   on the diagonal and complex off-diagonal. The GUE matrices can be used
   to model quantum statistical properties of classically chaotic
   Hamiltonians and have been first applied to studies of nuclear
   resonances. Second moments of a GUE matrix $V$ are normalised as
\begin{equation}
\ave{V_{jk}V_{lm}}_{\rm GUE}=\delta_{jm}\delta_{kl}/{\cal N},
\end{equation}
where the averaging is done over a GUE ensemble.
\par
Let us briefly describe the QFT algorithm. Basis qubit states in a Hilbert space of dimension ${\cal N}=2^n$ will be denoted by $\ket{k}, \ \ k=0,\ldots,2^n-1$. The 
unitary matrix $U_{\rm QFT}$ performs the following transformation on a state with 
expansion coefficients $x_k$
\begin{equation}
U_{\rm QFT} (\sum_{k=0}^{{\cal N}-1}{x_k \ket{k}})=\sum_{k=0}^{{\cal N}-1}{\tilde{x}_k\ket{k}},
\end{equation}
where $\tilde{x}_k=\frac{1}{\sqrt{{\cal N}}}\sum_{j=0}^{{\cal N}-1}{\exp{(2\pi \ii j k/{\cal N})} x_j}$. The resulting expansion coefficients $\tilde{x}_k$ are {\em Fourier transformed} input coefficients $x_k$. The ``dynamics'' of the QFT is decomposed into three kinds of unitary gates: One-qubit gates ${\rm A}_j$ acting on $j$-th 
qubit
\begin{equation}
{\rm A}_j=\frac{1}{\sqrt{2}} \pmatrix{ 1 & 1 \cr
  1 & -1 \cr},
\label{eq:Agate}
\end{equation}
where the basis is ordered as $(\ket{0},\ket{1})$, diagonal two-qubit gates ${\rm B}_{jk}={\rm diag}\{ 1,1,1,\exp{(i \theta_{jk})} \}$, 
with $\theta_{jk}=\pi/2^{k-j}$, and transposition gates ${\rm T}_{jk}$
which interchange the $j$-th and $k$-th qubit, $T_{jk}\ket{\ldots j\ldots k\ldots}=\ket{\ldots k \ldots j \ldots}$. There are $n$ ${\rm A}$-gates, $n(n-1)/2$ ${\rm B}$-gates 
and $[n/2]$ transposition gates, where $[x]$ is the integer part of $x$. 
The total number of gates for the algorithm is therefore $T=[n(n+2)/2]$. 
For instance, in the case of $n=4$ we have a sequence of $T=12$ gates (time runs 
from right to left)
\begin{equation}
U_{\rm QFT} = 
{\rm T}_{03} {\rm T}_{12} {\rm A}_0 {\rm B}_{01} {\rm B}_{02} {\rm B}_{03} {\rm A}_1 {\rm B}_{12} {\rm B}_{13} {\rm A}_2 {\rm B}_{23} {\rm A}_3.
\label{eq:plain4}
\end{equation}  
\par
For the GUE perturbation we can average the correlation function $C(t,t')$ (\ref{eq:corr}) over the GUE ensemble, resulting in
\begin{equation}
\ave{C(t,t')}_{\rm GUE} = \left|\frac{1}{{\cal N}} \tr{U(t,t')}\right|^2,
\label{eq:C_anal}
\end{equation}
where $U(t,t')$ is the unperturbed propagator from gate $t'+1$ to $t$, $U(t,t'):=U_t\cdots U_{t'+1}$ with the convention $U(t,t)\equiv \mathbbm{1}$. Averaging over the GUE is done only to ease analytical calculation and to yield a 
quantity that is independent of a particular realization of the perturbation. 
Qualitatively similar (numerical) results are obtained without the averaging. 
\par
\begin{figure}[h]
\begin{center}
\begin{minipage}[b]{7.5cm}
\includegraphics[width=7.5cm]{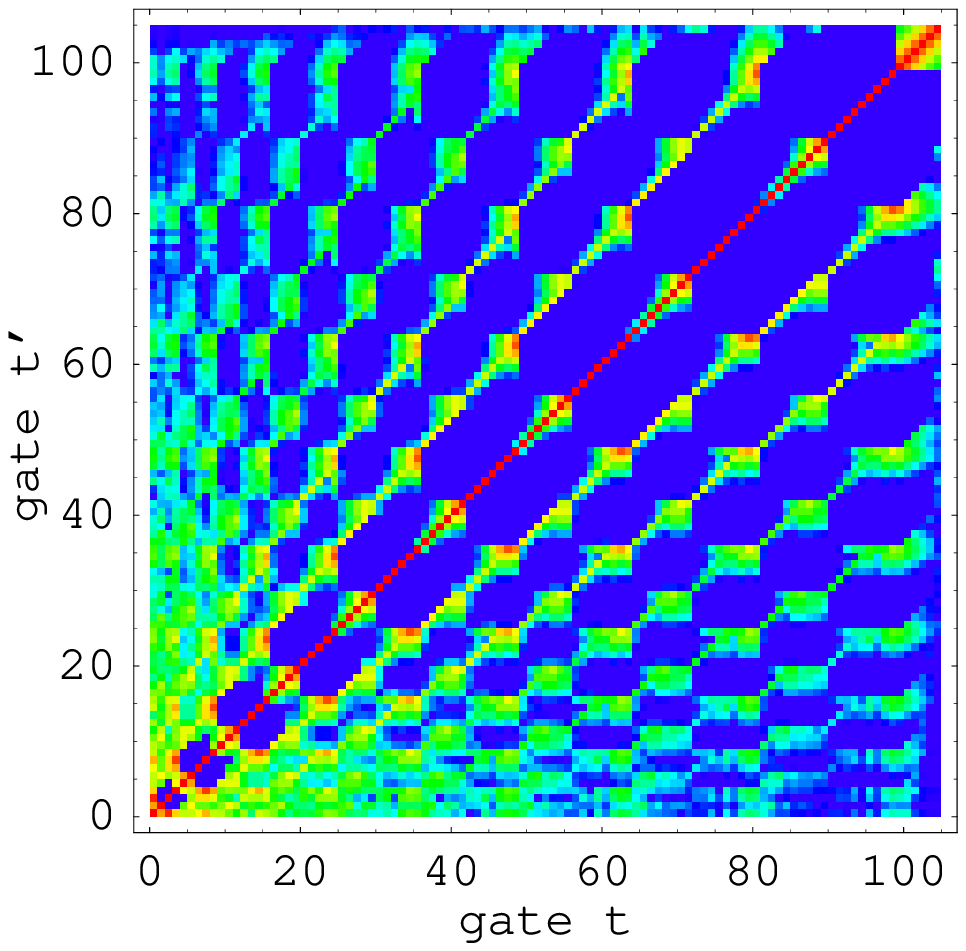}
\end{minipage}
\hspace{2mm}
\vspace{ -2mm}
\begin{minipage}[b]{6cm}
\includegraphics[width=6cm]{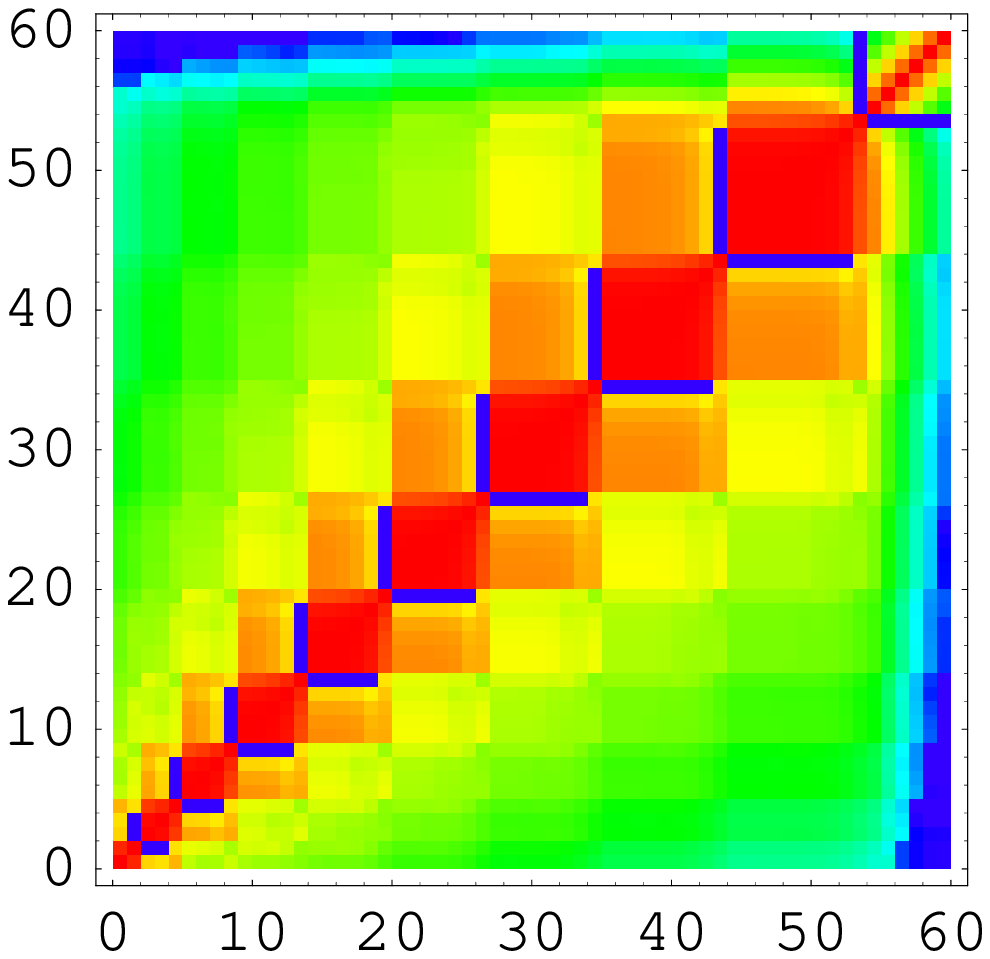}
\end{minipage}
\vspace{-5mm}
\begin{minipage}[b]{0.4in}
\includegraphics[height=6.2cm]{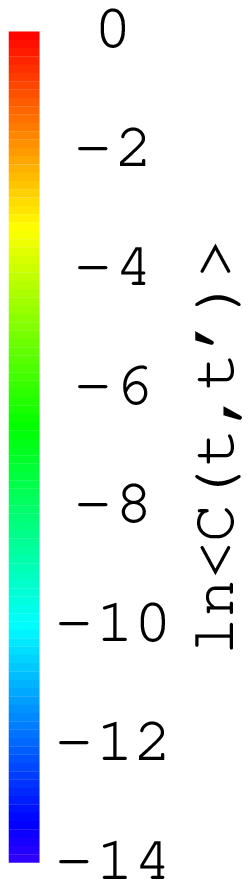} 
\end{minipage}
\end{center}
\caption{Correlation functions $\ave{C(t,t')}_{\rm GUE}$
(\ref{eq:C_anal}) for $n=10$ qubits and static GUE perturbation. The right figure shows the standard QFT (\ref{eq:plain4}) with $T=60$, while the left figure shows the IQFT with $T=105$ gates. Colour represents the size of 
elements in a log-scale from red ($e^{-0}$) to blue ($e^{-14}$ and less).}
\label{fig:corel}
\end{figure}

We have calculated the correlator $\ave{C(t,t')}$ for the QFT (\ref{eq:C_anal}) 
which is shown in the right Figure~\ref{fig:corel}. One can clearly see square red plateaus 
on the diagonal due to blocks of successive ${\rm B}$-gates. 
Similar square plateaus can also be seen off diagonal (from orange, yellow to green), so the correlation function has a staircase-like structure, with the ${\rm A}$-gates responsible for 
the drops and ${\rm B}$-gates responsible for the flat regions in between. 
This can be easily understood. For ``distant'' qubits $k-j \gg 1$ 
the gates ${\rm B}_{jk}$ are close to the identity and therefore cannot reduce the correlator. This slow correlation decay results in the correlation sum 
\begin{equation}
\nu:=\sum_{t,t'=1}^T C(t,t'),
\end{equation}
being proportional to $\nu \propto n^3$ (sum of the first $n$ squares) as 
compared to the theoretical minimum $\nu \propto T \propto n^2$. 
\par
\begin{figure}[h!]
\centerline{\includegraphics{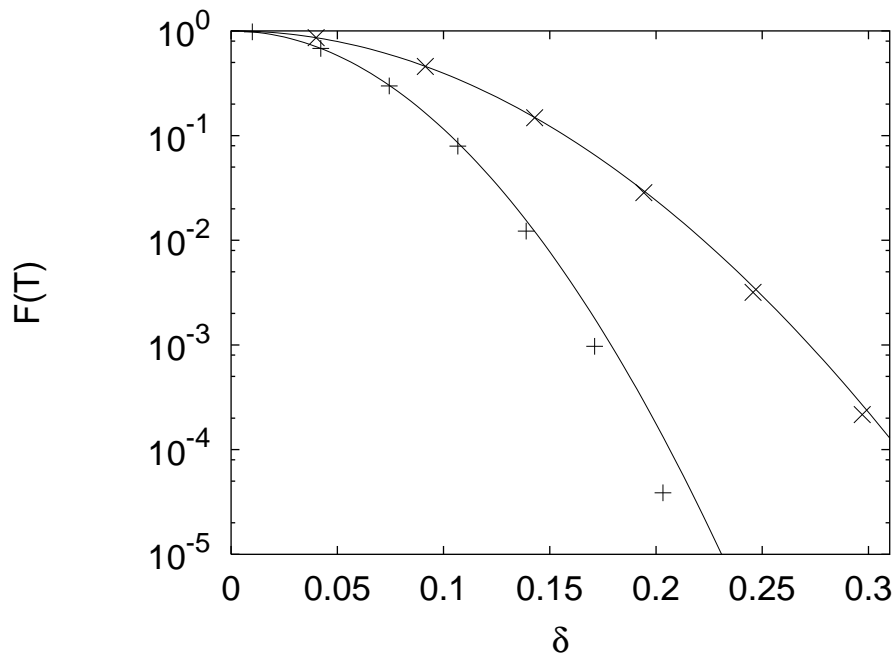}}
\caption{Dependence of the fidelity $F(T)$ on $\delta$ for the QFT (pluses) and the IQFT (crosses), for fixed
$n=8$. Solid curves are functions $\exp{(-\nu \delta^2)}$ (see text) with $\nu$ calculated analytically 
(\ref{eq:C_anal}) and equal to 
$\nu=216$ for the QFT and $93.2$ for the IQFT.}
\label{fig:gauss}
\end{figure}

In view of this, we will now try to rewrite the QFT with a goal to
achieve a smaller correlation sum, hopefully $\nu \propto n^2$. From (\ref{eq:C_anal}) we learn that the gates that are {\em traceless} 
(e.g. ${\rm A}$-gates) reduce the correlator very efficiently. In the plain QFT algorithm 
(\ref{eq:plain4}) we have $n-1$ blocks of ${\rm B}$-gates, where in each block all ${\rm B}$-gates 
act on the same first qubit, say $j$. In each such block, we propose to replace ${\rm B}_{jk}$ 
with a new gate ${\rm G}_{jk}={\rm R}_{jk}^\dagger {\rm B}_{jk}$, 
where a unitary gate ${\rm R}_{jk}$ will be chosen so as to 
commute with all diagonal gates ${\rm B}_{jl}$ in the block, whereas at the end of the 
block we will insert ${\rm R}_{jk}$ in order to ``annihilate'' ${\rm R}^\dagger_{jk}$ 
so as to preserve the evolution matrix of a whole block. 
The unitarity condition ${\rm R}^\dagger_{jk} {\rm R}_{jk}=1$ and 
$[{\rm R}_{jk},{\rm B}_{jl}]=0$ for all $j,k,l$ leaves us with a 6 parametric set 
of matrices ${\rm R}_{jk}$. 
By further enforcing $\tr{{\rm R}_{jk}}=0$ in order to maximally reduce the correlator, 
we end up with 4 free real parameters in ${\rm R}_{jk}$. One of the simplest choices, that turned out to be as suitable as any other, is the following
\begin{equation}
{\rm R}_{jk}=\pmatrix{ 0 & 0 & -1 & 0 \cr
		0 & 1 & 0 & 0 \cr
		1 & 0 & 0 & 0 \cr
		0 & 0 & 0 & -1 \cr},
\label{eq:R_gate}
\end{equation}
with the basis states $(\ket{jk})$ ordered as $(00,10,01,11)$. The ${\rm R}$ gate can be compactly written as ${\rm R}_{jk}\ket{\ldots a_j \ldots b_k\ldots}=(-1)^{b_k}\ket{\ldots a_j \ldots (\overline{a_j}\oplus b_k)\ldots}$, where $\oplus$ is an addition modulo $2$, bar denotes a negation and $a_j,b_k$ are $0$ or $1$. Furthermore, we find that ${\rm R}$-gates also commute among themselves, 
$[{\rm R}_{jk},{\rm R}_{jl}]=0$, which enables us to write a sequence of
${\rm R}$-gates whichever way we like, e.g. in the same order as a sequence of ${\rm G}$'s, so 
that pairs of gates ${\rm G}_{jk}$, ${\rm R}_{jk}$ operating on the same pair of qubits 
$(j,k)$, whose product is a {\em bad gate} ${\rm B}_{jk}$, are never neighbouring.
This is best illustrated by an example. For instance, the block 
${\rm B}_{01} {\rm B}_{02} {\rm B}_{03}$ will be replaced by 
${\rm R}_{01} {\rm R}_{02} {\rm R}_{03}
{\rm R}^\dagger_{01}{\rm B}_{01} {\rm R}_{02}^\dagger{\rm B}_{02} {\rm R}^\dagger_{03} {\rm B}_{03} 
= {\rm R}_{01} {\rm R}_{02} {\rm R}_{03} {\rm G}_{01} {\rm G}_{02} {\rm G}_{03}$. 
This is how we construct an {\em improved quantum Fourier transform algorithm} (IQFT). 
For the IQFT we need one additional type of gate, instead of diagonal ${\rm B}$-gates, 
we use nondiagonal ones ${\rm R}$ and ${\rm G}$.
To illustrate the obvious general procedure we write out the whole IQFT algorithm for $n=4$ qubits 
(compare with (\ref{eq:plain4}))
\begin{equation}
U_{\rm IQFT} = {\rm T}_{03} {\rm T}_{12} {\rm A}_0 {\rm R}_{01} {\rm R}_{02} {\rm R}_{03} {\rm G}_{01} {\rm G}_{02} {\rm G}_{03} {\rm A}_1 {\rm R}_{12} {\rm R}_{13} {\rm G}_{12} {\rm G}_{13} {\rm A}_2 {\rm R}_{23} {\rm G}_{23} {\rm A}_3.
\label{eq:IQFT4}
\end{equation}
\begin{figure}[h!]
\centerline{\includegraphics{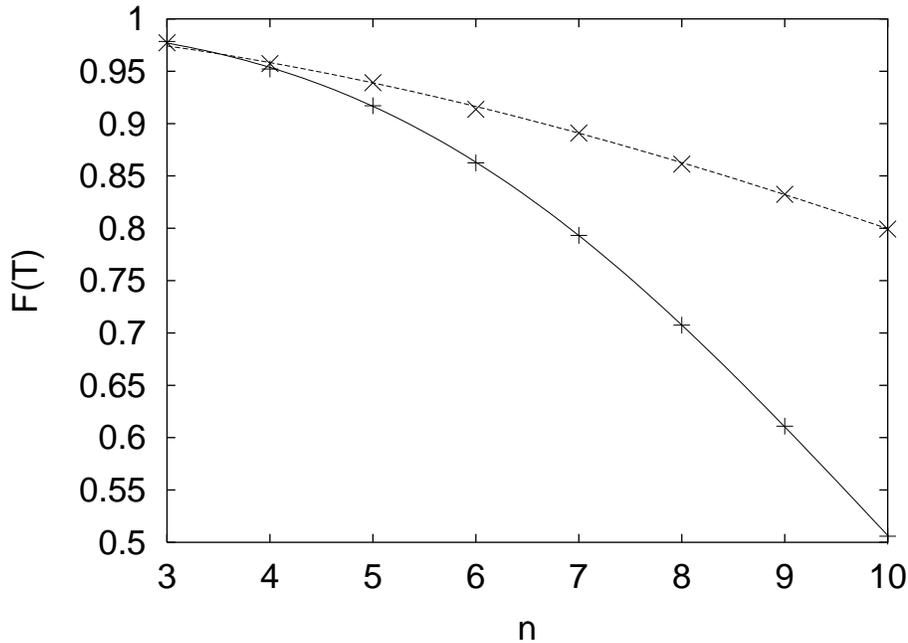}}
\caption{Dependence of the fidelity $F(T)$ on the number of 
qubits $n$ for the QFT (pluses) and the IQFT algorithms (crosses), for
fixed $\delta=0.04$. Numerical averaging over 50 GUE realizations is
performed. The full curve is 
$\exp{(-\delta^2 \{ 0.47 n^3 -0.76 n^2+2.90 n \})}$ and the dashed one 
$\exp{(-\delta^2 \{ 1.22 n^2+1.78 n\})}$. For $n=10$ the trace is approximated by an average 
over 200 Gaussian random register states.}
\label{fig:ft}
\end{figure}
Such IQFT algorithm consists in total of $T=[n(2n+1)/2]$ gates (note that it does not pay of 
to replace a block with a single ${\rm B}$ gate as we have done, so we could safely leave 
${\rm B}_{23} \equiv {\rm R}_{23} {\rm G}_{23}$). 
The correlation function for the IQFT algorithm is shown on the left
of Figure~\ref{fig:corel}. Almost all off-diagonal correlations are greatly reduced (to the level $\propto 1/{\cal N}^2$), leaving us only with a dominant diagonal. If we would have only diagonal elements, the fidelity would be $F(T)=1-\delta^2 T$,
(as in the case of noisy perturbation or decoherence, however, with a
different physical meaning of the strength scale $\delta$) where the
number of gates scales as $T \propto n^2$. From Figure~\ref{fig:corel}
it is clear that we have a very fast correlation decay for the
IQFT. Studying the scaling with $n$, the correlation sum $\nu$ has decreased from 
$\nu \propto n^3$ to $\nu \propto n^2$. To further illustrate this, 
we have numerically calculated fidelity by simulating the quantum algorithm
and applying the perturbation $\exp{(-\ii \delta V)}$ at each
gate. The results are shown in Figure~\ref{fig:ft} where one can see
much faster decay of the fidelity for the QFT than for the IQFT. Note that the IQFT has higher fidelity despite the perturbation being applied $\sim n^2$ times for the IQFT and only $\sim n^2/2$ times for the QFT. As we have argued before, the sum of 2-point correlator (\ref{eq:Fcor}) gives us only the 
first nontrivial order in the $\delta$-expansion. 
For dynamical systems, being either integrable or mixing and ergodic, we have shown in previous chapters that also higher orders of the fidelity can approximately be written as simple powers 
of the correlation sum $\nu$, so that the fidelity has the simple form 
\begin{equation}
F(T)\approx \exp(-\nu \delta^2).
\end{equation} 
Although quantum algorithm is quite inhomogeneous in time, we may still hope that $\exp{(-\nu \delta^2)}$ is a reasonable approximation of the fidelity also to higher orders in $\delta$. This is in fact the case as can be seen in Figure~\ref{fig:gauss}. 
Note also that the leading coefficient in the exponent for the IQFT fidelity, 
$\lim_{n\to\infty}\nu/n^2 = 1.22$, is close to the theoretical minimum of $1$.
As the definition of what is a fundamental single gate is somehow
arbitrary, the problem of minimising the sum $\nu$ depends on a given
technical realization of gates and the nature of the perturbation $V$
for an experimental setup. It has been shown~\citep{Celardo:03}, that
the IQFT algorithm improves stability against GUE perturbations also
for a more realistic model of a quantum computer, namely for an Ising quantum computer. 
\par
We should mention that the optimisation becomes harder if we consider 
{\em few-body} (e.g. two-body) random perturbation. This is connected with 
the fact that quantum gates are two-body operators and can perform only a very limited set 
of rotations on a full Hilbert space and consequently have a limited capability of reducing 
correlation function in a single step. For such errors the fidelity will typically decay with the square of the number of errors $T$ (i.e. gates), like $\sim \exp(-\delta^2 T^2)$, that is the same as for regular systems, see also~\citep{Banacloche:00} for a similar result. This means that the very fact that the algorithm is efficient, having a polynomial number of gates, makes it very hard to reduce the correlation function and therefore causes a fast fidelity decay. However, our simple approach based on $n$-body random matrices
seems reasonable, if the errors due to unwanted few-body qubit interactions can be eliminated by other 
methods.


\chapter{Conclusion}
\begin{flushright}
\baselineskip=13pt
\parbox{85mm}{\baselineskip=13pt
\sf I hate quotations. Tell me what you know.
}\medskip\\
---{\sf \itshape Ralph Waldo Emerson}\\\vspace{20pt}
\end{flushright}

We have studied the decay of quantum fidelity, of reduced fidelity and of purity in quantum systems. We considered two extreme cases of system's dynamics, mixing and regular. The dependence on the initial conditions, in particular for a random and coherent initial state, and the influence of the perturbation type on quantum stability has been analysed. 
\par
For a {\em general} type of {\em perturbation}, having a nonzero
diagonal matrix elements in the eigenbasis of the unperturbed system,
i.e. having a nonzero time average, the fidelity decay depends on the
mixing properties of system dynamics. For {\em mixing systems} in the
Fermi golden rule regime fidelity decay is exponential,
with the decay time given by the transport coefficient, which is in turn the
integral of the correlation function of the perturbation. The decay
time scales with Planck's
constant and with the perturbations strength as $\tau_{\rm m}\sim \hbar^2/\delta^2$. In this regime quantum
fidelity decay is much slower than classical fidelity decay and
moreover, it will in general be the {\em slower the more chaotic} the
corresponding classical system is. This surprising result does not
violate the quantum-classical correspondence though, as for large
perturbation strengths, in the so called Lyapunov regime, quantum
fidelity agrees with classical fidelity. Whether we observe the
quantum or the classical behaviour depends on the order of two
noncommuting limits, namely the semiclassical limit of vanishing
Planck constant $\hbar \to 0$ and the limit of vanishing
perturbation strength $\delta \to 0$. For sufficiently long times fidelity decay in mixing systems does not depend on the initial
state. In {\em regular systems} and for perturbations with a nonzero
time average the fidelity decay for wave packets is governed by a
ballistic separation of the packets. For coherent initial states the
resulting decay is Gaussian with the decay time scaling as $\tau_{\rm
r}\sim \sqrt{\hbar}/\delta$, and can be for sufficiently small
perturbations {\em smaller} than for chaotic systems. 
For random initial states in a regular regime quantum fidelity decays
according to a power law, $F \sim (\hbar/\delta t)^d$ in $d$ degrees of freedom system. In regular systems we also considered the decay averaged over
random positions of the initial coherent state, resulting in an algebraic decay but with the power being system specific. By a semiclassical method we theoretically calculated all decay times in both cases of regular and mixing dynamics in terms of classical quantities only, despite the fact that the quantum and the classical fidelity do not agree for mixing systems. 
\par
The quantum fidelity decay is markedly different for perturbations
with a {\em zero time average}, which can be written as a time
derivative of another operator. 
For such perturbations fidelity freezes at a constant plateau
regardless of the dynamics and starts to again decay only after a much
longer time, scaling as $\sim 1/\delta$. This freezing is a pure
quantum phenomenon as the correspondence with classical fidelity ends
before the plateau starts and classical fidelity decays much
faster. The only exception where the classical fidelity also exhibits
freezing and agrees with quantum fidelity is for a harmonic oscillator, for
which the plateau is higher than for other systems. We explicitly
calculate the plateau value for mixing and regular dynamics and
coherent and random initial states. For mixing dynamics a universal
relation holds between the plateau $F^{\rm RIS}_{\rm plat}$ for random
and $F^{\rm CIS}_{\rm plat}$ for coherent states, $(F^{\rm CIS}_{\rm
plat})^2=F^{\rm RIS}_{\rm plat}$. In the linear response regime the
scaling of the plateau is $1-F_{\rm plat}\sim \delta^2/\hbar^2$
regardless of the dynamics. The asymptotic decay of fidelity after the
plateau ends is also theoretically calculated in terms of a
``renormalised'' perturbation with strength $\delta^2/2$. For regular
systems this long time Gaussian decay happens on a time scale $\sim
\sqrt{\hbar}/\delta^2$ for coherent initial states whereas it is power
law with the prefactor scaling as $\sim \hbar/\delta^2$ for random initial states. For mixing dynamics the long time decay does not depend on the initial state and is exponential with the decay time $\sim \hbar^2/\delta^4$ or Gaussian with the decay time $\sim \hbar^{1-d/2}/\delta^2$. The crossover from the exponential to the Gaussian decay happens at the Heisenberg time. For one dimensional regular systems we also explain echo resonances, a sudden revivals of fidelity.
\par
We also study composite systems, composed of a central system and an
environment, and connect the decay of purity with the decay of
reduced fidelity. We prove a rigorous inequality between the fidelity
(characterising the stability of the whole system), the reduced
fidelity (characterising the stability of the central system) and the
purity (characterising the entanglement). For {\em mixing systems}
fidelity, reduced fidelity and purity all decay on the same time
scale. For {\em regular systems} though, reduced fidelity has a
Gaussian decay for coherent states whereas purity decays on an $\hbar$
independent time scale. We explicitly calculate the purity decay and
the power of the asymptotic algebraic decay depends on the
perturbation and can range between $1$ and $d_{\rm c}$, the $d_{\rm c}$
being number of degrees of freedom of the central system. All decay
constants are explicitly calculated. We also discuss an interesting
case where the time scales of the central system and the environment are vastly different and one can use averaging over the faster system to simplify the theory.  Decoherence for macroscopic superpositions of coherent states is derived and shown to be faster than for a single coherent state.
\par
Finally, we show an application of fidelity theory. By ``randomising''
the quantum Fourier transform algorithm we are able to make it more resistant against random perturbations from the environment.


\appendix
\renewcommand{\theequation}{\Alph{chapter}.\arabic{equation}}

\renewcommand{\chaptername}{\appendixname}      

\addappheadtotoc
\chapter{Spin Wigner functions}
\label{app:wig}
The Wigner function enables us to represent the quantum density
matrices in a phase space and thereby compare it with the classical
probability densities. If we have a one particle quantum system,
described by a canonical pair $[\hat{q},\hat{p}]=\ii \hbar$, the Wigner function $W_{\!\rho}$ of a quantum state given by a density matrix $\hat{\rho}$ is
\begin{equation}
W_{\!\rho}(q,p):=\frac{1}{2\pi \hbar} \int{{\rm d}x \bracket{q-x}{\rho}{q+x} \exp{(-\ii 2 p x/\hbar)}},
\end{equation}
if $\ket{x}$ is an eigenstate of operator $\hat{q}$ with an eigenvalue
$x$. In this appendix we will use a hat for quantum operators. For
spin state such a definition can not be used as the Hilbert space has different structure due to SU(2) commutation relations of spin operators.
\par
We would like to obtain a Weyl symbol of an arbitrary operator
$\hat{A}$ acting on a Hilbert space of size $2S+1$, i.e. on a state
space of spin of size $S$. In the special case when the operator
$\hat{A}$ is equal to a density matrix, the resulting Weyl symbol is called a Wigner
function. Such functions have been first proposed
by~\citet{Agarwal:81}. The Weyl symbol $W_A(\vartheta,\varphi)$ will
be a function of coordinates on a sphere. Furthermore, we would like
the standard trace dot product for operators to carry over to Weyl symbols, i.e. we demand that the following equality should hold
\begin{equation}
\tr{(\hat{A},\hat{B}^\dagger)}=\int{\!W_A W_B^* {\rm d}\Omega},
\label{eq:dot}
\end{equation}
with ${\rm d}\Omega={\rm d}\varphi \sin{\vartheta}{\rm d}\vartheta$. Spherical harmonic functions $Y_l^m$ constitute an orthonormal basis on a sphere,
\begin{equation}
\int{\!Y_l^m Y_{l'}^{m'*} {\rm d}\Omega}=\delta_{ll'}\delta_{mm'},
\end{equation}
and so we can expand the Weyl symbol over $Y_l^m$. On the other hand,
an arbitrary operator $\hat{A}$ can be in turn expanded over {\em
  multipole operators} $\hat{T}_l^m$ forming an orthogonal basis in
the Hilbert space of operators,
\begin{equation}
\tr{(\hat{T}_l^m \hat{T}_{l'}^{m'\dagger})}=\delta_{ll'}\delta_{mm'}.
\end{equation}
Therefore we can write the operator $\hat{A}$ as
\begin{equation}
\hat{A}:=\sum_{l=0}^{2S}{\sum_{m=-l}^{l}{a_{lm}\hat{T}_l^m }},\qquad a_{lm}=\tr{(\hat{A}\hat{T}_l^{m\dagger})}.
\end{equation}
From this and the orthogonality property of $Y_l^m$ we immediately see
that the Weyl symbol defined as
\begin{equation}
W_A(\vartheta,\varphi):=\sum_{l=0}^{2S}{\sum_{m=-l}^{l}{a_{lm}Y_l^m(\vartheta,\varphi) }},
\end{equation}
will satisfy dot property (\ref{eq:dot}) we demanded. The coefficients
$a_{lm}$ can be calculated using the explicit form of the multipole operators
\begin{equation}
\hat{T}_l^m:=\sum_{q=-S}^S{(-1)^{S-m-q}\sqrt{2l+1} \pmatrix{S & l & S\cr -m-q & m & q} \ket{m+q}\bra{q}},
\end{equation}
with $\pmatrix{S & l & S\cr -m-q & m & q}$ being the Wigner $3j$ symbol.
\par
The spin Wigner function for a state represented by a density matrix
$\hat{\rho}$ is obtained by taking $\hat{A}=\hat{\rho}$ in the above
formulas. As the density matrix is a Hermitian operator, the resulting
Wigner function is real. In the case of two pure states, $\rho_1=\ket{\psi_1}\bra{\psi_1}$, $\rho_2=\ket{\psi_2}\bra{\psi_2}$ the dot condition gives simply
\begin{equation}
|\braket{\psi_1}{\psi_2}|^2=\int{W_{\!\rho_1} W_{\!\rho_2} {\rm d}\Omega}.
\end{equation}
For examples of Wigner functions of some simple states see e.g.~\citep{Dowling:94}.

\chapter{Coherent State Expectation Values}
\label{app:expect}
\subsubsection{Boson coherent states}
Expectation values of expressions involving creation and annihilation
operators $a^+$ and $a$ for a harmonic oscillator coherent state $\ket{\alpha}$ are frequently needed. Using the definition of $\ket{\alpha}$ (\ref{eq:boson_coh}) it is easy to show the following equality
\begin{equation}
\bracket{\alpha}{g(a^+)f(a)}{\alpha}=g(\alpha^*)f(\alpha),
\end{equation}
with two polynomials $g$ and $f$. Let $p(a,a^+)$ be some polynomial
function and by $:p(a,a^+):$ we will denote a polynomial in a {\em
  normal} order which can be obtained from $p(a,a^+)$ by using the commutation relation $[a,a^+]=1$ to bring all terms involving $a^+$ to the left of terms with $a$. For instance, $:\!aa^+\!: \,=a^+ a+1$. Then we can write the expectation value of an arbitrary polynomial as
\begin{equation}
\bracket{\alpha}{p(a,a^+)}{\alpha}=\bracket{\alpha}{:p(a,a^+):}{\alpha}=:p(\alpha,\alpha^*):\, .
\end{equation}

\subsubsection{Spin coherent states}
For spin coherent states formulas are a bit more complicated due to different group structure. The easiest systematic method for the calculation of expectation values of polynomials in operators $S_{\rm x,y,z}$ in coherent state $\ket{\vartheta^*,\varphi^*}$ (\ref{eq:SU2_coh}) is using generating function formalism~\citep{Arecchi:72}. For brevity let us denote a spin coherent state with a complex parameter $\tau:={\rm e}^{\ii \varphi^*}\tan{(\vartheta^*/2)}$, i.e. $\ket{\tau}:=\ket{\vartheta^*,\varphi^*}$. The following expression holds,
\begin{equation}
\bracket{\tau_1}{S_+^a S_{\rm z}^b S_-^c}{\tau_2}=\left\{ \left( \frac{\partial}{\partial \xi}\right)^a \left( \frac{\partial}{\partial \eta} \right)^b \left( \frac{\partial}{\partial \zeta} \right)^c X(\xi,\eta,\zeta) \right\}_{\xi=\eta=\zeta=0},
\end{equation}
where
\begin{equation}
X(\xi,\eta,\zeta):=\bracket{\tau_1}{{\rm e}^{\xi S_+}{\rm e}^{\eta S_{\rm z}}{\rm e}^{\zeta S_-}}{\tau_2}=\frac{\left\{ {\rm e}^{\eta/2}+{\rm e}^{-\eta/2} (\tau_1^*+\xi) (\tau_2+\zeta)\right\}^{2S}}{(1+|\tau_1|^2)^S(1+|\tau_2|^2)^S} {\rm e}^{\ii (\varphi_1^*-\varphi_2^*)S}.
\end{equation}
This formula, together with the commutation relations, can be used to calculate some of the lowest powers
\begin{eqnarray}
\bracket{\tau}{S_{\rm z}}{\tau}&=& Sz \nonumber \\
\bracket{\tau}{S_{\rm z}^2}{\tau}&=& S^2 z^2+\frac{S}{2} (1-z^2) \nonumber\\
\bracket{\tau}{S_{\rm z}^3}{\tau}&=& S^3 z^3+\frac{3S^2}{2}z(1-z^2)-\frac{S}{2}z(1-z^2) \nonumber\\
\bracket{\tau}{S_{\rm z}^4}{\tau}&=& S^4 z^4+3S^3 z^2(1-z^2)+\frac{S^2}{4}(11z^4-14z^2+3)+\frac{S}{4}(-3z^4+4z^2-1),
\end{eqnarray}
where $z=\cos{\vartheta^*}$. The expressions for other spin operators $S_{\rm y}$ and $S_{\rm x}$ are obtained by replacing $z$ with $y$ or $x$ on the right hand side, respectively. Some other useful expectation values are
\begin{eqnarray}
\bracket{\tau}{S_+S_-}{\tau}&=& S^2(1-z^2)+\frac{S}{2}(1+z)^2 \nonumber\\
\bracket{\tau}{S_-S_+}{\tau}&=& S^2(1-z^2)+\frac{S}{2}(1-z)^2.
\end{eqnarray}  

\bibliographystyle{authordate3}
\bibliography{stability_eng}

\end{document}